\documentclass[english]{article}
\usepackage[utf8]{inputenc}
\usepackage{amsmath}
\usepackage{amssymb}
\usepackage{mathrsfs}
\usepackage{cite}
\usepackage{multirow}
\usepackage{framed}
\usepackage[colorlinks=true,urlcolor=blue,anchorcolor=blue,citecolor=blue,filecolor=blue,linkcolor=blue,menucolor=blue,linktocpage=true,pdfproducer=medialab,pdfa=true]{hyperref}
\usepackage{cancel}
\usepackage{diagbox} % for the diagonal box
\usepackage{array}   % optional, for better column formatting
\usepackage{tikz}

\makeatletter

\@addtoreset{equation}{section}

\def\draft{

\usepackage[showseconds=false,showzone=false,useregional]{datetime2}
\usepackage{draftwatermark}
\SetWatermarkText{\tt \DTMtoday\ (\DTMcurrenttime)}
\SetWatermarkColor[gray]{0.8}
\SetWatermarkFontSize{0.8cm}
\SetWatermarkAngle{90}
\SetWatermarkHorCenter{0.65cm}

\def\thtystars{******************************}
\def\sixtystars{\thtystars\thtystars}
\typeout{}
\typeout{\sixtystars**}
\typeout{* Draft mode!
         For final version remove \protect\draft\space in source 
         file *}
\typeout{\sixtystars**}
\typeout{}
\def\draftdate{\today}
\def\mua{\marginpar[\boldmath\hfil$\uparrow$]%
                   {\boldmath$\uparrow$\hfil}%
                    \typeout{marginpar: $\uparrow$}\ignorespaces}
\def\mda{\marginpar[\boldmath\hfil$\downarrow$]%
                   {\boldmath$\downarrow$\hfil}%
                    \typeout{marginpar: $\downarrow$}\ignorespaces}
\def\mla{\marginpar[\boldmath\hfil$\rightarrow$]%
                   {\boldmath$\leftarrow $\hfil}%
                    \typeout{marginpar: 
                    $\leftrightarrow$}\ignorespaces}
\overfullrule 5pt
\oddsidemargin -12mm
\marginparwidth 29mm
}

\def\stars{\strut\leaders\hbox{*}\hfill\strut}
\def\starline{\hfil\strut\hfil\hbox to \textwidth {\stars}\hfil}

\def\bleq#1{\begin{equation}
\label{eq:#1}
\end{equation}
\vspace{-40pt}
\begin{quote}\raggedright}
\def\eleq{\end{quote}}

\def\draftdate{\relax}
\def\mda{\relax}
\def\mua{\relax}
\def\mla{\relax}

\newcommand\refr[1]      {ref.~\cite{#1}}
\newcommand\refrs[1]    {refs.~\cite{#1}}

\newcommand\eqn[1]     {eq.~(\ref{#1})}
\newcommand\eqns[2]    {eqs.~(\ref{#1}) and~(\ref{#2})}
\newcommand\eqnss[2]   {eqs.~(\ref{#1})--(\ref{#2})}

\newcommand\Fig[1]     {Fig.~{\ref{#1}}}
\newcommand\fig[1]     {fig.~{\ref{#1}}}
\newcommand\figs[2]    {fig.~{\ref{#1}} and~{\ref{#2}}}

\newcommand\Sect[1]    {Sec.~{\ref{#1}}}

\newcommand\sect[1]    {sec.~{\ref{#1}}}

\newcommand\sectss[2]   {secs.~\ref{#1}--\ref{#2}}

\newcommand\appx[1]     {appendix~\ref{#1}}

\newcommand\tab[1]     {table~\ref{#1}}

\newcommand\nt         {\notag}
\newcommand\nn         {\nonumber}

\def\beq{\begin{equation}}
\def\eeq{\end{equation}}
\def\bsp#1\esp{\begin{split}#1\end{split}}
\def\bal#1\eal{\begin{align}#1\end{align}}

\def\beeq{\begin{eqnarray}}
\def\eeeq{\end{eqnarray}}
\newcommand\bom[1]     {{\mbox{\boldmath $#1$}}}

\newcommand\as         {\ensuremath{\alpha_{\mathrm{s}}}}

\newcommand{\CF}       {C_{\mathrm{F}}}
\newcommand{\CA}       {C_{\mathrm{A}}}
\newcommand{\TR}       {T_{\mathrm{R}}}
\newcommand{\Nc}       {N_{\mathrm{c}}}

\newcommand{\bT}       {\bom{T}}

\newcommand\qb         {{\bar q}}

\newcommand{\ep}       {\varepsilon}

\newcommand{\rd}       {{\mathrm{d}}}
\newcommand{\PS}[1]    {\rd\phi_{#1}}
\newcommand\tsig[1]    {\sigma^{\mathrm{#1}}}

\newcommand\dsig[1]    {\rd\sigma^{{\rm #1}}}
\newcommand\dsigk[4]    {\rd\sigma^{{\rm #1}_{#2} {\rm #3}_{#4}}}

\newcommand\dsiga[2]   {\rd\sigma^{{\rm #1,A}_{\scriptscriptstyle #2}}}

\newcommand\la         {\langle}
\newcommand\ra         {\rangle}

\newcommand{\cA}       {{\cal A}}
\newcommand{\cII}[1] {{\cal I}\kern-4pt *\kern-4pt{\cal I}_{#1}}
\newcommand{\cIJ}     {{\cal I}\kern-4pt *\kern-4pt{\cal J}}
\newcommand{\cJJ}[1] {{\cal J}\kern-4pt *\kern-4pt{\cal J}_{#1}}
\newcommand{\cJI}     {{\cal J}\kern-4pt *\kern-4pt{\cal I}}
\newcommand{\cJK}     {{\cal J}\kern-4pt *\kern-4pt{\cal K}}
\newcommand{\cKJ}[1] {{\cal K}\kern-4pt *\kern-4pt{\cal J}_{#1}}
\newcommand{\cKI}     {{\cal K}\kern-4pt *\kern-4pt{\cal I}}
\newcommand{\cKK}     {{\cal K}\kern-4pt *\kern-4pt{\cal K}}

\newcommand\SME[3]     {\big|{\cal M}_{#1}^{#2}{#3}\big|^2}

\newcommand\bra[3]     {\big\la {\cal M}_{#1}^{#2}#3\big|}
\newcommand\ket[3]     {\big|{\cal M}_{#1}^{#2}#3\big\ra}
\newcommand\braket[6]     {\big\la {\cal M}_{#1}^{#2}#3\big|{\cal M}_{#4}^{#5}#6\big\ra}

\newcommand{\mom}[1]   {\{p\}^{#1}}
\newcommand{\momb}[1]   {\{\bar{p}\}^{#1}}
\newcommand{\momt}[1]   {\{\ti{p}\}^{#1}}
\newcommand{\momh}[1]   {\{\ha{p}\}^{#1}}
\newcommand{\momhh}[1]   {\{\ha{\ha{p}}\}^{#1}}
\newcommand{\momht}[1]   {\{\ha{\ti{p}}\}^{#1}}
\newcommand{\momth}[1]   {\{\ti{\ha{p}}\}^{#1}}
\newcommand{\momtt}[1]   {\{\ti{\ti{p}}\}^{#1}}

\newcommand{\cmp}[2]   {C_{#1}^{#2}}
\newcommand{\smp}[1]   {S_{#1}}
\newcommand{\cmap}[2]   {\xrightarrow{C_{#1}^{#2}}}
\newcommand{\smap}[1]   {\xrightarrow{S_{#1}}}

\newcommand{\bSCS}[1]  {\bom{\mathrm C}\kern-2pt\bom{\mathrm S}_{#1}}

\def\hP{\ha{P}}

\newcommand{\calS}     {{\cal S}}
\newcommand{\cC}[2]    {{\cal C}_{#1}^{#2}}
\newcommand{\cS}[2]    {{\cal S}_{#1}^{#2}}

\newcommand{\cSCS}[2]  {{\cal C}\kern-2pt{\cal S}_{#1}^{#2}}

\newcommand{\IcC}[2]   {{\mathrm C}_{#1}^{#2}}
\newcommand{\IcS}[2]   {{\mathrm S}_{#1}^{#2}}

\newcommand{\IcSCS}[2]  {\mathrm{C}\kern-2pt\mathrm{S}_{#1}^{#2}}
\newcommand{\ba}[1]    {\bar{#1}}
\newcommand{\ti}[1]    {\tilde{#1}}

\newcommand{\ha}[1]    {\hat{#1}}
\newcommand{\wha}[1]   {\widehat{#1}}

\newcommand\kTm[1]     {k_{\perp}^{#1}}
\newcommand\kT[1]      {k_{\perp #1}}
\newcommand\kTt[1]     {\tilde{k}_{\perp #1}}
\newcommand\kTh[1]     {\hat{k}_{\perp #1}}

\def\s12{s_{12}}
\def\ab                 {Q}
\def\hahb                 {\ha{Q}}

\newcommand\al	{\alpha}
\newcommand\be	{\beta}
\newcommand\almin	{\alpha_{\mathrm{min}}}
\newcommand\almax {\alpha_{\mathrm{max}}}

\newcommand\lam	{\lambda}
\newcommand\lsq	{\lambda^2}
\newcommand\lammin	{\lambda_{\mathrm{min}}}
\newcommand\lammax	{\lambda_{\mathrm{max}}}
\newcommand\hlammin	{\ha{\lambda}_{\mathrm{min}}}
\newcommand\hlammax	{\ha{\lambda}_{\mathrm{max}}}
\newcommand\tlammin	{\ti{\lambda}_{\mathrm{min}}}
\newcommand\tlammax	{\ti{\lambda}_{\mathrm{max}}}

\begin{document}

%%%
%%% Title page
%%%

\begin{titlepage}
%\renewcommand{\thefootnote}{\fnsymbol{footnote}}

% Date

\noindent
\hfill December 2025
\vspace{0.6cm}

% Title

\begin{center}
{\LARGE \bf 
CoLoRFulNNLO for hadron collisions:\\[0.5em] 
regularizing initial-state double real emissions}

\vspace{1.0cm}

% Authors

\large
V. Del Duca$^{\, a}$,  
G. Somogyi$^{\, b}$, 
F. Tramontano$^{\, c}$\\
\vspace{0,5cm}
\normalsize
{\it $^{\, a}$INFN, Laboratori Nazionali di Frascati, 
00044 Frascati (RM), Italy}\\
{\it $^{\, b}$HUN-REN Wigner Research Centre for Physics, 
1121 Budapest, Konkoly-Thege Mikl\'os \'ut 29-33, Hungary}\\
{\it $^{\, c}$Dipartimento di Fisica Ettore Pancini, 
Universit\`a di Napoli Federico II and 
INFN - Sezione di Napoli, Complesso Universitario di 
Monte Sant’Angelo Ed. 6, Via Cintia, 80126 Napoli, Italy}
\vspace{1.4cm}

%Abstract

{\large \bf Abstract}
\vspace{-0.2cm}
\end{center}
We present the extension of the completely local subtraction 
scheme CoLoRFulNNLO to color-singlet production in hadron 
collisions. We provide explicit momentum mappings and the complete 
set of double-real counterterms required for this class of 
processes. The counterterms are systematically derived from the 
known infrared limit formulae of QCD matrix elements, and 
particular care has been taken to ensure their analytic 
integrability. The resulting construction involves a relatively 
small number of counter-events, preserving the locality and 
efficiency of the scheme. All formulae have been implemented 
within the publicly available {\tt NNLOCAL} Monte Carlo program, 
and we explicitly validate all IR limits using arbitrary-precision 
computer algebra and present representative results. The 
counterterms presented here constitute a self-contained subset 
applicable to general hadronic processes within the CoLoRFulNNLO 
approach.

\end{titlepage}
\clearpage

%%%
%%% TOC here
%%%

\tableofcontents

%%%
%%% Introduction
%%%

\section{Introduction}

Precision calculations in Quantum Chromodynamics (QCD) for hadron
collider processes are indispensable for exploiting the full 
potential of the LHC and future experiments. The steadily
improving experimental precision requires theoretical predictions 
at next-to-next-to-leading order (NNLO) in the perturbative 
expansion in order to reduce theory uncertainties and to provide 
reliable background estimates for searches of new physics. The 
explicit construction of NNLO corrections is, however, technically 
challenging because of the intricate pattern of overlapping soft 
and collinear infrared (IR) singularities that appear in the 
double real, real–virtual and double virtual contributions.

A productive strategy to tame these singularities is furnished by 
local subtraction methods, which introduce local counterterms that 
reproduce the singular limits of the QCD matrix elements 
point-by-point in phase space and thus permit stable numerical 
integration of the finite remainders. Representative local approaches 
include antenna subtraction~\cite{Gehrmann-DeRidder:2005btv}, 
sector-improved residue subtraction~\cite{Czakon:2014oma}, local 
analytic sector subtraction~\cite{Magnea:2018hab}, nested 
soft-collinear subtractions~\cite{Caola:2017dug}, the 
projection-to-Born method~\cite{Cacciari:2015jma}, geometric IR 
subtraction~\cite{Herzog:2018ily} and the CoLoRFulNNLO 
framework~\cite{DelDuca:2016ily}. Alternative strategies based on the 
definition of non-local counterterms and slicing parameters have also 
been successfully pursued, including $q_T$ 
subtraction~\cite{Catani:2007vq} and $N$-jettiness 
subtraction~\cite{Gaunt:2015pea}. Although the local subtraction 
methods differ in technical implementation, the class of processes for 
which they are fully worked out, as well as in the specific treatment 
of phase space (e.g., mappings of momenta and whether or not phase 
space is partitioned), they share the common objective of achieving 
exact local cancellation of infrared divergences while enabling 
efficient Monte Carlo integration.

The CoLoRFulNNLO framework~\cite{Somogyi:2005xz, Somogyi:2006db, 
Somogyi:2006da, DelDuca:2015zqa, DelDuca:2016csb, DelDuca:2016ily, 
Somogyi:2020mmk} is distinguished by its systematic construction of 
local counterterms that explicitly account for all spin and color 
degrees of freedom and by the partial availability of analytic results 
for the corresponding integrated subtraction terms. In particular, 
closed analytic expressions for the integrated counterterms have been 
obtained for classes of configurations relevant to final-state 
radiation off massive partons~\cite{DelDuca:2015zqa, Somogyi:2020mmk} 
and to the initial-state radiation discussed here. For final-state 
radiation from massless partons, integrated counterterms have been 
computed analytically up to and including the $\mathcal{O}(\ep^{-1})$ 
poles in the dimensional-regularization expansion; at $\mathcal{O}
(\ep^{0})$ all but some contributions are available in closed 
analytic form whereas the remaining pieces have been evaluated 
numerically and incorporated through 
interpolation~\cite{DelDuca:2016csb, DelDuca:2016ily}. The 
availability of fully analytic integrated subtraction terms where 
possible is a significant practical advantage: such expressions 
simplify the combination of integrated counterterms with virtual 
contributions and can often be used directly in Monte Carlo 
programs. This in turn can improve numerical stability, since one 
need not worry about ensuring a sufficiently small interpolation 
error. Moreover, the availability of analytic integrated 
counterterms increases the portability of the subtraction 
framework.

An important structural feature of CoLoRFulNNLO is the freedom to 
redefine local counterterms by exploiting various degrees of 
freedom such as the choice of momentum mappings, constraining the 
subtractions to the neighborhood of singular regions and 
reorganizations of subtraction kernels. This flexibility can be 
used to optimize numerical behavior, to reduce the number of 
momentum mappings, and hence the number of counter-events, and to 
adapt the scheme to different classes of processes. In particular, 
these freedoms will be instrumental when extending the framework 
to accommodate more complex final states, for example by adding an 
observed jet to the color-singlet final state in hadronic 
collisions.

In this paper we extend the CoLoRFulNNLO methodology to treat 
double initial-state radiation in processes that produce a 
color-singlet final state at hadron colliders. Our construction 
provides a local subtraction scheme that cancels the infrared 
singularities associated with single and double emissions from the 
incoming partons, while preserving the desirable practical 
properties of the original framework:
\begin{itemize}
\item numerical stability, achieved by exact local cancellation of 
soft and collinear divergences at integrand level;
\item computational efficiency, through a design that aims to keep 
the number of required momentum mappings, and thus counter-events, 
under control;
\item analytic integrated counterterms, enabling a transparent 
treatment of pole cancellation with virtual contributions;
\item modularity and flexibility, which facilitate portability to 
new processes and future extensions, for instance, to final states 
containing jets.
\end{itemize}
The results we present constitute both a concrete realization of 
double initial-state subtraction within CoLoRFulNNLO and a step 
towards a more general, practical local subtraction framework for 
NNLO QCD calculations at hadron colliders. The remainder of the 
paper is organized as follows.

In \sect{sec:setup} we present the basic layout of our  
subtraction scheme, while \sect{sec:colorful} reviews the elements 
of the CoLoRFulNNLO construction relevant to our extension. 
\Sect{sec:doublereal} introduces the approximate cross sections 
that regularize the double real contribution. The new counterterms 
and momentum mappings for double initial-state radiation are then 
presented in full detail in \sectss{sec:A1}{sec:A12}, where we 
also define the integrated counterterms. \Sect{sec:validation} 
discusses the numerical validation of our construction. Finally, 
\sect{sec:conclusions} summarizes our findings and outlines 
directions for future work, including possible refinements of the 
counterterm definitions in view of adding jets to the final state.

%%%
%%% The subtraction scheme
%%%

\section{The subtraction scheme}
\label{sec:setup}

We consider the production of $m$ jets plus any number of 
colorless particles collectively denoted by $X$ in the collision 
of hadrons $A$ and $B$ of momenta $p_A$ and $p_B$, $A+ B \to m 
\mbox{ jets} + X$. The cross section for this process is given by
\beq
\sigma(p_A,p_B) = \sum_{a,b}
	\int_0^1 \rd x_a\, f_{a/A}(x_a,\mu_F^2) \int_0^1 \rd x_b\, 
    f_{b/B}(x_b,\mu_F^2) \tsig{}_{ab}(p_a,p_b;\mu_F^2)\,,
\eeq
i.e., it is a convolution of the partonic cross section with 
parton density functions (PDFs). Here $p_a = x_a p_A$ and $p_b = 
x_b p_B$, while the summations are over parton flavors. The 
partonic cross section can be computed in perturbation theory and 
its perturbative expansion up to NNLO reads
\beq
\tsig{}_{ab}(p_a,p_b;\mu_F^2) = 
	\tsig{LO}_{ab}(p_a,p_b;\mu_R^2,\mu_F^2) 
	+\tsig{NLO}_{ab}(p_a,p_b;\mu_R^2,\mu_F^2)
	+\tsig{NNLO}_{ab}(p_a,p_b;\mu_R^2,\mu_F^2)+\ldots\,.
\eeq
In the following, when no confusion can arise, the dependence of 
the cross sections on partonic momenta as well as the 
renormalization and the factorization scales will be suppressed. 

The leading order (LO) partonic cross section is just the integral 
of the fully differential Born cross section over the phase space 
of the produced final state,
\beq
\tsig{LO}_{ab} = \int_{m+X} \dsig{B}_{ab}\, J_{m+X}\,.
\label{eq:sigLO}
\eeq
Above, $J_{m+X}$ is the value of some infrared and collinear-safe 
physical observable $J$ evaluated on the phase space of $m$ 
partons plus the colorless particles in $X$. By virtue of the 
infrared and collinear  safety of $J$, the phase space integral in 
\eqn{eq:sigLO} is finite and can be computed by standard numerical 
methods directly in four spacetime dimensions.

In contrast, higher-order corrections are given by sums of 
real-emission and/or virtual contributions which are separately IR 
divergent and require regularization.\footnote{In this work, we 
employ dimensional regularization in $d=4-2\ep$ dimensions in the  
CDR scheme~\cite{Collins:1984xc} to regulate IR singularities.} In 
particular, at N$^k$LO there are precisely $k$ extra emissions 
which may become unresolved and each of these may be either real 
or virtual. E.g., at NLO only a single extra emission is allowed, 
hence the NLO correction is the just the sum of the (single) 
real-emission contribution $\dsig{R}_{ab}$ and the virtual 
contribution $\dsig{V}_{ab}$,
\beq
\tsig{NLO}_{ab} = \int_{m+X+1} \dsig{R}_{ab}\, J_{m+X+1}
	+ \int_{m+X} \left(\dsig{V}_{ab} 
    +  \dsigk{C}{1}{}{}_{ab} \right) J_{m+X}\,.
\label{eq:sigNLO}
\eeq
Note that due to the renormalization of the PDFs, the collinear 
remnant $\dsigk{C}{1}{}{}_{ab}$ must also be included to obtain a 
finite NLO cross section. The IR singularities appearing in 
\eqn{eq:sigNLO} can be handled by any number of well-established 
methods~\cite{Frixione:1995ms,Catani:1996vz}.

At NNLO, two extra emissions are allowed which may both be real or 
virtual. Thus the NNLO correction is a sum of the double real 
contribution $\dsig{RR}_{ab}$, the real-virtual contribution 
$\dsig{RV}_{ab}$ and the double virtual contribution 
$\dsig{VV}_{ab}$,
\beq
\tsig{NNLO}_{ab} = \int_{m+X+2} \dsig{RR}_{ab}\, J_{m+X+2}
	+ \int_{m+X+1} \left(\dsig{RV}_{ab} 
    + \dsigk{C}{1}{}{}_{ab} \right) J_{m+X+1}
	+ \int_{m+X} \left(\dsig{VV}_{ab} 
    + \dsigk{C}{2}{}{}_{ab} \right) J_{m+X}\,,
\label{eq:sigNNLO}
\eeq
where once more, $\dsigk{C}{1}{}{}_{ab}$ and 
$\dsigk{C}{2}{}{}_{ab}$ are collinear remnants coming from PDF 
renormalization. In the CoLoRFulNNLO scheme, the IR singularities 
in \eqn{eq:sigNNLO} associated with one or both of the extra 
emissions becoming unresolved are regularized by subtracting and 
adding back various approximate cross sections. In the double real 
part, up to two partons can become unresolved. This is reflected 
in the structure of subtractions,
\beq
\tsig{RR}_{ab, \mathrm{reg.}} = 
\int_{m+X+2}
\bigg\{
    \dsig{RR}_{ab} J_{m+X+2} 
    - \dsiga{RR}{1}_{ab} J_{m+X+1} 
    - \dsiga{RR}{2}_{ab} J_{m+X} 
    + \dsiga{RR}{12}_{ab} J_{m+X} 
    \bigg\}\,,
\label{eq:sigRRreg}
\eeq
where $\dsiga{RR}{1}_{ab}$ and $\dsiga{RR}{2}_{ab}$ are the single 
and double unresolved approximate cross sections that match the 
singular behavior of $\dsig{RR}_{ab}$ point-wise in IR limits 
where one and two partons become unresolved. The last term,  
$\dsiga{RR}{12}_{ab}$, is introduced in order to avoid double 
subtraction in regions of phase space where single and double 
limits overlap, and hence plays a dual role. Conceptually, it 
regularizes $\dsiga{RR}{1}_{ab}$ in double unresolved limits and 
$\dsiga{RR}{2}_{ab}$ in single unresolved ones. However, the true 
structure of cancellations is somewhat more involved. In fact, in 
single unresolved limits, in addition to regularizing 
$\dsiga{RR}{2}_{ab}$, this approximate cross section must also 
cancel all spurious singularities (i.e., singularities of 
subtraction terms unrelated to the specific single unresolved 
limit being considered) that develop in $\dsiga{RR}{1}_{ab}$. We 
will refer to $\dsiga{RR}{12}_{ab}$ as the iterated single 
unresolved approximate cross section. Turning to the real-virtual 
contribution, we recall that in this case at most one parton can 
become unresolved, hence the subtractions take the form
\beq
\bsp
\tsig{RV}_{ab, \mathrm{reg.}} = 
    \int_{m+X+1}\bigg\{
    &\bigg[\dsig{RV}_{ab} 
    + \dsigk{C}{1}{}{}_{ab} + \int_1 \dsiga{RR}{1}_{ab}\bigg] 
    J_{m+X+1}
\\    
    -&\bigg[\dsiga{RV}{1}_{ab} 
    + \dsigk{C}{1,}{A}{1}_{ab} 
    + \left(\int_1 \dsiga{RR}{1}_{ab}\right)^{\!{\rm A}_1}\bigg] 
    J_{m+X}
    \bigg\}\,.
\esp
\label{eq:sigRVreg}
\eeq
Notice in particular that the integrated version of 
$\dsiga{RR}{1}_{ab}$, which is subtracted in \eqn{eq:sigRRreg}, is 
added back here and together with $\dsigk{C}{1}{}{}_{ab} $ it 
cancels the explicit $\ep$-poles of $\dsig{RV}_{ab}$ coming from 
the extra loop. However, all three terms on the first line also 
have kinematic singularities as a parton becomes unresolved, which 
are regularized by the three single unresolved approximate cross 
sections on the second line. Note that the latter must be 
constructed in such a way that their sum is also free of explicit 
$\ep$-poles. Thus, $\ep$-poles cancel within the  brackets, while 
IR singularities cancel between them. Finally, the double virtual 
contribution is free of kinematic singularities since $J$ is 
infrared and collinear safe, but due to the two extra loops, it 
contains explicit IR $\ep$-poles. These poles are then canceled by 
the integrated forms of the various subtraction terms that we have 
not yet added back,
\beq
\bsp
\tsig{VV}_{ab, \mathrm{reg.}} = 
    \int_{m+X} \bigg\{&
    \dsig{VV}_{ab} 
    + \dsigk{C}{2}{}{}_{ab} 
    + \int_2\bigg[\dsiga{RR}{2}_{ab} 
    - \dsiga{RR}{12}_{ab}\bigg]     
\\&
    + \int_1\bigg[\dsiga{RV}{1}_{ab} 
    + \dsigk{C}{1,}{A}{1}_{ab} 
    + \left(\int_1 \dsiga{RR}{1}_{ab}\right)^{\!{\rm A}_1}
    \bigg] \bigg\} J_{m+X}\,.
\esp
\label{eq:sigVVreg}
\eeq
As a result of these manipulations, the five terms in 
\eqn{eq:sigNNLO} are rearranged into the three contributions in 
\eqnss{eq:sigRRreg}{eq:sigVVreg}, each separately finite in four 
dimensions.

Of course, the formal approximate cross sections must be precisely 
defined before \eqnss{eq:sigRRreg}{eq:sigVVreg} can be applied in 
practical calculations. In the following, we focus on double real 
emission and present explicit formulae for the approximate cross 
sections $\dsiga{RR}{1}_{ab}$, $ \dsiga{RR}{2}_{ab}$ and 
$\dsiga{RR}{12}_{ab}$ for the case of color-singlet production in 
hadronic collisions (i.e., $m=0$). We emphasize that all 
subtraction terms remain valid for the more general case when 
final-state radiation is present already at Born level. For such 
processes, the subtraction terms presented here will need to be 
supplemented by additional ones regulating those unresolved 
configurations that do not occur for color-singlet production.

%%%
%%% Building subtraction terms
%%%

\section{Building subtraction terms}
\label{sec:colorful}

In the CoLoRFulNNLO framework, approximate cross sections are 
built from subtraction terms that are in turn constructed starting 
from QCD IR factorization formulae. These formulae capture the 
universal behavior of QCD squared matrix elements as some number 
of partons becomes unresolved (i.e., soft and/or collinear) and 
are completely known up to NNLO accuracy~\cite{Campbell:1997hg, 
Catani:1998nv, Catani:1999ss, DelDuca:1999iql, Kosower:2002su, 
Czakon:2011ve, Bern:1994zx, Bern:1998sc, Kosower:1999rx, 
Bern:1999ry, Catani:2000pi, Braun-White:2022rtg, Gehrmann:2025xab} 
and in some cases beyond~\cite{DelDuca:2019ggv, DelDuca:2020vst, 
Catani:2003vu, Badger:2015cxa, Czakon:2022fqi, Duhr:2014nda, 
Li:2013lsa, Duhr:2013msa, Dixon:2019lnw, Zhu:2020ftr, 
Catani:2021kcy, Catani:2019nqv, DelDuca:2022noh, Chen:2024hvp, 
Guan:2024hlf, Herzog:2023sgb, Czakon:2022dwk, Catani:2022hkb}. 
Their general structure can be given as follows. Let us denote the 
$k$-fold real emission squared matrix element for some specific 
partonic subprocess as\footnote{If the Born process is loop-induced, 
the $k$-fold real emission squared matrix element will not 
literally correspond to the square of zero-loop matrix elements. 
In this case, the notation and in particular the $(0)$ superscript 
should be understood to refer to the squared matrix element with 
zero-loop corrections compared to the Born process. 
\label{fn:loop}} 
$\SME{ab,m+X+k}{(0)}{(p_a, p_b; \mom{}_{m+X+k})}$. Here the 
subscripts $a$ and $b$ denote the {\em flavors} of the incoming 
partons, while the subscript on the set of final-state momenta in 
the argument gives the {\em multiplicity}. For later convenience, 
the dependence of the matrix element on initial-state momenta is 
shown explicitly. Furthermore, we denote by $\bom{U}_{\!j}$ the 
formal operator that takes some specific $j$-fold unresolved limit 
($j\le k$). Then, the IR factorization formula can be written in 
the following symbolic form
\beq
\bom{U}_{\!j} \SME{ab,m+X+k}{(0)}{(p_a, p_b; \mom{}_{m+X+k})} = 
\left(\frac{\as}{2\pi}\right)^{j} \mathrm{Sing}_{j}^{(0)} 
\SME{\ba{a}\ba{b},m+X+k-j}{(0)}
{(\ba{p}_a,\ba{p}_b; \momb{}_{m+X+k-j})}\,,
\label{eq:IRlimit}
\eeq
i.e., it is a product of a tree-level, $j$-fold unresolved 
singular structure $\mathrm{Sing}_{j}^{(0)} $ times a reduced 
matrix element with $j$ partons removed. The singular structures 
involve Altarelli-Parisi splitting functions, eikonal factors and 
their multi-emission generalizations and are generally matrices in 
color and/or spin space. Thus, the product in \eqn{eq:IRlimit} is 
to be understood accordingly, see \appx{appx:spincolor} for 
details. Moreover, the parton flavors in the reduced matrix 
element may be different from the original ones as indicated by 
the $\ba{a}$ and $\ba{b}$ subscripts. We note in passing that up 
to NNLO accuracy, the singular structures 
$\mathrm{Sing}_{j}^{(0)} $ are known to be universal, i.e., they 
do not depend on the specific process being considered. This opens 
the door to the construction of general NNLO subtraction schemes. 
However, strict process-independent factorization as implied by 
\eqn{eq:IRlimit} may be violated beyond NNLO for initial-state 
radiation~\cite{Catani:2011st, Cieri:2024ytf}. This issue is not 
relevant for our immediate purposes, but of course must be 
addressed by any calculational setup for computing perturbative 
corrections beyond NNLO accuracy.

As is well-known, the IR limit formulae of \eqn{eq:IRlimit} cannot 
be used directly as subtraction terms because of the following 
reasons. First, at any given order beyond LO, the regions of phase 
space corresponding to the various IR limits generally overlap. 
Indeed, already at NLO, the momentum of a parton can be soft and 
collinear to another momentum, hence the soft and collinear 
singular regions are not distinct. Clearly, the structure of 
overlaps becomes more elaborate at higher orders. This then 
implies that care must be taken to avoid multiple subtraction in 
regions of phase space where singular limits overlap. Second, IR 
factorization formulae are only well-defined in the strict limit 
which they describe. To appreciate this, it is enough to examine 
the prescription for the set of momenta $\momb{}$ entering the 
factorized matrix element on the right hand side of 
\eqn{eq:IRlimit}. For example in soft limits (single or multiple), 
the relevant IR factorization theorem instructs us to simply drop 
all soft momenta from the original set of momenta $\mom{}$ in 
order to obtain $\momb{}$. However, this prescription clearly 
violates momentum conservation unless all soft momenta are 
strictly zero, i.e., the right hand side is only well-defined in 
the strict soft limit. In collinear limits, according to the IR 
factorization theorem, $\momb{}$ is obtained simply by replacing 
the collinear momenta with their sum. This prescription however 
does not respect the mass-shell condition for the parent momentum, 
unless the combined momenta are strictly collinear. Moreover, the 
momentum fractions entering the Altarelli-Parisi splitting 
functions are also only well-defined if the collinear momenta are 
strictly proportional to the parent momentum, i.e., in the strict 
collinear limit. Thus, proper extensions of the IR limit formulae 
must be defined away from the limits.

%
% Dealing with overlapping limits
%

\subsection{Dealing with overlapping limits}
\label{sec:overlaps}

Let us begin by formalizing the concept of overlapping limit, 
i.e., a limit in which several unresolved regions are approached 
simultaneously, as follows. An overlapping limit refers simply to 
the successive application of limits and we say that some $j$-fold 
unresolved limits overlap if their subsequent application leads to 
a configuration that is also $j$-fold unresolved. It should be 
noted that taking two $j$-fold unresolved limits one after the 
other can lead to a configuration that is more than $j$-fold 
unresolved.\footnote{A simple example is the successive 
application of two single unresolved $(j=1)$ soft limits $p_r^\mu 
\to 0$ and $p_s^\mu \to 0$, which obviously produces a double soft 
configuration in which $j=2$ momenta are unresolved.} In these 
cases, we do not consider the limits as overlapping. It can be 
shown~\cite{Somogyi:2005xz} that the basic structure of IR limit 
formulae for overlapping limits is the same as in \eqn{eq:IRlimit}. 
Hence, in the following, the formal limit operators $\bom{U}_{\!j}$ 
will refer to any generic $j$-fold unresolved limit, including 
overlapping limits.

The issue of over-counting of overlapping singular regions can 
then be dealt with easily by using the inclusion-exclusion 
principle: we start by summing all basic limits, then subtract the 
pairwise overlaps of limits, add the triple overlaps and so 
on.\footnote{Alternatively, one can choose to construct 
subtraction terms that smoothly interpolate between various limits 
as, e.g., in the case of dipole subtraction~\cite{Catani:1996vz} 
or antenna subtraction~\cite{Gehrmann-DeRidder:2005btv}.} E.g., at 
NLO we encounter only single collinear ($p_i^\mu \parallel 
p_r^\mu$) and single soft ($p_r^\mu \to 0$) limits. Then, in order 
construct a single unresolved approximation that counts each 
singular region once, the inclusion-exclusion principle instructs 
us to sum all collinear and all soft limits and subtract the 
collinear-soft overlaps. To do this, we introduce the formal 
single unresolved limit operator $\bom{A}_1$ as follows,
\beq
\bom{A}_1 = \sum\Big(\bom{C}_{ir} + \bom{S}_{r} 
    - \bom{C}_{ir} \cap \bom{S}_{r} \Big)\,,
\label{eq:A1symb}
\eeq
where $\bom{C}_{ir}$ and $\bom{S}_{r}$ are generic operators that 
take the single collinear and single soft limits, while the 
overlapping limit is denoted by the $\cap$ symbol. We can then 
apply this formal operator to any specific object for which we 
wish to build a single unresolved approximation, such as 
$\dsig{RR}_{ab}$ and $\dsig{RV}_{ab}$ but also 
$\dsiga{RR}{2}_{ab}$, $\dsigk{C}{1}{}{}_{ab}$ and 
$\int_1 \dsiga{RR}{1}_{ab}$. This construction generalizes to NNLO 
easily. Now we have four types of basic singular configurations: 
triple collinear ($p_i^\mu \parallel p_r^\mu \parallel p_s^\mu$), 
double collinear ($p_i^\mu \parallel p_r^\mu$ and $p_j^\mu 
\parallel p_s^\mu$), soft-collinear ($p_i^\mu \parallel p_r^\mu$ 
and $p_s^\mu \to 0$) and finally double soft ($p_r^\mu \to 0$ and 
$p_s^\mu \to 0$). Accordingly, the double unresolved limit 
operator $\bom{A}_2$ counting each singular region once is given by
\beq
\bsp
\bom{A}_2 = \sum\Big[&\bom{C}_{irs} 
    + \bom{C}_{ir,js} + \bom{C\!S}_{ir,s} + \bom{S}_{rs} 
\\
&
-\Big(\bom{C}_{irs} \cap \bom{C\!S}_{ir,s} 
    + \bom{C}_{irs} \cap \bom{S}_{rs}
    + \bom{C}_{ir,js} \cap \bom{C\!S}_{ir,s} 
    + \bom{C}_{ir,js} \cap \bom{S}_{rs}
    + \bom{C\!S}_{ir,s} \cap \bom{S}_{rs}
\Big)
\\
&
+\Big(
\bom{C}_{irs} \cap \bom{C\!S}_{ir,s} \cap \bom{S}_{rs}
+ \bom{C}_{ir,js} \cap \bom{C\!S}_{ir,s} \cap \bom{S}_{rs}
\Big)
\Big]\,.
\esp
\label{eq:A2symb}
\eeq
Here $\bom{C}_{irs}$, $\bom{C}_{ir,js}$, $\bom{C\!S}_{ir,s}$ and 
$\bom{S}_{rs}$ are limit operators for the four basic types of 
limits and overlaps are denoted as before. In writing 
\eqn{eq:A2symb} we have used that the overlap of a triple and 
double collinear limit leads to a configuration where at least 
three partons are unresolved\footnote{Consider the $\bom{C}_{irs} 
\cap \bom{C}_{ir,js}$ overlap, i.e., the limit where the $p_i^\mu 
\parallel p_r^\mu \parallel p_s^\mu$ and $p_i^\mu \parallel 
p_r^\mu$, $p_j^\mu \parallel p_s^\mu$ regions are approached 
simultaneously. In this region clearly $p_i^\mu \parallel p_r^\mu 
\parallel p_s^\mu \parallel p_j^\mu$, which is a quadruple 
collinear limit where three partons are unresolved.} so at NNLO 
we do not need to consider this overlap. Moreover, it is not 
necessarily true that each overlapping term in \eqn{eq:A2symb} 
corresponds to a distinct limit formula. In particular, explicit 
calculations~\cite{Somogyi:2005xz} show that at the level of 
symbolic operators, $\bom{C}_{ir,js} \cap \bom{C\!S}_{ir,s} \cap 
\bom{S}_{rs} = \bom{C}_{ir,js} \cap \bom{S}_{rs}$. Nevertheless, 
we cannot simply cancel these terms altogether, since the 
summation in \eqn{eq:A2symb} is symbolic and it is not immediately 
clear if they both appear with the same coefficient once the 
summation is properly defined. In fact, as we will see in 
\sect{sec:doubleunresolv}, a complete cancellation does not occur.

Obviously \eqns{eq:A1symb}{eq:A2symb} are symbolic since neither 
the summations nor the meaning of the symbol $\cap$ are precisely 
defined. Defining the summations properly is not difficult: one 
simply has to make sure that all basic unresolved configurations 
are counted precisely once. However, the meaning of the $\cap$ 
symbol is somewhat more involved, since it turns out that at the 
symbolic operator level, the overlap of limits as defined above is 
non-commutative. This non-commutativity arises from the formal 
definition of the soft and collinear limits and was investigated 
in detail in~\cite{Somogyi:2005xz}. There it was also shown that 
the $\cap$ symbol in \eqns{eq:A1symb}{eq:A2symb} should be 
understood to refer to the overlap of limits in the sense 
introduced above, i.e., as successive applications of limits. 
Hence, e.g., $\bom{C}_{ir} \cap \bom{S}_{r}$ is to be understood 
as an operator that computes the collinear limit of the soft 
limit, in this order.

%
% From limit formulae to subtraction terms: general considerations
%

\subsection{From limit formulae to subtraction terms: 
general considerations}
\label{sec:extension}

The extension of the IR limit formulae (both for direct and 
overlapping limits) over the full phase space requires two steps. 
First, the momenta entering the factorized squared matrix elements 
must be specified. Second, all kinematic quantities appearing in 
the IR factorization formulae, such as momentum fractions and 
transverse momenta in collinear parton splitting, as well as 
eikonal factors in soft emission, must be explicitly defined as 
functions of the original momenta of the event. Clearly, both the 
momentum mappings and the definitions of momentum fractions, etc., 
must be chosen in a way that the structure of cancellations in all 
overlapping limits is respected. This is of course a non-trivial 
constraint on the entire construction. Here we provide some 
general considerations which motivate our approach to the
\begin{itemize}
    \item definition of momentum mappings;
    \item choice of momentum fractions;
    \item parametrization of transverse momenta;
    \item treatment of soft eikonal factors.
\end{itemize}
%

%
% Definition of momentum mappings
%

\subsubsection{Definition of momentum mappings}

Starting with the momentum mappings, we first of all require 
transformations of the original set of momenta, 
$(p_a,p_b;\mom{}_{m+X+k})$, to sets of reduced momenta, $(\ti{p}_a, 
\ti{p}_b;\momt{}_{m+X+k-j})$ (here $j\le k$ is the number of 
unresolved particles) which respect momentum conservation and the 
mass-shell conditions.\footnote{This set of tilded momenta, 
$(\ti{p}_a, \ti{p}_b;\momt{}_{m+X+k-j})$, is {\em different} from the 
set of barred momenta, $(\ba{p}_a, \ba{p}_b;\momb{}_{m+X+k-j})$, that 
appears in the IR factorization formula of \eqn{eq:IRlimit}. As 
explained above, the latter may violate momentum conservation and/or 
the mass-shell conditions away from the strict limit.} In addition, 
the mapped momenta must also have the proper behavior in IR limits as 
spelled out in the IR factorization theorems. Obviously, the choice of 
momentum mappings is not unique, and we opt to use mappings where the 
recoil is distributed over the entire event rather than being 
absorbed by some specific spectator parton(s). This choice makes 
it more straightforward to construct a single unresolved 
approximate cross section such that both $\dsiga{RR}{1}_{ab}$ as 
well as its integrated form $\int_1 \dsiga{RR}{1}_{ab}$ have 
universal IR limits. In fact, this feature is not guaranteed by 
QCD factorization alone~\cite{Somogyi:2009ri}, and relies on the 
specific definitions of the subtraction terms. We note that for 
initial-state radiation at NNLO, we find it is enough to introduce 
just five elementary momentum mappings, whose precise definitions 
will be given below. (These definitions are also collected in 
\appx{appx:mommaps} for the readers' convenience.) In particular, as 
we will see, subtraction terms corresponding to overlapping limits do 
not require dedicated momentum mappings, while subtraction terms for 
iterated single unresolved limits can be defined using convolutions of 
single unresolved ones. 

%
% Choice of momentum fractions
%

\subsubsection{Choice of momentum fractions}

Turning to the definition of kinematic quantities that appear in 
the IR limit formulae, let us begin with the momentum fractions in 
collinear splitting. Consider first the case of final-final single 
collinear splitting, $(ir) \to i+r$, where we seek to define the 
momentum fractions $z_i$ and $z_r$ of the daughter partons. In the 
strict collinear limit, the parent momentum is simply $p_{ir}^\mu 
= p_i^\mu+p_r^\mu$ and $p_i^\mu$ as well as $p_r^\mu$ are 
strictly proportional to $p_{ir}^\mu$. The factors of 
proportionality define the momentum fractions $z_i$ and $z_r$. In 
the strictly collinear configuration, these may be computed by 
contracting the parent and daughter momenta with some auxiliary 
four-vector $v^\mu$ (with $v\cdot p_{ir}\ne 0$) and taking the 
appropriate ratios, i.e., $z_i = (p_i\cdot v)/[(p_i+p_r)\cdot v]$ 
and $z_r = (p_r\cdot v)/[(p_i+p_r)\cdot v]$. Then, a rather 
obvious way of specifying the momentum fractions away from the 
strict limit is to adopt these expressions as the 
{\em definitions} of momentum fractions. These definitions can be 
made Lorentz-invariant by choosing $v^\mu$ to be the total 
(partonic) four-momentum of the event, $v^\mu=Q^\mu=
(p_a+p_b)^\mu$. Here we have introduced $Q^\mu=p_a^\mu+p_b^\mu$ to 
denote the total incoming {\em partonic} momentum. Thus, for {\em 
final-state} partons $i$ and $r$ we {\em define} the quantity 
\beq
z_{i,r} = \frac{s_{i\ab}}{s_{(ir)\ab}}\,,
\label{eq:zjk-def}
\eeq
which is then interpreted as the momentum fraction of parton $i$ 
in the $(ir)\to i+r$ final-state splitting. Obviously, the 
momentum fraction for parton $r$ is then $z_{r,i}=1-z_{i,r}$. In 
\eqn{eq:zjk-def} we have taken the occasion to introduce some 
notation that we will use throughout. First, $s_{jk}$ will denote 
twice the dot-product of any two momenta $p_j$ and $p_k$. Second, 
indices in parentheses denote sums of the corresponding momenta so 
that, e.g.,
\beq
s_{jk} = 2p_j\cdot p_k\,,
\qquad
s_{j(k\ell)} = 2p_j\cdot (p_k+p_\ell)\,,
\qquad
s_{(jk)(\ell m)} = 2(p_j+p_k)\cdot (p_\ell+p_m)\,,
\eeq
and so on. Importantly, \eqn{eq:zjk-def} should be understood to 
define the {\em function} $z_{i,r}$ whose arguments are the two 
final-state momenta with indices $i$ and $r$, i.e., $p_i^\mu$ and 
$p_r^\mu$. Thus, any expression of this form is to be interpreted 
as the ratio of the invariants $s_{i\ab}$ and $s_{(ir)\ab}$, where 
$i$ and $r$ are indices of final-state momenta from some 
particular set. This is significant, since we will often need to 
define momentum factions for momenta obtained after some mapping, 
i.e., when the momenta are elements of the set $(\ti{p}_a, 
\ti{p}_b;\momt{}_{m+X+k-j})$ discussed above. In such cases, the 
indices will inherit the tilde (or corresponding) notation. E.g., 
an expression of the form$z_{\ti{i},r}$ will be understood to be 
defined by 
\eqn{eq:zjk-def},
\beq
z_{\ti{i},r} = \frac{s_{\ti{i}\ab}}{s_{(\ti{i}r)\ab}}
	= \frac{2\ti{p}_i\cdot (p_a+p_b)}
    {2(\ti{p}_i+p_r)\cdot (p_a+p_b)}\,,
\eeq
where for full clarity, we have also resolved our notation for 
invariants. Notice in particular that the dot-products are always 
computed with the total partonic momentum $Q^\mu=(p_a+p_b)^\mu$, 
irrespective of the fact that $\ti{p}_i^\mu$ is an element of a 
set of mapped momenta. We will make heavy use of this notation 
throughout.

The definition in \eqn{eq:zjk-def} has several nice features. 
First, the momenta $p_i$ and $p_r$ appear symmetrically and the 
momentum fractions sum to one by construction, $z_{i,r}+z_{r,i} = 
1$. Second, the generalization to final-state $n$-fold splitting 
is obvious. E.g., for final-state triple collinear $(irs)\to 
i+r+s$ splitting, we define
\beq
z_{i,rs} = \frac{s_{i\ab}}{s_{(irs)\ab}}\,.
\label{eq:zizrzs-def}
\eeq
This relation is again to be understood as defining the function 
$z_{i,rs}$ of the three final-state momenta whose indices appear 
in the subscript.

Turning to initial-final splitting, let us begin once more by 
considering the case of the single collinear limit $a\to (ar) + 
r$. We now look to define the momentum fractions $x_a$ and $x_r$ 
of the parton entering the reduced matrix element and the parton 
in the final-state. Note the slight abuse of notation in using 
$x_a$ to denote the momentum fraction of parton $(ar)$. In the 
strict collinear limit, the momentum of parton $(ar)$ is 
$p_{ar}^\mu=p_a^\mu-p_r^\mu$ and $p_{ar}^\mu$ as well as 
$p_r^\mu$ are strictly proportional to $p_a^\mu$. The factors of 
proportionality are the momentum fractions $x_a$ and $x_r$. As in 
the final-state case, in the strictly collinear configuration, 
these may be computed by contracting the momenta with some 
auxiliary four-vector $v^\mu$ (with $v\cdot p_a\ne 0$) and taking 
appropriate ratios, i.e., $x_r = (p_r\cdot v)/(p_a\cdot v)$ and 
$x_a = [(p_a-p_r)\cdot v]/(p_a\cdot v) = 1-x_r$. Thus, a rather 
straightforward way of specifying the momentum fractions away from 
the strict limit is to {\em define} them by these expressions. 
Once again, we can choose $v^\mu$ to be the total (partonic) 
four-momentum of the event, $v^\mu = Q^\mu=(p_a+p_b)^\mu$, which 
leads to a Lorentz-invariant definition of momentum fractions. 
Thus, for an {\em initial-state} parton $a$ and {\em final-state} 
parton $r$, we {\em define} the quantities
\beq
x_{r,a} = \frac{s_{r\ab}}{s_{a\ab}}\,,
\qquad
x_{a,r} = 1 - x_{r,a}\,,
\label{eq:xja-def}
\eeq
which are then interpreted as momentum fractions for partons $r$ 
and $(ar)$ in the $a\to (ar)+r$ initial-state splitting. Once more 
we stress that \eqn{eq:xja-def} is to be understood as defining 
the {\em functions} $x_{r,a}$ and $x_{a,r}$ whose arguments are 
the final-state momentum $p_r^\mu$ and the initial-state momentum 
$p_a^\mu$. Notice that the definitions in \eqn{eq:xja-def} are not 
symmetric between initial-state and final-state momenta and the 
precise definition depends on whether the first index is that of a 
final-state parton or an initial-state parton. We emphasize 
furthermore that although $s_{a\ab} = s_{ab}$, the right-hand side 
of $x_{r,a}$ in \eqn{eq:xja-def} must be written precisely as 
given above and the subscripts refer to the indices of momenta in 
the numerator and denominator which {\em always} multiply the 
four-vector $Q^\mu=(p_a+p_b)^\mu$. This becomes significant when 
defining momentum fractions for mapped momenta. As with the 
final-state momentum fractions discussed above, in such cases the 
indices will inherit the tilde (or corresponding) notation of the 
mapped momenta, but dot-products must always be computed with 
$Q^\mu=(p_a+p_b)^\mu$, such that, e.g., $x_{r,\ti{a}}$ is
\beq
x_{r,\ti{a}} = \frac{s_{r\ab}}{s_{\ti{a}\ab}} 
	= \frac{2p_r\cdot(p_a+p_b)}{2\ti{p}_a\cdot(p_a+p_b)}\,,
\eeq
where we have spelled out all definitions explicitly for complete 
clarity. Such notation will also be used prominently. 

The generalization to $n$-fold initial-state splitting is again 
simple. E.g., for initial-state $a\to (ars)+r+s$ triple collinear 
splitting, the obvious generalization of \eqn{eq:xja-def} reads,
\beq
x_{r,a} = \frac{s_{r\ab}}{s_{a\ab}}\,,
\qquad
x_{s,a} = \frac{s_{s\ab}}{s_{a\ab}}\,,
\qquad
x_{a,rs} = 1 - x_{r,a} - x_{s,a}\,.
\label{eq:xajk-def}
\eeq
At the risk of belaboring the point, we stress one more time that 
\eqn{eq:xajk-def} defines {\em functions} of the initial-state 
momentum $p_a^\mu$ and the final-state momenta $p_r^\mu$ and 
$p_s^\mu$, and dot-products are always computed with the 
four-vector $Q^\mu=(p_a+p_b)^\mu$. Thus, the number of indices in 
the subscript reflects the number of arguments of these functions 
and in particular the definitions of the momentum fractions of the 
final-state partons $r$ and $s$ are unchanged from to 
\eqn{eq:xja-def}. As discussed above, when one or more of the 
arguments is from a set which was obtained through some momentum 
mapping, the corresponding index inherits the notation of the 
mapped set. As for initial-final single collinear splitting, the 
definitions of momentum fractions in the $n$-fold case are not 
symmetric between initial-state and final-state partons, however 
the final-state partons do appear symmetrically.

We note in passing that the definitions in \eqn{eq:xja-def} can 
also be obtained from \eqn{eq:zjk-def} by a crossing 
transformation, $x_{a,r} = 1/z_{i,r}|_{p_i \to -p_a}$ and $x_{r,a} 
= -z_{r,i}/z_{i,r}|_{p_i \to -p_a}$, see the discussion in 
\appx{appx:AP-functions}. This transformation also 
relates momentum fractions for multiple-collinear splitting. 
Indeed, it is easy to check that $x_{a,rs}$ of \eqn{eq:xajk-def} 
can be obtained from \eqn{eq:zizrzs-def} as $x_{a,rs} = 
1/z_{i,rs}|_{p_i \to -p_a}$ and moreover $x_{r,a} = -
z_{r,is}/z_{i,rs}|_{p_i \to -p_a}$ and $x_{s,a} = -
z_{s,ir}/z_{i,rs}|_{p_i \to -p_a}$. With regards to the last two 
relations, notice that the ratio $z_{r,is}/z_{i,rs}$ depends only 
on the momenta $p_r$ and $p_i$, but not on $p_s$ (similarly 
$z_{s,ir}/z_{i,rs}$ depends only on the momenta $p_s$ and $p_i$, 
but not on $p_r$). Thus, ratios of momentum fractions for 
final-state triple collinear splitting actually define functions 
of just two momenta. Hence, the number of indices in the 
subscripts differ on the two sides of these particular crossing 
relations.

We finish the discussion of momentum fractions by observing that 
somewhat surprisingly, the definition in \eqn{eq:xajk-def} raises 
the following subtlety. It is not very difficult to see that 
$x_{a,rs}$ can take both positive and negative values. This 
implies that it vanishes at certain points of the double real 
emission phase space and these points turn out not to correspond 
to any IR limit (i.e., they are ordinary points inside the double 
real emission phase space). This alone would not be an issue, but 
it can be checked that the initial-final-final triple collinear 
splitting formulae have pieces with $x_{a,rs}$ in the 
denominator.\footnote{In fact, negative powers of $x_{a,rs}$ up to 
$1/x_{a,rs}^2$ can appear.} Thus, this definition of $x_{a,rs}$ 
introduces spurious non-integrable singularities. One way of 
addressing this issue is of course to search for new definitions 
of momentum fractions that do not vanish inside the real emission 
phase space. This is possible, but the obtained formulae are quite 
cumbersome and lead to complications when computing the integrated 
versions of the triple collinear subtraction terms. However, we 
may also deal with the spurious singularities by introducing 
appropriate damping functions around these singularities. In this 
work, we have chosen to pursue the latter option and keep the 
definition of the momentum fraction in \eqn{eq:xajk-def}.

%
% {Parametrization of transverse momenta
%

\subsubsection{Parametrization of transverse momenta}

Next we discuss the choice of transverse momenta. We start by 
observing that at the level of IR factorization formulae, there is 
a relation between momentum fractions, invariants and the squares 
of transverse momenta. E.g., for final-final single collinear 
$(ir)\to i+r$ splitting, standard Sudakov parametrization implies 
(see, e.g.,~\refr{Catani:1996vz})
\beq
\kT{}^2 = -z_i z_r s_{ir}\,,
\eeq
while the appropriate relations for triple parton splitting can be 
found in~\refr{Catani:1999ss} and are recalled in 
\appx{appx:AP-functions}, see 
\eqns{eq:ktkt2ktktonkt2_1}{eq:ktkt2ktktonkt2_2}. Thus, the 
collinear splitting functions can be written in different ways 
that are nonetheless equivalent in the IR limit, e.g., 
\beq
P^{\mu\nu} = -A g^{\mu\nu} 
    + B \kT{}^\mu \kT{}^\nu
\qquad\mbox{and}\qquad
P^{\mu\nu} = -A g^{\mu\nu} 
    - B z_i z_r s_{ir} \frac{\kT{}^\mu \kT{}^\nu}{\kT{}^2}
\eeq
are equivalent for any $A$ and $B$. However, the two forms will 
not necessarily be equal once some explicit definitions of the 
momentum fractions and transverse momenta are adopted. We find it 
most convenient  to always write the azimuthally correlated terms 
in the splitting functions as $\kT{}^\mu \kT{}^\nu/\kT{}^2$ such that 
the overall scale of $\kT{}$ cancels and substitute the definitions of 
momentum fractions and transverse momenta into those expressions. This 
choice renders the computation of azimuthally averaged splitting 
function trivial and simplifies their integration.

Regarding the actual definition of transverse momenta, let us 
consider first the case of final-final collinear splitting 
$(ir)\to i+r$. The momenta naturally available to construct the 
transverse momentum are $p_i^\mu$, $p_r^\mu$ and the mapped 
momentum $\ha{p}_{ir}^\mu$ which enters the reduced matrix element 
(to be defined precisely in \eqn{eq:cirFFmap} below). Thus for 
{\em final-state} partons $i$ and $r$ we define\footnote{We have 
chosen the notation for the coefficients and signs for later 
convenience.}
\beq
\kT{i,r}^\mu = \zeta_{i,r} p_r^\mu - \zeta_{r,i} p_i^\mu 
    + Z_{ir} \ha{p}_{ir}^\mu\,.
\label{eq:ktjk-ansatz}
\eeq
The coefficients $\zeta_{i,r}$, $\zeta_{r,i}$ and $Z_{ir}$ in 
\eqn{eq:ktjk-ansatz} can be fixed (up to an overall factor which 
is irrelevant as explained above) by requiting that $\kT{i,r}^\mu$ 
be orthogonal to $\ha{p}_{ir}^\mu$ and the total partonic momentum 
$Q^\mu=(p_a+p_b)^\mu$. These choices simplify the integration of 
the collinear subtraction terms. We will spell out our specific 
definitions below in \sect{sec:CirFF00}. As before, 
\eqn{eq:ktjk-ansatz} should be considered to define the 
{\em function} $\kT{i,r}^\mu$ of final-state momenta $p_i$ and 
$p_r$ and we will understand expressions such as $\kT{r,s}^\mu$ as 
being given by \eqn{eq:ktjk-ansatz} in a fashion very similar to 
the definition of momentum fractions with arbitrary indices above. 
We note that \eqn{eq:ktjk-ansatz} can be generalized to $n$-fold 
collinear splitting in a reasonably straightforward way. We will 
not use this generalization in this paper, however the explicit 
definition for the case of final-state triple collinear splitting 
was presented in~\refr{Somogyi:2006da}.

Turning to the case of initial-final collinear splitting $a \to 
(ar)+r$, it is most straightforward to define the transverse 
momentum of the final-state parton to be simply orthogonal to the 
direction of $p_a^\mu$. Thus for a general {\em final-state} 
parton $r$ we define
\beq
\kT{r,a}^\mu = p_{r,\perp}^\mu\,,
\qquad
\kT{a,r}^\mu = -\kT{r,a}^\mu\,,
\label{eq:kj-def}
\eeq
where the `transverse momentum' of the initial-state hard parton 
was formally defined by requiring that the transverse momenta in 
the splitting sum to zero.\footnote{Although the definition of 
$\kT{r,a}^\mu$, \eqn{eq:kj-def}, appears to involve only the momentum 
of parton $r$, notice that in fact $p_a^\mu$ must also be known in 
order to define the perpendicular direction. Hence, $a$ also appears 
in the subscript.} Similarly to the case of initial-state momentum 
fractions, $\kT{a,r}^\mu$ actually denotes the transverse momentum of 
parton $(ar)$. Clearly, \eqn{eq:kj-def} generalizes immediately to 
initial-state $n$-fold collinear splitting. E.g., for the 
initial-final-final $a\to (ars)+r+s$ triple collinear splitting, the 
transverse momenta of the final-state partons $r$ and $s$ are still 
defined as in \eqn{eq:kj-def}, while the `transverse momentum' of the 
initial-state hard parton $(ars)$ is once again given by the 
requirement that all transverse momenta sum to zero,
\beq
\kT{r,a}^\mu = p_{r,\perp}^\mu\,,
\qquad
\kT{s,a}^\mu = p_{s,\perp}^\mu\,,
\qquad
\kT{a,rs}^\mu = -\kT{r,a}^\mu-\kT{s,a}^\mu\,.
\label{eq:kcjk-def}
\eeq
At this point, it is hopefully clear that 
\eqns{eq:kj-def}{eq:kcjk-def} should once again be viewed as 
{\em functions} of the momenta whose indices appear in the 
subscripts, such that expressions like $\kT{\ti{r},\ti{a}}^\mu$ 
are understood to be defined by these equations. We note in passing 
that for a final-state momentum $p_r^\mu$ it is possible to give a 
Lorentz-invariant definition of $p_{r,\perp}^\mu$ as follows,
\beq
p_{r,\perp}^\mu = p_{r}^\mu - \frac{s_{br}}{s_{ab}} p_a^\mu 
    - \frac{s_{ar}}{s_{ab}} p_b^\mu\,.
\eeq
However, in actual numerical calculations, the direction of 
$p_a^\mu$ is usually fixed and there is no need to use this 
representation.

%
% Treatment of soft eikonal factors
%

\subsubsection{Treatment of soft eikonal factors}

Last we briefly comment on the choice of eikonal functions in soft 
formulae. Clearly, the eikonal function for the emission of a soft 
gluon $r$,
\beq
\calS_{jk}(r) = \frac{2s_{jk}}{s_{jr} s_{kr}}\,,
\label{eq:eik-def}
\eeq
is well-defined as it stands over the full real emission phase 
space in terms of the original momenta of the event. However, we 
are free to use also the mapped hard momenta, i.e., make the 
replacement
\beq
\frac{2s_{jk}}{s_{jr} s_{kr}} \to \frac{2s_{\ti{j}\ti{k}}}{s_{\ti{j}r} s_{\ti{k}r}}\,.
\label{eq:eik-def-mapped}
\eeq
At the level of IR limits, the two expressions are identical, as 
the transformed hard momenta go to the original ones in the soft 
limit. On the other hand, we find that using the form involving 
the mapped momenta leads to easier integrals when computing the 
integrated subtractions and at the same time does not lead to any 
complications for defining the unintegrated subtraction terms. 

We have collected the definitions of all kinematic quantities 
introduced above in \appx{appx:kinquant} for the readers' 
convenience.

%
% The general structure of subtraction terms
%

\subsection{The general structure of subtraction terms}

The upshot of the previous discussion is that we can find 
definitions of momentum mappings, momentum fractions, and so on 
that allow us promote the IR limit formulae of \eqn{eq:IRlimit} to 
true subtraction terms that are unambiguously defined over the 
full real emission phase space,
\beq
\bom{U}_{\!j} \SME{ab,m+X+k}{(0)}{(p_a,p_b;\mom{}_{m+X+k})} \to 
    {\cal U}_{j}^{(0,0)}\,,
\label{eq:IRct}
\eeq
where we have set
\beq
{\cal U}_{j}^{(0,0)} = \left(\frac{\as}{2\pi}\right)^{j} 
    \widetilde{\mathrm{Sing}}_{j}^{(0)} 
    \SME{\ti{a}\ti{b},m+X+k-j}{(0)}
    {(\ti{p}_a,\ti{p}_b;\momt{}_{m+X+k-j})}\,.
\label{eq:IRct1}
\eeq
Notice that the reduced matrix element is evaluated over the set 
of mapped momenta $(\ti{p}_a,\ti{p}_b;\momt{}_{m+X+k-j})$, while 
$\widetilde{\mathrm{Sing}}_{j}^{(0)}$ denotes the expression for 
the appropriate singular structure, incorporating the precise 
definitions of momentum fractions, eikonal factors and so on. The 
$(0,0)$ superscript reminds us that these subtraction terms were 
derived from IR limit formulae that involve zero-loop singular 
structures and zero-loop squared matrix elements (as compared to 
the Born matrix element, see footnote~\ref{fn:loop} above). 

Finally, the subtraction terms of \eqn{eq:IRct1} can be used to 
build approximate cross sections. These are essentially sums over 
subtraction terms, constructed such that all relevant singular 
limits, including overlapping ones, are counted precisely once as 
dictated by the inclusion-exclusion principle. In 
\sect{sec:doublereal}, we present the precise definitions of all 
approximate cross sections that appear in \eqn{eq:sigRRreg} in 
terms of subtraction terms which are in turn defined in 
\sectss{sec:A1}{sec:A12}.

%%%
%%% Approximate cross sections for double real emission
%%%

\section{Approximate cross sections for double real emission}
\label{sec:doublereal}

We start by defining precisely the Born, real emission and double 
real emission partonic cross sections. Thus, let us consider the 
partonic process $a + b \to m\mbox{ jets} + X$ (recall $X$ denotes 
collectively the set of all colorless final-state particles). 
Clearly each cross section involves a sum over the various 
partonic subprocesses, hence
\bal
\dsig{B}_{ab}(p_a, p_b) &= 
    \sum_{\{m\}} \frac{1}{S_{\{m\}}} 
    \dsig{B}_{ab,m+X}(p_a, p_b)\,,
\label{eq:Bdef}
\\
\dsig{R}_{ab}(p_a, p_b) &= 
    \sum_{\{m+1\}} \frac{1}{S_{\{m+1\}}} 
    \dsig{R}_{ab,m+X+1}(p_a, p_b)\,,
\label{eq:Rdef}
\\
\dsig{RR}_{ab}(p_a, p_b) &= 
    \sum_{\{m+2\}} \frac{1}{S_{\{m+2\}}} 
    \dsig{RR}_{ab,m+X+2}(p_a, p_b)\,,
\label{eq:RRdef}
\eal
where $\displaystyle \sum_{\{n\}}$ denotes the summation over the 
$a+b\to n \mbox{ partons} + X$ partonic subprocesses and 
$S_{\{n\}}$ is the corresponding Bose symmetry factor for 
identical particles in the final state. Then the cross sections 
for the individual Born, real and double real subprocesses read
\bal
\dsig{B}_{ab,m+X}(p_a,p_b) &= 
    {\cal N} \frac{1}{\omega(a)\omega(b)\Phi(p_a\cdot p_b)} 
\label{eq:Bdefsub}
\\&\times
    \PS{m+X}(\mom{}_{m+X};p_a+p_b) 
    \SME{ab,m+X}{(0)}{(p_a,p_b;\mom{}_{m+X})}\,,
\nt
\\
\dsig{R}_{ab,m+X+1}(p_a,p_b) &= 
    {\cal N} \frac{1}{\omega(a)\omega(b)\Phi(p_a\cdot p_b)} 
\label{eq:Rdefsub}
\\&\times
    \PS{m+X+1}(\mom{}_{m+X+1};p_a+p_b) 
    \SME{ab,m+X+1}{(0)}{(p_a,p_b;\mom{}_{m+X+1})}\,,
\nt
\intertext{and}
\dsig{RR}_{ab,m+X+2}(p_a,p_b) &= 
    {\cal N} \frac{1}{\omega(a)\omega(b)\Phi(p_a\cdot p_b)} 
\label{eq:RRdefsub}
\\&\times
    \PS{m+X+2}(\mom{}_{m+X+2};p_a+p_b) 
    \SME{ab,m+X+2}{(0)}{(p_a,p_b;\mom{}_{m+X+2})}\,.
\nt
\eal
Here ${\cal N}$ collects all factors that are independent of QCD 
while $\Phi(p_a\cdot p_b)$ is the flux factor and 
$\mom{}_{m+X+k}$, ($k=0,1,2$) denotes the set of momenta of the 
final-state particles. The $(m+X+k)$-particle phase space 
appearing in \eqnss{eq:Bdefsub}{eq:RRdefsub} reads
\beq
\bsp
&
\PS{m+X+k}(\mom{}_{m+X+k};p_a+p_b) =
\\&\qquad = 
	\prod_{i=1}^{m+k} \frac{\rd^d p_i}{(2\pi)^{d-1}} 
    \delta_+(p_i^2)
	\frac{\rd^d p_X}{(2\pi)^{d-1}} \delta_+(p_X^2-M_X^2) 
	(2\pi)^{d} \delta^{(d)}\left( p_a + p_b 
    - \sum_{i=1}^{m+k} p_i -p_X\right)\,.
\esp
\label{eq:psmx2}
\eeq
Finally, $\omega(a)$ and $\omega(b)$ count the number of colors 
and spins of the incoming partons and account for the necessary 
averaging over these in the cross section. They depend only on 
parton flavor and in conventional dimensional regularization we 
have
\beq
\omega(q) = \omega(\qb) = 2 \Nc\,, 
\qquad 
\omega(g) = 2 (1-\ep) (\Nc^2-1)\,.
\label{eq:omegafac}
\eeq
We can now spell out the precise definitions of the approximate 
cross sections relevant for regularizing the IR divergences 
present in the double real emission contribution.

%%%
%%% Single unresolved emission
%%%

\subsection{Single unresolved emission}
\label{sec:singleunresolv}

The approximate cross section regularizing single unresolved 
emission, $\dsiga{RR}{1}_{ab}$ reads
\beq
\dsiga{RR}{1}_{ab} = 
    \sum_{\{m+2\}} \frac{1}{S_{\{m+2\}}} 
    \dsiga{RR}{1}_{ab,m+X+2}\,,
\eeq
where the approximate cross section for any particular subprocess 
can be written symbolically as
\beq
\dsiga{RR}{1}_{ab,m+X+2} = 
	{\cal N} \frac{1}{\omega(a)\omega(b)\Phi(p_a\cdot p_b)}
    \PS{m+X+2}(\mom{}_{m+X+2};p_a+p_b) \cA_{1}^{(0)}\,,
\label{eq:dsigRRA1}
\eeq
with
\beq
\cA_{1}^{(0)} =
	\sum_{r \in F}\bigg[
	\cS{r}{(0,0)} 
	+ \sum_{\substack{i \in F \\ i \ne r}} 
    \bigg(\frac12\, \cC{ir}{FF (0,0)} 
    - \cC{ir}{FF}\cS{r}{(0,0)}\bigg)
	+ \sum_{c \in I} \bigg(\cC{cr}{IF (0,0)} 
    - \cC{cr}{IF}\cS{r}{(0,0)}\bigg)\bigg]\,.
\label{eq:A1}
\eeq
Here $I$ and $F$ denote the sets of initial-state and final-state 
partons. The various subtraction terms are explicit realizations of 
the generic subtraction term ${\cal U}_{j}^{(0,0)}$ of \eqn{eq:IRct1} 
with $j=1$ and have the following physical origin: $\cS{r}{(0,0)}$ 
denotes the subtraction term that regularizes the emission of a single 
soft gluon, while $\cC{ir}{FF (0,0)}$ and $\cC{cr}{IF (0,0)}$ are 
subtraction terms regularizing final-final and initial-final collinear 
singularities. Finally, as explained in detail in \sect{sec:overlaps}, 
the overlapping terms $\cC{ir}{FF}\cS{r}{(0,0)}$ and $\cC{cr}{IF}\cS{r}
{(0,0)}$ remove the double subtraction in collinear-soft regions. Thus 
\eqn{eq:A1} is simply the explicit realization of \eqn{eq:A1symb}, 
where we have carefully spelled out the summations such that each 
singular region is counted precisely once. Indeed, the factor of 
$\frac12$ appears with $\cC{ir}{FF (0,0)}$ for exactly this reason: it 
accounts for the fact that the double summation over $i,r \in F$ in 
\eqn{eq:A1} counts this term twice. Finally, we remark that the 
subscripts on the concrete subtraction terms do not simply give 
the number of unresolved partons ($j=1$) in contrast to the 
generic notation in \eqn{eq:IRct1}. Instead they specify the 
actual limit from which the term derives. For example, a single 
unresolved collinear limit is specified by the indices of the two 
partons whose momenta become collinear, and these then appear in 
the subscripts of the corresponding subtraction terms in 
\eqn{eq:A1}. We will follow similar conventions for the rest of 
the subtraction terms throughout this paper. 

We emphasize that the momentum mappings used to define the five 
types of subtraction terms in \eqn{eq:A1} are not all different. 
Hence, the number of distinct configurations of reduced momenta, 
i.e., the number of counter-events, will be less than the number 
of subtraction terms. In fact, we find that the subtraction terms 
for overlapping limits can be defined using the momentum mapping 
appropriate to the {\em innermost} limit, which can be read off as 
the rightmost one in our notation for the subtraction terms. Thus, 
the subtraction terms corresponding to collinear-soft overlaps, 
$\cC{ir}{FF}\cS{r}{(0,0)}$ and $\cC{ar}{IF}\cS{r}{(0,0)}$, both 
employ the soft momentum mapping. This fact can be used to 
optimize the implementation of the approximate cross section, 
hence we present the grouping of the subtraction terms according 
to momentum mappings in \tab{tab:A1maps}. The precise definitions 
of all momentum mappings and subtraction terms are provided in 
\sect{sec:A1}.
\begin{table}
\setlength{\tabcolsep}{10pt}
\renewcommand{\arraystretch}{2}
\begin{center}
\begin{tabular}{|c|ccc|}
\hline
\parbox{\widthof{Momentum}}
{\centering Momentum mapping}\vphantom{$\Bigg|$} 
& \multicolumn{3}{c|}{Subtraction terms}
\\
\hline\hline
$\cmp{ir}{FF}$
	& $\cC{ir}{FF(0,0)}$
	& 
	&
\\
\hline
$\cmp{cd,r}{II,F}$
	& $\cC{cr}{IF(0,0)}$
	&
	&
\\
\hline	
$\smp{r}$
	& $\cS{r}{(0,0)}$
  	& $\cC{ir}{FF}\cS{r}{(0,0)}$
	& $\cC{cr}{IF}\cS{r}{(0,0)}$
\\
\hline
\end{tabular}
\end{center}
\caption{\label{tab:A1maps}
Single unresolved subtraction terms grouped according to momentum 
mappings. Note that collinear-soft overlapping terms are defined 
using the single soft momentum mapping.
}
\end{table}

%%%
%%% Double unresolved emission
%%%

\subsection{Double unresolved emission}
\label{sec:doubleunresolv}

The approximate cross section $\dsiga{RR}{2}_{ab}$ regularizing 
double unresolved emission for initial state radiation is
\beq
\dsiga{RR}{2}_{ab} = 
    \sum_{\{m+2\}} \frac{1}{S_{\{m+2\}}} 
    \dsiga{RR}{2}_{ab,m+X+2}\,,
\eeq
where the approximate cross section for any given subprocess can 
be written symbolically as 
\beq
\dsiga{RR}{2}_{ab,m+X+2} = 
	{\cal N} \frac{1}{\omega(a)\omega(b)\Phi(p_a\cdot p_b)} 
    \PS{m+X+2}(\mom{}_{m+X+2};p_a+p_b) \cA_{2}^{(0)}\,,
\label{eq:dsigRRA2}
\eeq
with
\beq
\bsp
\cA_{2}^{(0)} &= 
	\sum_{r \in F}\sum_{\substack{s \in F \\ s \ne r}}\bigg\{
	\frac12\, \cS{rs}{(0,0)} 
\\ & +
	\sum_{c \in I}\bigg[
	\frac12\, \cC{crs}{IFF (0,0)} 
	+ \cSCS{cr,s}{IF (0,0)}
	- \cC{crs}{IFF}\cSCS{cr,s}{IF (0,0)}
	- \frac12\,\cC{crs}{IFF}\cS{rs}{(0,0)}
	- \cSCS{cr,s}{IF}\cS{rs}{(0,0)}
	+ \cC{crs}{IFF}\cSCS{cr,s}{IF}\cS{rs}{(0,0)}
\\ &\qquad +	
	\sum_{\substack{d \in I \\ d \ne c}}\bigg(
	\frac12\, \cC{cr,ds}{IF,IF (0,0)}	
	- \cC{cr,ds}{IF,IF} \cSCS{cr,s}{IF (0,0)}
	- \frac12\, \cC{cr,ds}{IF,IF} \cS{rs}{(0,0)}
	+ \cC{cr,ds}{IF,IF} \cSCS{cr,s}{IF}\cS{rs}{(0,0)}
	\bigg)
	\bigg]\bigg\}\,.
\label{eq:A2}
\esp
\eeq
The various subtraction terms appearing above correspond to the 
limits implied by the notation and in fact \eqn{eq:A2} is just the 
explicit realization of the symbolic expression in 
\eqn{eq:A2symb}. As in \eqn{eq:A1}, various symmetry factors of 
$\frac12$ have been inserted to ensure that each singular limit is 
counted precisely once. However, the expression in \eqn{eq:A2} can 
be simplified, since the equivalence of the $\bom{C}_{ir,js} \cap 
\bom{C\!S}_{ir,s} \cap \bom{S}_{rs}$ and $\bom{C}_{ir,js} \cap 
\bom{S}_{rs}$ limits discussed below \eqn{eq:A2symb} implies 
\beq
\cC{cr,ds}{IF,IF} \cSCS{cr,s}{IF}\cS{rs}{(0,0)} = 
    \cC{cr,ds}{IF,IF} \cS{rs}{(0,0)}\,.
\label{eq:CCSSeqCS}
\eeq
Hence we can write
\beq
\bsp
\cA_{2}^{(0)} &= 
	\sum_{r \in F}\sum_{\substack{s \in F \\ s \ne r}}\bigg\{
	\frac12\, \cS{rs}{(0,0)} 
\\ & +
	\sum_{c \in I}\bigg[
	\frac12\, \cC{crs}{IFF (0,0)} 
	+ \cSCS{cr,s}{IF (0,0)}
	- \cC{crs}{IFF}\cSCS{cr,s}{IF (0,0)}
	- \frac12\,\cC{crs}{IFF}\cS{rs}{(0,0)}
	- \cSCS{cr,s}{IF}\cS{rs}{(0,0)}
	+ \cC{crs}{IFF}\cSCS{cr,s}{IF}\cS{rs}{(0,0)}
\\ &\qquad +	
	\sum_{\substack{d \in I \\ d \ne c}}\bigg(
	\frac12\, \cC{cr,ds}{IF,IF (0,0)}	
	- \cC{cr,ds}{IF,IF} \cSCS{cr,s}{IF (0,0)}
	+ \frac12\, \cC{cr,ds}{IF,IF} \cS{rs}{(0,0)}
	\bigg)
	\bigg]\bigg\}\,.
\label{eq:A2short}
\esp
\eeq
Obviously this is a more economical representation of 
$\cA_{2}^{(0)}$. Nevertheless, the original form in \eqn{eq:A2} 
has the advantage that it makes certain further simplifications 
which emerge specifically in the color-singlet case more apparent. 
Indeed, as we will now discuss, all terms in \eqn{eq:A2} that 
involve the soft-collinear limit  cancel among each other for 
color-singlet production. To see this, consider first the 
soft-collinear IR limit formula for the configuration $p_c^\mu 
\parallel p_r^\mu$ and $p_s^\mu \to 0$, where $c$ is an initial-
state parton ($c=a,b$),
\beq
\bsp
\bom{CS}^{IF}_{cr,s}\SME{ab,m+X+2}{(0)}{(p_a,p_b;\mom{}_{m+X+2})} 
	&= -(8\pi\as \mu^{2\ep})^2 
    \sum_{\substack{j,k \in I \cup F \\ j, k \ne r, s}} 
	\frac{1}{2}\calS_{jk}(s) \bT_j \bT_k 
	\frac{1}{x_{c,r}} \frac{1}{s_{cr}} 
\\& \times	    
    \hP_{(cr) r}(x_{c,r},\kT{r,c};\ep) 
	\SME{(cr)d,m+X}{(0)}{(\ha{p}_c,\ha{p}_d;\momh{}_{m+X})}\,,
\label{eq:CSars-limit}
\esp
\eeq
where $\hP_{(cr) r}$ is the Altarelli-Parisi splitting kernel for 
the initial-state splitting $c \to (cr)+r$ and $d\ne c$ is the 
other initial-state parton. In \eqn{eq:CSars-limit}, whenever $j$ 
or $k$ is equal to $c$ in the sum, we must evaluate the 
appropriate term with the color charge operator $\bT_{(cr)} = 
\bT_c+\bT_r$, while the eikonal factor must be computed with the 
reduced momentum $\ha{p}_c$ that appears in the factorized matrix 
element. However, in this limit $\ha{p}_c$ is strictly 
proportional to $p_c$, while the eikonal factor is degree-zero 
homogeneous in hard momenta (i.e., $\calS_{\ha{c}k}(s) = \calS_{ck}
(s)$), thus in the color-singlet case ($j,k = a,b$ only), we find
\beq
\bsp
\bom{CS}^{IF}_{cr,s}\SME{ab,m+X+2}{(0)}{(p_a,p_b;\mom{}_{m+X+2})} 
	&= -(8\pi\as \mu^{2\ep})^2 \frac{2s_{ab}}{s_{as} s_{bs}}
	\bT_a \bT_b 
	\frac{1}{x_{c,r}} \frac{1}{s_{cr}} 
\\&\times     
    \hP_{(cr) r}(x_{c,r},\kT{r,c};\ep) 
	\SME{(cr)d,m+X}{(0)}{(\ha{p}_c,\ha{p}_d;\momh{}_{m+X})}\,.
\label{eq:CSars-limit-cs}
\esp
\eeq
Let us now compute the triple collinear limit of 
\eqn{eq:CSars-limit-cs}, when $p_r \to x_{r,c} p_c$ and 
$p_s \to x_{s,c} p_c$. A quick computation yields
\beq
\bsp
\bom{C}^{IFF}_{crs}\bom{CS}^{IF}_{cr,s}
\SME{ab,m+X+2}{(0)}{(p_a,p_b;\mom{}_{m+X+2})} 
	&= (8\pi\as \mu^{2\ep})^2 \frac{1}{s_{cs}} \frac{2}{x_{s,c}}
	\bT_c^2 	
	\frac{1}{x_{c,r}} \frac{1}{s_{cr}} 
\\&\times    
    \hP_{(cr) r}(x_{c,r},\kT{r,c};\ep) 
	\SME{(cr)d,m+X}{(0)}{(\ha{p}_c,\ha{p}_d;\momh{}_{m+X})}\,,
\label{eq:CarsCSars-limit-cs}
\esp
\eeq
where we have used color conservation ($\bT_a + \bT_b = 0$). 
Similarly, in the double collinear limit, when $p_r = x_{r,c} p_c$ 
and $p_s = x_{s,d} p_d$ (where $c,d=a,b$ with $c\ne d$) we find
\beq
\bsp
\bom{C}^{IF,IF}_{cr,ds}\bom{CS}^{IF}_{cr,s}
\SME{ab,m+X+2}{(0)}{(p_a,p_b;\mom{}_{m+X+2})} 
	&= (8\pi\as \mu^{2\ep})^2 \frac{1}{s_{ds}} \frac{2}{x_{s,d}}
	\bT_d^2 
	\frac{1}{x_{c,r}} \frac{1}{s_{cr}} 
\\&\times	    
    \hP_{(cr) r}(x_{c,r},\kT{r,c};\ep) 
	\SME{(cr)d,m+X}{(0)}{(\ha{p}_c,\ha{p}_d;\momh{}_{m+X})}\,.
\label{eq:CarbsCSars-limit-cs}
\esp
\eeq
But the three terms in 
\eqnss{eq:CSars-limit-cs}{eq:CarbsCSars-limit-cs} arise in 
\eqn{eq:A2} in the combination
\beq
\bsp
&
\left[\bom{CS}^{IF}_{cr,s} 
- \bom{C}^{IFF}_{crs}\bom{CS}^{IF}_{cr,s} 
- \bom{C}^{IF,IF}_{cr,ds}\bom{CS}^{IF}_{cr,s}\right]
\SME{ab,m+X+2}{(0)}{(p_a,p_b;\mom{}_{m+X+2})} 
\\ & \qquad =
    (8\pi\as \mu^{2\ep})^2    
	\left[-\frac{2s_{ab}}{s_{as} s_{bs}} \bT_a \bT_b 
	- \frac{1}{s_{cs}} \frac{2}{x_{s,c}} \bT_c^2
	- \frac{1}{s_{ds}} \frac{2}{x_{s,d}} \bT_d^2
	\right]
	\frac{1}{x_{c,r}} \frac{1}{s_{cr}} 
    \hP_{(cr) r}(x_{c,r},\kT{r,c};\ep) 
\\ & \qquad \times
	\SME{(cr)d,m+X}{(0)}{(\ha{p}_c,\ha{p}_d;\momh{}_{m+X})}\,.
\esp
\label{eq:CSterms}
\eeq
The color-charge algebra is trivial in the color-singlet case, 
$\bT_{a}\bT_b = -\bT_a^2 = -\bT_b^2$, (recall $c,d\in I=\{a,b\}$) 
so the color-charge operators in \eqn{eq:CSterms} can be factored 
and the combination of terms becomes proportional to 
\beq
\left[\frac{s_{ab}}{s_{as} s_{bs}}
	- \frac{1}{s_{cs}} \frac{1}{x_{s,c}}
	- \frac{1}{s_{ds}} \frac{1}{x_{s,d}}
	\right]\,.
\label{eq:CSterms-expr}
\eeq
However, if we define the momentum fractions $x_{s,c}$ and 
$x_{s,d}$ as in \eqn{eq:xja-def}, then the expression in 
\eqn{eq:CSterms-expr} vanishes for both $c=a$, $d=b$ and $c=b$, 
$d=a$. Thus, with this choice of momentum fractions, we find the 
following cancellation among the subtraction terms appearing in 
\eqn{eq:A2},\footnote{Notice that since $I=\{a,b\}$, the sum over 
$d$ in \eqn{eq:CScancel1} constitutes only a single term: if $c=a$ 
then $d=b$, while if $c=b$ then $d=a$.}
\beq
\cSCS{cr,s}{IF (0,0)} 
- \cC{crs}{IFF}\cSCS{cr,s}{IF (0,0)} 
- \sum_{\substack{d \in I \\ d \ne c}}
\cC{cr,ds}{IF,IF} \cSCS{cr,s}{IF (0,0)} = 0\,.
\label{eq:CScancel1}
\eeq
Very similar arguments also establish that
\beq
\cSCS{cr,s}{IF} \cS{rs}{(0,0)} 
- \cC{crs}{IFF}\cSCS{cr,s}{IF} \cS{rs}{(0,0)}
- \sum_{\substack{d \in I \\ d \ne c}}
\cC{cr,ds}{IF,IF} \cSCS{cr,s}{IF} \cS{rs}{(0,0)} = 0\,,
\label{eq:CScancel2}
\eeq
hence all terms involving the soft-collinear limit cancel in 
\eqn{eq:A2} as promised and we find the simplified form of 
$\cA_{2}^{(0)}$ from~\refr{DelDuca:2024ovc}, appropriate for 
color-singlet production,
\beq
\cA_{2}^{(0)} = 
	\frac12 \sum_{r \in F}
    \sum_{\substack{s \in F \\ s \ne r}}\bigg\{
	\cS{rs}{(0,0)} 
+
	\sum_{c \in I}\bigg[
	\cC{crs}{IFF (0,0)} 
	- \cC{crs}{IFF}\cS{rs}{(0,0)}
+	
	\sum_{\substack{d \in I \\ d \ne c}}\bigg(
	\cC{cr,ds}{IF,IF (0,0)}	
	- \cC{cr,ds}{IF,IF} \cS{rs}{(0,0)}
	\bigg)
	\bigg]\bigg\}\,.
\label{eq:A2cs}
\eeq
Obviously, this result can also be reached starting from 
\eqn{eq:A2short} by using \eqn{eq:CCSSeqCS}. However, the nature 
of the cancellations is then somewhat obscured.

As in the case of \eqn{eq:A1}, not all momentum mappings used to 
define the five types of subtraction terms in \eqn{eq:A2cs} are 
distinct, since once again, subtraction terms corresponding to 
overlapping limits can be defined using the momentum mapping 
appropriate to the innermost limit, i.e., the rightmost one in our 
notation. The grouping of subtraction terms according to momentum 
mappings is presented in \tab{tab:A2maps} for the color-singlet 
case. The precise definition of each momentum mapping and 
subtraction term is given in \sect{sec:A2}.
\begin{table}
\setlength{\tabcolsep}{10pt}
\renewcommand{\arraystretch}{2}
\begin{center}
\begin{tabular}{|c|ccc|}
\hline
\parbox{\widthof{Momentum}}
{\centering Momentum mapping}\vphantom{$\Bigg|$} & 
\multicolumn{3}{c|}{Subtraction terms}
\\
\hline\hline
$\cmp{cd,rs}{II,FF}$
	& $\cC{crs}{IFF(0,0)}$
	& $\cC{cr,ds}{IF,IF(0,0)}$
	&
\\
\hline	
$\smp{rs}$
	& $\cS{rs}{(0,0)}$
  	& $\cC{crs}{IFF}\cS{rs}{(0,0)}$
	& $\cC{cr,ds}{IF,IF}\cS{rs}{(0,0)}$
\\
\hline
\end{tabular}
\end{center}
\caption{\label{tab:A2maps}
Double unresolved subtraction terms grouped according to momentum 
mappings. Note that collinear-soft overlapping terms are defined 
using the double soft momentum mapping.
}
\end{table}

%%%
%%% Iterated single unresolved emission
%%%

\subsection{Iterated single unresolved emission}
\label{sec:itersingleunresolv}

The approximate cross section regularizing iterated single 
unresolved emission, $\dsiga{RR}{12}_{ab}$, reads
\beq
\dsiga{RR}{12}_{ab} = 
    \sum_{\{m+2\}} \frac{1}{S_{\{m+2\}}} 
    \dsiga{RR}{12}_{ab,m+X+2}\,,
\eeq
where for any particular subprocess the approximate cross section 
can be written symbolically as 
\beq
\dsiga{RR}{12}_{ab,m+X+2} = 
	{\cal N} \frac{1}{\omega(a)\omega(b)\Phi(p_a\cdot p_b)} 
    \PS{m+X+2}(\mom{}_{m+X+2};p_a+p_b) \cA_{12}^{(0)}\,,
\label{eq:dsigRRA12}
\eeq
with
\beq
\cA_{12}^{(0)} = 
	\sum_{s \in F}\bigg[
	\cA_{2}^{(0)} \cS{s}{} 
	+ \sum_{\substack{r \in F \\ r \ne s}} 
    \bigg(\frac12\,\cA_{2}^{(0)}\cC{rs}{FF}
	- \cA_{2}^{(0)} \cC{rs}{FF}\cS{s}{}\bigg)
	+ \sum_{c \in I} \bigg( \cA_{2}^{(0)} \cC{cs}{IF}
	- \cA_{2}^{(0)}\cC{cs}{IF}\cS{s}{} \bigg)\bigg]\,.
\label{eq:A12}
\eeq
As anticipated above in \sect{sec:overlaps}, $\cA_{12}^{(0)}$ is 
formally obtained by applying the same construction as in 
\eqn{eq:A1} to $\cA_{2}^{(0)}$. It is then clearly more 
advantageous to consider the form of $\cA_{2}^{(0)}$ in 
\eqn{eq:A2short} in order to decrease the number of terms we must 
consider and we find
\bal
\cA_{2}^{(0)} \cS{s}{} &= 
	\sum_{\substack{r \in F \\ r \ne s}}\bigg[
	\cS{rs}{(0)} \cS{s}{}
	+ \sum_{c \in I}\bigg(	
	\cC{crs}{IFF (0)} \cS{s}{}	
	+ \cSCS{cr,s}{IF (0)} \cS{s}{}
	- \cC{crs}{IFF}\cSCS{cr,s}{IF (0)} \cS{s}{}	
\nn\\ & \qquad
	- \cC{crs}{IFF}\cS{rs}{(0)}	  \cS{s}{}
	- \cSCS{cr,s}{IF}\cS{rs}{(0)} \cS{s}{}	
	+ \cC{crs}{IFF}\cSCS{cr,s}{IF}\cS{rs}{(0)} \cS{s}{}
	\bigg)
	\bigg]	
	\,, \label{eq:SsA2}\\
\cA_{2}^{(0)} \cC{rs}{FF} &= 
	\cS{rs}{(0)} \cC{rs}{FF}
	+ \sum_{c \in I}\bigg(
	\cC{crs}{IFF (0)} \cC{rs}{FF}
	-\cC{crs}{IFF}\cS{rs}{(0)} \cC{rs}{FF}
	\bigg)
	\,, \label{eq:CrsA2}\\
\cA_{2}^{(0)} \cC{rs}{FF}\cS{s}{} &= 
	\cS{rs}{(0)} \cC{rs}{FF}\cS{s}{}
	+ \sum_{c \in I}\bigg(
	\cC{crs}{IFF (0)} \cC{rs}{FF}\cS{s}{}
	-\cC{crs}{IFF}\cS{rs}{(0)} \cC{rs}{FF}\cS{s}{}
	\bigg)
	\,, \label{eq:CrsFFSsA2}\\
\cA_{2}^{(0)} \cC{cs}{IF}  &= 
	\sum_{\substack{r \in F \\ r \ne s}} 
    \bigg[ \cC{csr}{IFF (0)} \cC{cs}{IF}
	+ \cSCS{cs,r}{IF (0)} \cC{cs}{IF}
	- \cC{csr}{IFF}\cSCS{cs,r}{IF (0)} \cC{cs}{IF}
\nn\\ & \qquad
	+ \sum_{\substack{d \in I \\ d \ne c}}\bigg(
	\cC{cs,dr}{IF,IF (0)} \cC{cs}{IF}
	-\cC{cs,dr}{IF,IF (0)}\cSCS{cs,r}{IF (0)} \cC{cs}{IF}
	\bigg) \bigg]
	\,, \label{eq:CasA2}\\
\cA_{2}^{(0)} \cC{cs}{IF}\cS{s}{} &= 
	\sum_{\substack{r \in F \\ r \ne s}}\bigg[
	\cS{rs}{(0)} \cC{cs}{IF}\cS{s}{}
	+ \cC{csr}{IFF (0)}  \cC{cs}{IF}\cS{s}{}
	- \cC{csr}{IFF}\cS{rs}{(0)} \cC{cs}{IF}\cS{s}{}
\nn\\ & \qquad
	+ \sum_{\substack{d \in I \\ d \ne c}}\bigg(
	\cSCS{dr,s}{IF (0)} \cC{cs}{IF}\cS{s}{}\
	- \cSCS{dr,s}{IF}\cS{rs}{(0)} \cC{cs}{IF}\cS{s}{}
	\bigg)
	\bigg]	
	\,.
\label{eq:CasIFSsA2}
\eal
We note that in writing \eqnss{eq:SsA2}{eq:CasIFSsA2}, we have 
made use of various cancellations that are manifest already at the 
level of iterated IR factorization formulae, since the direct 
calculation of the terms on the right-hand side of \eqn{eq:A12} 
yields significantly larger expressions than 
these~\cite{Somogyi:2005xz}. Concentrating on color-singlet 
production, it is obvious that the cancellations which lead from 
\eqns{eq:A2}{eq:A2short} to \eqn{eq:A2cs} imply several relations 
among the terms in \eqnss{eq:SsA2}{eq:CasIFSsA2}. In particular, 
we find 
\bal
&
\cS{rs}{(0)} \cC{rs}{FF}\cS{s}{}
- \sum_{c \in I} 
\cC{crs}{IFF}\cS{rs}{(0)} \cC{rs}{FF}\cS{s}{} = 0\,,
\label{eq:rel1}
\\ &
\sum_{c \in I} \bigg(
\cSCS{cs,r}{IF,(0)}  \cC{cs}{IF}
- \cC{csr}{IFF}\cSCS{cs,r}{IF,(0)} \cC{cs}{IF}
- \sum_{\substack{d \in I \\ d \ne c}} 
\cC{cs,dr}{IF,IF}\cSCS{cs,r}{IF,(0)} \cC{cs}{IF} \bigg)= 0\,,
\label{eq:rel11}
\\ &
\sum_{c \in I} \bigg( 
\cSCS{cr,s}{IF (0)}  \cS{s}{}
- \cC{crs}{IFF}\cSCS{cr,s}{IF (0)} \cS{s}{}\
- \sum_{\substack{d \in I \\ d \ne c}} 
\cSCS{dr,s}{IF (0)} \cC{cs}{IF}\cS{s}{} \bigg) = 0\,,
\label{eq:rel2}
\\ &
\sum_{c \in I} \bigg(
\cSCS{cr,s}{IF}\cS{rs}{(0)} \cS{s}{}
- \cC{crs}{IFF}\cSCS{cr,s}{IF}\cS{rs}{(0)} \cS{s}{}
- \sum_{\substack{d \in I \\ d \ne c}}
\cSCS{dr,s}{IF}\cS{rs}{(0)} \cC{cs}{IF}\cS{s}{} \bigg) = 0\,,
\label{eq:rel3}
\eal
which allows us to rewrite the various terms in \eqn{eq:A12} as
\bal
\cA_{2}^{(0)} \cS{s}{} &=
	\sum_{\substack{r \in F \\ r \ne s}}\bigg[
	\cS{rs}{(0,0)} \cS{s}{}
	+ \sum_{c \in I}\bigg(	
	\cC{crs}{IFF (0,0)} \cS{s}{}	
	- \cC{crs}{IFF}\cS{rs}{(0,0)} \cS{s}{}
	\bigg)
	\bigg]	
	\,, \label{eq:SsA2new}
\\
\cA_{2}^{(0)} \cC{rs}{FF} &=
	\cS{rs}{(0,0)} \cC{rs}{FF}
	+ \sum_{c \in I}\bigg(
	\cC{crs}{IFF (0,0)} \cC{rs}{FF}
	-\cC{crs}{IFF}\cS{rs}{(0,0)} \cC{rs}{FF}
	\bigg)
	\,, \label{eq:CrsA2new}
\\
\cA_{2}^{(0)} \cC{rs}{FF}\cS{s}{} &=
    \sum_{c \in I}
	\cC{crs}{IFF (0,0)} \cC{rs}{FF}\cS{s}{}
	\,, \label{eq:CrsFFSsA2new}  
\\
\cA_{2}^{(0)} \cC{cs}{IF}  &=
	\sum_{\substack{r \in F \\ r \ne s}} 
    \bigg( \cC{csr}{IFF (0,0)} \cC{cs}{IF}
	+ \sum_{\substack{d \in I \\ d \ne c}}
	\cC{cs,dr}{IF,IF (0,0)} \cC{cs}{IF}
	 \bigg)
	\,, \label{eq:CasA2new}
\\
\cA_{2}^{(0)} \cC{cs}{IF}\cS{s}{} &=
	\sum_{\substack{r \in F \\ r \ne s}}\bigg(
	\cS{rs}{(0,0)} \cC{cs}{IF}\cS{s}{}
	+ \cC{csr}{IFF (0,0)}  \cC{cs}{IF}\cS{s}{}
	- \cC{csr}{IFF}\cS{rs}{(0,0)} \cC{cs}{IF}\cS{s}{}
	\bigg)
	\,.
\label{eq:CasIFSsA2new}
\eal
Clearly, \eqnss{eq:SsA2new}{eq:CasIFSsA2new} can also be derived 
starting from the simplified form of $\cA_{2}^{(0)}$ in 
\eqn{eq:A2cs}, appropriate to color-singlet production.

Once more, the number of independent momentum mappings used to 
define the subtraction terms in \eqn{eq:A12} is less than the 
total number of subtraction terms and the grouping of terms 
according to momentum mappings is presented in \tab{tab:A12maps} 
for the color-singlet case.\footnote{In \tab{tab:A12maps}, we use 
the standard notation for the convolution of maps, so that, e.g., 
$\smp{\ti{r}} \circ \smp{s}$ implies that the $\smp{s}$ mapping is 
performed first, followed by the $\smp{\ti{r}}$ mapping.} In fact, 
as anticipated above, iterated single unresolved subtraction terms 
can be defined using convolutions of single unresolved momentum 
mappings. The precise convolution for each term can be determined 
as follows. First, we perform a mapping of momenta appropriate to 
the single unresolved limit involved in the definition of the 
specific term. Thus, collinear limits imply that the first mapping 
is of single collinear type, while soft and collinear-soft limits 
indicate that the first mapping is a single soft one. We have 
chosen our notation such that the single limits stand to the right 
of the double limits in iterated single unresolved subtraction 
terms specifically so that this mapping can be read off as the 
rightmost limit, which then corresponds to the same prescription 
as for the overlaps of limits. The next mapping can then be 
inferred from the rightmost double unresolved limit appearing in 
the notation: it is of collinear or soft type depending on the 
nature of this double unresolved limit. By way of example, 
consider e.g., the subtraction term 
$\cC{csr}{IFF}\cS{rs}{(0,0)}\cC{cs}{IF}\cS{s}{}$. The rightmost 
limit is the single soft limit $\cS{s}{}$, so the first mapping is 
going to be the corresponding single soft one, which we denote by 
$\smp{s}$. Then, the first double unresolved limit from the right 
is the double soft limit $\cS{rs}{(0,0)}$. Thus, the second 
mapping is also of soft type. As the first mapping removes the 
soft momentum $p_s$ and remaps the other soft momentum $p_r\to 
\ti{p}_r$, this second single soft momentum mapping will 
correspond to removing the remapped momentum $\ti{p}_r$ and is 
denoted by $\smp{\ti{r}}$. Thus the momentum mapping for this 
subtraction term is obtained by the convolution of these two 
mappings, $\smp{\ti{r}} \circ \smp{s}$, as shown in 
\tab{tab:A12maps}. Hence, in the case of both overlapping as well 
as iterated single unresolved subtraction terms, the appropriate 
mappings can be inferred from the notation by reading from right 
to left. The explicit definition of each subtraction term in 
\eqnss{eq:SsA2new}{eq:CasIFSsA2new} is given in \sect{sec:A12}.
\begin{table}
\setlength{\tabcolsep}{10pt}
\renewcommand{\arraystretch}{2}
\begin{center}
\begin{tabular}{|c|cccc|}
\hline
\parbox{\widthof{$\cmp{II,F}{\ha{c}\ha{d},\ha{r}} 
\circ \cmp{II,F}{cd,s}\vphantom{\bigg|}$}}
{\centering Momentum mapping}\vphantom{$\Bigg|$} & 
\multicolumn{4}{c|}{Subtraction terms}
\\
\hline\hline
$\cmp{\ha{c}\ha{d},\wha{rs}}{II,F} 
\circ \cmp{rs}{FF}\vphantom{\bigg|}$
	& $\cC{crs}{IFF(0,0)}\cC{rs}{FF}$
	&
	&
	&
\\
\hline
$\cmp{\ha{c}\ha{d},\ha{r}}{II,F} 
\circ \cmp{cd,s}{II,F}\vphantom{\bigg|}$
	& $\cC{csr}{IFF(0,0)}\cC{cs}{IF}$
	& $\cC{cs,dr}{IF,IF(0,0)}\cC{cs}{IF}$
	&
	&
\\	
\hline
$\cmp{\ti{c}\ti{d},\ti{r}}{II,F} \circ \smp{s}\vphantom{\bigg|}$
	& $\cC{crs}{IFF(0,0)}\cS{s}{}$
	& $\cC{crs}{IFF(0,0)}\cC{rs}{FF}\cS{s}{}$
	& $\cC{crs}{IFF(0,0)}\cC{cs}{IF}\cS{s}{}$
	&
\\
\hline
$\smp{\wha{rs}} \circ \cmp{rs}{FF}\vphantom{\bigg|}$
	& $\cS{rs}{(0,0)}\cC{rs}{FF}$
	& $\cC{crs}{IFF}\cS{rs}{(0,0)}\cC{rs}{FF}$
	&
	&
\\
\hline
$\smp{\ti{r}} \circ \smp{s}\vphantom{\bigg|}$
	& $\cS{rs}{(0,0)}\cS{s}{}$
  	& $\cC{crs}{IFF}\cS{rs}{(0,0)}\cS{s}{}$
	& $\cS{rs}{(0,0)}\cC{cs}{IF}\cS{s}{}$
	& $\cC{crs}{IFF}\cS{rs}{(0,0)}\cC{cs}{IF}\cS{s}{}$
\\
\hline
\end{tabular}
\end{center}
\caption{\label{tab:A12maps}
Iterated single unresolved subtraction terms grouped according to 
momentum mappings. The symbol $\circ$ denotes the standard 
convolution of maps.
}
\end{table}

%%%
%%% Subtraction terms for single unresolved emission
%%%

\section{Subtraction terms for single unresolved emission}
\label{sec:A1}

In sec.~\ref{sec:singleunresolv}, we introduced the subtraction 
terms for single unresolved emission. In this section we 
provide their precise definitions. 

%
% Final-state single collinear-type subtraction term
%

\subsection{Final-state single collinear-type subtraction term}

%
% $\cC{ir}{FF (0,0)}$
%

\subsubsection{\texorpdfstring{$\cC{ir}{FF (0,0)}$}{CirFF00}}
\label{sec:CirFF00}

%%%%%
\paragraph{Subtraction term:} 

\begin{figure}
\setlength{\tabcolsep}{10pt}
\renewcommand{\arraystretch}{1.2}
\begin{center}
\parbox{18em}{\includegraphics[scale=1]{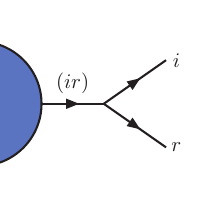}}
\begin{tabular}{|c|c|c|}
\hline
$i$ & $r$ & $(ir)$ 
\\ 
\hline\hline
$q/\qb$ & $\qb/q$ & $g$
\\
$q/\qb$ & $g$ & $q/\qb$
\\
$g$ & $q/\qb$ & $q/\qb$
\\
$g$ & $g$ & $g$
\\
\hline
\end{tabular}
\end{center}
\caption{\label{fig:Pirflav}
The splitting configuration for $(ir)\to i+r$ final-state 
splitting. The arrows in the figure denote momentum and flavor 
flow. The precise flavor mapping for the $\cC{ir}{FF(0,0)}$ 
counterterm is given in the table.
}
\end{figure}

For final-state partons $i$ and $r$ we define the final-final 
single collinear subtraction term
\beq
\cC{ir}{FF (0,0)}(p_a,p_b;\mom{}_{m+X+2}) = 
	8\pi\as\mu^{2\ep} 
	\frac{1}{s_{ir}}
	\hP_{i r}^{(0)}(z_{i,r},\kT{i,r};\ep) 
	\SME{ab,m+X+1}{(0)}{(\ha{p}_a,\ha{p}_b;\momh{}_{m+X+1})} \,,
\label{eq:CirFF00}
\eeq
where the reduced matrix element is obtained by removing the 
final-state partons $i$ and $r$ and replacing them with a single 
parton $(ir)$ whose flavor is determined by the flavor summation 
rules
\beq
q+g=q\,,\qquad 
\qb+g=\qb\,,\qquad 
q+\qb = g\,,\qquad
g+g=g\,.
\label{eq:flsum}
\eeq
\Fig{fig:Pirflav} shows the corresponding final-state splitting 
configuration and gives the precise flavor mapping for all cases. 
The set of mapped momenta entering the reduced matrix element is 
defined below in \eqn{eq:cirFFmap}. Furthermore, $\hP_{i r}^{(0)}$ 
are the final-final tree-level Altarelli-Parisi splitting 
functions for $(ir) \to i+r$ splitting, whose explicit expressions 
are recalled in \eqnss{eq:Pqg0FF}{eq:Pgg0FF}. The momentum 
fraction $z_{i,r}$ is given in \eqn{eq:zjk-def}. The transverse 
momentum $\kT{i,r}^\mu$ is defined as in \eqn{eq:ktjk-ansatz}. We 
choose to fix the coefficients appearing there as follows,
\beq
\zeta_{i,r} = z_{i,r} - \frac{s_{ir}}{\al_{ir} s_{(ir)\ab}}\,,
\qquad\qquad
\zeta_{r,i} = z_{r,i} - \frac{s_{ir}}{\al_{ir} s_{(ir)\ab}}\,,
\label{eq:zirri}
\eeq
and
\beq
Z_{ir} = \frac{s_{ir}}{\al_{ir} s_{\wha{ir}\ab}} 
    (z_{r,i} - z_{i,r})\,,
\label{eq:Zir}
\eeq
where $\al_{ir}$ is defined by \eqn{eq:Cir0FF_al} below, while $s_{\wha{ir}\ab} = 2\ha{p}_{ir}\cdot (p_a+p_b)$, with 
$\ha{p}_{ir}^\mu$ given in \eqn{eq:cirFFmap}. These choices ensure 
that $\kT{i,r}^\mu$ is orthogonal to both $\ha{p}_{ir}^\mu$ and to 
$Q^\mu=(p_a+p_b)^\mu$.

%%%%%
\paragraph{Momentum mapping:}

The final-state single collinear (FF) momentum mapping,
\beq
\cmp{ir}{FF}:\; (p_a, p_b; \mom{}_{m+X+2}) 
    \cmap{ir}{FF} (\ha{p}_a, \ha{p}_b; \momh{}_{m+X+1})\,,
\eeq
is defined as ($i,r \in F$)
\beq
\bsp
\ha{p}_a^\mu &= (1-\al_{ir})p_a^\mu\,,
\\
\ha{p}_b^\mu &= (1-\al_{ir})p_b^\mu\,,
\\
\ha{p}_{ir}^\mu &= p_i^\mu + p_r^\mu - \al_{ir} (p_a+p_b)^\mu\,,
\\
\ha{p}_n^\mu &= p_n^\mu\,, \qquad n\in F\,,\quad n \ne i,r\,,
\\
\ha{p}_X^\mu &= p_X^\mu\,.
\esp
\label{eq:cirFFmap}
\eeq
The value of $\al_{ir}$ is fixed by requiring that the parent 
momentum, $\ha{p}_{ir}^\mu$, be massless, $\ha{p}_{ir}^2 = 0$, 
which gives
\beq
\al_{ir} = \frac{1}{2}\left[\frac{s_{(ir)\ab}}{s_{ab}} 
    - \sqrt{\frac{s_{(ir)\ab}^2}{s_{ab}^2} 
    - \frac{4s_{ir}}{s_{ab}}}\,\right]\,.
\label{eq:Cir0FF_al}
\eeq
Clearly, as $p_i^\mu$ and $p_r^\mu$ become collinear, $\al_{ir} 
\to 0$ and so $\ha{p}_{ir}^\mu \to p_i^\mu+p_r^\mu$ while the rest 
of the momenta remain unchanged. This corresponds precisely to the 
$p_i^\mu \parallel p_r^\mu$ final-final single collinear kinematic 
configuration. Throughout, we will denote momenta obtained via a 
collinear-type momentum mapping with hats, as in \eqn{eq:cirFFmap}.

%%%%%
\paragraph{Phase space convolution:}

By inserting the identity,
\beq
1 = \frac{\rd^d K}{(2\pi)^{d}} (2\pi)^{d} 
    \delta^{(d)} ( K - p_i - p_r )\, 
    \rd K^2\, \delta_+(K^2 - s_{ir})\,,
\eeq
the $(m+X+2)$-particle phase space of \eqn{eq:psmx2} (with $k=2$) 
can be factored on partons $i$ and $r$ through the convolution 
formula,
\beq
\PS{m+X+2}(\mom{}_{m+X+2};p_a+p_b)
    = \frac{ \rd K^2}{2\pi}\, 
    \PS{m+X+1}(p_a,\ldots, K, \ldots, p_{m+2}, p_X; p_a+p_b)\, 
    \PS{2}(p_i,p_r; K)\,.
\label{eq:genconv}
\eeq
Moreover, using then the identity,
\beq
1 = \rd\alpha\, \delta(\alpha-\alpha_{ir})\, \rd^d\ha{p}_{ir}\, 
    \delta^{(d)}\left( \ha{p}_{ir} - K + \alpha (p_a+p_b)\right)\,,
\label{eq:idir}
\eeq
and exchanging the momenta $\{p_n\}$ and $p_X$ with the momenta 
$\{\ha{p}_n\}$ and $\ha{p}_X$ through the mapping in 
\eqn{eq:cirFFmap}, we can write the convolution of the 
$(m+X+2)$-particle phase space in \eqn{eq:genconv} as,
\beq
\bsp
&
\PS{m+X+2}(\mom{}_{m+X+2};p_a+p_b) =
\\&\qquad = 
	\int_{\almin}^{\almax} \rd\al\,
	\PS{m+X+1}\big(\momh{}_{m+X+1}; (1-\al)(p_a+p_b)\big)
	\frac{s_{\wha{ir}\ab}}{2\pi}
	\PS{2}(p_i,p_r; K)\,,
\esp
\label{eq:CirFF-map-PSconv} 
\eeq
where $K$ is fixed by \eqn{eq:idir}, i.e., 
$K=\ha{p}_{ir}+\al(p_a+p_b)$. The limits of the $\al$ integration 
are
\beq
\almin = 0
\qquad\mbox{and}\qquad
\almax = 1 - \frac{M}{\sqrt{s_{ab}}}\,,
\label{eq:alminmax}
\eeq
where $M$ is the sum of the masses of all final-state particles 
(both partons and non-QCD particles). Finally, we note that the flux 
factor $\Phi(p_a\cdot p_b)$ written in terms of mapped momenta reads
\beq
\Phi(p_a\cdot p_b) = 
    \frac{\Phi(\ha{p}_a\cdot \ha{p}_b)}{(1-\al)^2}\,,
\label{eq:cirFFflux}
\eeq
where we have exploited \eqn{eq:idir} to set $\alpha_{ir} = 
\alpha$.

%%%%%
\paragraph{Integrated subtraction term:} 

Using the definitions of the subtraction term and the phase space 
convolution, \eqns{eq:CirFF00}{eq:CirFF-map-PSconv}, we write the 
integrated counterterm as
\beq
\bsp
&
\int_1{\cal N} \frac{1}{\omega(a)\omega(b)\Phi(p_a\cdot p_b)} 
	\PS{m+X+2}(\mom{}_{m+X+2};p_a+p_b)\,
	\cC{ir}{FF (0,0)}(p_a,p_b;\mom{}_{m+X+2}) 
	\\ &\qquad
	{\cal N} \int_{\almin}^{\almax} \rd\al\,
	\frac{1}{\omega(a)\omega(b)\Phi(\ha{p}_a\cdot \ha{p}_b)}
	\PS{m+X+1}\big(\momh{}_{m+X+1};(1-\al)(p_a+p_b)\big) 
\\ &\qquad \times
	\frac{\as}{2\pi} S_\ep \left(\frac{\mu^2}{s_{ab}}\right)^\ep
	\left[\IcC{ir}{FF (0,0)}(\al;\ep)\right]
	\SME{ab,m+X+1}{(0)}{(\ha{p}_a,\ha{p}_b;\momh{}_{m+X+1})}\,.
\esp
\label{eq:Int_CirFF00}
\eeq
Here and in the following, the notation $\int_1$ indicates that we 
are integrating only over the phase space measure of the single 
unresolved emission. Furthermore, using the delta function in 
\eqn{eq:idir}, the initial-state momenta in the reduced matrix 
element above are written as $\ha{p}_a=(1-\al)p_a$ and $\ha{p}_b=
(1-\al)p_b$ and we introduced the integrated counterterm
\beq
\left[\IcC{ir}{FF (0,0)}(\al;\ep)\right] = 
	\frac{(4\pi)^2 }{S_\ep} s_{ab}^\ep
	\int \PS{2}(p_i,p_r; K) (1-\al)^2
	\frac{s_{\wha{ir}\ab}}{2\pi}
	\frac{1}{s_{ir}}
	\hP_{i r}^{(0)}(z_{i,r},\kT{i,r};\ep)\,.
\label{eq:ICirFF00}
\eeq
Recall that $K = \ha{p}_{ir}+\al(p_a+p_b)$ and we set
\beq
S_\ep = \frac{(4\pi)^\ep}{\Gamma(1-\ep)}\,.
\label{eq:seps}
\eeq
Notice that on the right hand side of \eqn{eq:Int_CirFF00}, we 
have rewritten the flux factor in terms of mapped momenta using 
\eqn{eq:cirFFflux} and hence a factor of $\Phi(\ha{p}_a\cdot 
\ha{p}_b)/\Phi(p_a\cdot p_b) = (1-\al)^2$ appears explicitly in 
the definition of the integrated counterterm in \eqn{eq:ICirFF00}. 
With this rewriting, we readily recognize the expression of the 
real emission cross section for the appropriate partonic 
subprocess, $\dsig{R}_{ab,m+X+1}(\ha{p}_a,\ha{p}_b)$ defined in 
\eqn{eq:Rdefsub}, on the right hand side of \eqn{eq:Int_CirFF00}. 
Thus we can write
\beq
\bsp
&
\int_1{\cal N} \frac{1}{\omega(a)\omega(b)\Phi(p_a\cdot p_b)} 
	\PS{m+X+2}(\mom{}_{m+X+2};p_a+p_b)\,
	\cC{ir}{FF (0,0)}(p_a,p_b;\mom{}_{m+X+2}) 
\\ &\qquad=
	\frac{\as}{2\pi} S_\ep \left(\frac{\mu^2}{s_{ab}}\right)^\ep
	\Big(
	\left[\IcC{ir}{FF (0,0)}\right] \otimes \dsig{R}_{ab,m+X+1}
	\Big) 
\esp
\label{eq:Int_CirFF00-fin}
\eeq
where the symbol $\otimes$ denotes the phase space convolution 
\beq
\left[\IcC{ir}{FF (0,0)}\right] \otimes \dsig{R}_{ab,m+X+1}
= \int_{\almin}^{\almax} \rd\al\,
	\left[\IcC{ir}{FF (0,0)}(\al;\ep)\right]
	\dsig{R}_{ab,m+X+1}(\ha{p}_a, \ha{p}_b)\,.
\label{eq:CirRR00xdsigR}
\eeq
As noted previously, the unintegrated subtraction term is 
generally a matrix in color and/or spin space and this property is 
inherited by the integrated counterterm.\footnote{In practice it 
turns out that all integrated counterterms become proportional to 
the unit matrix in spin space. However, the same is not true in 
color space.} Thus, the product of the integrated counterterm and 
the real emission cross section in \eqn{eq:CirRR00xdsigR} is to be 
understood accordingly. In the rest of the paper, we will present 
all integrated counterterms immediately in the more compact form 
corresponding to \eqns{eq:Int_CirFF00-fin}{eq:CirRR00xdsigR}.
In order to evaluate $\left[\IcC{ir}{FF (0,0)}(\al;\ep)\right]$, 
we start by using $\kT{i,r}\cdot\ha{p}_{ir} = 0$ and perform the 
azimuthal integration by simply passing in \eqn{eq:ICirFF00} to 
the azimuthally averaged splitting functions, $\hP_{i r}^{(0)}
(z_{i,r},\kT{i,r};\ep) \to P_{i r}^{(0)}(z_{i,r};\ep)$, which are 
given in \eqnss{eq:Pqg-ave}{eq:Pgg-ave}.\footnote{Thus, as noted 
above, the non-trivial spin correlations generally present in 
\eqn{eq:CirFF00} disappear upon integration.} Hence we have
\beq
\left[\IcC{ir}{FF (0,0)}(\al;\ep)\right] =
	\frac{8\pi}{S_\ep} s_{ab}^{\ep}
	\int \PS{2} (p_i,p_r; K) (1-\al)^2
	\frac{s_{\wha{ir}\ab} }{s_{ir}}
	P_{i r}^{(0)}(z_{i,r};\ep)\,.
\label{eq:ICirFF00-2}
\eeq
In this paper, we will limit ourselves to setting up the 
definitions of the integrated subtraction terms as in 
\eqn{eq:ICirFF00-2} and defer their explicit calculation to 
separate dedicated publications.

%
% Initial-state single collinear-type subtraction term
%

\subsection{Initial-state single collinear-type subtraction term}

%
% \cC{ar}{IF (0,0)}
%
\subsubsection{\texorpdfstring{$\cC{ar}{IF (0,0)}$}{CarIF00}}
\label{sec:CarIF00}

%%%%%
\paragraph{Subtraction term:} 

\begin{figure}
\setlength{\tabcolsep}{10pt}
\renewcommand{\arraystretch}{1.2}
\begin{center}
\parbox{20em}{\includegraphics[scale=1]{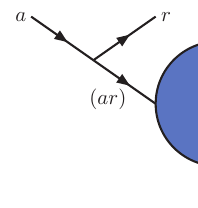}}
\begin{tabular}{|c|c|c|}
\hline
$a$ & $r$ & $(ar)$ 
\\ 
\hline\hline
$q/\qb$ & $q/\qb$ & $g$
\\
$q/\qb$ & $g$ & $q/\qb$
\\
$g$ & $q/\qb$ & $\qb/q$
\\
$g$ & $g$ & $g$
\\
\hline
\end{tabular}
\end{center}
\caption{\label{fig:Parflav}
The splitting configuration for $a\to (ar)+r$ initial-state 
splitting. The arrows in the figure denote momentum and flavor 
flow. The precise flavor mapping for the $\cC{ar}{IF(0,0)}$ 
counterterm is given in the table.
}
\end{figure}

For an initial-state parton $a$ and a final-state parton $r$ we 
define the initial-final single collinear subtraction term
\beq
\cC{ar}{IF (0,0)}(p_a,p_b;\mom{}_{m+X+2}) = 
	8\pi\as\mu^{2\ep}
	\frac{1}{x_{a,r} s_{ar}}
	\hP_{(ar) r}^{(0)}(x_{a,r},\kT{r,a};\ep) 
	\SME{(ar)b,m+X+1}{(0)}{(\ha{p}_a,\ha{p}_b;\momh{}_{m+X+1})}\,,
\label{eq:CarIF00}
\eeq
where the reduced matrix element is obtained by removing the 
final-state parton $r$ and replacing the initial-state parton $a$ 
by the parton $(ar)$ whose flavor is determined by the flavor 
summation rules of \eqn{eq:flsum} by requiring that $a=(ar)+r$. 
Clearly the flavor of the initial-state parton that is involved in 
the splitting is generally different in the original double real 
emission matrix element and in the reduced matrix element. The 
initial-state splitting configuration and precise flavor mapping 
are shown in \fig{fig:Parflav}. The set of momenta entering the 
reduced matrix element is defined below in \eqn{eq:cabrIIFmap}. 
The tree-level initial-final Altarelli-Parisi splitting functions 
for $a \to (ar) + r$ splitting, $\hP_{(ar) r}^{(0)}$, can be 
obtained from those for final-final splitting by the crossing 
relation discussed in \appx{appx:AP-functions}, see 
\eqn{eq:Pifcross}, and are given explicitly in 
\eqnss{eq:Pqg0IF}{eq:Pgg0IF}. As usual, we have chosen to label 
the splitting function with the flavor of the initial-state parton 
in the {\em reduced} matrix element, $(ar)$, and the flavor of the 
final-state parton $r$. The momentum fraction $x_{a,r}$ and 
transverse momentum $\kT{r,a}$ are defined by 
\eqns{eq:xja-def}{eq:kj-def}.

%%%%%
\paragraph{Momentum mapping:}

The initial-state single collinear (IF) momentum mapping,
\beq
\cmp{ab,r}{II,F}:\; (p_a, p_b; \mom{}_{m+X+2}) 
    \cmap{ab,r}{II,F} (\ha{p}_a, \ha{p}_b; \momh{}_{m+X+1})\,,
\eeq
is defined as ($a,b \in I$ and $r \in F$)
\beq
\bsp
\ha{p}_a^\mu &= \xi_{a,r} p_a^\mu\,,
\\
\ha{p}_b^\mu &= \xi_{b,r} p_b^\mu\,,
\\
\ha{p}_n^\mu &= {\Lambda(P,\ha{P})^\mu}_{\!\nu}\, p_n^\nu\,, 
\qquad n \in F, \quad n\ne r\,,
\\
\ha{p}_X^\mu &= {\Lambda(P,\ha{P})^\mu}_{\!\nu}\, p_X^\nu\,.
\esp
\label{eq:cabrIIFmap}
\eeq
Here ${\Lambda(P,\ha{P})^\mu}_{\!\nu}$ is a proper Lorentz 
transformation that takes the massive momentum $P^\mu$ into a 
momentum $\ha{P}^\mu$ of the same mass, with
\beq
P^\mu = (p_a+p_b)^\mu - p_r^\mu\,,
\qquad\qquad
\ha{P}^\mu = (\ha{p}_a + \ha{p}_b)^\mu 
    = (\xi_{a,r}p_a + \xi_{b,r} p_b)^\mu\,.
\label{eq:phap}
\eeq
One specific representation of ${\Lambda(P,\ha{P})^\mu}_{\!\nu}$ is
\beq
{\Lambda(P,\ha{P})^\mu}_{\!\nu} = 
	{g^\mu}_{\!\nu} - 2\frac{(P+\ha{P})^\mu (P+\ha{P})_\nu}
    {(P+\ha{P})^2} + 2\frac{\ha{P}^\mu P_\nu}{P^2}\,.
\label{eq:Lambda_munu}
\eeq
Requiring that $P^2 = \ha{P}^2$ only fixes the value of the 
product of $\xi_{a,r}$ and $\xi_{b,r}$,
\beq
\xi_{a,r} \xi_{b,r} = 1 - \frac{s_{r\ab}}{s_{ab}}\,,
\label{eq:prod12r}
\eeq
so there is some freedom in their explicit definition. We find the 
choice of~\refr{Daleo:2006xa} convenient and we set
\beq
\xi_{a,r} = \sqrt{\frac{s_{ab} - s_{br}}{s_{ab} - s_{ar}}
	\frac{s_{ab} - s_{r\ab}}{s_{ab}}}\,,
\qquad
\xi_{b,r} = \sqrt{\frac{s_{ab} - s_{ar}}{s_{ab} - s_{br}}
	\frac{s_{ab} - s_{r\ab}}{s_{ab}}}\,,
\label{eq:daleo}
\eeq
which fulfills \eqn{eq:prod12r}. Notice that $\xi_{a,r}$ and 
$\xi_{b,r}$ are related by the exchange $a \leftrightarrow b$. It 
is not difficult to see that as, e.g., $p_a^\mu$ and $p_r^\mu$ 
become collinear such that $p_r^\mu \to x_r p_a^\mu$, we have 
$\xi_{a,r} \to 1-x_r$ while $\xi_{b,r} \to 1$, hence $\ha{P}^\mu 
\to P^\mu$ and the Lorentz-transformation in \eqn{eq:Lambda_munu} 
becomes the identity. Thus we obtain precisely the $p_a^\mu 
\parallel p_r^\mu$ initial-final single collinear kinematic 
configuration. Due to the $a \leftrightarrow b$ symmetry of the 
mapping, the mapped momenta also have the correct behavior to 
describe the $p_b^\mu \parallel p_r^\mu$ initial-final collinear 
limit.

%%%%%
\paragraph{Phase space convolution:}

Using the identity,
\beq
1 = \rd\xi_a\, \delta(\xi_a - \xi_{a,r})\, 
    \rd\xi_b\, \delta(\xi_b - \xi_{b,r}) \,,
\label{eq:idIF}
\eeq
and shifting the momenta as in \eqn{eq:cabrIIFmap}, the 
$(m+X+2)$-particle phase space of \eqn{eq:psmx2} (with $k=2$) can 
be written as the convolution,
\beq
\PS{m+X+2}(\mom{}_{m+X+2};p_a+p_b) = 
	 \int \rd\xi_a\, \rd\xi_b\,
	\PS{m+X+1}\big(\momh{}_{m+X+1};\xi_a p_a + \xi_b p_b\big) \,
	\PS{II,F}(p_r,\xi_a,\xi_b)\,,
\label{eq:CarIF-map-PSconv}
\eeq
with 
\beq
\PS{II,F}(p_r,\xi_a,\xi_b) = \frac{\rd^d p_r}{(2\pi)^{d-1}}\, 
    \delta_+(p_r^2)\, \delta(\xi_a - \xi_{a,r})\, 
    \delta(\xi_b - \xi_{b,r})\,,
\label{eq:ps2ar}
\eeq
where in \eqn{eq:CarIF-map-PSconv} the limits of integration for 
both $\xi_a$ and $\xi_b$ are 0 and 1, with the constraints
\beq
(\xi_a\xi_b)_{\rm min} = \frac{M^2}{s_{ab}}\,,
\qquad
(\xi_a\xi_b)_{\rm max} = 1\,,
\label{eq:xi1xi2lims}
\eeq
where $M$ was defined below \eqn{eq:alminmax}. We note that the phase 
space measure $\PS{II,F}(p_r,\xi_a,\xi_b)$ can also be expressed in 
the following form,
\beq
\PS{II,F}(p_r,\xi_a,\xi_b) = 
	\PS{2}(p_r,K;p_a+p_b)\,
	{\cal J}_{II,F}\, 
	\delta[\xi_a(s_{ab}-s_{ar}) - \xi_b(s_{ab}-s_{br})]\,,
\label{eq:PSif-PS2}
\eeq
with the Jacobian
\beq
{\cal J}_{II,F} = 
\frac{s_{ab}[\xi_a(s_{ab}-s_{ar}) + \xi_b(s_{ab}-s_{br})]}{2\pi}\,,
\eeq
where the momentum $K$ appearing in the two-particle phase space 
above is massive with $K^2 = \xi_a \xi_b s_{ab}$. Finally, we note 
that the flux factor $\Phi(p_a\cdot p_b)$ can be written in terms of 
mapped momenta as
\beq
\Phi(p_a\cdot p_b) = 
    \frac{\Phi(\ha{p}_a\cdot \ha{p}_b)}{\xi_a \xi_b}\,,
\label{eq:carIFflux}
\eeq
where we have used \eqn{eq:idIF} to set $\xi_{a,r} = \xi_a$ and 
$\xi_{b,r} = \xi_b$.

%%%%%
\paragraph{Integrated subtraction term:} 

Using the definitions of the subtraction term and the phase space 
convolution, \eqns{eq:CarIF00}{eq:CarIF-map-PSconv}, we write the 
integrated counterterm as
\beq
\bsp
&
\int_1{\cal N} \frac{1}{\omega(a)\omega(b)\Phi(p_a\cdot p_b)} 
	\PS{m+X+2}(\mom{}_{m+X+2};p_a+p_b)
	\cC{ar}{IF (0,0)}(p_a,p_b;\mom{}_{m+X+2}) 
	\\ &\qquad=
	\frac{\as}{2\pi} S_\ep \left(\frac{\mu^2}{s_{ab}}\right)^\ep
	\Big( 
	\left[\IcC{ar}{IF (0,0)}\right] \otimes \dsig{R}_{(ar)b,m+X+1}
	\Big)\,,
\esp
\label{eq:Int_CarIF00-fin}
\eeq
where 
\beq
\left[\IcC{ar}{IF (0,0)}\right] \otimes \dsig{R}_{(ar)b,m+X+1} = 
	\int \rd \xi_a\,\rd \xi_b\, 
	\left[\IcC{ar}{IF (0,0)}(\xi_a,\xi_b;\ep)\right]
	\dsig{R}_{(ar)b,m+X+1}(\ha{p}_a,\ha{p}_b)\,.
\label{eq:CarIF00xdsigR}
\eeq
In \eqn{eq:CarIF00xdsigR} the initial-state momenta entering the 
real emission cross section are written as $\ha{p}_a = \xi_a p_a$ 
and $\ha{p}_b = \xi_b p_b$ using the delta functions in 
\eqn{eq:ps2ar} and we introduced the integrated counterterm
\beq
\left[\IcC{ar}{IF (0,0)}(\xi_a,\xi_b;\ep)\right] = 
	\frac{(4\pi)^2 }{S_\ep} s_{ab}^\ep
	\int \PS{II,F}(p_r,\xi_a,\xi_b)\, \xi_a \xi_b
	 \frac{1}{x_{a,r} s_{ar}}
	\frac{\omega(ar)}{\omega(a)}
	\hP_{(ar) r}^{(0)}(x_{a,r},\kT{r,a};\ep)\,.
\label{eq:ICarIF00}
\eeq
Note in particular the factor of $\omega(ar)/\omega(r)$ in 
\eqn{eq:ICarIF00}. This is present because, similarly to the flux 
factor, the initial-state averaging factor $\omega(a)$ on the 
left-hand side of \eqn{eq:Int_CarIF00-fin} must be rewritten in 
terms of $\omega(ar)$, since it is the latter that appears in 
$\dsig{R}_{(ar)b,m+X+1}(\ha{p}_a,\ha{p}_b)$, see \eqn{eq:Rdefsub}. 
Hence, the integrated counterterm includes the appropriate ratio 
of $\omega$s. Such factors appear in integrated counterterms 
whenever the flavor of the initial-state partons is different in 
the original double real emission matrix element and in the 
reduced matrix element in the subtraction term. In order to evaluate 
$\left[\IcC{ar}{IF (0,0)}(\xi_a,\xi_b;\ep)\right]$, we use  
$\kT{r,a}\cdot \ha{p}_a=0$ to perform the azimuthal integration by 
simply passing to the azimuthally averaged splitting functions, 
$\hP_{(ar) r}^{(0)}(x_{a,r},\kT{r,a};\ep) \to P_{(ar) r}^{(0)}
(x_{a,r};\ep)$, given in \eqnss{eq:Pqg-ave-IF}{eq:Pgg-ave-IF}. 
However, after azimuthal integration, the rest of the phase space 
measure $\PS{II,F}(p_r,\xi_a,\xi_b)$ is completely fixed by the delta 
functions in \eqn{eq:ps2ar}. In fact, inverting the relations in 
\eqn{eq:daleo}, we find
\beq
s_{ar} = s_{ab} \frac{\xi_{a,r} 
    (1-\xi_{b,r}^2)}{\xi_{a,r} + \xi_{b,r}}\,, \qquad 
s_{br} = s_{ab} \frac{\xi_{b,r} 
    (1-\xi_{a,r}^2)}{\xi_{a,r} + \xi_{b,r}}\,,
\label{eq:invertdaleo}
\eeq
which implies
\beq
x_{a,r} = \xi_{a,r}\xi_{b,r}\,.
\label{eq:xareval}
\eeq
Moreover, working out the Jacobian of the transformation from the 
remaining components of $p_r$ to $\xi_{a,r}$ and $\xi_{b,r}$, the 
phase space measure $\PS{II,F} (p_r,\xi_a,\xi_b)$ can be written 
as~\cite{Daleo:2006xa}
\beq
\bsp
&\PS{II,F}(p_r,\xi_a,\xi_b) 
\\&\qquad= 
	\frac{S_\ep}{8\pi^2}
	\frac{\rd \Omega_{d-2}}{\Omega_{d-2}}
	\rd \xi_{a,r}\, \rd \xi_{b,r}\,
	s_{ab}^{1-\ep} 
	\left[\frac{\xi_a \xi_b (1-\xi_a^2) (1-\xi_b^2)}
    {(\xi_a + \xi_b)^2} \right]^{-\ep}
	\frac{\xi_a \xi_b (1+\xi_a \xi_b)}{(\xi_a + \xi_b)^2} 
	\delta(\xi_{a,r} - \xi_a) \delta(\xi_{b,r} - \xi_b)
	\,,
\esp
\label{eq:ps2arexp}
\eeq
where $\rd \Omega_{d-2}$ represents the measure over angles on 
which the integrand does not depend and $S_\ep$ is given in 
\eqn{eq:seps}. Then, we can write the integrated subtraction term 
in \eqn{eq:ICarIF00} as 
\beq
\left[\IcC{ar}{IF (0,0)}(\xi_a,\xi_b;\ep)\right] = 
	2 s_{ab} \left[ \frac{\xi_a \xi_b (1-\xi_a^2) (1-\xi_b^2)}
    {(\xi_a + \xi_b)^2} \right]^{-\ep}
	\frac{\xi_a^2 \xi_b^2 (1+\xi_a \xi_b)}{(\xi_a + \xi_b)^2}
	\frac{1}{x_{a,r} s_{ar}}
	\frac{\omega(ar)}{\omega(a)}
	P_{(ar) r}^{(0)}(x_{a,r};\ep)\,,
\label{eq:integcar}
\eeq
where we have used the delta functions in \eqn{eq:ps2arexp} to 
perform the integrations over $\xi_{a,r}$ and $\xi_{b,s}$. In 
particular, this implies that in \eqn{eq:integcar} above, 
$x_{a,r}$ and $s_{ar}$ are expressed in terms of $\xi_{a}$ and 
$\xi_{b}$ as in \eqns{eq:invertdaleo}{eq:xareval} with the 
replacements $\xi_{a,r} \to \xi_a$ and $\xi_{b,r} \to \xi_b$.

%
% Single soft-type subtraction terms
%

\subsection{Single soft-type subtraction terms}

%
% \cS{r}{(0,0)}
%
\subsubsection{\texorpdfstring{$\cS{r}{(0,0)}$}{Sr00}}
\label{sec:Sr00}

%%%%%
\paragraph{Subtraction term:} 

For a final-state gluon $r$, we define the single soft subtraction 
term as follows,
\beq
\cS{r}{(0,0)}(p_a,p_b;\mom{}_{m+X+2}) = 
	-8\pi\as\mu^{2\ep}
	\sum_{\substack{j,k \in I \cup F \\ j,k \ne r}}
	\frac{1}{2}\calS_{\ti{j}\ti{k}}(r)
	\bT_j \bT_k 
	\SME{ab,m+X+1}{(0)}{(\ti{p}_a,\ti{p}_b;\momt{}_{m+X+1})}\,,
\label{eq:Sr00}
\eeq
where the reduced matrix element is obtained simply by removing 
the final-state gluon $r$. The set of mapped momenta entering the 
reduced matrix element is defined below in \eqn{eq:srmap}. 
Further, $\calS_{\ti{j}\ti{k}}(r)$ denotes the eikonal factor, 
evaluated with the mapped hard momenta, as explained in 
\sect{sec:extension},
\beq
\calS_{\ti{j}\ti{k}}(r) = 
    \frac{2s_{\ti{j}\ti{k}}}{s_{\ti{j}r}s_{\ti{k}r}}\,,
    \qquad j,k \in I \cup F\,, \quad j,k \ne r\,.
\label{eq:eikfact}
\eeq
The explicit factor of $1/2$ in \eqn{eq:Sr00} comes from the sum 
over $j$ and $k$ not being ordered, hence each term is counted 
twice.

%%%%%
\paragraph{Momentum mapping:}

The single soft (S) mapping,
\beq
\smp{r}:\; (p_a,p_b;\mom{}_{m+X+2}) 
    \to (\ti{p}_a, \ti{p}_b; \momt{}_{m+X+1})\,,
\eeq
is defined as in~\refr{DelDuca:2019ctm} ($r \in F$) 
\beq
\bsp
\ti{p}_a^\mu &= \lambda_r p_a^\mu\,,
\\
\ti{p}_b^\mu &= \lambda_r p_b^\mu\,,
\\
\ti{p}_n^\mu &= {\Lambda(P,\ti{P})^\mu}_{\!\nu}\, p_n^\nu\,, 
\qquad n \in F, \quad n\ne r\,,
\\
\ti{p}_X^\mu &= {\Lambda(P,\ti{P})^\mu}_{\!\nu}\, p_X^\nu\,,
\esp
\label{eq:srmap}
\eeq
where ${\Lambda(P,\ti{P})^\mu}_{\!\nu}$ is the proper Lorentz 
transformation of \eqn{eq:Lambda_munu} that takes the massive 
momentum $P^\mu$ into a momentum $\ti{P}^\mu$ of the same mass, with
\beq
P^\mu = (p_a+p_b)^\mu - p_r^\mu
\qquad\mbox{and}\qquad
\ti{P}^\mu = (\ti{p}_a + \ti{p}_b)^\mu = \lambda_r (p_a+p_b)^\mu\,.
\eeq
The value of $\lambda_{r}$ is fixed by requiring that $P^2 = 
\ti{P}^2$,
\beq
\lambda_{r} = \sqrt{1 - \frac{s_{r\ab}}{s_{ab}}}\,.
\label{eq:lamr-def}
\eeq
Since the initial-state momenta $p_a^\mu$ and $p_b^\mu$ are 
rescaled by $\lambda_{r}$ in \eqn{eq:srmap}, so is the total 
momentum. The advantage of this mapping over the 
momentum-conserving soft mapping of~\refr{Somogyi:2006cz} is that 
\eqn{eq:srmap} can also be used in the presence of massive 
final-state particles, as well as in the color-singlet case, for 
which the mapped set $\momt{}_{X}$ consists of only massive 
non-QCD particles in the final state~\cite{DelDuca:2019ctm}. In 
the $p_r^\mu \to 0$ limit clearly $\lambda_r \to 1$ which implies 
$\ti{P}^\mu \to P^\mu$, so the Lorentz-transformation goes to the 
identity and we recover the single soft $p_r^\mu \to 0$ kinematic 
configuration. We will use tildes to denote momenta obtained via 
soft-type momentum mappings throughout, as in \eqn{eq:srmap}.

%%%%%
\paragraph{Phase space convolution:}

By inserting the identity,
\beq
1 = \frac{\rd^d K}{(2\pi)^{d}} (2\pi)^{d} 
    \delta^{(d)}\left( K+ p_r - p_a - p_b\right)\, 
    \rd K^2\, \delta_+(K^2 - s_{ab} + s_{r\ab})\,,
\eeq
with $\delta_+(K^2 - s_{ab} + s_{r\ab}) = \delta_+(K^2 - 
\lambda_r^2 s_{ab})$, the $(m+X+2)$-particle phase space in 
\eqn{eq:psmx2} (with $k=2$) can be factored on parton $r$ and 
momentum $K$ through the convolution formula,
\beq
\PS{m+X+2}(\mom{}_{m+X+2};p_a+p_b) 
= \int \frac{ \rd K^2}{2\pi}\, 
    \PS{m+X+1}(p_a,\ldots, p_{m+2}, p_X; K)\, 
    \PS{2}(p_r, K; p_a+p_b)\,.
\label{eq:genconvsoft}
\eeq
Then using the identity,
\beq
1 = \rd\lambda\, \delta(\lambda - \lambda_r) \,,
\label{eq:idlambdar}
\eeq
and exchanging the momenta $\{p_n\}$ and $p_X$ with the momenta 
$\{\ti{p}_n\}$ and $\ti{p}_X$ through the mapping of 
\eqn{eq:srmap} lets us write the convolution in 
\eqn{eq:genconvsoft} as,
\beq
\PS{m+X+2}(\mom{}_{m+X+2};p_a+p_b) = 
	\int_{\lammin}^{\lammax} \rd\lam\,
	\PS{m+X+1}\big(\momt{}_{m+X+1};\lam (p_a+p_b)\big) \,
	\frac{s_{ab}}{\pi} \lam\,
	\PS{2}(p_r,K;p_a+p_b)\,,
\label{eq:Sr-map-PSconv}
\eeq
where momentum $K$ is massive with $K^2 = \lam^2 s_{ab}$. The 
limits of the $\lam$ integration are
\beq
\lammin = \frac{M}{\sqrt{s_{ab}}}\,,
\qquad
\lammax = 1\,,
\label{eq:limitslam}
\eeq
where $M$ has been defined below \eqn{eq:alminmax}. Finally, we note 
that the flux factor $\Phi(p_a\cdot p_b)$ written in terms of mapped 
momenta reads
\beq
\Phi(p_a\cdot p_b) = \frac{\Phi(\ti{p}_a\cdot \ti{p}_b)}{\lam^2}\,,
\label{eq:srflux}
\eeq
where we have exploited \eqn{eq:idlambdar} to set $\lam_{r} = 
\lam$.

%%%%%
\paragraph{Integrated subtraction term:} 

Using the definitions of the subtraction term and the phase space 
convolution, \eqns{eq:Sr00}{eq:Sr-map-PSconv}, we write the 
integrated counterterm as
\beq
\bsp
&
\int_1{\cal N} \frac{1}{\omega(a)\omega(b)\Phi(p_a\cdot p_b)} 
	\PS{m+X+2}(\mom{}_{m+X+2};p_a+p_b)
	\cS{r}{(0,0)}(p_a,p_b;\mom{}_{m+X+2}) 	
	\\ &\quad=
	\frac{\as}{2\pi} S_\ep \left(\frac{\mu^2}{s_{ab}}\right)^\ep
	\Big(\left[\IcS{r}{(0,0)}\right] \otimes 
    \dsig{R}_{ab,m+X+1}\Big)
	\,,
\label{eq:Int_Sr00-fin}
\esp
\eeq
where 
\beq
\left[\IcS{r}{(0,0)}\right] \otimes \dsig{R}_{ab,m+X+1} = 
	\int_{\lammin}^{\lammax} \rd \lam\, 
	\sum_{\substack{j,k \in I \cup F \\ j,k \ne r}}
	\frac{1}{2}
    \left[\IcS{r}{(0,0)}(\lam;\ep)\right]^{(j,k)}
	\bT_j \bT_k 
	\dsig{R}_{ab,m+X+1}(\ti{p}_a,\ti{p}_b)\,.
\label{eq:Sr00xdsigR}
\eeq
The initial-state momenta entering the real emission cross section 
in \eqn{eq:Sr00xdsigR} are $\ti{p}_a = \lam p_a$ and $\ti{p}_b = 
\lam p_b$, where we have exploited the delta function in 
\eqn{eq:idlambdar}, and we introduced the integrated counterterm
\beq
\left[\IcS{r}{(0,0)}(\lam;\ep)\right]^{(j,k)} =
	- \frac{(4\pi)^2 }{S_\ep} s_{ab}^\ep
	\int \PS{2}(p_r,K;p_a+p_b)\, 
	\frac{s_{ab}}{\pi} \lam^3 
    \calS_{\ti{j}\ti{k}}(r)\,.
\label{eq:ISr00}
\eeq

%
% \cC{ir}{FF}\cS{r}{(0,0)}
%
\subsubsection{
\texorpdfstring{$\cC{ir}{FF}\cS{r}{(0,0)}$}{CirFFSr00}}
\label{sec:CirFFSr00}

%%%%%
\paragraph{Subtraction term:} 

For a final-state parton $i$ and a final-state gluon $r$ we define 
the single final-final collinear-soft overlapping subtraction term 
as follows,
\beq
\cC{ir}{FF}\cS{r}{(0,0)}(p_a,p_b;\mom{}_{m+X+2}) = 
	8\pi\as\mu^{2\ep}
	\frac{1}{s_{\ti{i}r}} \frac{2 z_{\ti{i},r}}{z_{r,\ti{i}}} 
    \bT_i^2
    \SME{ab,m+X+1}{(0)}{(\ti{p}_a,\ti{p}_b;\momt{}_{m+X+1})}\,,
\label{eq:CirFFSr00}
\eeq
where the reduced matrix element is obtained by removing the 
final-state gluon $r$ and the momenta of the partons in this 
reduced matrix element are given in \eqn{eq:srmap}.\footnote{Note 
that the flavor of parton $i$ remains unchanged in the reduced 
matrix element. This is of course consistent with the cancellation 
between $\cC{ir}{FF}\cS{r}{(0,0)}$ and $\cC{ir}{FF(0,0)}$ in the 
soft limit, because whenever parton $r$ is a gluon, the flavor of 
the emitting parton $(ir)$ in $\cC{ir}{FF(0,0)}$ is the same as 
that of the original parton $i$. Similar comments apply to the 
rest of the collinear-soft type terms that appear in  
$\cA_1^{(0)}$,  $\cA_2^{(0)}$ as well as $\cA_{12}^{(0)}$.} The 
momentum fractions $z_{\ti{i},r}$ and $z_{r,\ti{i}}$ appearing in 
\eqn{eq:CirFFSr00} are given by \eqn{eq:zjk-def}. Notice that the 
momentum fractions $z_{\ti{i},r}$ and $z_{r,\ti{i}}$, as well as 
the invariant  $s_{\ti{i}r}$ in \eqn{eq:CirFFSr00} must be defined 
with the mapped momentum $\ti{p}_i^\mu$ in order to correctly cancel 
the $\ti{p}_i^\mu \parallel \ti{p}_r^\mu$ collinear limit of the 
soft subtraction term $\cS{r}{(0,0)}$. This is because in this 
collinear limit $\ti{p}_i^\mu$ does not go to $p_i^\mu$ and so the 
mapped momenta must be retained in the definition of $\cC{ir}
{FF}\cS{r}{(0,0)}$. However, in the $p_r \to 0$ soft limit we have 
$\ti{p}_i^\mu \to p_i^\mu$, hence $\cC{ir}{FF}\cS{r}{(0,0)}$ 
correctly cancels $\cC{ir}{FF (0,0)}$ in this limit.

%%%%%
\paragraph{Momentum mapping and phase space factorization:}

This subtraction term is defined using the single soft momentum 
mapping of~\sect{sec:Sr00}, where the appropriate phase space 
convolution is also presented.

%%%%%
\paragraph{Integrated subtraction term:} 

Using the definitions of the phase space convolution and the 
subtraction term, \eqns{eq:Sr-map-PSconv}{eq:CirFFSr00}, we write 
the integrated counterterm as
\beq
\bsp
&
\int_1{\cal N} \frac{1}{\omega(a)\omega(b)\Phi(p_a\cdot p_b)} 
	\PS{m+X+2}(\mom{}_{m+X+2};p_a+p_b)
	\cC{ir}{FF}\cS{r}{(0,0)}(p_a,p_b;\mom{}_{m+X+2}) 	
	\\ &\quad=
	\frac{\as}{2\pi} S_\ep \left(\frac{\mu^2}{s_{ab}}\right)^\ep
	\Big(\left[\IcC{ir}{FF}\IcS{r}{(0,0)}\right] \otimes 
    \dsig{R}_{ab,m+X+1}\Big)
	\,,
\label{eq:Int_CirFFSr00-fin}
\esp
\eeq
where 
\beq
\left[\IcC{ir}{FF}\IcS{r}{(0,0)}\right] \otimes 
\dsig{R}_{ab,m+X+1} = 
	\int_{\lammin}^{\lammax} \rd \lam\, 
	\left[\IcC{ir}{FF}\IcS{r}{(0,0)}(\lam;\ep)\right]
	\dsig{R}_{ab,m+X+1}(\ti{p}_a,\ti{p}_b)\,.
\label{eq:CirFFSr00xdsigR}
\eeq
The initial-state momenta entering the real emission cross section 
in \eqn{eq:CirFFSr00xdsigR} are once again $\ti{p}_a = \lam p_a$ 
and $\ti{p}_b = \lam p_b$ and we introduced the integrated 
counterterm
\beq
\left[\IcC{ir}{FF}\IcS{r}{(0,0)}(\lam;\ep)\right] = 
	\frac{(4\pi)^2 }{S_\ep} s_{ab}^\ep
	\int \PS{2}(p_r,K;p_a+p_b)\, 
	\frac{s_{ab}}{\pi} \lam^3
	 \frac{1}{s_{\ti{i}r}}\frac{2z_{\ti{i},r}}{z_{r,\ti{i}}}
     \bT_i^2\,.
\label{eq:ICirFFSr00}
\eeq

%
% \cC{ar}{IF}\cS{r}{(0,0)}
%

\subsubsection{
\texorpdfstring{$\cC{ar}{IF}\cS{r}{(0,0)}$}{CarIFSr00}}
\label{sec:CarIFSr00}

%%%%%
\paragraph{Subtraction term:} 

For an initial-state parton $a$ and a final-state gluon $r$ we 
define the initial-final collinear-soft overlapping subtraction 
term as follows,
\beq
\cC{ar}{IF}\cS{r}{(0,0)}(p_a,p_b;\mom{}_{m+X+2}) = 
	8\pi\as\mu^{2\ep}
	\frac{1}{s_{\ti{a}r}} \frac{2}{x_{r,\ti{a}}} \bT_a^2
	\SME{ab,m+X+1}{(0)}{(\ti{p}_a,\ti{p}_b;\momt{}_{m+X+1})}\,,
\label{eq:CarIFSr00}
\eeq
where the reduced matrix element is obtained once more by removing 
the final-state gluon $r$, and the momenta entering the reduced 
matrix element are given in \eqn{eq:srmap}. The momentum fraction 
$x_{r,\ti{a}}$ appearing in \eqn{eq:CarIFSr00} is defined by 
\eqn{eq:xja-def}. Using this definition of $x_{r,\ti{a}}$ and the 
fact that the single soft momentum mapping simply rescales $p_a$ 
to obtain $\ti{p}_a$, we find the equality
\beq
\frac{1}{s_{\ti{a}r}} \frac{2}{x_{r,\ti{a}}} = 
    \frac{1}{s_{ar}} \frac{2}{x_{r,a}}\,.
\eeq
Nevertheless, we choose to retain the definition as given in 
\eqn{eq:CarIFSr00} for the following reason. Using 
\eqns{eq:zjk-def}{eq:xja-def}, we have immediately that
\beq
\frac{1}{s_{\ti{i}r}}\frac{2 z_{\ti{i},r}}{z_{r,\ti{i}}} = 
\frac{1}{s_{\ti{i}r}}\frac{2s_{\ti{i}\ab}}{s_{r\ab}}
\qquad\mbox{and}\qquad
\frac{1}{s_{\ti{a}r}}\frac{2}{x_{r,\ti{a}}} = 
\frac{1}{s_{\ti{a}r}}\frac{2s_{\ti{a}\ab}}{s_{r\ab}}\,.
\eeq
Hence, the collinear-soft counterterms of 
\eqns{eq:CirFFSr00}{eq:CarIFSr00} have the same form once the 
momentum fractions are written explicitly in terms of mapped 
momenta. This fact can be exploited to optimize the implementation 
of these subtraction terms. We will follow this convention of 
defining initial-final collinear-soft factors in terms of mapped 
momenta throughout.

%%%%%
\paragraph{Momentum mapping and phase space factorization:}

This subtraction term is defined using the single soft momentum 
mapping of~\sect{sec:Sr00}, where the appropriate phase space 
convolution is also presented.

%%%%%
\paragraph{Integrated subtraction term:} 

Using the definitions of the phase space convolution and the 
subtraction term, \eqns{eq:Sr-map-PSconv}{eq:CarIFSr00}, we write 
the integrated counterterm as
\beq
\bsp
&
\int_1{\cal N} \frac{1}{\omega(a)\omega(b)\Phi(p_a\cdot p_b)} 
	\PS{m+X+2}(\mom{}_{m+X+2};p_a+p_b)
	\cC{ar}{IF}\cS{r}{(0,0)}(p_a,p_b;\mom{}_{m+X+2}) 	
	\\ &\quad=
	\frac{\as}{2\pi} S_\ep \left(\frac{\mu^2}{s_{ab}}\right)^\ep
	\Big(\left[\IcC{ar}{IF}\IcS{r}{(0,0)}\right] \otimes 
    \dsig{R}_{ab,m+X+1}\Big)
	\,,
\label{eq:Int_CarIFSr00-fin}
\esp
\eeq
where 
\beq
\left[\IcC{ar}{IF}\IcS{r}{(0,0)}\right] \otimes 
\dsig{R}_{ab,m+X+1} = 
	\int_{\lammin}^{\lammax} \rd \lam\, 
	\left[\IcC{ar}{IF}\IcS{r}{(0,0)}(\lam;\ep)\right]
	\dsig{R}_{ab,m+X+1}(\ti{p}_a,\ti{p}_b)\,.
\label{eq:CarIFSr00xdsigR}
\eeq
Recall that in \eqn{eq:CarIFSr00xdsigR} we have $\ti{p}_a = \lam 
p_a$ and $\ti{p}_b = \lam p_b$ and we introduced the integrated 
subtraction term
\beq
\left[\IcC{ar}{IF}\IcS{r}{(0,0)}(\lam;\ep)\right] = 
	\frac{(4\pi)^2 }{S_\ep} s_{ab}^\ep
	\int \PS{2}(p_r,K;p_a+p_b)\, 
	\frac{s_{ab}}{\pi} \lam^3
	 \frac{1}{s_{\ti{a}r}}\frac{2}{x_{r,\ti{a}}} \bT_a^2\,.
\label{eq:ICarIFSr00}
\eeq

%%%
%%% Subtraction terms for double unresolved emission
%%%

\section{Subtraction terms for double unresolved emission}
\label{sec:A2}

In this section, we provide the explicit definitions of the 
individual double unresolved subtraction terms introduced in 
\sect{sec:doubleunresolv}.

%
% Initial-state double collinear-type subtraction terms
%
\subsection{Initial-state double collinear-type subtraction terms}

%
% \cC{ars}{IFF (0,0)}
%
\subsubsection{\texorpdfstring{$\cC{ars}{IFF (0,0)}$}{CarsIFF00}}
\label{sec:CarsIFF00}

%%%%%
\paragraph{Subtraction term:} 

\begin{figure}
\setlength{\tabcolsep}{10pt}
\renewcommand{\arraystretch}{1.2}
\begin{center}
\parbox{15em}{\includegraphics[scale=1]{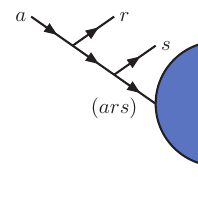}}
\begin{tabular}{|c|c|c|c|}
\hline
$a$ & $r$ & $s$ & $(ars)$ 
\\ 
\hline\hline
$q/\qb$ & $q/\qb$ & $g$ & $g$
\\
$q/\qb$ & $g$ & $q/\qb$ & $g$
\\
$q/\qb$ & $q/\qb$ & $\qb/q$ & $q/\qb$
\\
$q/\qb$ & $g$ & $g$ & $q/\qb$
\\
\hline
\end{tabular}
%%%%%
\begin{tabular}{|c|c|c|c|}
\hline
$a$ & $r$ & $s$ & $(ars)$ 
\\ 
\hline\hline
$g$ & $q/\qb$ & $g$ & $\qb/q$
\\
$g$ & $g$ & $q/\qb$ & $\qb/q$
\\
$g$ & $q/\qb$ & $\qb/q$ & $g$
\\
$g$ & $g$ & $g$ & $g$
\\
\hline
\end{tabular}
\end{center}
\caption{\label{fig:Parsflav}
The splitting configuration for $a\to (ars)+r+s$ initial-state 
triple collinear splitting. The arrows in the figure denote 
momentum and flavor flow. The precise flavor mapping for the 
$\cC{ars}{IFF(0,0)}$ counterterm is given in the table.
}
\end{figure}

For an initial-state parton $a$ and final-state partons $r$ and 
$s$ we define the initial-final-final triple collinear subtraction 
term as
\beq
\bsp
\cC{ars}{IFF (0,0)}(p_a,p_b;\mom{}_{m+X+2}) &=
	(8\pi\as\mu^{2\ep})^2 
    \frac{1}{x_{a,rs}(s_{ar}+s_{as}-s_{rs})^2}
\\ &\times
	\hP_{(ars) r s}^{(0)}(x_{a,rs}, x_{r,a}, x_{s,a}, s_{ar}, 
    s_{as}, s_{rs},\kT{a,rs}, \kT{r,a}, \kT{s,a}; \ep)
\\ &\times 
	\SME{(ars)b,m+X}{(0)}{(\ha{p}_a,\ha{p}_b;\momh{}_{m+X})}
	{\cal F}(x_{a,rs},\xi_{a,rs}\xi_{b,rs}) \,, 
\esp
\label{eq:CarsIFF00}
\eeq
where the reduced matrix element is obtained by removing the 
final-state partons $r$ and $s$ and replacing the initial-state 
parton $a$ with a parton $(ars)$ whose flavor is determined by the 
flavor summation rules of \eqn{eq:flsum} by the requirement that 
$a=(ars)+r+s$. Hence, the flavor of the initial-state parton that 
is involved in the splitting is in general different in the 
original double real emission matrix element and in the reduced 
matrix element. The initial-state triple collinear splitting 
configuration is shown in \fig{fig:Parsflav}, together with the 
precise flavor mapping. The set of momenta entering the reduced 
matrix element is defined below in \eqn{eq:cabrsIIFFmap}. The 
tree-level initial-final-final splitting functions for $a\to (ars) 
+ r + s$ splitting,  $\hP_{(ars) r s}^{(0)}$, can be obtained from 
the final-state triple collinear splitting functions 
of~\refrs{Catani:1998nv,Catani:1999ss}, recalled here in 
\eqnss{eq:Pqqbpqp0FFF}{eq:Pggg0FFF}, by the crossing relation 
discussed in \appx{appx:AP-functions}, see \eqn{eq:Piffcross}. We 
have once more chosen to label the splitting function with the 
flavor of the initial-state parton in the {\em reduced} matrix 
element, $(ars)$, and the flavors of the final-state partons $r$ 
and $s$. The momentum fractions $x_{a,rs}$, $x_{r,a}$ and 
$x_{s,a}$ are defined by \eqn{eq:xajk-def}, while the transverse 
momenta $\kT{a,rs}$, $\kT{r,a}$ and $\kT{s,a}$ are given by 
\eqn{eq:kcjk-def}. Finally, the function ${\cal F}$ in 
\eqn{eq:CarsIFF00} is defined as
\beq
{\cal F}(x,y) =\left(\frac{x}{y}\right)^2\,,
\label{eq:Ffunc}
\eeq
so that
\beq
{\cal F}(x_{a,rs},\xi_{a,rs}\xi_{b,rs}) 
	= \left( \frac{x_{a,rs}}{\xi_{a,rs}\xi_{b,rs}} \right)^2
	= \left( \frac{s_{ab} - s_{(rs)\ab}}
    {s_{ab} - s_{(rs)\ab} + s_{rs}} \right)^2\,,
\label{eq:Fiff}
\eeq
where $\xi_{a,rs}$ and $\xi_{b,rs}$ are given in 
\eqn{eq:xiars_xibrs_def} below and their product is recorded in 
\eqn{eq:arsprod}. Clearly in the triple collinear $p_a^\mu 
\parallel p_r^\mu \parallel p_s^\mu$ configuration ${\cal F}
(x_{a,rs},\xi_{a,rs}\xi_{b,rs}) \to 1$, so the subtraction term in 
\eqn{eq:CarsIFF00} is a correct regulator of the double real 
emission matrix element in this limit. The reason for including 
${\cal F}(x_{a,rs},\xi_{a,rs}\xi_{b,rs})$ in \eqn{eq:CarsIFF00} 
was discussed in \sect{sec:extension}: its role is to cancel the 
unphysical singularity associated with the vanishing of $x_{a,rs}$ 
inside the double real emission phase space. Indeed, using the 
expressions in \appx{appx:AP-functions}, one can check that the 
subtraction term contains at most factors of $1/x_{a,rs}^2$, so 
the spurious singularities may be canceled by simply including the 
factor of ${\cal F}(x_{a,rs},\xi_{a,rs}\xi_{b,rs})$ in 
\eqn{eq:CarsIFF00}. On the other hand, the rest of the subtraction 
terms, especially the iterated single unresolved ones, can be 
defined in a way such that all internal cancellation between the 
various subtraction terms are maintained in every unresolved limit.

%%%%%
\paragraph{Momentum mapping:}

The initial-state double collinear momentum mapping,
\beq
\cmp{ab,rs}{II,FF}:\; (p_a, p_b; \mom{}_{m+X+2}) 
\cmap{ab,rs}{II,FF} (\ha{p}_a,\ha{p}_b;\momh{}_{m+X})\,,
\eeq
is defined as ($a,b \in I$ and $r,s \in F$)
\beq
\bsp
\ha{p}_a^\mu &= \xi_{a,rs} p_a^\mu\,,
\\
\ha{p}_b^\mu &= \xi_{b,rs} p_b^\mu\,,
\\
\ha{p}_n^\mu &= {\Lambda(P,\ha{P})^\mu}_{\!\nu}\, p_n^\nu\,, 
\qquad n \in F, \quad n\ne r\,,
\\
\ha{p}_X^\mu &= {\Lambda(P,\ha{P})^\mu}_{\!\nu}\, p_X^\nu\,,
\esp
\label{eq:cabrsIIFFmap}
\eeq
where ${\Lambda(P,\ha{P})^\mu}_{\!\nu}$ is the proper Lorentz 
transformation of \eqn{eq:Lambda_munu} that takes the massive 
momentum $P^\mu$ into a momentum $\ha{P}^\mu$ of the same mass, with
\beq
P^\mu = (p_a+p_b)^\mu - p_r^\mu - p_s^\mu \,,
\qquad
\ha{P}^\mu = (\ha{p}_a + \ha{p}_b)^\mu 
    = (\xi_{a,rs}p_a + \xi_{b,rs} p_b)^\mu\,.
\label{eq:pptilda}
\eeq
The value of the product $\xi_{a,rs} \xi_{b,rs}$ is fixed by 
requiring that $P^2 = \ha{P}^2$,
\beq
\xi_{a,rs} \xi_{b,rs} = 
	1 - \frac{s_{(rs)\ab}}{s_{ab}} 
	+ \frac{s_{rs}}{s_{ab}} 
\,,
\label{eq:arsprod}
\eeq
but this does not fix $\xi_{a,rs}$ and $\xi_{b,rs}$ uniquely. 
Following~\refr{Daleo:2006xa}, we set
\beq
\xi_{a,rs} = \sqrt{\frac{s_{ab} - s_{b(rs)}}{s_{ab} - s_{a(rs)}}
	\frac{s_{ab} - s_{(rs)\ab} + s_{rs}}{s_{ab}}}\,,
\qquad\quad
\xi_{b,rs} = \sqrt{\frac{s_{ab} - s_{a(rs)}}{s_{ab} - s_{b(rs)}}
	\frac{s_{ab} - s_{(rs)\ab} + s_{rs}}{s_{ab}}}\,,
\label{eq:xiars_xibrs_def}
\eeq
which fulfills \eqn{eq:arsprod}. Clearly this mapping is simply a 
generalization of the initial-state single collinear momentum 
mapping of \eqn{eq:cabrIIFmap}. Note that $\xi_{a,rs}$ and 
$\xi_{b,rs}$ are related by exchanging $a\leftrightarrow b$. 
Furthermore, they are symmetric under the $r\leftrightarrow s$ 
interchange. The mapping of \eqn{eq:cabrsIIFFmap} has the 
advantage that it reproduces both the initial-final-final triple 
collinear kinematic configuration, as well as the initial-final, 
initial-final double collinear kinematic configuration in the 
appropriate limit. Indeed, when e.g., $p_a^\mu \parallel p_r^\mu 
\parallel p_s^\mu$ so that $p_r^\mu \to x_r p_a^\mu$ and $p_s^\mu 
\to x_s p_a^\mu$ we find $\xi_{a,rs} \to 1 - x_r - x_s$ and 
$\xi_{b,rs} \to 1$. Then, $\ha{P}^\mu \to P^\mu$  implying that 
the Lorentz transformation becomes the identity, and we recover 
the $p_a^\mu \parallel p_r^\mu \parallel p_s^\mu$ 
initial-final-final triple collinear kinematic configuration. 
Likewise, in the $p_a^\mu \parallel p_r^\mu$, $p_b^\mu \parallel 
p_s^\mu$ limit when $p_r^\mu \to x_r p_a^\mu$ and $p_s^\mu \to x_s 
p_b^\mu$, we obtain $\xi_{a,rs} \to 1 - x_r $ and $\xi_{b,rs} \to 
1 - x_s$. This again leads to $\ha{P}^\mu \to P^\mu$ and the 
Lorentz transformation going to the identity, and we obtain the 
$p_a^\mu \parallel p_r^\mu$, $p_b^\mu \parallel p_s^\mu$ 
initial-final, initial-final double collinear kinematic 
configuration. Due to the $a\leftrightarrow b$ symmetry of the 
mapping, the mapped momenta have the correct behavior also in the 
$p_b^\mu \parallel p_r^\mu \parallel p_s^\mu$, as well as the 
$p_b^\mu \parallel p_r^\mu$, $p_a^\mu \parallel p_s^\mu$ limits.

%%%%%
\paragraph{Phase space convolution:}

Using the identity,
\beq
1 = \rd\xi_a\, \delta(\xi_a - \xi_{a,rs})\, \rd\xi_b\, \delta(\xi_b - \xi_{b,rs}) \,,
\label{eq:idIFF}
\eeq
and shifting the momenta as in \eqn{eq:cabrsIIFFmap}, the 
$(m+X+2)$-particle phase space of \eqn{eq:psmx2} (with $k=2$) can 
be written as the convolution,
\beq
\PS{m+X+2}(\mom{}_{m+X+2};p_a+p_b) = 
	 \int \rd\xi_a\, \rd\xi_b\,
	\PS{m+X}\big(\momh{}_{m+X};\xi_a p_a+ \xi_b p_b\big) \,
	\PS{II,FF}(p_r,p_s,\xi_a,\xi_b)\,,
\label{eq:CarIFF-map-PSconv}
\eeq
with 
\beq
\PS{II,FF}(p_r,p_s,\xi_a,\xi_b) = 
	\frac{\rd^d p_r}{(2\pi)^{d-1}}\, \delta_+(p_r^2)\, 
    \frac{\rd^d p_s}{(2\pi)^{d-1}}\, \delta_+(p_s^2)\,
	\delta(\xi_a - \xi_{a,rs})\, \delta(\xi_b - \xi_{b,rs})\,.
\label{eq:ps2ars}
\eeq
The limits of the $\xi_a$ and $\xi_b$ integrations are as in 
\eqn{eq:xi1xi2lims}. We note that the phase space measure 
$\PS{II,FF}(p_r,p_s,\xi_a,\xi_b) $ can also be expressed in the 
following form,
\beq
\PS{II,FF}(p_r,p_s,\xi_a,\xi_b) = 
	\PS{3}(p_r,p_s,K;p_a+p_b)\,
	{\cal J}_{II,FF}\,
	\delta[\xi_a(s_{ab}-s_{a(rs)}) - \xi_b(s_{ab}-s_{b(rs)})]\,,
\label{eq:PSiff-PS3}
\eeq
with the Jacobian
\beq
{\cal J}_{II,FF} = \frac{s_{ab}[\xi_a(s_{ab}-s_{a(rs)}) 
    + \xi_b(s_{ab}-s_{b(rs)})]}{2\pi}\,.
\eeq
The momentum $K$ appearing in the three-particle phase space above is 
massive with $K^2 = \xi_a \xi_b s_{ab}$. Finally, we note that the 
flux factor $\Phi(p_a\cdot p_b)$ can be written in terms of mapped 
momenta as
\beq
\Phi(p_a\cdot p_b) 
    = \frac{\Phi(\ha{p}_a\cdot \ha{p}_b)}{\xi_a \xi_b}\,,
\label{eq:carsIFFflux}
\eeq
where we have used \eqn{eq:idIFF} to set $\xi_{a,rs} = \xi_a$ and 
$\xi_{b,rs} = \xi_b$. Note that \eqn{eq:carsIFFflux} is formally 
identical to \eqn{eq:carIFflux}, although the precise meaning of 
the mapped momenta in these equations is different.

%%%%%
\paragraph{Integrated subtraction term:} 

Using the definitions of the subtraction term and the phase space 
convolution, \eqns{eq:CarsIFF00}{eq:CarIFF-map-PSconv}, we write 
the integrated counterterm as
\beq
\bsp
&
\int_2{\cal N} \frac{1}{\omega(a)\omega(b)\Phi(p_a\cdot p_b)} 
	\PS{m+X+2}(\mom{}_{m+X+2};p_a+p_b)\,
	\cC{ars}{IFF (0,0)}(p_a,p_b;\mom{}_{m+X+2})
\\ &\qquad=
	\left[\frac{\as}{2\pi} S_\ep 
    \left(\frac{\mu^2}{s_{ab}}\right)^\ep\right]^2
	\Big( \left[\IcC{ars}{IFF (0,0)}\right] \otimes 
    \dsig{B}_{(ars)b,m+X}\Big)\,,
\esp
\label{eq:Int_CarsIFF00-fin}
\eeq
where
\beq
\left[\IcC{ars}{IFF (0,0)}\right] \otimes \dsig{B}_{(ars)b,m+X} = 
	\int \rd \xi_a\,\rd \xi_b\,
	\left[\IcC{ars}{IFF (0,0)}(\xi_a,\xi_b;\ep)\right]
	\dsig{B}_{(ars)b,m+X}(\ha{p}_a,\ha{p}_b)\,.
\eeq
Here and below, the notation $\int_2$ indicates that the phase 
space integration is performed only over the two unresolved 
emissions. Furthermore, exploiting the delta functions in 
\eqn{eq:idIFF}, the initial-state momenta entering the Born cross 
section are written as $\ha{p}_a = \xi_a p_a$ and $\ha{p}_b = 
\xi_b p_b$ and we introduced the integrated counterterm
\beq
\bsp
&
\left[\IcC{ars}{IFF (0,0)}(\xi_a,\xi_b;\ep)\right] = 
	\left( \frac{(4\pi)^2 }{S_\ep} s_{ab}^\ep \right)^2
	\int \PS{II,FF}(p_r,p_s,\xi_a,\xi_b)\, \xi_a \xi_b
	\frac{1}{x_{a,rs} (s_{ar}+s_{as}-s_{rs})^2}
\\ &\qquad\times	
	\frac{\omega(ars)}{\omega(a)}
	\hP_{(ars) r s}^{(0)}(x_{a,rs}, x_{r,a}, x_{s,a}, s_{ar}, 
    s_{as}, s_{rs},\kT{a,rs}, \kT{r,a}, \kT{s,a}; \ep)
	{\cal F}(x_{a,rs},\xi_a\xi_b)\,.
\esp
\label{eq:Int_Cars0IFF}
\eeq
In order to evaluate $\left[\IcC{ars}{IFF (0,0)}
(\xi_a,\xi_b;\ep)\right]$, we use the representation of the 
factorized phase space measure in \eqn{eq:PSiff-PS3} and exploit 
$\kT{a,rs}\cdot \ha{p}_a = \kT{r,a}\cdot \ha{p}_a = \kT{s,a}\cdot 
\ha{p}_a = 0$ to perform the azimuthal integration by passing to 
the azimuthally averaged splitting functions, 
\[
\hP_{(ars) r s}^{(0)}(x_{a,rs}, x_{r,a}, x_{s,a}, s_{ar}, s_{as}, 
s_{rs},\kT{a,rs}, \kT{r,a}, \kT{s,a}; \ep) 
\to P_{(ars) r s}^{(0)}(x_{a,rs}, x_{r,a}, x_{s,a}, s_{ar}, 
s_{as}, s_{rs}; \ep)\,,
\]
see the discussion in \appx{appx:AP-functions}. We then find
\beq
\bsp
\left[\IcC{ars}{IFF (0,0)}(\xi_a,\xi_b;\ep)\right] &= 
	\left( \frac{(4\pi)^2 }{S_\ep} s_{ab}^\ep \right)^2
	\int \PS{3}(p_r,p_s,K;p_a+p_b)\,
	{\cal J}_{II,FF}\, 
    \delta[\xi_a(s_{ab}-s_{a(rs)}) - \xi_b(s_{ab}-s_{b(rs)})]
\\ &\times
	\frac{\xi_a \xi_b }{x_{a,rs} (s_{ar}+s_{as}-s_{rs})^2}
    \frac{\omega(ars)}{\omega(a)}
	P_{(ars) r s}^{(0)}(x_{a,rs}, x_{r,a}, x_{s,a}, s_{ar}, 
    s_{as}, s_{rs}; \ep) {\cal F}(x_{a,rs},\xi_a\xi_b)\,.
\esp
\label{eq:ICars00IFF-2}
\eeq
In contrast to the case of the integrated single unresolved 
initial-final subtraction term $\left[\IcC{ar}{IF (0,0)}
(\xi_a,\xi_b;\ep)\right]$ of \eqn{eq:ICarIF00}, this time the 
integration is no longer fully fixed by the delta functions in 
\eqn{eq:ps2ars}. 

%
% \cC{ar,bs}{IF,IF (0,0)}
%
\subsubsection{
\texorpdfstring{$\cC{ar,bs}{IF,IF (0,0)}$}{CarbsIFIF00}}
\label{sec:CarbsIFIF00}

%%%%%
\paragraph{Subtraction term:} 

For initial-state partons $a$ and $b$ and final-state partons $r$ 
and $s$ we define the initial-final, initial-final double 
collinear subtraction term as
\beq
\bsp
\cC{ar,bs}{IF,IF (0,0)}(p_a,p_b;\mom{}_{m+X+2}) &=
	(8\pi\as\mu^{2\ep})^2  
    \frac{1}{x_{a,r} s_{ar}} \frac{1}{x_{b,s} s_{bs}}
	\hP_{(ar) r}^{(0)}(x_{a,r},\kT{r,a}; \ep)
	\hP_{(bs) s}^{(0)}(x_{b,s},\kT{s,b}; \ep)
\\ & \times
	\SME{(ar)(bs),m+X}{(0)}{(\ha{p}_a,\ha{p}_b;\momh{}_{m+X})}
	{\cal F}(x_{a,r}x_{b,s},\xi_{a,rs}\xi_{b,rs}) \,, 
\esp
\label{eq:CarbsIFIF00}
\eeq
where the reduced matrix element is obtained by removing the 
final-state partons $r$ and $s$ and replacing the initial-state 
partons $a$ and $b$ by partons of flavors $(ar)$ and $(bs)$ using 
the flavor summation rules in \eqn{eq:flsum}. The flavors of 
partons $(ar)$ and $(bs)$ are fixed by requiring that $a=(ar)+r$ 
and $b=(bs)+s$, see \fig{fig:Parflav}. Clearly, the flavors of 
both initial-state partons in the reduced matrix element are 
generally different from those in the original double real 
emission matrix element. The set of momenta entering the reduced 
matrix element was defined in \eqn{eq:cabrsIIFFmap}, while the 
tree-level initial-final Altarelli-Parisi splitting functions 
$\hP_{(ar) r}^{(0)}$ and $\hP_{(bs) s}^{(0)}$ are the same as 
those appearing in the single unresolved initial-final collinear 
subtraction term of \eqn{eq:CarIF00}. Their explicit expressions 
are given in \eqnss{eq:Pqg0IF}{eq:Pgg0IF}. The momentum fractions 
and transverse momenta entering \eqn{eq:CarbsIFIF00} are defined 
by \eqns{eq:xja-def}{eq:kj-def}. Note that these definitions are 
identical to the ones in the single unresolved initial-final 
collinear subtraction term of \sect{sec:CarIF00} for the $(ar)$ 
pair, while the corresponding expressions for the $(bs)$ pair are 
obtained by the simple relabeling of $a\to b$ and $r\to s$. 
Finally, the function ${\cal F}$ which appears in 
\eqn{eq:CarbsIFIF00} was defined in \eqn{eq:Ffunc}, hence
\beq
{\cal F}(x_{a,r}x_{b,s},\xi_{a,rs}\xi_{b,rs}) 
	= \left( \frac{x_{a,r}x_{b,s}}{\xi_{a,rs}\xi_{b,rs}} \right)^2
	= \left( \frac{(s_{ab} - s_{r\ab})(s_{ab} - s_{s\ab})}
    {s_{ab}(s_{ab} - s_{(rs)\ab} + s_{rs})} \right)^2\,.
\label{eq:Fifif}
\eeq
In the double collinear $p_a^\mu \parallel p_r^\mu$, $p_b^\mu 
\parallel p_s^\mu$ configuration ${\cal F}
(x_{a,r}x_{b,s},\xi_{a,rs}\xi_{b,rs}) \to 1$, so the subtraction 
term in \eqn{eq:CarbsIFIF00} is a correct regulator of the double 
real emission matrix element in this limit. The function ${\cal F}
(x_{a,r}x_{b,s},\xi_{a,rs}\xi_{b,rs})$ is included in 
\eqn{eq:CarbsIFIF00} in order to preserve the structure of 
internal cancellations between the subtraction terms in all 
unresolved limits, see the discussion in 
\sect{sec:CasbrIFIF00CasIF}.

%%%%%
\paragraph{Momentum mapping and phase space factorization:}

This subtraction term is defined using the initial-state double 
collinear momentum mapping of~\sect{sec:CarsIFF00}, where the 
appropriate phase space convolution is also presented.

%%%%%
\paragraph{Integrated subtraction term:} 

Using the definitions of the phase space convolution and the 
subtraction term, \eqns{eq:CarIFF-map-PSconv}{eq:CarbsIFIF00}, we 
write the integrated counterterm as
\beq
\bsp
&
\int_2{\cal N} \frac{1}{\omega(a)\omega(b)\Phi(p_a\cdot p_b)} 
	\PS{m+X+2}(\mom{}_{m+X+2};p_a+p_b)\,
	\cC{ar,bs}{IF,IF (0,0)}(p_a,p_b;\mom{}_{m+X+2})
\\ &\qquad=
	\left[\frac{\as}{2\pi} S_\ep 
    \left(\frac{\mu^2}{s_{ab}}\right)^\ep\right]^2
	\Big( \left[\IcC{ar,bs}{IF,IF (0,0)}\right] \otimes 
    \dsig{B}_{(ar)(bs),m+X}\Big)\,,
\esp
\label{eq:Int_CarbsIFIF00-fin}
\eeq
where
\beq
\left[\IcC{ar,bs}{IF,IF (0,0)}\right] \otimes 
\dsig{B}_{(ar)(bs),m+X} = 
	\int \rd \xi_a\,\rd \xi_b\,
	\left[\IcC{ar,bs}{IF,IF (0,0)}(\xi_a,\xi_b;\ep)\right]
	\dsig{B}_{(ar)(bs),m+X}(\ha{p}_a,\ha{p}_b)\,.
\label{eq:CarbsIFIF00xB}
\eeq
In \eqn{eq:CarbsIFIF00xB}, we once again have $\ha{p}_a = \xi_a 
p_a$ and $\ha{p}_b = \xi_b p_b$ and we introduced  the integrated 
counterterm
\beq
\bsp
\left[\IcC{ar,bs}{IF,IF (0,0)}(\xi_a,\xi_b;\ep)\right] &= 
	\left( \frac{(4\pi)^2 }{S_\ep} s_{ab}^\ep \right)^2
	\int \PS{II,FF}(p_r,p_s,\xi_a,\xi_b)\, \xi_a \xi_b
	\frac{1}{x_{a,r} s_{ar}}\frac{1}{x_{b,s} s_{bs}}
\\ &\times	
	\frac{\omega(ar)}{\omega(a)} \frac{\omega(bs)}{\omega(b)}
	\hP_{(ar) r}^{(0)}(x_{a,r}, \kT{r,a}; \ep)
	\hP_{(bs) s}^{(0)}(x_{b,s}, \kT{s,b}; \ep)
	{\cal F}(x_{a,r}x_{b,s},\xi_a\xi_b)\,.
\esp
\label{eq:Int_Carbs00IFIF}
\eeq
In order to evaluate $\left[\IcC{ar,bs}{IF,IF (0,0)}
(\xi_a,\xi_b;\ep)\right]$, we use the representation of the 
factorized phase space measure in \eqn{eq:PSiff-PS3} and exploit 
$\kT{r,a}\cdot \ha{p}_a=0$ and $\kT{s,b}\cdot \ha{p}_b=0$ to 
perform the azimuthal integration by passing to the azimuthally 
averaged splitting functions, $\hP_{(ar) r}^{(0)}(x_{a,r}, 
\kT{r,a}; \ep) \to P_{(ar) r}^{(0)}(x_{a,r}; \ep)$ and $\hP_{(bs) 
s}^{(0)}(x_{b,s}, \kT{s,b}; \ep)\to P_{(bs) s}^{(0)}(x_{b,s}; 
\ep)$, given in \eqnss{eq:Pqg-ave-IF}{eq:Pgg-ave-IF}. We then find
\beq
\bsp
\left[\IcC{ar,bs}{IF,IF (0,0)}(\xi_a,\xi_b;\ep)\right] &= 
	\left( \frac{(4\pi)^2 }{S_\ep} s_{ab}^\ep \right)^2
	\int \PS{3}(p_r,p_s,K;p_a+p_b) \,
	{\cal J}_{II,FF} \, 
    \delta[\xi_a(s_{ab}-s_{a(rs)}) - \xi_b(s_{ab}-s_{b(rs)})]
\\ &\times
    \xi_a \xi_b
	\frac{1}{x_{a,r} s_{ar}} \frac{1}{x_{b,s} s_{bs}}
	\frac{\omega(ar)}{\omega(a)} \frac{\omega(bs)}{\omega(b)}
	P_{(ar) r}^{(0)}(x_{a,r}; \ep)	
	P_{(bs) s}^{(0)}(x_{b,s}; \ep)
	{\cal F}(x_{a,r}x_{b,s},\xi_a\xi_b)\,.
\esp
\label{eq:ICarbs00IFIF-2}
\eeq
Similarly to the case of the integrated triple collinear 
subtraction term $\left[\IcC{ars}{IFF (0,0)}
(\xi_a,\xi_b;\ep)\right]$ of \eqn{eq:ICars00IFF-2}, the 
integration is not fully fixed by the delta functions in 
\eqn{eq:ps2ars}. 

%
% Double soft-type subtraction terms
%

\subsection{Double soft-type subtraction terms}

%
% \cS{rs}{(0,0)}
%
\subsubsection{\texorpdfstring{$\cS{rs}{(0,0)}$}{Srs00}}
\label{sec:Srs00}

%%%%%
\paragraph{Subtraction term:} 

For two final-state partons $r$ and $s$, either two gluons or a 
same-flavor quark-antiquark pair, we define the double soft 
subtraction term as follows,
\beq
\bsp
&
\cS{r_g s_g}{(0,0)}(p_a,p_b;\mom{}_{m+X+2}) = (8\pi\as\mu^{2\ep})^2
\\&\qquad\times
	\bigg[\frac18 
	\sum_{\substack{i,k,j,\ell \in I \cup F \\ 
    i, k, j, \ell \ne r, s}} 
	\calS_{\ti{i}\ti{k}}(r) \calS_{\ti{j}\ti{\ell}}(s)
	\{\bT_i \bT_k,\bT_j \bT_\ell\}
	\SME{ab,m+X}{(0)}{(\ti{p}_a,\ti{p}_b;\momt{}_{m+X})}
\\& \qquad\qquad
	-\frac14 \CA
	\sum_{\substack{i,k \in I \cup F \\ i, k \ne r, s}} 
	\calS_{\ti{i}\ti{k}}(r,s)
	\bT_i \bT_k
	\SME{ab,m+X}{(0)}{(\ti{p}_a,\ti{p}_b;\momt{}_{m+X})} \bigg]\,,
\esp
\label{eq:Srsgg00}
\eeq
and
\beq
\bsp
&
\cS{r_q s_\qb}{(0,0)}(p_a,p_b;\mom{}_{m+X+2}) = 
(8\pi\as\mu^{2\ep})^2
\\ &\qquad\times
	\frac{1}{s_{rs}^2}\TR
	\sum_{\substack{i,k \in I \cup F \\ i, k \ne r, s}} 
	\left(
	\frac{s_{\ti{i}r} s_{\ti{k}s} + s_{\ti{i}s} s_{\ti{k}r} 
	- s_{\ti{i}\ti{k}} s_{rs}}{s_{\ti{i}(rs)} s_{\ti{k}(rs)}}
	- \frac{s_{\ti{i}r}s_{\ti{i}s}}{s_{\ti{i}(rs)}^2}
	- \frac{s_{\ti{k}r}s_{\ti{k}s}}{s_{\ti{k}(rs)}^2}
	\right)	
	\bT_i \bT_k
	\SME{ab,m+X}{(0)}{(\ti{p}_a,\ti{p}_b;\momt{}_{m+X})}\,,
\esp
\label{eq:Srsqq00}
\eeq
where the reduced matrix elements are obtained by removing the 
final-state partons $r$ and $s$. The set of mapped momenta 
entering the reduced matrix elements is defined below in 
\eqn{eq:srsmap}. In \eqn{eq:Srsgg00}, the eikonal factors 
$\calS_{\ti{i}\ti{k}}(r)$ and $\calS_{\ti{j}\ti{\ell}}(s)$ are 
again written in terms of mapped hard momenta as in 
\eqn{eq:eikfact},
\beq
\calS_{\ti{i}\ti{k}}(r) 
    = \frac{2 s_{\ti{i}\ti{k}}}{s_{\ti{i}r} s_{\ti{k}r}}\,,
\qquad
\calS_{\ti{j}\ti{\ell}}(s) 
    = \frac{2 s_{\ti{j}\ti{\ell}}}{s_{\ti{j}s} s_{\ti{\ell}s}}\,,
\qquad i,k,j,\ell \in I \cup F\,, \quad i, k, j, \ell \ne r, s\,.
\label{eq:eikfact-2}
\eeq 
The non-abelian two-gluon soft function in \eqn{eq:Srsgg00} is given 
by ($i,k\in I \cup F$, $i,k\ne r,s$)
\beq
\calS_{\ti{i}\ti{k}}(r,s) = 
	\calS_{\ti{i}\ti{k}}^{(\mathrm{s.o.})}(r,s)
	+ 4 \frac{s_{\ti{i}r} s_{\ti{k}s} + s_{\ti{i}s} s_{\ti{k}r}}
	{s_{\ti{i}(rs)} s_{\ti{k}(rs)}}
	\left[\frac{1-\ep}{s_{rs}^2} 
	- \frac18 \calS_{\ti{i}\ti{k}}^{(\mathrm{s.o.})}(r,s)\right]
	- \frac{8s_{\ti{i}\ti{k}}}{s_{rs}s_{\ti{i}(rs)}s_{\ti{k}(rs)}}
    \,,
\label{eq:NABsoft}
\eeq
where 
\beq
\calS_{\ti{i}\ti{k}}^{(\mathrm{s.o.})}(r,s) = 
	\calS_{\ti{i}\ti{k}}(s)\Big(
	\calS_{\ti{i}s}(r)
	+ \calS_{\ti{k}s}(r)
	- \calS_{\ti{i}\ti{k}}(r)\Big)
\eeq
is the form of $\calS_{\ti{i}\ti{k}}(r,s)$ in the strongly-ordered 
limit (gluon $s$ is much softer than gluon $r$). Note that for 
$i=k$, \eqn{eq:NABsoft} does not vanish. In fact,
\beq
\calS_{\ti{i}\ti{i}}(r,s) 
    = 8 \frac{s_{\ti{i}r}s_{\ti{i}s}}{s_{\ti{i}(rs)}^2} 
    \frac{1-\ep}{s_{rs}^2}\,.
\eeq
We note that in the color-singlet case ($m=0$), all sums in 
\eqns{eq:Srsgg00}{eq:Srsqq00} run only over the two initial-state 
partons $a$ and $b$. Moreover, the color-charge algebra is 
trivial, since color conservation ($\bT_a+\bT_b=0$) implies 
\beq
\bT_a^2 = \bT_b^2 \equiv \bT^2_{\mathrm{ini}}\,,
\qquad
\bT_a\bT_b = -\bT^2_{\mathrm{ini}}
\label{eq:TaTb}
\eeq
and
\beq
\{\bT_a\bT_b,\bT_a\bT_b\} = 2\left(\bT^2_{\mathrm{ini}}\right)^2\,.
\label{eq:TaTbTaTb}
\eeq
Thus, in the color-singlet case, we find
\beq
\bsp
&
\cS{r_g s_g}{(0,0)}(p_a,p_b;\mom{}_{X+2}) = (8\pi\as\mu^{2\ep})^2
\\&\qquad\times
	\bigg\{ 
	\calS_{\ti{a}\ti{b}}(r) \calS_{\ti{a}\ti{b}}(s)
    \bT^2_{\mathrm{ini}}
	+\frac12
	 \bigg[
     \calS_{\ti{a}\ti{b}}(r,s)
     -\frac12\Big(\calS_{\ti{a}\ti{a}}(r,s)
     +\calS_{\ti{b}\ti{b}}(r,s)\Big)
     \bigg] \CA\bigg\}\bT^2_{\mathrm{ini}}
	\SME{ab,X}{(0)}{(\ti{p}_a,\ti{p}_b;\momt{}_{X})}\,,
\esp
\label{eq:Srsgg00-colorsinglet}
\eeq
and
\beq
\bsp
&
\cS{r_q s_\qb}{(0,0)}(p_a,p_b;\mom{}_{X+2}) = -
(8\pi\as\mu^{2\ep})^2
\\ &\qquad\times
	\frac{2}{s_{rs}^2}\TR
	\left(
	\frac{s_{\ti{a}r} s_{\ti{b}s} + s_{\ti{a}s} s_{\ti{b}r} 
	- s_{\ti{a}\ti{b}} s_{rs}}{s_{\ti{a}(rs)} s_{\ti{b}(rs)}}
	- \frac{s_{\ti{a}r}s_{\ti{a}s}}{s_{\ti{a}(rs)}^2}
	- \frac{s_{\ti{b}r}s_{\ti{b}s}}{s_{\ti{b}(rs)}^2}
	\right)	
    \bT^2_{\mathrm{ini}}
	\SME{ab,X}{(0)}{(\ti{p}_a,\ti{p}_b;\momt{}_{X})}\,.
\esp
\label{eq:Srsqq00-colorsinglet}
\eeq

%%%%%
\paragraph{Momentum mapping:}

The double soft mapping,
\beq
\smp{rs}:\; (p_a, p_b; \mom{}_{m+X+2}) 
    \smap{rs} (\ti{p}_a,\ti{p}_b;\momt{}_{m+X})\,,
\eeq
is defined as ($r,s \in F$),
\beq
\bsp
\ti{p}_a^\mu &= \lambda_{rs} p_a^\mu\,,
\\
\ti{p}_b^\mu &= \lambda_{rs} p_b^\mu\,,
\\
\ti{p}_n^\mu &= {\Lambda(P,\ha{P})^\mu}_{\!\nu}\, p_n^\nu\,,  
\qquad n \in F, \quad n\ne r\,,
\\
\ti{p}_X^\mu &= {\Lambda(P,\ha{P})^\mu}_{\!\nu}\, p_X^\nu\,,
\esp
\label{eq:srsmap}
\eeq
where ${\Lambda(P,\ha{P})^\mu}_{\!\nu}$ is the proper Lorentz 
transformation of \eqn{eq:Lambda_munu} that takes the massive 
momentum $P^\mu$ into a momentum $\ti{P}^\mu$ of the same mass, 
where
\beq
P^\mu = (p_a+p_b)^\mu - p_r^\mu - p_s^\mu\,, 
\qquad
\ti{P}^\mu = (\ti{p}_a + \ti{p}_b)^\mu 
    = \lambda_{rs} (p_a+p_b)^\mu\,.
\label{eq:pphatsoft}
\eeq
The value of $\lambda_{rs}$ is fixed by requiring that 
$P^2 = \ti{P}^2$,
\beq
\lambda_{rs} = \sqrt{1 - \frac{s_{(rs)\ab}}{s_{ab}}  
    + \frac{s_{rs}}{s_{ab}}} \,.
\label{eq:lambdars}
\eeq
Clearly this mapping is simply a generalization of the single soft 
mapping of \eqn{eq:srmap}. As in the single soft case, 
\eqn{eq:srsmap} can be used in the presence of massive final-state 
particles as well as in the color-singlet case. Moreover, in the 
$p_r^\mu \to 0$, $p_s^\mu \to 0$ limit we have $\lambda_{rs} \to 
1$ which implies $\ti{P}^\mu \to P^\mu$. Then, the Lorentz 
transformation becomes the identity and we obtain the $p_r^\mu \to 
0$, $p_s^\mu \to 0$ double soft kinematic configuration.

%%%%%
\paragraph{Phase space convolution:}

Using the identity,
\beq
1 = \frac{\rd^d K}{(2\pi)^{d}} (2\pi)^{d} 
    \delta^{(d)}\left( K+ p_r + p_s - p_a - p_b\right)\, 
    \rd K^2\, \delta_+(K^2 - s_{ab} + s_{(rs)\ab} - s_{rs})\,,
\eeq
with $\delta_+(K^2 - s_{ab} + s_{(rs)\ab} - s_{rs}) = \delta_+(K^2 
- \lambda_{rs}^2 s_{ab})$, the $(m+X+2)$-particle phase space of 
\eqn{eq:psmx2} (with $k=2$) can be factored on partons $r$ and $s$ 
and momentum $K$ through the convolution formula,
\beq
\PS{m+X+2}(\mom{}_{m+X+2};p_a+p_b) 
= \int \frac{ \rd K^2}{2\pi}\, 
    \PS{m+X}(p_a,\ldots, p_{m+2}, p_X; K)\, 
    \PS{3}(p_r, p_s, K; p_a+p_b)\,.
\label{eq:genconvdoubsoft}
\eeq
Inserting then the identity,
\beq
1 = \rd\lambda\, \delta(\lambda - \lambda_{rs}) \,,
\label{eq:idlambdars}
\eeq
and exchanging the momenta $\{p_n\}$ and $p_X$ with the momenta 
$\{\ti{p}_n\}$ and $\ti{p}_X$ through the mapping of 
\eqn{eq:srsmap} lets us write the convolution in 
\eqn{eq:genconvdoubsoft} as,
\beq
\PS{m+X+2}(\mom{}_{m+X+2};p_a+p_b) = 
	\int_{\lammin}^{\lammax} \rd\lam\,
	\PS{m+X}\big(\momt{}_{m+X};\lam (p_a+p_b)\big) \,
	\frac{s_{ab}}{\pi} \lam\,
	\PS{3}(p_r,p_s, K;p_a+p_b)\,,
\label{eq:Srs-map-PSconv}
\eeq
where momentum $K$ is massive with $K^2 = \lam^2 s_{ab}$, and the 
integration limits are given in \eqn{eq:limitslam}. Finally, we note 
that the flux factor $\Phi(p_a\cdot p_b)$ written in terms of mapped 
momenta reads
\beq
\Phi(p_a\cdot p_b) = \frac{\Phi(\ti{p}_a\cdot \ti{p}_b)}{\lam^2}\,,
\label{eq:srsflux}
\eeq
where we have exploited \eqn{eq:idlambdars} to set $\lam_{rs} = 
\lam$. Notice that \eqn{eq:srsflux} is formally the same as 
\eqn{eq:srflux}, however the precise meaning of the mapped momenta 
is different in the two equations.

%%%%%
\paragraph{Integrated subtraction term for double soft gluon 
emission:} 

Using the definitions of the subtraction term and the phase space 
convolution, \eqns{eq:Srsgg00}{eq:Srs-map-PSconv}, we write the 
integrated counterterm for double soft gluon emission as
\beq
\bsp
& 
\int_2{\cal N} \frac{1}{\omega(a)\omega(b)\Phi(p_a\cdot p_b)} 
	\PS{m+X+2}(\mom{}_{m+X+2};p_a+p_b)
	\cS{r_g s_g}{(0,0)}(p_a,p_b;\mom{}_{m+X+2})
\\ &\quad=
	\left[\frac{\as}{2\pi} S_\ep 
    \left(\frac{\mu^2}{s_{ab}}\right)^{\ep}\right]^2
	\Big(\left[\IcS{g g}{(0,0)}\right] \otimes 
    \dsig{B}_{ab,m+X}\Big)\,,
\esp
\label{eq:Int_Srsgg00-fin}
\eeq
where 
\beq
\bsp
\left[\IcS{g g}{(0,0)}\right] \otimes \dsig{B}_{ab,m+X} = 
	\int_{\lammin}^{\lammax} \rd \lam\,
	\bigg[&
	\frac18 \sum_{\substack{i,k,j,\ell \in I \cup F 
    \\ i, k, j, \ell \ne r, s}} 
	\left[\IcS{g g}{(0,0)}(\lam;\ep)\right]^{(ik,j\ell)} 
    \{\bT_i \bT_k,\bT_j \bT_\ell\}
	\dsig{B}_{ab,m+X}(\ti{p}_a, \ti{p}_b)
\\ & 
	-\frac14
	\sum_{\substack{i,k \in I \cup F \\ i, k \ne r, s}} 
	\left[\IcS{g g}{(0,0)}(\lam;\ep)\right]^{(i,k)} 
    \CA \bT_i \bT_k 
	\dsig{B}_{ab,m+X}(\ti{p}_a, \ti{p}_b)\bigg]\,.
\esp
\label{eq:ISgg00}
\eeq
The initial-state momenta entering the Born cross sections in 
\eqn{eq:ISgg00} are $\ti{p}_a = \lam p_a$ and $\ti{p}_b = \lam 
p_b$ (using the delta function in \eqn{eq:idlambdars}) and we have 
introduced the integrated counterterms
\bal
\left[\IcS{g g}{(0,0)}(\lam;\ep)\right]^{(ik,j\ell)} = &
	\left( \frac{(4\pi)^2 }{S_\ep} s_{ab}^\ep \right)^2 
	\int  \PS{3}(p_r, p_s, K; p_a+p_b)\, \frac{s_{ab}}{\pi} 
    \lam^3 \calS_{\ti{i}\ti{k}}(r) \calS_{\ti{j}\ti{\ell}}(s)\,,
\label{eq:ISrs0ggikjl}
\\
\left[\IcS{g g}{(0,0)}(\lam;\ep)\right]^{(i,k)} = &
	\left( \frac{(4\pi)^2 }{S_\ep} s_{ab}^\ep \right)^2 
	\int  \PS{3}(p_r, p_s, K; p_a+p_b)\, \frac{s_{ab}}{\pi} 
    \lam^3 \calS_{\ti{i}\ti{k}}(r,s)\,.
\label{eq:ISrs0ggik}
\eal
Of course, for color-singlet production, we only need to evaluate 
\eqns{eq:ISrs0ggikjl}{eq:ISrs0ggik} for $i,j,k,\ell = a,b$. Using 
\eqns{eq:TaTb}{eq:TaTbTaTb}, in the color-singlet case 
\eqn{eq:ISgg00} becomes
\beq
\bsp
&
\left[\IcS{g g}{(0,0)}\right] \otimes \dsig{B}_{ab,X} = 
	\int_{\lammin}^{\lammax} \rd \lam\, 
	\bigg\{
	\left[\IcS{g g}{(0,0)}(\lam;\ep)\right]^{(ab,ab)} 
    \bT^2_{\mathrm{ini}} 
\\ &\qquad
	+ \frac{1}{2}\bigg(
	\left[\IcS{g g}{(0,0)}(\lam;\ep)\right]^{(a,b)}
	- \frac{1}{2}
	\left[\IcS{g g}{(0,0)}(\lam;\ep)\right]^{(a,a)}
    - \frac{1}{2}
    \left[\IcS{g g}{(0,0)}(\lam;\ep)\right]^{(b,b)}
	\bigg) \CA
	\bigg\}\bT^2_{\mathrm{ini}} 
	\dsig{B}_{ab,X}(\ti{p}_a, \ti{p}_b)\,.
\esp
\eeq
%

%%%%%
\paragraph{Integrated subtraction term for double soft 
quark-antiquark emission:} 

Using the definitions of the subtraction term and the phase space 
convolution, \eqns{eq:Srsqq00}{eq:Srs-map-PSconv}, we write the 
integrated counterterm for double soft quark-antiquark emission as
\beq
\bsp
& 
\int_2{\cal N} \frac{1}{\omega(a)\omega(b)\Phi(p_a\cdot p_b)} 
	\PS{m+X+2}(\mom{}_{m+X+2};p_a+p_b)
	\cS{r_q s_\qb}{(0,0)}(p_a,p_b;\mom{}_{m+X+2})
\\ &\quad=
	\left[\frac{\as}{2\pi} S_\ep 
    \left(\frac{\mu^2}{s_{ab}}\right)^{\ep}\right]^2
	\Big(\left[\IcS{q \qb}{(0,0)}\right] \otimes 
    \dsig{B}_{ab,m+X}\Big)\,,
\esp
\label{eq:Int_Srsqq00-fin}
\eeq
where 
\beq
\left[\IcS{q \qb}{(0,0)}\right] \otimes \dsig{B}_{ab,m+X} =
	\int_{\lammin}^{\lammax} \rd \lam\,
	\sum_{\substack{i,k \in I \cup F \\ i, k \ne r, s}} 
	\left[\IcS{q \qb}{(0,0)}(\lam;\ep)\right]^{(i,k)}
    \TR \bT_i \bT_k
	\dsig{B}_{ab,m+X}(\ti{p}_a,\ti{p}_b)\,.
\eeq
Once more, the initial-state momenta are $\ti{p}_a = \lam p_a$ and 
$\ti{p}_b = \lam p_b$ and we introduced the integrated counterterm
\beq
\bsp
\left[\IcS{q \qb}{(0,0)}(\lam;\ep)\right]^{(i,k)} &=
	\left( \frac{(4\pi)^2 }{S_\ep} s_{ab}^\ep \right)^2 
	\int  \PS{3}(p_r, p_s, K; p_a+p_b)\, \frac{s_{ab}}{\pi} 
    \lam^3
\\ &\times
	\frac1{s_{rs}^2}
	\left(
	\frac{s_{\ti{i}r} s_{\ti{k}s} + s_{\ti{i}s} s_{\ti{k}r} 
	- s_{\ti{i}\ti{k}} s_{rs}}{s_{\ti{i}(rs)} s_{\ti{k}(rs)}}
	- \frac{s_{\ti{i}r}s_{\ti{i}s}}{s_{\ti{i}(rs)}^2}
	- \frac{s_{\ti{k}r}s_{\ti{k}s}}{s_{\ti{k}(rs)}^2}
	\right) \,.
\label{eq:ISrsqq00}
\esp
\eeq
In the color-singlet case, we need to evaluate \eqn{eq:ISrsqq00} 
only for $i,k = a,b$ and exploiting \eqn{eq:TaTb} we find
\beq
\bsp
\left[\IcS{q \qb}{(0,0)}\right] \otimes \dsig{B}_{ab,X} &= 
	-2 \int_{\lammin}^{\lammax} \rd \lam\, 
	\left[\IcS{q \qb}{(0,0)}(\lam;\ep)\right]^{(a,b)} 
    \TR \bT^2_{\mathrm{ini}} 
	\dsig{B}_{ab,X}(\ti{p}_a, \ti{p}_b)\,.
\esp
\eeq
%

%
% \cC{ars}{IFF}\cS{rs}{(0,0)}
%

\subsubsection{
\texorpdfstring{$\cC{ars}{IFF}\cS{rs}{(0,0)}$}{CarsIFFSrs00}}
\label{sec:CarsIFFSrs00}

%%%%%
\paragraph{Subtraction term:} 

For an initial-state parton $a$ and final-state partons $r$ and 
$s$ which can be either two gluons or a same-flavor 
quark-antiquark pair, we define the triple collinear-double soft 
overlapping subtraction term as follows,
\beq
\bsp
\cC{ars}{IFF}\cS{rs}{(0,0)}(p_a,p_b;\mom{}_{m+X+2}) &= 
(8\pi\as\mu^{2\ep})^2
	C_{ars}^{IFF}\!S_{rs}^{(0)}(x_{\ti{a},rs}, x_{r,\ti{a}}, 
    x_{s,\ti{a}}, s_{\ti{a}r}, s_{\ti{a}s}, s_{rs};\ep)
\\ & \times    
	\bT_a^2
	 \SME{ab,m+X}{(0)}{(\ti{p}_a,\ti{p}_b;\momt{}_{m+X})}\,,
\label{eq:CarsIFFSrs00}
\esp
\eeq
where the reduced matrix element is obtained by removing the 
final-state partons $r$ and $s$, while the set of mapped momenta 
is defined in \eqn{eq:srsmap}. The function 
$C_{ars}^{IFF}\!S_{rs}^{(0)}(x_{\ti{a},rs}, x_{r,\ti{a}}, 
x_{s,\ti{a}}, s_{\ti{a}r}, s_{\ti{a}s}, s_{rs};\ep)$ in 
\eqn{eq:CarsIFFSrs00} depends on the flavors of partons $r$ and 
$s$. When both $r$ and $s$ are gluons, we have
\beq
\bsp
&
C_{a r_g s_g}^{IFF}\!S_{r_g s_g}^{(0)}
(x_{\ti{a},rs}, x_{r,\ti{a}}, x_{s,\ti{a}}, s_{\ti{a}r}, 
s_{\ti{a}s}, s_{rs};\ep) =
	\bT_a^2 
	\frac{4}{s_{\ti{a}r} s_{\ti{a}s} x_{r,\ti{a}} x_{s,\ti{a}}}
\\&\qquad
	+ \CA \bigg[
	\frac{(1 - \ep)}{s_{\ti{a}(rs)} s_{rs}} 
	\frac{(s_{\ti{a}r} x_{s,\ti{a}} - s_{\ti{a}s} x_{r,\ti{a}})^2}
	{s_{\ti{a}(rs)} s_{rs} (x_{r,\ti{a}} + x_{s,\ti{a}})^2}
	- \frac{1}{s_{\ti{a}(rs)} s_{rs}} 
	\bigg(\frac{4}{x_{r,\ti{a}} + x_{s,\ti{a}}} 
    - \frac{1}{x_{r,\ti{a}}}\bigg)
	- \frac{1}{s_{\ti{a}(rs)} s_{\ti{a}r}} 
	\frac{2}{x_{r,\ti{a}} (x_{r,\ti{a}} + x_{s,\ti{a}}) }
\\& \qquad\qquad\quad
	- \frac{1}{s_{\ti{a}(rs)} s_{\ti{a}s}}
	\frac{1}{x_{r,\ti{a}} (x_{r,\ti{a}} + x_{s,\ti{a}})} 
	+ \frac{1}{s_{\ti{a}r} s_{rs}} 
	\bigg(\frac{1}{x_{s,\ti{a}}} 
		+ \frac{1}{x_{r,\ti{a}} + x_{s,\ti{a}}}\bigg)
	+ (r \leftrightarrow s)\bigg] \,, 
\esp
\label{eq:CarsIFFSrs0gg} 
\eeq
while for a same-flavor quark-antiquark pair we find
\beq
C_{a r_q s_\qb}^{IFF}\!S_{r_q s_\qb}^{(0)}
(x_{\ti{a},rs}, x_{r,\ti{a}}, x_{s,\ti{a}}, s_{\ti{a}r}, 
s_{\ti{a}s}, s_{rs};\ep) = 
	\TR
	\frac{2}{s_{\ti{a}(rs)} s_{rs}} \bigg[
	\frac{1}{x_{r,\ti{a}} + x_{s,\ti{a}}}
	- \frac{(s_{\ti{a}r} x_{s,\ti{a}} 
    - s_{\ti{a}s} x_{r,\ti{a}})^2}
	{s_{\ti{a}(rs)} s_{rs} (x_{r,\ti{a}} + x_{s,\ti{a}})^2}
    \bigg]\,.
\label{eq:CarsIFFSrs0qq}
\eeq
The momentum fractions $x_{\ti{a},rs}$, $x_{r,\ti{a}}$ and 
$x_{s,\ti{a}}$ are defined by \eqn{eq:xajk-def}.

%%%%%
\paragraph{Momentum mapping and phase space factorization:}

This subtraction term is defined using the double soft momentum 
mapping of~\sect{sec:Srs00}, where the appropriate phase space 
convolution is also presented.

%%%%%
\paragraph{Integrated subtraction term:}

Using the definitions of the phase space convolution and the 
subtraction term, \eqns{eq:Srs-map-PSconv}{eq:CarsIFFSrs00}, we 
write the integrated counterterm as
\beq
\bsp
& 
\int_2{\cal N} \frac{1}{\omega(a)\omega(b)\Phi(p_a\cdot p_b)} 
	\PS{m+X+2}(\mom{}_{m+X+2};p_a+p_b)
	\cC{ars}{IFF}\cS{rs}{(0,0)}(p_a,p_b;\mom{}_{m+X+2})
\\ &\quad=
	\left[\frac{\as}{2\pi} S_\ep 
    \left(\frac{\mu^2}{s_{ab}}\right)^{\ep}\right]^2
	\Big(\left[\IcC{ars}{IFF}\IcS{rs}{(0,0)}\right] \otimes 
    \dsig{B}_{ab,m+X}\Big)\,,
\esp
\label{eq:Int_CarsIFFSrs00-fin}
\eeq
where 
\beq
\left[\IcC{ars}{IFF}\IcS{rs}{(0,0)}\right] \otimes 
\dsig{B}_{ab,m+X} = 
	\int_{\lammin}^{\lammax} \rd \lam\,
	\left[\IcC{ars}{IFF}\IcS{rs}{(0,0)}(\lam;\ep)\right]
	\dsig{B}_{ab,m+X}(\ti{p}_a,\ti{p}_b)\,.
\eeq
Once more, in the Born cross section $\ti{p}_a = \lam p_a$ and 
$\ti{p}_b = \lam p_b$ and the integrated counterterm is defined as
\beq
\bsp
\left[\IcC{ars}{IFF}\IcS{rs}{(0,0)}(\lam;\ep)\right] &= 
	\left( \frac{(4\pi)^2 }{S_\ep} s_{ab}^\ep \right)^2 
	\int \PS{3}(p_r, p_s, K; p_a+p_b)\, \frac{s_{ab}}{\pi} \lam^3
\\ &\times
	C_{ars}^{IFF}\!S_{rs}^{(0)}(x_{\ti{a},rs}, x_{r,\ti{a}}, 
    x_{s,\ti{a}}, s_{\ti{a}r}, s_{\ti{a}s}, s_{rs};\ep)
	\bT_a^2\,.
\esp
\label{eq:ICarsIFFSrs00}
\eeq
%

%
% \cC{ar,bs}{IF,IF}\cS{rs}{(0,0)}
%

\subsubsection{
\texorpdfstring{$\cC{ar,bs}{IF,IF}\cS{rs}{(0,0)}$}{CarbsIFIFSrs00}}
\label{sec:CarbsIFIFSrs00}

%%%%%
\paragraph{Subtraction term:} 

For initial-state partons $a$ and $b$ and final-state gluons $r$ 
and $s$, we define the double collinear-double soft overlapping 
subtraction term as follows,\footnote{Note that the double soft 
quark-antiquark subtraction term $\cS{r_q s_\qb}{(0,0)}$ in 
\eqn{eq:Srsqq00} does not have a leading singularity in the 
$p_a^\mu \parallel p_r^\mu$, $p_b^\mu \parallel p_s^\mu$ double 
collinear limit.}
\beq
\cC{ar,bs}{IF,IF}\cS{rs}{(0,0)}(p_a,p_b;\mom{}_{m+X+2}) = 
	(8\pi\as\mu^{2\ep})^2
	\frac{2}{s_{\ti{a}r} x_{r,\ti{a}}} \bT_a^2
	\frac{2}{s_{\ti{b}s} x_{s,\ti{b}}} \bT_b^2
	 \SME{ab,m+X}{(0)}{(\ti{p}_a,\ti{p}_b;\momt{}_{m+X})}\,,
\label{eq:CarbsIFIFSrs00}
\eeq
where the reduced matrix element is yet again obtained by dropping 
the final-state partons $r$ and $s$ and the set of mapped momenta 
is given in \eqn{eq:srsmap}. The momentum fractions $x_{r,\ti{a}}$ 
and $x_{s,\ti{b}}$ are defined by \eqn{eq:xja-def}.

%%%%%
\paragraph{Momentum mapping and phase space factorization:}

This subtraction term is defined using the double soft momentum 
mapping of~\sect{sec:Srs00}, where the appropriate phase space 
convolution is also presented.

%%%%%
\paragraph{Integrated subtraction term:}

Using the definitions of the phase space convolution and the 
subtraction term, \eqns{eq:Srs-map-PSconv}{eq:CarbsIFIFSrs00}, we 
write the integrated counterterm as
\beq
\bsp
& 
\int_2{\cal N} \frac{1}{\omega(a)\omega(b)\Phi(p_a\cdot p_b)} 
	\PS{m+X+2}(\mom{}_{m+X+2};p_a+p_b)
	\cC{ar,bs}{IF,IF}\cS{rs}{(0,0)}(p_a,p_b;\mom{}_{m+X+2})
\\ &\quad=
	\left[\frac{\as}{2\pi} S_\ep 
    \left(\frac{\mu^2}{s_{ab}}\right)^{\ep}\right]^2
	\Big(\left[\IcC{ar,bs}{IF,IF}\IcS{rs}{(0,0)}\right] \otimes 
    \dsig{B}_{ab,m+X}\Big)\,,
\esp
\label{eq:Int_CarbsSrsgg00-fin}
\eeq
where 
\beq
\left[\IcC{ar,bs}{IF,IF}\IcS{rs}{(0,0)}\right] \otimes 
\dsig{B}_{ab,m+X} = 
	\int_{\lammin}^{\lammax} \rd \lam\,
	\left[\IcC{ar,bs}{IF,IF}\IcS{rs}{(0,0)}(\lam;\ep)\right]
	\dsig{B}_{ab,m+X}(\ti{p}_a,\ti{p}_b)\,.
\eeq
As before, $\ti{p}_a = \lam p_a$ and $\ti{p}_b = \lam p_b$ and the 
integrated subtraction term is defined as
\beq
\left[\IcC{ar,bs}{IF,IF}\IcS{rs}{(0,0)}(\lam;\ep)\right] = 
	\left( \frac{(4\pi)^2 }{S_\ep} s_{ab}^\ep \right)^2 
	\int \PS{3}(p_r, p_s, K; p_a+p_b)\, \frac{s_{ab}}{\pi} \lam^3
	\frac{2}{s_{\ti{a}r} x_{r,\ti{a}}} \bT_a^2
	\frac{2}{s_{\ti{b}s} x_{s,\ti{b}}} \bT_b^2\,.
\label{eq:ICarbsIFIFSrs00}
\eeq

%%%
%%% Subtraction terms for iterated single unresolved emission
%%%

\section{Subtraction terms for iterated single unresolved emission}
\label{sec:A12}

In this section, we provide the explicit definitions of all 
subtraction terms entering ${\cal A}_{12}^{(0)}$, which was 
defined in \sect{sec:itersingleunresolv}. 

%
% Final-state collinear limit of double collinear-type 
% subtractions
%

\subsection{Final-state collinear limit of double collinear-type 
subtractions}

%
% \cC{ars}{IFF (0,0)}\cC{rs}{FF}
%
\subsubsection{
\texorpdfstring{$\cC{ars}{IFF (0,0)}\cC{rs}{FF}$}{CarsIFF00CrsFF}}
\label{sec:CarsIFF00CrsFF}

%%%%%
\paragraph{Subtraction term:} 

For an initial-state parton $a$ and final-state partons $r$ and 
$s$, we define 
\beq
\bsp
&
\cC{ars}{IFF (0,0)}\cC{rs}{FF}(p_a,p_b;\mom{}_{m+X+2}) = 
(8\pi\as\mu^{2\ep})^2
	\frac{1}{s_{rs}} 
    \frac{1}{x_{\ha{a},\wha{rs}} s_{\ha{a}\wha{rs}}}
\\ &\qquad\times
	\hP^{\mathrm{(C)},(0)}_{(ars)rs}
	(z_{r,s}, \kT{r,s}, \kTh{r,s}, x_{\ha{a},\wha{rs}}, 
    \kT{\wha{rs},\ha{a}}, \ha{p}_a;\ep) 
	\SME{(ars)b,m+X}{(0)}
    {(\ha{\ha{p}}_a,\ha{\ha{p}}_b;\momhh{}_{m+X})} \,,
\esp
\label{eq:CarsIFF00CrsFF}
\eeq
where the reduced matrix element is obtained by removing the 
final-state partons $r$ and $s$ and replacing the initial-state 
parton $a$ with a parton of flavor $(ars)$ using the flavor 
summation rules of \eqn{eq:flsum} and requiring that $a=
(ars)+r+s$, see \fig{fig:Parsflav}. Hence, the flavor of the 
initial-state parton involved in the splitting is generally 
different in the original double real emission matrix element and 
in the reduced matrix element. The set of mapped momenta  entering 
the reduced matrix element is defined below in \eqn{eq:IF-FFmap}. 
The strongly-ordered initial-final-final splitting functions for 
the $a \to a + (rs) \to a + r + s$ splitting, $\hP^{\mathrm{(C)},
(0)}_{(ars) r s}$, can be obtained from the corresponding final-
state strongly-ordered splitting functions introduced 
in~\refr{Somogyi:2005xz} and recalled in \eqnss{eq:CFFPqqbq0FFF}
{eq:CFFPggg0FFF} by the crossing relation discussed in 
\appx{appx:SO-AP-functions}, see \eqn{eq:PCars-crossing}. The 
splitting function is labeled with the flavor $(ars)$ of the 
initial-state parton in the reduced matrix element and the flavors 
$r$ and $s$ of the final-state partons. In particular, the last 
two indices correspond to those of the strongly-ordered pair. The 
momentum fractions entering \eqn{eq:CarsIFF00CrsFF} are defined by 
\eqns{eq:zjk-def}{eq:xja-def}. The various transverse momenta are 
defined as follows. First, $\kT{r,s}^\mu$ is defined as in 
\eqn{eq:ktjk-ansatz}, with coefficients given in 
\eqns{eq:zirri}{eq:Zir} with the relabeling $i\to r$ and $r\to s$. 
Hence, $\kT{r,s}^\mu$ is orthogonal to both $\ha{p}_{rs}^\mu$ and 
$(p_a+p_b)^\mu$. Note that $\kT{r,s}$ appears in 
\eqnss{eq:CFFPqqbq0FFF}{eq:CFFPggg0FFF} only in the invariant 
$s_{\ha{i}\kT{r,s}} = 2\ha{p}_i\cdot \kT{r,s}$ or squared as 
$\kT{r,s}^2$, but never with a free index to be contracted with 
the factorized matrix element. After crossing parton $i$ to the 
initial state, $\hP^{\mathrm{(C)},(0)}_{(ars) r s}$ will depend on 
the invariant $s_{\ha{a}\kT{r,s}} = 2\ha{p}_a\cdot \kT{r,s}$. This 
is of note, because $\kT{r,s}^\mu$ is not orthogonal to the parent 
momentum $\ha{\ha{p}}_a^\mu$ in the factorized matrix element. 
Hence, whenever it would appear in the strongly ordered splitting 
function of~\refr{Somogyi:2005xz} with open indices\footnote{The 
strongly ordered triple collinear splitting functions as 
originally derived in~\cite{Somogyi:2005xz} naturally depend only 
on two transverse momenta: the one of the two partons that 
constitute the strongly ordered pair and the one of the strongly 
ordered pair with respect to the third parton.}, it is replaced by 
$\kTh{r,s}^\mu$, which is orthogonal to $\ha{\ha{p}}_a$ by 
construction,
\beq
\kTh{r,s}^\mu = \kT{r,s}^\mu 
	- \frac{\kT{r,s}\cdot \ha{\ha{p}}_a}
    {\ha{\ha{p}}_a \cdot (p_a+p_b)} (p_a+p_b)^\mu\,.
\eeq
Such a modification was justified in detail 
in~\refr{Bolzoni:2010bt} (see their appendix D.1) for the case of 
final-state splitting and the arguments presented there remain 
valid also after crossing. Finally, $\kT{\wha{rs},\ha{a}}^\mu$ is 
given by \eqn{eq:kj-def}. 

%%%%%
\paragraph{Momentum mapping:} 

The IF--FF iterated momentum mapping is defined by the successive 
application of the final-state single collinear momentum mapping 
of \eqn{eq:cirFFmap} and the initial-state single collinear 
momentum mapping of \eqn{eq:cabrIIFmap},
\beq
\cmp{\ha{a}\ha{b},\wha{rs}}{II,F} \circ \cmp{rs}{FF}:\; 
	(p_a,p_b;\mom{}_{m+X+2}) \cmap{rs}{FF} 
	(\ha{p}_a, \ha{p}_b; \momh{}_{m+X+1}) 
    \cmap{\ha{a}\ha{b},\wha{rs}}{II,F} 
	(\ha{\ha{p}}_a,\ha{\ha{p}}_b;\momhh{}_{m+X})\,.
\eeq
Specifically, we first define the set of momenta $(\ha{p}_a, 
\ha{p}_b;\momh{}_{m+X+1})$ via the $\cmp{rs}{FF}$ mapping of 
\eqn{eq:cirFFmap}
\beq
\bsp
\ha{p}_a^\mu &= (1-\al_{rs})p_a^\mu\,,
\\
\ha{p}_b^\mu &= (1-\al_{rs})p_b^\mu\,,
\\
\ha{p}_{rs}^\mu &= p_r^\mu + p_s^\mu - \al_{rs} (p_a+p_b)^\mu\,,
\\
\ha{p}_n^\mu &= p_n^\mu\,, \qquad n\in F\,,\quad n \ne r,s\,,
\\
\ha{p}_X^\mu &= p_X^\mu\,,
\esp
\label{eq:crsFFmap}
\eeq
with $\al_{rs}$ given in \eqn{eq:Cir0FF_al} with the relabeling 
$i\to r$ and $r\to s$.
Then, the set $(\ha{\ha{p}}_a,\ha{\ha{p}}_b;\momhh{}_{m+X})$ is 
constructed using the $\cmp{\ha{a}\ha{b},\wha{rs}}{II,F}$ mapping 
of \eqn{eq:cabrIIFmap}, starting from the momenta in 
\eqn{eq:crsFFmap},
\beq
\bsp
\ha{\ha{p}}_a^\mu &= \xi_{\ha{a},\wha{rs}} \ha{p}_a^\mu\,,
\\
\ha{\ha{p}}_b^\mu &= \xi_{\ha{b},\wha{rs}} \ha{p}_b^\mu\,,
\\
\ha{\ha{p}}_n^\mu &= {\Lambda(\ha{P},\ha{\ha{P}})^\mu}_{\!\nu}\, 
\ha{p}_n^\nu\,, \qquad n \in F, \quad n\ne \wha{rs}\,,
\\
\ha{\ha{p}}_X^\mu &= {\Lambda(\ha{P},\ha{\ha{P}})^\mu}_{\!\nu}\, 
\ha{p}_X^\nu\,,
\esp
\label{eq:IF-FFmap}
\eeq
where ${\Lambda(\ha{P},\ha{\ha{P}})^\mu}_{\!\nu}$ is the proper 
Lorentz transformation in \eqn{eq:Lambda_munu}, while
\beq
\ha{P}^\mu = (\ha{p}_a+\ha{p}_b)^\mu - \ha{p}_{rs}^\mu\,,
\qquad
\ha{\ha{P}}^\mu = (\ha{\ha{p}}_a+\ha{\ha{p}}_b)^\mu 
	= (\xi_{\ha{a},\wha{rs}} \ha{p}_a + \xi_{\ha{b},\wha{rs}} 
    \ha{p}_b)^\mu
	= (1-\al_{rs})(\xi_{\ha{a},\wha{rs}} p_a 
    + \xi_{\ha{b},\wha{rs}} p_b)^\mu\,.
\eeq
The quantities $\xi_{\ha{a},\wha{rs}}$ and $\xi_{\ha{b},\wha{rs}}$ 
are computed as in \eqn{eq:daleo} but with hatted momenta, i.e., 
with the replacement $p_a \to \ha{p}_a$, $p_b \to \ha{p}_b$, 
$Q\to \ha{Q}=\ha{p}_a+\ha{p}_b$ and $p_r \to \ha{p}_{rs}$. Notice that 
we use the double hat notation to indicate that the final set of 
momenta are obtained via the successive application of two 
collinear-type mappings.

%%%%%
\paragraph{Phase space convolution:} 

Using the appropriate convolution formulae for the two mappings, 
\eqns{eq:CirFF-map-PSconv}{eq:CarIF-map-PSconv}, we obtain the 
following representation of the $(m+X+2)$-particle phase space of 
\eqn{eq:psmx2} (with $k=2$),
\beq
\bsp
&
\PS{m+X+2}(\mom{}_{m+X+2};p_a+p_b) =
\\&\qquad = 
	\int_{\almin}^{\almax} \rd\al\, 
	 \int \rd\ha{\xi}_a\, \rd\ha{\xi}_b\,
	\PS{m+X}\big(\momhh{}_{m+X}; (1-\al)
    (\ha{\xi}_{a}p_a+\ha{\xi}_{b}p_b)\big)
	\PS{II,F}(\ha{p}_{rs},\ha{\xi}_a,\ha{\xi}_b)
	\frac{s_{\wha{rs}\ab}}{2\pi}
	\PS{2}(p_r,p_s; K)\,,
\esp
\label{eq:IF-FFmap-PSconv} 
\eeq
where $\PS{II,F}$ is defined by \eqn{eq:ps2ar} and $K = 
\ha{p}_{rs}+\al(p_a+p_b)$. The limits for the $\al$ integration 
are the same as in \eqn{eq:alminmax}, while the $\ha{\xi}_a$ and 
$\ha{\xi}_b$ integrations run from 0 to 1 with the constraints
\beq
(\ha{\xi}_a \ha{\xi}_b)_{\rm min} = \frac{M^2}{(1-\al)^2s_{ab}}\,,
\qquad
(\ha{\xi}_a \ha{\xi}_b)_{\rm max} = 1\,.
\eeq
Finally, the flux factor $\Phi(p_a\cdot p_b)$ can be written in 
terms of the mapped momenta as follows,
\beq
\Phi(p_a\cdot p_b) = \frac{\Phi(\ha{\ha{p}}_a\cdot \ha{\ha{p}}_b)}
{(1-\alpha)^2\ha{\xi}_a\ha{\xi}_b}\,.
\label{eq:cFFcIFFflux}
\eeq
%

%%%%%
\paragraph{Integrated subtraction term:} 

Using the definitions of the subtraction term and the phase space 
convolution, \eqns{eq:CarsIFF00CrsFF}{eq:IF-FFmap-PSconv}, we 
write the integrated counterterm as 
\beq
\bsp
&
\int_2{\cal N} \frac{1}{\omega(a)\omega(b)\Phi(p_a\cdot p_b)} 
	\PS{m+X+2}(\mom{}_{m+X+2};p_a+p_b)\,
	\cC{ars}{IFF (0,0)}\cC{rs}{FF}(p_a,p_b;\mom{}_{m+X+2})
\\ &\qquad=
	\left[\frac{\as}{2\pi} S_\ep 
    \left(\frac{\mu^2}{s_{ab}}\right)^\ep\right]^2
	\Big( \left[\IcC{ars}{IFF (0,0)} \IcC{rs}{FF}\right] \otimes 
    \dsig{B}_{(ars)b,m+X}\Big)\,,
\esp
\label{eq:Int_CarsIFF00CrsFF-fin}
\eeq
where 
\beq
\left[\IcC{ars}{IFF (0,0)} \IcC{rs}{FF}\right] \otimes 
\dsig{B}_{(ars)b,m+X} = 
	\int \rd \al\, \int \rd \ha{\xi}_a\,\rd \ha{\xi}_b\,
	\left[\IcC{ars}{IFF (0,0)} \IcC{rs}{FF}(\al,\ha{\xi}_a, 
    \ha{\xi}_b;\ep)\right] 
    \dsig{B}_{(ars)b,m+X}(\ha{\ha{p}}_a,\ha{\ha{p}}_b)\,.
\eeq
The initial-state momenta in the Born cross section are 
$\ha{\ha{p}}_a = (1-\al)\ha{\xi}_a p_a$ and $\ha{\ha{p}}_b = (1-
\al)\ha{\xi}_b p_b$ and we defined the integrated counterterm, 
\beq
\bsp
\left[\IcC{ars}{IFF (0,0)} \IcC{rs}{FF}(\al,\ha{\xi}_a, 
\ha{\xi}_b;\ep)\right] &=  
	\left( \frac{(4\pi)^2 }{S_\ep} s_{ab}^\ep \right)^2
	\int
	\PS{II,F}(\ha{p}_{rs},\ha{\xi}_a,\ha{\xi}_b)
	\frac{s_{\wha{rs}\ab}}{2\pi}
	\PS{2}(p_r,p_s; K)\,	
\\& \times 
	\frac{(1-\alpha)^2 \ha{\xi}_a\ha{\xi}_b}{s_{rs} 
    x_{\ha{a},\wha{rs}} s_{\ha{a}\wha{rs}}} 	
	\frac{\omega(ars)}{\omega(a)}
	\hP^{\mathrm{(C)},(0)}_{(ars) r s}
	(z_{r,s}, \kT{r,s}, \kTh{r,s}, x_{\ha{a},\wha{rs}}, 
    \kT{\wha{rs},\ha{a}}, \ha{p}_a;\ep)\,.
\esp
\label{eq:I_cIFFcFF}	
\eeq
In order to evaluate $\left[\IcC{ars}{IFF (0,0)} \IcC{rs}{FF}
(\al,\ha{\xi}_a, \ha{\xi}_b;\ep)\right]$, we exploit that 
$\kTh{r,s}\cdot \ha{\ha{p}}_a = \kT{\wha{rs},\ha{a}}\cdot 
\ha{\ha{p}}_a =0$ and  perform the azimuthal integration by 
passing to the azimuthally averaged splitting 
functions,\footnote{As noted above, $\kT{r,s}\cdot \ha{\ha{p}}_a 
\ne 0$, so the integration of terms involving $s_{\ha{a}\kT{r,s}}$ 
requires care. Nevertheless, the free indices can all be treated 
in the usual way, through azimuthal averaging.} 
\[
\hP^{\mathrm{(C)},(0)}_{(ars) r s}
	(z_{r,s}, \kT{r,s}, \kTh{r,s}, x_{\ha{a},\wha{rs}}, 
    \kT{\wha{rs},\ha{a}}, \ha{p}_a;\ep)
\to
P^{\mathrm{(C)},(0)}_{(ars) r s}(z_{r,s},   
    x_{\ha{a},\wha{rs}}, s_{\ha{a}\kT{r,s}};\ep)\,,
\]
see the discussion in \appx{appx:SO-AP-functions}. After azimuthal 
integration, the phase space measure $\PS{II,F}
(\ha{p}_{rs},\ha{\xi}_a,\ha{\xi}_b)$ is fully fixed by the delta 
functions in \eqn{eq:ps2ar} and we have (with an obvious 
relabeling of \eqn{eq:ps2arexp})
\beq
\bsp
&
\PS{II,F}(\ha{p}_{rs},\ha{\xi}_a,\ha{\xi}_b)
\\&\qquad= 
	\frac{S_\ep}{8\pi^2}
	\frac{\rd \Omega_{d-2}}{\Omega_{d-2}} 
    \rd \xi_{\ha{a},\wha{rs}}\,\rd \xi_{\ha{b},\wha{rs}}\,
	s_{\ha{a}\ha{b}}^{1-\ep}
	\left[ \frac{\ha{\xi}_a \ha{\xi}_b (1-\ha{\xi}_a^2) 
    (1-\ha{\xi}_b^2)}
	{(\ha{\xi}_a + \ha{\xi}_b)^2} \right]^{-\ep}
	\frac{\ha{\xi}_a \ha{\xi}_b (1+\ha{\xi}_a \ha{\xi}_b)}
    {(\ha{\xi}_a + \ha{\xi}_b)^2}
	\delta(\xi_{\ha{a},\wha{rs}} - \ha{\xi}_a) 
    \delta(\xi_{\ha{b},\wha{rs}} - \ha{\xi}_b)\,.
\esp
\label{eq:CarsCrs_measure}
\eeq
Then, using $s_{\ha{a}\ha{b}} = (1-\al)^2 s_{ab}$, we find
\beq
\bsp
&
\left[\IcC{ars}{IFF (0,0)} \IcC{rs}{FF}(\al,\ha{\xi}_a, 
\ha{\xi}_b;\ep)\right] =  
	\frac{(4\pi)^2 }{S_\ep} s_{ab}^{1+\ep}
	\int \PS{2}(p_r,p_s; K)\, \frac{s_{\wha{rs}\ab}}{\pi} 
    (1-\al)^{4-2\ep}
\\&\qquad\times 
	\left[\frac{\ha{\xi}_a \ha{\xi}_b (1-\ha{\xi}_a^2) 
    (1-\ha{\xi}_b^2)}
	{(\ha{\xi}_a + \ha{\xi}_b)^2} \right]^{-\ep}
	\frac{\ha{\xi}_a^2 \ha{\xi}_b^2 (1+\ha{\xi}_a \ha{\xi}_b)}
    {(\ha{\xi}_a + \ha{\xi}_b)^2}
	\frac{1}{s_{rs} x_{\ha{a},\wha{rs}} s_{\ha{a}\wha{rs}}}
	\frac{\omega(ars)}{\omega(a)}
	P^{\mathrm{(C)},(0)}_{(ars) r s}(z_{r,s}, 
    x_{\ha{a},\wha{rs}}, s_{\ha{a}\kT{r,s}};\ep)\,,
\esp
\label{eq:I_cFFcIFF2}
\eeq
where we have used the delta functions in \eqn{eq:CarsCrs_measure} 
to perform the integrations over $\xi_{\ha{a},\wha{rs}}$ and 
$\xi_{\ha{b},\wha{rs}}$. This implies that $x_{\ha{a},\wha{rs}}$ 
and $s_{\ha{a}\wha{rs}}$ in \eqn{eq:I_cFFcIFF2} above are 
understood to be expressed with $\ha{\xi}_a$ and $\ha{\xi}_b$. In 
particular, we find
\beq
s_{\ha{a}\wha{rs}} = (1-\al)^2 s_{ab} 
    \frac{\ha{\xi}_{a} (1-\ha{\xi}_{b}^2)}
    {\ha{\xi}_{a} + \ha{\xi}_{b}}\,, 
\qquad 
s_{\ha{b}\wha{rs}} = (1-\al)^2 s_{ab} 
    \frac{\ha{\xi}_{b} (1-\ha{\xi}_{a}^2)}
    {\ha{\xi}_{a} + \ha{\xi}_{b}}\,,
\eeq
which implies
\beq
x_{\ha{a},\wha{rs}} = \ha{\xi}_a \ha{\xi}_b\,.
\eeq
%

%
% Initial-state collinear limit of double 
% collinear-type subtractions
%

\subsection{Initial-state collinear limit of double 
collinear-type subtractions}

%
% \cC{asr}{IFF (0,0)}\cC{as}{IF}
%

\subsubsection{
\texorpdfstring{$\cC{asr}{IFF (0,0)}\cC{as}{IF}$}{CasrIFF00CasIF}}
\label{sec:CasrIFF00CasIF}

%%%%%
\paragraph{Subtraction term:} 

For an initial-state parton $a$ and final-state partons $r$ and 
$s$, we define
\beq
\bsp
\cC{asr}{IFF (0,0)}\cC{as}{IF}(p_a,p_b;\mom{}_{m+X+2}) &= 
	(8\pi\as\mu^{2\ep})^2
	\frac{1}{x_{a,s} s_{as} x_{\ha{a},\ha{r}} s_{\ha{a}\ha{r}}}
	\hP^{\mathrm{(C)},(0)}_{(ars) (as)s}
	(x_{a,s}, \kT{s,a}, \kTh{s,a}, x_{\ha{a},\ha{r}}, 
    \kT{\ha{r},\ha{a}}, \ha{p}_r; \ep)
\\&\times
    \SME{(ars)b,m+X}{(0)}
    {(\ha{\ha{p}}_a,\ha{\ha{p}}_b;\momhh{}_{m+X})} {\cal F}
    (x_{\ha{a},\ha{r}},\xi_{\ha{a},\ha{r}}\xi_{\ha{b},\ha{r}}) 
	\,,
\esp
\label{eq:CasrIFF00CasIF} 
\eeq
where once again, the reduced matrix element is obtained by 
dropping the final-state partons $r$ and $s$ and replacing the 
initial-state parton $a$ by a parton of flavor $(ars)$, given by 
the flavor summation rules of \eqn{eq:flsum} such that $a=
(ars)+r+s$, see \fig{fig:Parsflav}. Thus, the flavor of the 
initial-state parton involved in the splitting is generally not 
the same in the original double real emission matrix element and 
the reduced matrix element. The momenta entering the reduced 
matrix element are defined below in \eqn{eq:IF-IFmap}. The 
strongly-ordered initial-final-final splitting functions for the 
$a \to (as) + r \to a + s + r$ splitting, $\hP^{\mathrm{(C)},
(0)}_{(ars) (as) s}$, can be obtained from the corresponding 
final-state strongly-ordered splitting functions introduced 
in~\refr{Somogyi:2005xz} and recalled in 
\eqnss{eq:CFFPqqbq0FFF}{eq:CFFPggg0FFF} by the crossing relation 
discussed in \appx{appx:SO-AP-functions}, see 
\eqn{eq:PCars-crossing-IF}. The strongly-ordered splitting 
function is indexed by the flavors of the initial-state partons 
$(ars)$ and $(as)$, as well as the final-state parton $s$. Once 
more, the last two indices correspond to those of the 
strongly-ordered pair. The momentum fractions entering 
\eqn{eq:CasrIFF00CasIF} are defined by \eqn{eq:xja-def}, while the 
transverse momenta are defined as follows. First, $\kT{s,a}$ and 
$\kT{\ha{r},\ha{a}}$ are given by \eqn{eq:kj-def}. Then, since the 
mapping in \eqn{eq:IF-IFmap} is such that $\ha{\ha{p}}_a$ is 
related to $\ha{p}_a$ by a simple rescaling, $\kT{s,a}$ is also 
orthogonal to $\ha{\ha{p}}_a$. Hence we can simply set
\beq
\kTh{s,a}^\mu = \kT{s,a}^\mu\,.
\eeq
Finally, the function ${\cal F}$ in \eqn{eq:CasrIFF00CasIF} is 
defined as in \eqn{eq:Ffunc}, so that we find
\beq
{\cal F}(x_{\ha{a},\ha{r}},\xi_{\ha{a},\ha{r}}\xi_{\ha{b},\ha{r}})
	= \left( \frac{x_{\ha{a},\ha{r}}}
    {\xi_{\ha{a},\ha{r}}\xi_{\ha{b},\ha{r}}} \right)^2
	= \left( \frac{s_{\ha{a}\hahb}(s_{\ha{a}\ab} - s_{\ha{r}\ab})}
	{s_{\ha{a}\ab}(s_{\ha{a}\hahb} - s_{\ha{r}\hahb})} \right)^2\,.
\label{eq:FCiif}
\eeq
This function must be included in \eqn{eq:CasrIFF00CasIF} for two 
reasons. On the one hand, it is needed in order to reproduce the 
$p_a^\mu \parallel p_s^\mu$ collinear behavior of the triple 
collinear subtraction term $\cC{ars}{IFF (0,0)}$ of 
\eqn{eq:CarsIFF00}. Indeed, it is easy to see that in this limit, 
the ${\cal F}(x_{a,rs},\xi_{a,rs}\xi_{b,rs})$ function in 
\eqn{eq:Fiff} does not go to one. However, it is not difficult to 
show that when $p_a^\mu \parallel p_s^\mu$, the values of ${\cal F}
(x_{\ha{a},\ha{r}},\xi_{\ha{a},\ha{r}}\xi_{\ha{b},\ha{r}})$ and 
${\cal F}(x_{a,rs},\xi_{a,rs}\xi_{b,rs})$ approach the same limit. 
Hence $\cC{asr}{IFF (0,0)}\cC{as}{IF}$ in \eqn{eq:CasrIFF00CasIF} 
is a correct regulator of the triple collinear subtraction term 
$\cC{ars}{IFF (0,0)}$ in the $p_a^\mu \parallel p_s^\mu$ collinear 
limit. On the other hand, ${\cal F}
(x_{\ha{a},\ha{r}},\xi_{\ha{a},\ha{r}}\xi_{\ha{b},\ha{r}})$ also 
plays a role in cancelling an unphysical singularity, this time 
associated with the vanishing of $x_{\ha{a},\ha{r}}$ inside the 
double real emission phase space. Using 
\eqnss{eq:CFFPqqbq0FFF}{eq:CFFPggg0FFF}, one can check that this 
subtraction term contains at most factors of 
$1/x_{\ha{a},\ha{r}}^2$, so the spurious singularities may be 
canceled by including the factor of ${\cal F}
(x_{\ha{a},\ha{r}},\xi_{\ha{a},\ha{r}}\xi_{\ha{b},\ha{r}})$ in 
\eqn{eq:CasrIFF00CasIF}. We note in passing that as $\ha{p}_a^\mu 
\parallel \ha{p}_r^\mu$, we find ${\cal F}
(x_{\ha{a},\ha{r}},\xi_{\ha{a},\ha{r}}\xi_{\ha{b},\ha{r}})\to 1$, 
and $\cC{asr}{IFF (0,0)}\cC{as}{IF}$ correctly regulates the 
single unresolved subtraction term $\cC{as}{IF (0,0)}$ in this 
limit.

%%%%%
\paragraph{Momentum mapping:} 

The IF--IF iterated momentum mapping is defined by successive 
application of the initial-state single collinear momentum mapping 
of \eqn{eq:cabrIIFmap},
\beq
\cmp{\ha{a}\ha{b},\ha{r}}{II,F} \circ \cmp{ab,s}{II,F}:\; 
	(p_a, p_b; \mom{}_{m+X+2}) \cmap{ab,s}{II,F} 
	(\ha{p}_a, \ha{p}_b; \momh{}_{m+X+1}) 
    \cmap{\ha{a}\ha{b},\ha{r}}{II,F} 
	(\ha{\ha{p}}_a,\ha{\ha{p}}_b;\momhh{}_{m+X}) \,.
\label{eq:casIFcarIFmap}
\eeq
Specifically, we first define the set of momenta $(\ha{p}_a, 
\ha{p}_b;\momh{}_{m+X+1})$ via the $\cmp{ab,s}{II,F}$ mapping,
\beq
\bsp
\ha{p}_a^\mu &= \xi_{a,s} p_a^\mu\,,
\\
\ha{p}_b^\mu &= \xi_{b,s} p_b^\mu\,,
\\
\ha{p}_n^\mu &= {\Lambda(P,\ha{P})^\mu}_{\!\nu}\, p_n^\nu\,, 
\qquad n \in F, \quad n\ne s\,,
\\
\ha{p}_X^\mu &= {\Lambda(P,\ha{P})^\mu}_{\!\nu}\, p_X^\nu\,,
\esp
\label{eq:casIFmap}
\eeq
where ${\Lambda(P,\ha{P})^\mu}_{\!\nu}$ is the proper Lorentz 
transformation in \eqn{eq:Lambda_munu} and
\beq
P^\mu = (p_a+p_b)^\mu - p_s^\mu\,,
\qquad
\ha{P}^\mu = (\ha{p}_a+\ha{p}_b)^\mu 
	= (\xi_{a,s} p_a + \xi_{b,s} p_b)^\mu\,,
\eeq
with $\xi_{a,s}$ and $\xi_{b,r}$ defined by \eqn{eq:daleo} with 
the relabeling $r\to s$. Then, the set 
$(\ha{\ha{p}}_a,\ha{\ha{p}}_b;\momhh{}_{m+X})$ is constructed 
using the $\cmp{\ha{a}\ha{b},\ha{r}}{II,F}$ mapping of 
\eqn{eq:cabrIIFmap}, starting from the momenta in 
\eqn{eq:casIFmap},
\beq
\bsp
\ha{\ha{p}}_a^\mu &= \xi_{\ha{a},\ha{r}} \ha{p}_a^\mu\,,
\\
\ha{\ha{p}}_b^\mu &= \xi_{\ha{b},\ha{r}} \ha{p}_b^\mu\,,
\\
\ha{\ha{p}}_n^\mu &= {\Lambda(\ha{P},\ha{\ha{P}})^\mu}_{\!\nu}\, 
\ha{p}_n^\nu\,, \qquad n \in F, \quad n\ne \ha{r}\,,
\\
\ha{\ha{p}}_X^\mu &= {\Lambda(\ha{P},\ha{\ha{P}})^\mu}_{\!\nu}\, 
\ha{p}_X^\nu\,,
\esp
\label{eq:IF-IFmap}
\eeq
where now
\beq
\ha{P}^\mu = (\ha{p}_a+\ha{p}_b)^\mu - \ha{p}_{r}^\mu\,,
\qquad
\ha{\ha{P}}^\mu = (\ha{\ha{p}}_a+\ha{\ha{p}}_b)^\mu 
    = (\xi_{\ha{a},\ha{r}} \ha{p}_a + \xi_{\ha{b},\ha{r}} 
    \ha{p}_b)^\mu
	= (\xi_{\ha{a},\ha{r}} \xi_{a,s} p_a + \xi_{\ha{b},\ha{r}} 
    \xi_{b,s} p_b)^\mu\,.
\eeq
The quantities $\xi_{\ha{a},\ha{r}}$ and $\xi_{\ha{b},\ha{r}}$ are 
computed as in \eqn{eq:daleo} but with hatted momenta, i.e., with 
the replacement $p_a \to \ha{p}_a$, $p_b \to \ha{p}_b$, 
$Q\to \ha{Q}=\ha{p}_a+\ha{p}_b$ and $p_r \to \ha{p}_{r}$.

%%%%%
\paragraph{Phase space convolution:} 

Using the convolution formula for the initial-state single 
collinear mapping, \eqn{eq:CarIF-map-PSconv}, we obtain the 
following representation of the $(m+X+2)$-particle phase space of 
\eqn{eq:psmx2} (with $k=2$),
\beq
\bsp
&
\PS{m+X+2}(\mom{}_{m+X+2};p_a+p_b) =
\\&\qquad = 
	\int \rd\xi_a\, \rd\xi_b\,
	 \int \rd\ha{\xi}_a\, \rd\ha{\xi}_b\,
	\PS{m+X}\big(\momhh{}_{m+X}; 
    \ha{\xi}_a \xi_a p_a+\ha{\xi}_b \xi_b p_b\big)\,
	\PS{II,F}(\ha{p}_{r},\ha{\xi}_a,\ha{\xi}_b)\,
	\PS{II,F}(p_s,\xi_a,\xi_b)\,,
\esp
\label{eq:IF-IFmap-PSconv} 
\eeq
where $\PS{II,F}$ is defined by \eqn{eq:ps2ar}. The limits of 
integration for $\xi_a$ and $\xi_b$ are the same as in 
\eqn{eq:xi1xi2lims}, while the $\ha{\xi}_a$ and $\ha{\xi}_b$ 
integrations run from 0 to 1 with the constraints
\beq
(\ha{\xi}_a \ha{\xi}_b)_{\rm min} 
    = \frac{M^2}{\xi_a \xi_b s_{ab}}\,,
\qquad
(\ha{\xi}_a \ha{\xi}_b)_{\rm max} = 1\,.
\eeq
Finally, the flux factor $\Phi(p_a\cdot p_b)$ can be written in 
terms of the mapped momenta as follows,
\beq
\Phi(p_a\cdot p_b) = \frac{\Phi(\ha{\ha{p}}_a\cdot \ha{\ha{p}}_b)}
{\xi_a\xi_b \ha{\xi}_{a} \ha{\xi}_{b}}\,.
\label{eq:cIFcIFFflux}
\eeq
%

%%%%%
\paragraph{Integrated subtraction term:}

Using the definitions of the subtraction term and the phase space 
convolution, \eqns{eq:CasrIFF00CasIF}{eq:IF-IFmap-PSconv}, we 
write the integrated counterterm as
\beq
\bsp
&
\int_2{\cal N} \frac{1}{\omega(a)\omega(b)\Phi(p_a\cdot p_b)} 
	\PS{m+X+2}(\mom{}_{m+X+2};p_a+p_b)\,
	\cC{asr}{IFF (0,0)}\cC{as}{IF}(p_a,p_b;\mom{}_{m+X+2})
\\ &\qquad=
	\left[\frac{\as}{2\pi} S_\ep 
    \left(\frac{\mu^2}{s_{ab}}\right)^\ep\right]^2
	\Big( \left[\IcC{asr}{IFF (0,0)} \IcC{as}{IF}\right] \otimes 
    \dsig{B}_{(ars)b,m+X}\Big)\,,
\esp
\label{eq:Int_CarsIFF00CasIF-fin}
\eeq
where
\beq
\bsp
&
\left[\IcC{asr}{IFF (0,0)} \IcC{as}{IF}\right] \otimes 
\dsig{B}_{(ars)b,m+X} 
\\ &\qquad=
	\int \rd \xi_a\, \rd \xi_b\, \int \rd \ha{\xi}_a\,\rd 
    \ha{\xi}_b\,
	\left[\IcC{asr}{IFF (0,0)} \IcC{as}{IF}
    (\xi_a,\xi_b,\ha{\xi}_a, \ha{\xi}_b;\ep)\right]
	\dsig{B}_{(ars)b,m+X}(\ha{\ha{p}}_a, \ha{\ha{p}}_b)\,.
\esp
\eeq
Above, the initial-state momenta appearing in the Born cross 
section are given by $\ha{\ha{p}}_a = \xi_a \ha{\xi}_a p_a$ and 
$\ha{\ha{p}}_b = \xi_b \ha{\xi}_b p_b$ and we introduced the 
integrated counterterm,
\beq
\bsp
\left[\IcC{asr}{IFF (0,0)} \IcC{as}{IF}
(\xi_a,\xi_b,\ha{\xi}_a, \ha{\xi}_b;\ep)\right] &=  
	\left( \frac{(4\pi)^2 }{S_\ep} s_{ab}^\ep \right)^2  
	\int \PS{II,F}(\ha{p}_r,\ha{\xi}_a,\ha{\xi}_b)\, \PS{II,F}
    (p_s,\xi_a,\xi_b)
	\frac{\xi_a \xi_b \ha{\xi}_{a} \ha{\xi}_{b}}{x_{a,s} s_{as} 
    x_{\ha{a},\ha{r}} s_{\ha{a}\ha{r}}}
\\&\times 
	\frac{\omega(ars)}{\omega(a)}
	\hP^{\mathrm{(C)} (0)}_{(ars) (as) s}(x_{a,s}, \kT{s,a}, 
    \kTh{s,a}, 
	x_{\ha{a},\ha{r}}, \kT{\ha{r},\ha{a}}, \ha{p}_r;\ep){\cal F}
    (x_{\ha{a},\ha{r}},\xi_{\ha{a},\ha{r}}\xi_{\ha{b},\ha{r}}) \,.
\esp
\label{eq:I_cIFcIFF}
\eeq
In order to evaluate $\left[\IcC{asr}{IFF (0,0)} \IcC{as}{IF}
(\xi_a,\xi_b,\ha{\xi}_a, \ha{\xi}_b;\ep)\right]$, we exploit that 
$\kTh{s,a}\cdot \ha{\ha{p}}_a = \kT{\ha{r},\ha{a}}\cdot 
\ha{\ha{p}}_a = 0$ and perform the azimuthal integration by 
passing to the azimuthally averaged splitting 
functions,\footnote{The transverse momentum, $\kT{s,a}^\mu$ 
appears in the integrand in the invariant $s_{\ha{r}\kT{s,a}}$ and 
the integration of the corresponding terms once more requires 
care. However, the free indices can all be treated in the usual 
way, through azimuthal averaging.}
\[
\hP^{\mathrm{(C)} (0)}_{(ars) (as) s}(x_{a,s}, \kT{s,a}, 
\kTh{s,a}, x_{\ha{a},\ha{r}}, \kT{\ha{r},\ha{a}}, \ha{p}_r;\ep)
\to
P^{\mathrm{(C)} (0)}_{(ars) (as) s}
(x_{a,s}, x_{\ha{a},\ha{r}}, s_{\ha{r}\kT{s,a}};\ep)\,,
\]
see the discussion in \appx{appx:SO-AP-functions}. After azimuthal 
integration, the phase space measures $\PS{II,F}(p_s,\xi_a,\xi_b)$ 
and $\PS{II,F}(\ha{p}_r,\ha{\xi}_{a},\ha{\xi}_{b})$ are fully 
fixed by the delta functions in \eqn{eq:ps2ar} as in 
\eqn{eq:ps2arexp} (with appropriate relabeling when necessary) and 
we can write the complete integration measure appearing in 
\eqn{eq:I_cIFcIFF} above as 
\beq
\bsp
&
\PS{II,F}(p_s,\xi_a,\xi_b)\, 
\PS{II,F}(\ha{p}_r,\ha{\xi}_a,\ha{\xi}_b) = 
	\left(\frac{S_\ep}{8\pi^2}\right)^2 
	\frac{\rd \Omega_{d-2}}{\Omega_{d-2}} \rd \xi_a\,\rd\xi_b\, 
	\frac{\rd \Omega_{d-2}}{\Omega_{d-2}} 
    \rd\ha{\xi}_a\, \rd\ha{\xi}_b\,
	s_{ab}^{1-\ep} s_{\ha{a}\ha{b}}^{1-\ep}
\\&\qquad\times 
 	\left[ \frac{\xi_a \xi_b (1-\xi_a^2) (1-\xi_b^2)}
    {(\xi_a + \xi_b)^2} \right]^{-\ep}
	\frac{\xi_a \xi_b (1+\xi_a \xi_b)}{(\xi_a + \xi_b)^2}
	\left[ \frac{\ha{\xi}_a \ha{\xi}_b (1-\ha{\xi}_a^2) (1-
    \ha{\xi}_b^2)}
	{(\ha{\xi}_a + \ha{\xi}_b)^2} \right]^{-\ep}
	\frac{\ha{\xi}_a \ha{\xi}_b (1+\ha{\xi}_a \ha{\xi}_b)}
    {(\ha{\xi}_a + \ha{\xi}_b)^2} 
\\&\qquad\times
	\delta(\xi_{a,s} - \xi_a) \delta(\xi_{b,s} - \xi_b)
	\delta(\xi_{\ha{a},\ha{r}} - \ha{\xi}_a) 
    \delta(\xi_{\ha{b},\ha{r}} - \ha{\xi}_b)\,.
\esp
\label{eq:CarsCas_measure}
\eeq
Then, using $s_{\ha{a}\ha{b}} = \xi_{a,s}\xi_{b,s} s_{ab}$, we 
find 
\beq
\bsp
&
\left[\IcC{asr}{IFF (0,0)} \IcC{as}{IF}
(\xi_a,\xi_b,\ha{\xi}_a, \ha{\xi}_b;\ep)\right] = 
	4 s_{ab}^2
	\left[ \frac{\xi_a^2 \xi_b^2 (1-\xi_a^2) (1-\xi_b^2)}
    {(\xi_a + \xi_b)^2} \right]^{-\ep}
	\frac{\xi_a^3 \xi_b^3 (1+\xi_a \xi_b)}{(\xi_a + \xi_b)^2} 
	\left[ \frac{\ha{\xi}_a \ha{\xi}_b (1-\ha{\xi}_a^2) (1-
    \ha{\xi}_b^2)}
	{(\ha{\xi}_a + \ha{\xi}_b)^2} \right]^{-\ep}
\\&\qquad\times
	\frac{\ha{\xi}_a^2 \ha{\xi}_b^2 (1+\ha{\xi}_a \ha{\xi}_b)}
    {(\ha{\xi}_a + \ha{\xi}_b)^2} 
	\frac{1}{x_{a,s} s_{as} x_{\ha{a},\ha{r}} s_{\ha{a}\ha{r}}} 
	\frac{\omega(ars)}{\omega(a)}
	P^{\mathrm{(C)} (0)}_{(ars) (as) s}(x_{a,s}, 
    x_{\ha{a},\ha{r}}, s_{\ha{r}\kT{s,a}};\ep)  
	{\cal F}(x_{\ha{a},\ha{r}},\ha{\xi}_a \ha{\xi}_b) \,,
\esp
\label{eq:I_cIFcIFF2}
\eeq
where we have used the delta functions in \eqn{eq:CarsCas_measure} 
to perform the integrations over $\xi_{a,s}$, $\xi_{b,s}$ as well 
as $\xi_{\ha{a},\ha{r}}$ and $\xi_{\ha{b},\ha{r}}$. Thus, 
$s_{as}$, $x_{a,s}$, $s_{\ha{a}\ha{r}}$ and $x_{\ha{a},\ha{r}}$ 
are all understood to be expressed with $\xi_a$, $\xi_b$, 
$\ha{\xi}_a$ and $\ha{\xi}_b$ in \eqn{eq:CarsCas_measure}. In 
particular, $s_{as}$ and $x_{a,s}$ are given by 
\eqns{eq:invertdaleo}{eq:xareval} after a trivial $r\to s$ 
relabeling. The corresponding relations for $s_{\ha{a}\ha{r}}$ and 
$x_{\ha{a},\ha{r}}$ can be obtained through a straightforward, if 
somewhat tedious calculation and we find
\beq
s_{\ha{a}\ha{r}} = \xi_a \xi_b s_{ab} 
\frac{\ha{\xi}_a (1-\ha{\xi}_b^2)}{\ha{\xi}_a + \ha{\xi}_b}\,,
\qquad
s_{\ha{b}\ha{r}} = \xi_a \xi_b s_{ab} 
\frac{\ha{\xi}_b (1-\ha{\xi}_a^2)}{\ha{\xi}_a + \ha{\xi}_b}
\label{eq:invertdaleohat}
\eeq
and
\beq
x_{\ha{a},\ha{r}} = 
    \frac{\xi_a (\ha{\xi}_a+\ha{\xi}_b)
    -\ha{\xi}_a \xi_b(1-\ha{\xi}_b^2)
    -\xi_a \ha{\xi}_b(1-\ha{\xi}_a^2)}
    {\xi_a (\ha{\xi}_a+\ha{\xi}_b)}\,.
\label{eq:xarevalhat}
\eeq
Using this representation, it is not difficult to see that 
$x_{\ha{a},\ha{r}}$ does in fact vanish at certain points inside 
the double real emission phase space (i.e., $x_{\ha{a},\ha{r}}$ 
becomes zero for values of $\xi_a$, $\xi_b$, $\ha{\xi}_a$ and 
$\ha{\xi}_b$ that are strictly greater than zero and less than 
one).

%
%\cC{as,br}{IF,IF (0,0)} \cC{as}{IF}
%

\subsubsection{
\texorpdfstring{$\cC{as,br}{IF,IF (0,0)} \cC{as}{IF}$}
{CasbrIFIF00CasIF}}
\label{sec:CasbrIFIF00CasIF}

%%%%%
\paragraph{Subtraction term:} 

For initial-state partons $a$ and $b$ and final-state partons $r$ 
and $s$ we define 
\beq
\bsp
\cC{as,br}{IF,IF (0,0)} \cC{as}{IF}(p_a,p_b;\mom{}_{m+X+2}) &= 
	(8\pi\as\mu^{2\ep})^2 
    \frac{1}{x_{a,s} s_{as} x_{\ha{b},\ha{r}} s_{\ha{b}\ha{r}}}
	\hP_{(as) s}^{(0)}(x_{a,s},\kT{s,a};\ep)
	\hP_{(br) r}^{(0)}(x_{\ha{b},\ha{r}},\kT{\ha{r},\ha{b}};\ep) 
\\&\times
	\SME{(as)(br),m+X}{(0)}
    {(\ha{\ha{p}}_a,\ha{\ha{p}}_b;\momhh{}_{m+X})} {\cal F}
    (x_{\ha{b},\ha{r}},\xi_{\ha{a},\ha{r}}\xi_{\ha{b},\ha{r}})\,,
\esp
\label{eq:CcsdrIFIF00CcsIF}
\eeq
where the reduced matrix element is obtained by removing the 
final-state partons $r$ and $s$ and replacing the initial-state 
partons $a$ and $b$ by partons of flavors $(as)$ and $(br)$. The 
flavors of partons $(as)$ and $(br)$ are fixed by requiring that 
$a=(as)+s$ and $b=(br)+r$ as in \fig{fig:Parflav}. The set of 
mapped momenta entering the reduced matrix element is defined in 
\eqn{eq:IF-IFmap}. The tree-level initial-final Altarelli-Parisi 
splitting functions are given in \eqnss{eq:Pqg0IF}{eq:Pgg0IF}, while 
the momentum fractions and transverse momenta entering 
\eqn{eq:CcsdrIFIF00CcsIF} are defined by \eqns{eq:xja-def}{eq:kj-def}. 
Finally, the function ${\cal F}$ in \eqn{eq:CcsdrIFIF00CcsIF} is given 
in \eqn{eq:Ffunc}, so that
\beq
{\cal F}(x_{\ha{b},\ha{r}},\xi_{\ha{a},\ha{r}}\xi_{\ha{b},\ha{r}}) 
	= \left(\frac{x_{\ha{b},\ha{r}}}
    {\xi_{\ha{a},\ha{r}} \xi_{\ha{b},\ha{r}}}\right)^2 
	= \left( \frac{s_{\ha{b}\hahb}(s_{\ha{b}\ab} - s_{\ha{r}\ab})}
	{s_{\ha{b}\ab}(s_{\ha{b}\hahb} - s_{\ha{r}\hahb})} \right)^2\,.
\label{eq:FCifif}
\eeq
In fact, ${\cal F}
(x_{\ha{b},\ha{r}},\xi_{\ha{a},\ha{r}}\xi_{\ha{b},\ha{r}})$ is the 
same as the function ${\cal F}
(x_{\ha{a},\ha{r}},\xi_{\ha{a},\ha{r}}\xi_{\ha{b},\ha{r}})$ 
appearing in \eqn{eq:FCiif}, up to a $\ha{a} \leftrightarrow 
\ha{b}$ exchange, and serves the same role. On the one hand, it 
ensures that $\cC{as,br}{IF,IF (0,0)} \cC{as}{IF}$ in 
\eqn{eq:CcsdrIFIF00CcsIF} is a proper regulator of the double 
collinear subtraction term $\cC{as,br}{IF,IF (0,0)}$ of 
\eqn{eq:CarbsIFIF00} (with $r$ and $s$ exchanged). Indeed in the 
$p_a^\mu \parallel p_s^\mu$ limit, the ${\cal F}
(x_{a,s}x_{b,r},\xi_{a,rs}\xi_{b,rs})$ function in $\cC{as,br}
{IF,IF (0,0)}$ (see \eqn{eq:Fifif} with $r$ and $s$ exchanged) 
does not go to one. However, it is straightforward to show that 
its value coincides with the value of ${\cal F}
(x_{\ha{b},\ha{r}},\xi_{\ha{a},\ha{r}}\xi_{\ha{b},\ha{r}})$ in 
this limit. On the other hand,  the inclusion of ${\cal F}
(x_{\ha{b},\ha{r}},\xi_{\ha{a},\ha{r}}\xi_{\ha{b},\ha{r}})$ in 
\eqn{eq:CcsdrIFIF00CcsIF} is necessary to cancel the unphysical 
singularity due to the vanishing of $x_{\ha{b},\ha{r}}$ inside the 
double real emission phase space. We also note that in the 
$\ha{p}_b^\mu \parallel \ha{p}_r^\mu$ limit we have ${\cal F}
(x_{\ha{b},\ha{r}},\xi_{\ha{a},\ha{r}}\xi_{\ha{b},\ha{r}}) \to 1$, 
and $\cC{as,br}{IF,IF (0,0)} \cC{as}{IF}$ correctly regulates the 
single unresolved subtraction term $\cC{as}{IF (0,0)}$ in this 
limit.

%%%%%
\paragraph{Momentum mapping and phase space factorization:}

This subtraction term is defined using the IF--IF iterated 
momentum mapping of~\sect{sec:CasrIFF00CasIF}, where the 
appropriate phase space convolution is also presented.

%%%%%
\paragraph{Integrated subtraction term:}

Using the definitions of the phase space convolution and the 
subtraction term, \eqns{eq:IF-IFmap-PSconv}{eq:CcsdrIFIF00CcsIF}, 
we write the integrated counterterm as
\beq
\bsp
&
\int_2{\cal N} \frac{1}{\omega(a)\omega(b)\Phi(p_a\cdot p_b)} 
	\PS{m+X+2}(\mom{}_{m+X+2};p_a+p_b)\,
	\cC{as,br}{IF,IF (0,0)}\cC{as}{IF}(p_a,p_b;\mom{}_{m+X+2})
\\ &\qquad=
	\left[\frac{\as}{2\pi} S_\ep 
    \left(\frac{\mu^2}{s_{ab}}\right)^\ep\right]^2
	\Big( \left[\IcC{as,br}{IF,IF (0,0)} \IcC{as}{IF}\right] 
    \otimes \dsig{B}_{(as)(br),m+X}\Big)\,,
\esp
\label{eq:Int_CarbsIFIF00CasIF-fin}
\eeq
where
\beq
\bsp
&
\left[\IcC{as,br}{IF,IF (0,0)} \IcC{as}{IF}\right] \otimes 
\dsig{B}_{(ar)(bs),m+X} 
\\&\qquad= 
\int \rd \xi_a\, \rd \xi_b\, \int \rd \ha{\xi}_a\,\rd \ha{\xi}_b\, 
	\left[\IcC{as,br}{IF,IF (0,0)} \IcC{as}{IF}
    (\xi_a,\xi_b,\ha{\xi}_a, \ha{\xi}_b;\ep)\right]
	\dsig{B}_{(as)(br),m+X}(\ha{\ha{p}}_a, \ha{\ha{p}}_b)\,.
\esp
\eeq
Once again, the initial-state momenta in the Born cross section 
are $\ha{\ha{p}}_a = \xi_a \ha{\xi}_a p_a$ and $\ha{\ha{p}}_b = 
\xi_b \ha{\xi}_b p_b$ and we introduced the integrated counterterm,
\beq
\bsp
\left[\IcC{as,br}{IF,IF (0,0)} \IcC{as}{IF}
(\xi_a,\xi_b,\ha{\xi}_a, \ha{\xi}_b;\ep)\right] &=
	\left( \frac{(4\pi)^2 }{S_\ep} s_{ab}^\ep \right)^2  
	\int \PS{II,F}(\ha{p}_r,\ha{\xi}_a,\ha{\xi}_b)\, 
    \PS{II,F}(p_s,\xi_a,\xi_b)\,
	\frac{\xi_a \xi_b \ha{\xi}_a\ha{\xi}_b}{x_{a,s} s_{as} 
    x_{\ha{b},\ha{r}} s_{\ha{b}\ha{r}}}
\\&\times 
	\frac{\omega(as)}{\omega(a)} \frac{\omega(br)}{\omega(b)} 	
	\hP_{(as) s}^{(0)}(x_{a,s},\kT{s,a};\ep)
	\hP_{(br) r}^{(0)}(x_{\ha{b},\ha{r}},\kT{\ha{r},\ha{b}};\ep)
	{\cal F}
    (x_{\ha{b},\ha{r}},\xi_{\ha{a},\ha{r}},\xi_{\ha{b},\ha{r}}) \,.
\esp
\label{eq:I_cIFcIFIF}
\eeq
In order to evaluate $\left[\IcC{as,br}{IF,IF (0,0)} \IcC{as}{IF}
(\xi_a,\xi_b,\ha{\xi}_a, \ha{\xi}_b;\ep)\right]$, we exploit that 
$\kT{s,a}\cdot \ha{\ha{p}}_a = \kT{\ha{r},\ha{b}}\cdot 
\ha{\ha{p}}_b = 0$ and perform the azimuthal integration by 
passing to the azimuthally averaged splitting functions, 
$\hP_{(as) s}^{(0)}(x_{a,s},\kT{s,a};\ep) \to P_{(as) s}^{(0)}
(x_{a,s};\ep)$ and $\hP_{(br) r}^{(0)}
(x_{\ha{b},\ha{r}},\kT{\ha{r},\ha{b}};\ep) \to P_{(br) r}^{(0)}
(x_{\ha{b},\ha{r}};\ep)$, see \eqnss{eq:Pqg-ave-IF}{eq:Pgg-ave-IF}. 
After azimuthal integration, the phase space measures $\PS{II,F}
(p_s,\xi_a,\xi_b)$ and $\PS{II,F}(\ha{p}_r,\ha{\xi}_a,\ha{\xi}_b)$ 
are fully fixed by the delta functions in \eqn{eq:ps2ar} as in 
\eqn{eq:ps2arexp} (with appropriate relabeling when necessary) and 
we can write the complete integration measure appearing in 
\eqn{eq:I_cIFcIFIF} above as in \eqn{eq:CarsCas_measure}. Then, 
using $s_{\ha{a}\ha{b}}=\xi_{a,s}\xi_{b,s}s_{ab}$, we find
\beq
\bsp
&
\left[\IcC{as,br}{IF,IF (0,0)} \IcC{as}{IF}
(\xi_a,\xi_b,\ha{\xi}_a, \ha{\xi}_b;\ep)\right] =
	4 s_{ab}^2 
	\left[ \frac{\xi_a^2 \xi_b^2 (1-\xi_a^2) (1-\xi_b^2)}
    {(\xi_a + \xi_b)^2} \right]^{-\ep}
	\frac{\xi_a^3 \xi_b^3 (1+\xi_a \xi_b)}{(\xi_a + \xi_b)^2} 
	\left[ \frac{\ha{\xi}_a \ha{\xi}_b (1-\ha{\xi}_a^2) (1-
    \ha{\xi}_b^2)}
	{(\ha{\xi}_a + \ha{\xi}_b)^2} \right]^{-\ep}
\\ &\qquad\times    
	\frac{\ha{\xi}_a^2 \ha{\xi}_b^2 (1+\ha{\xi}_a \ha{\xi}_b)}
    {(\ha{\xi}_a + \ha{\xi}_b)^2} 
	\frac{1}{x_{a,s} s_{as} x_{\ha{b},\ha{r}} s_{\ha{b}\ha{r}}} 
	\frac{\omega(as)}{\omega(a)} \frac{\omega(br)}{\omega(b)} 
	P_{(as) s}^{(0)}(x_{a,s};\ep)P_{(br) r}^{(0)}
    (x_{\ha{b},\ha{r}};\ep)
	{\cal F}(x_{\ha{b},\ha{r}},\ha{\xi}_a\ha{\xi}_b)\,,
\esp
\label{eq:I_cIFcIFIF2}
\eeq
where the delta functions in \eqn{eq:CarsCas_measure} have been 
used to perform the integrations over $\xi_{a,s}$, $\xi_{b,s}$ as 
well as $\xi_{\ha{a},\ha{r}}$ and $\xi_{\ha{b},\ha{r}}$. Hence, in 
\eqn{eq:CarsCas_measure}, $x_{a,s}$, $s_{as}$, $x_{\ha{b},\ha{r}}$ 
and $s_{\ha{b}\ha{r}}$ are all expressed in terms of $\xi_a$, 
$\xi_b$, $\ha{\xi}_a$ and $\ha{\xi}_b$. In particular, $s_{as}$ 
and $x_{a,s}$ can be obtained from 
\eqns{eq:invertdaleo}{eq:xareval} after a trivial $r\to s$ 
relabeling, while the corresponding relations for 
$s_{\ha{b}\ha{r}}$ and $x_{\ha{b},\ha{r}}$ are given in 
\eqns{eq:invertdaleohat}{eq:xarevalhat} after the replacement 
$a\to b$. Because of this, it is evident that in addition to 
$x_{\ha{a},\ha{r}}$, $x_{\ha{b},\ha{r}}$ can also vanish inside 
the double real emission phase space.

%
% Soft-type limits of double collinear-type subtractions
%

\subsection{Soft-type limits of double collinear-type subtractions}

%
% \cC{ars}{IFF (0,0)}\cS{s}{}
%

\subsubsection{
\texorpdfstring{$\cC{ars}{IFF (0,0)}\cS{s}{}$}{CarsIFF00Ss}}
\label{sec:CarsIFF00Ss}

%%%%%
\paragraph{Subtraction term:} 

For an initial-state parton $a$, final-state parton $r$ and 
final-state gluon $s$ we define
\beq
\bsp
\cC{ars}{IFF (0,0)}\cS{s}{}(p_a,p_b;\mom{}_{m+X+2}) &= 
	(8\pi\as\mu^{2\ep})^2
	P^{(\mathrm S), (0)}_{(ar) r s}
	(x_{\ti{a},\ti{r}s}, x_{\ti{r},\ti{a}}, x_{s,\ti{a}},
	s_{\ti{a}\ti{r}}, s_{\ti{a}s}, s_{\ti{r}s};\ep)
\\ &\times	
	\frac{1}{x_{\ti{a},\ti{r}} s_{\ti{a}\ti{r}}}
	\hP_{(ar) r}^{(0)}(x_{\ti{a},\ti{r}},\kT{\ti{r},\ti{a}};\ep) 
	\SME{(ar)b,m+X}{(0)}
    {(\ha{\ti{p}}_a,\ha{\ti{p}}_b;\momht{}_{m+X})} \,,
\esp
\label{eq:CarsIFF00Ss}
\eeq
where the reduced matrix element is defined by removing the 
final-state parton $r$ and gluon $s$ and replacing the 
initial-state parton $a$ by a parton of flavor $(ar)$ such that 
$a=(ar)+r$, see \fig{fig:Parflav}. The set of momenta entering the 
reduced matrix element is defined below in \eqn{eq:IF-Smap}. In 
\eqn{eq:CarsIFF00Ss}, $\hP_{(ar) r}^{(0)}$ are once more the 
tree-level initial-final Altarelli-Parisi splitting functions, 
whose explicit expressions are given in 
\eqnss{eq:Pqg0IF}{eq:Pgg0IF}. Moreover, the initial-final-final 
soft functions $P^{(\mathrm S), (0)}_{(ar) r s}$ can be obtained 
from the corresponding final-final-final soft functions introduced 
in~\refr{Somogyi:2005xz} and recalled here in 
\eqnss{eq:SsPfifrq0FFF}{eq:SsPggg0FFF} by the crossing relation 
discussed in \appx{appx:Soft-AP-functions}, see 
\eqn{eq:sapspliIFF}. The momentum fractions entering 
\eqn{eq:CarsIFF00Ss} are defined by 
\eqns{eq:xja-def}{eq:xajk-def}, while the transverse momentum 
$\kT{\ti{r},\ti{a}}$ is given by \eqn{eq:kj-def}.

%%%%%
\paragraph{Momentum mapping:} 

The IF--S iterated momentum mapping is defined by the successive 
application of the single soft momentum mapping of \eqn{eq:srmap} 
and the initial-state single collinear momentum mapping of 
\eqn{eq:cabrIIFmap},
\beq
\cmp{\ha{a}\ha{b},\ha{r}}{II,F} \circ \smp{s}{}:\; 
	(p_a, p_b; \mom{}_{m+X+2}) 
	\smap{s}{} 
	(\ti{p}_a, \ti{p}_b; \momt{}_{m+X+1}) 
	\cmap{\ha{a}\ha{b},\ha{r}}{II,F} 
	(\ha{\ti{p}}_a, \ha{\ti{p}}_b;\momht{}_{m+X}) \,.
\label{eq:sscarIFmap}
\eeq
Specifically, we first define the set of momenta $(\ti{p}_a, 
\ti{p}_b;\momt{}_{m+X+1})$ via the $\smp{s}{}$ mapping,
\beq
\bsp
\ti{p}_a^\mu &= \lam_s p_a^\mu\,,
\\
\ti{p}_b^\mu &= \lam_s p_b^\mu\,,
\\
\ti{p}_n^\mu &= {\Lambda(P,\ti{P})^\mu}_{\!\nu}\, p_n^\nu\,, 
\qquad n \in F, \quad n\ne s\,,
\\
\ti{p}_X^\mu &= {\Lambda(P,\ti{P})^\mu}_{\!\nu}\, p_X^\nu\,,
\esp
\label{eq:ssmap}
\eeq
where ${\Lambda(P,\ti{P})^\mu}_{\!\nu}$ is the proper Lorentz 
transformation in \eqn{eq:Lambda_munu} and
\beq
P^\mu = (p_a+p_b)^\mu - p_s^\mu\,,
\qquad
\ti{P}^\mu = (\ti{p}_a+\ti{p}_b)^\mu 
	= \lam_s(p_a + p_b)^\mu\,,
\eeq
with $\lam_s$ defined by \eqn{eq:lamr-def} with the relabeling 
$r\to s$. Then, the set 
$(\ha{\ti{p}}_a, \ha{\ti{p}}_b;\momht{}_{m+X})$ is constructed 
using the $\cmp{\ha{a}\ha{b},\ha{r}}{II,F}$ mapping of 
\eqn{eq:cabrIIFmap}, starting from the momenta in \eqn{eq:ssmap},
\beq
\bsp
\ha{\ti{p}}_a^\mu &= \xi_{\ti{a},\ti{r}} \ti{p}_a^\mu\,,
\\
\ha{\ti{p}}_b^\mu &= \xi_{\ti{b},\ti{r}} \ti{p}_b^\mu\,,
\\
\ha{\ti{p}}_n^\mu &= {\Lambda(\ti{P},\ha{\ti{P}})^\mu}_{\!\nu}\, 
\ti{p}_n^\nu\,, \qquad n \in F, \quad n\ne \ti{r}\,,
\\
\ha{\ti{p}}_X^\mu &= {\Lambda(\ti{P},\ha{\ti{P}})^\mu}_{\!\nu}\, 
\ti{p}_X^\nu\,,
\esp
\label{eq:IF-Smap}
\eeq
where once again ${\Lambda(\ti{P},\ha{\ti{P}})^\mu}_{\!\nu}$ is 
the proper Lorentz transformation in \eqn{eq:Lambda_munu}, while
\beq
\ti{P}^\mu = (\ti{p}_a+\ti{p}_b)^\mu - \ti{p}_{r}^\mu\,,
\qquad
\ha{\ti{P}}^\mu = (\ha{\ti{p}}_a+\ha{\ti{p}}_b)^\mu 
	= (\xi_{\ti{a},\ti{r}} \ti{p}_a + \xi_{\ti{b},\ti{r}} 
    \ti{p}_b)^\mu
	= \lam_s(\xi_{\ti{a},\ti{r}} p_a + \xi_{\ti{b},\ti{r}} 
    p_b)^\mu\,.
\eeq
The quantities $\xi_{\ti{a},\ti{r}}$ and $\xi_{\ti{b},\ti{r}}$ are 
computed as in \eqn{eq:daleo} but with the tilded momenta of 
\eqn{eq:ssmap}, i.e., with the replacement $p_a \to \ti{p}_a$, 
$p_b \to \ti{p}_b$, $Q\to \ti{Q}=\ti{p}_a+\ti{p}_b$ and $p_r \to 
\ti{p}_{r}$. We use the tilde-hat notation to indicate that the final 
set of momenta is obtained by applying first a soft mapping, followed 
by a collinear one.

%%%%%
\paragraph{Phase space convolution:} 

Using the appropriate convolution formulae for the two mappings, 
\eqns{eq:CarIF-map-PSconv}{eq:Sr-map-PSconv}, we obtain the 
following representation of the $(m+X+2)$-particle phase space of 
\eqn{eq:psmx2} (with $k=2$),
\beq
\bsp
&
\PS{m+X+2}(\mom{}_{m+X+2};p_a+p_b) =
\\&\qquad = 
	\int \rd \lam\,
	 \int \rd\ti{\xi}_a\, \rd\ti{\xi}_b\,
	\PS{m+X}\big(\momht{}_{m+X}; 
    \lam(\ti{\xi}_a p_a+\ti{\xi}_b p_b)\big)
	\PS{II,F}(\ti{p}_{r},\ti{\xi}_a,\ti{\xi}_b)
	\frac{s_{ab}}{\pi} \lam\,
	\PS{2}(p_s,K;p_a+p_b)\,,
\esp
\label{eq:IF-Smap-PSconv} 
\eeq
where $\PS{II,F}$ is defined by \eqn{eq:ps2ar}, while $K$ is a 
massive momentum with mass $K^2 = \lam^2 s_{ab}$. The limits of 
the $\lam$ integration are given in \eqn{eq:limitslam}, while the 
$\ti{\xi}_a$ and $\ti{\xi}_b$ integrations run from 0 to 1 with 
the constraints
\beq
(\ti{\xi}_a \ti{\xi}_b)_{\rm min} = \frac{M^2}{\lam^2 s_{ab}}\,,
\qquad
(\ti{\xi}_a \ti{\xi}_b)_{\rm max} = 1\,.
\eeq
Finally, the flux factor $\Phi(p_a\cdot p_b)$ can be written in 
terms of the mapped momenta as follows,
\beq
\Phi(p_a\cdot p_b) = \frac{\Phi(\ha{\ti{p}}_a\cdot \ha{\ti{p}}_b)}
    {\lsq \ti{\xi}_a \ti{\xi}_b}\,.
\label{eq:ScIFFflux}
\eeq
%

%%%%%
\paragraph{Integrated counterterm:}

Using the definitions of the subtraction term and the phase space 
convolution, \eqns{eq:CarsIFF00Ss}{eq:IF-Smap-PSconv}, we write 
the integrated counterterm as
\beq
\bsp
&
\int_2{\cal N} \frac{1}{\omega(a)\omega(b)\Phi(p_a\cdot p_b)} 
	\PS{m+X+2}(\mom{}_{m+X+2};p_a+p_b)\,
	\cC{ars}{IFF (0,0)}\cS{s}{}(p_a,p_b;\mom{}_{m+X+2})
\\ &\qquad=
	\left[\frac{\as}{2\pi} S_\ep 
    \left(\frac{\mu^2}{s_{ab}}\right)^\ep\right]^2
	\Big( \left[\IcC{ars}{IFF (0,0)} \IcS{s}{}\right] \otimes 
    \dsig{B}_{(ar)b,m+X}\Big)\,,
\esp
\label{eq:Int_CarsIFF00SsFF-fin}
\eeq
where
\beq
\left[\IcC{ars}{IFF (0,0)} \IcS{s}{}\right] \otimes 
\dsig{B}_{(ar)b,m+X} =
\int_{\lammin}^{\lammax} \rd \lam\, \int \rd \ti{\xi}_a\,
    \rd \ti{\xi}_b\, 
	\left[\IcC{ars}{IFF (0,0)} \IcS{s}{}(\lam,\ti{\xi}_a, 
    \ti{\xi}_b;\ep)\right]
	\dsig{B}_{(ar)b,m+X}(\ha{\ti{p}}_a, \ha{\ti{p}}_b)\,.
\eeq
The initial-state momenta in the Born cross section are given by 
$\ha{\ti{p}}_a = \lam \ti{\xi}_a p_a$ and $\ha{\ti{p}}_b = \lam 
\ti{\xi}_b p_b$ and we define the integrated counterterm
\beq
\bsp
\left[\IcC{ars}{IFF (0,0)} \IcS{s}{}
(\lam,\ti{\xi}_a, \ti{\xi}_b; \ep)\right] &=
	\left( \frac{(4\pi)^2 }{S_\ep} s_{ab}^\ep \right)^2 
	\int \PS{II,F}(\ti{p}_{r},\ti{\xi}_a,\ti{\xi}_b)
	\frac{s_{ab}}{\pi}
	\PS{2}(p_s,K;p_a+p_b)
	\frac{\lam^3 \ti{\xi}_a \ti{\xi}_b}
    {x_{\ti{a},\ti{r}} s_{\ti{a}\ti{r}}}
\\&\times 	
	\frac{\omega(ar)}{\omega(a)}
	P^{(\mathrm S), (0)}_{(ar) r s}
	(x_{\ti{a},\ti{r}s}, x_{\ti{r},\ti{a}}, x_{s,\ti{a}},
	s_{\ti{a}\ti{r}}, s_{\ti{a}s}, s_{\ti{r}s};\ep)
	\hP_{(ar) r}^{(0)}(x_{\ti{a},\ti{r}},\kT{\ti{r},\ti{a}};\ep)\,.
\esp
\label{eq:I_ScIFF}
\eeq
In order to evaluate $\left[\IcC{ars}{IFF (0,0)} \IcS{s}{}
(\lam,\ti{\xi}_a, \ti{\xi}_b; \ep)\right]$, we exploit that 
$\kT{\ti{r},\ti{a}}\cdot \ha{\ti{p}}_a = 0$ and perform the 
azimuthal integration by passing to the azimuthally averaged 
splitting functions $\hP_{(ar) r}^{(0)}
(x_{\ti{a},\ti{r}},\kT{\ti{r},\ti{a}};\ep) \to P_{(ar) r}^{(0)}
(x_{\ti{a},\ti{r}};\ep)$ of \eqnss{eq:Pqg-ave-IF}{eq:Pgg-ave-IF}. 
After azimuthal integration, the phase space measure $\PS{II,F}
(\ti{p}_{r},\ti{\xi}_a,\ti{\xi}_b)$ is fully fixed by the delta 
functions in \eqn{eq:ps2ar} and can be written (with a trivial 
relabeling of \eqn{eq:ps2arexp}) as
\beq
\bsp
&
\PS{II,F}(\ti{p}_{r},\ti{\xi}_a,\ti{\xi}_b)
\\&\qquad= 
	\frac{S_\ep}{8\pi^2} 
    \frac{\rd\Omega_{d-2}}{\Omega_{d-2}}
    \rd \xi_{\ti{a},\ti{r}}\, 
    \rd \xi_{\ti{b},\ti{r}}\,
	s_{\ti{a}\ti{b}}^{1-\ep}  
	\left[\frac{\ti{\xi}_a \ti{\xi}_b (1-\ti{\xi}_a^2) 
    (1-\ti{\xi}_b^2)}
	{(\ti{\xi}_a + \ti{\xi}_b)^2} \right]^{-\ep}
	\frac{\ti{\xi}_a \ti{\xi}_b (1+\ti{\xi}_a \ti{\xi}_b)}
    {(\ti{\xi}_a + \ti{\xi}_b)^2} 
	\delta(\xi_{\ti{a},\ti{r}} - \ti{\xi}_a) 
    \delta(\xi_{\ti{b},\ti{r}} - \ti{\xi}_b)\,.
\esp
\label{eq:int2pssiff}
\eeq
We then find 
\beq
\bsp
&
\left[\IcC{ars}{IFF (0,0)} \IcS{s}{}
(\lam,\ti{\xi}_a, \ti{\xi}_b; \ep)\right] =
	2 \frac{(4\pi)^2 }{S_\ep} s_{ab}^{1+\ep} \int 
    \PS{2}(p_s,K;p_a+p_b)\, 
	\frac{s_{ab}}{\pi} \lam^{5-2\ep}
	\left[\frac{\ti{\xi}_a \ti{\xi}_b (1-\ti{\xi}_a^2) 
    (1-\ti{\xi}_b^2)}
	{(\ti{\xi}_a + \ti{\xi}_b)^2} \right]^{-\ep} 
\\&\qquad\times
	\frac{\ti{\xi}_a^2 \ti{\xi}_b^2 (1+\ti{\xi}_a \ti{\xi}_b)}
    {(\ti{\xi}_a + \ti{\xi}_b)^2}
	\frac{1}{x_{\ti{a},\ti{r}} s_{\ti{a}\ti{r}}}
	\frac{\omega(ar)}{\omega(a)}
	P^{(\mathrm S), (0)}_{(ar) r s}
	(x_{\ti{a},\ti{r}s}, x_{\ti{r},\ti{a}}, x_{s,\ti{a}},
	s_{\ti{a}\ti{r}}, s_{\ti{a}s}, s_{\ti{r}s};\ep)
	P_{(ar) r}^{(0)}(x_{\ti{a},\ti{r}};\ep)\,,
\esp
\label{eq:I_ScIFF2}
\eeq
where we used that $s_{\ti{a}\ti{b}} = \lam^2 s_{ab}$ and 
performed the integrations over $\xi_{\ti{a},\ti{r}}$ and 
$\xi_{\ti{b},\ti{r}}$ using the delta functions in 
\eqn{eq:int2pssiff}. This means that in \eqn{eq:I_ScIFF2} above, 
$x_{\ti{a},\ti{r}}$ and $s_{\ti{a}\ti{r}}$ are understood to be 
expressed with $\ti{\xi}_a$ and $\ti{\xi}_b$. In particular, we 
find
\beq
s_{\ti{a}\ti{r}} 
= \lam^2 s_{ab} \frac{\ti{\xi}_{a} (1-\ti{\xi}_{b}^2)}
    {\ti{\xi}_{a} + \ti{\xi}_{b}}\,, 
\qquad 
s_{\ti{b}\ti{r}} 
= \lam^2 s_{ab} \frac{\ti{\xi}_{b} (1-\ti{\xi}_{a}^2)}
    {\ti{\xi}_{a} + \ti{\xi}_{b}}\,,
\label{eq:invertdaleotilde}
\eeq
which gives
\beq
x_{\ti{a},\ti{r}} = \ti{\xi}_a \ti{\xi}_b\,.
\label{eq:xarevaltilde}
\eeq
%

%
% \cC{ars}{IFF (0,0)}\cC{rs}{FF}\cS{s}{}
%

\subsubsection{
\texorpdfstring{$\cC{ars}{IFF (0,0)}\cC{rs}{FF}\cS{s}{}$}
{CarsIFF00CrsFFSs}}
\label{sec:CarsIFF00CrsFFSs}

%%%%%
\paragraph{Subtraction term:} 

For an initial-state parton $a$, final-state parton $r$ and 
final-state gluon $s$ we define
\beq
\bsp
\cC{ars}{IFF (0,0)} \cC{rs}{FF}\cS{s}{}(p_a,p_b;\mom{}_{m+X+2}) &= 
	(8\pi\as\mu^{2\ep})^2
	\frac{2z_{\ti{r},s}}{s_{\ti{r}s} z_{s,\ti{r}}} \bT_r^2
\\&\times
	\frac{1}{x_{\ti{a},\ti{r}} s_{\ti{a}\ti{r}}}
	\hP_{(ar) r}^{(0)}(x_{\ti{a},\ti{r}},\kT{\ti{r},\ti{a}};\ep)
	\SME{(ar)b,m+X}{(0)}
    {(\ha{\ti{p}}_a,\ha{\ti{p}}_b;\momht{}_{m+X})}\,,
\label{eq:CrsFFSsCars0IFF}
\esp
\eeq
where again the reduced matrix element is obtained by removing the 
final-state parton $r$ and gluon $s$ and replacing the 
initial-state parton $a$ with a parton of flavor $(ar)$, where 
$a=(ar)+r$ as in \fig{fig:Parflav}. The mapped momenta entering 
the factorized matrix element are given by \eqn{eq:IF-Smap}. In 
\eqn{eq:CrsFFSsCars0IFF}, $\hP_{(ar) r}^{(0)}$ are once again the 
tree-level initial-final Altarelli-Parisi splitting functions, 
whose explicit expressions are given in 
\eqnss{eq:Pqg0IF}{eq:Pgg0IF}. The momentum fractions entering 
\eqn{eq:CarsIFF00Ss} are defined by \eqns{eq:zjk-def}{eq:xja-def}, 
while the transverse momentum $\kT{\ti{r},\ti{a}}$ is given by 
\eqn{eq:kj-def}.

%%%%%
\paragraph{Momentum mapping and phase space factorization:}

This subtraction term is defined using the IF--S iterated momentum 
mapping of~\sect{sec:CarsIFF00Ss}, where the appropriate phase 
space convolution is also presented.

%%%%%
\paragraph{Integrated subtraction term:}

Using the definitions of the phase space convolution and the 
subtraction term, \eqns{eq:IF-Smap-PSconv}{eq:CrsFFSsCars0IFF}, we 
write the integrated counterterm as 
\beq
\bsp
&
\int_2{\cal N} \frac{1}{\omega(a)\omega(b)\Phi(p_a\cdot p_b)} 
	\PS{m+X+2}(\mom{}_{m+X+2};p_a+p_b)\,
	\cC{ars}{IFF (0,0)}\cC{rs}{} \cS{s}{}(p_a,p_b;\mom{}_{m+X+2})
\\ &\qquad=
	\left[\frac{\as}{2\pi} S_\ep 
    \left(\frac{\mu^2}{s_{ab}}\right)^\ep\right]^2
	\Big( \left[\IcC{ars}{IFF (0,0)} \IcC{rs}{} \IcS{s}{}\right] 
    \otimes \dsig{B}_{(ar)b,m+X}\Big)\,,
\esp
\label{eq:Int_CarsIFF00CrsFFSs-fin}
\eeq
where
\beq
\left[\IcC{ars}{IFF (0,0)} \IcC{rs}{} \IcS{s}{}\right] \otimes 
\dsig{B}_{(ar)b,m+X} = 
\int_{\lammin}^{\lammax} \rd \lam\, \int \rd \ti{\xi}_a\,
    \rd \ti{\xi}_b\, 
	\left[\IcC{ars}{IFF (0,0)} \IcC{rs}{} \IcS{s}{}
    (\lam,\ti{\xi}_a, \ti{\xi}_b;\ep)\right]
	\dsig{B}_{(ar)b,m+X}(\ha{\ti{p}}_a, \ha{\ti{p}}_b)\,.
\eeq
Once again, the initial-state momenta in the Born cross section 
are $\ha{\ti{p}}_a = \lam \ti{\xi}_a p_a$ and $\ha{\ti{p}}_b = 
\lam \ti{\xi}_b p_b$ and we introduced the integrated counterterm
\beq
\bsp
\left[\IcC{ars}{IFF (0,0)} \IcC{rs}{} \IcS{s}{}
(\lam,\ti{\xi}_a, \ti{\xi}_b; \ep)\right] &=
	\left( \frac{(4\pi)^2 }{S_\ep} s_{ab}^\ep \right)^2 
	\int \PS{II,F}(\ti{p}_{r},\ti{\xi}_a,\ti{\xi}_b)
	\frac{s_{ab}}{\pi}
	\PS{2}(p_s,K;p_a+p_b)
	\frac{1}{s_{\ti{r}s}} 
    \frac{2z_{\ti{r},s}}{z_{s,\ti{r}}}\bT_r^2 
\\&\times
	\frac{\lam^3 \ti{\xi}_a \ti{\xi}_b}
    {x_{\ti{a},\ti{r}} s_{\ti{a}\ti{r}}}
	\frac{\omega(ar)}{\omega(a)}
	\hP_{(ar) r}^{(0)}(x_{\ti{a},\ti{r}},\kT{\ti{r},\ti{a}};\ep)\,.
\esp
\label{eq:I_CrsFFSsCars0IFF}
\eeq
In order to evaluate $\left[\IcC{ars}{IFF (0,0)} \IcC{rs}{} \IcS{r}
{}(\lam,\ti{\xi}_a, \ti{\xi}_b; \ep)\right]$, we exploit 
$\kT{\ti{r},\ti{a}}\cdot \ha{\ti{p}}_a=0$ and perform the 
azimuthal integration by passing to the azimuthally averaged 
splitting functions, $\hP_{(ar) r}^{(0)}
(x_{\ti{a},\ti{r}},\kT{\ti{r},\ti{a}};\ep) \to P_{(ar) r}^{(0)}
(x_{\ti{a},\ti{r}};\ep)$, see 
\eqnss{eq:Pqg-ave-IF}{eq:Pgg-ave-IF}. After azimuthal integration, 
the phase space measure $\PS{II,F}
(\ti{p}_{r},\ti{\xi}_a,\ti{\xi}_b)$ is fully fixed by the delta 
functions in \eqn{eq:ps2ar} and is given in \eqn{eq:int2pssiff}. 
Then, using $s_{\ti{a}\ti{b}} = \lam^2 s_{ab}$, we find
\beq
\bsp
\left[\IcC{ars}{IFF (0,0)} \IcC{rs}{} \IcS{s}{}
(\lam,\ti{\xi}_a, \ti{\xi}_b; \ep)\right] &=
	2 \frac{(4\pi)^2 }{S_\ep} s_{ab}^{1+\ep}
	\int \PS{2}(p_s,K;p_a+p_b)\, \frac{s_{ab}}{\pi} 
    \lam^{5-2\ep}
	\left[\frac{\ti{\xi}_a \ti{\xi}_b (1-\ti{\xi}_a^2) 
    (1-\ti{\xi}_b^2)}{(\ti{\xi}_a + \ti{\xi}_b)^2} \right]^{-\ep} 
\\&\times
	\frac{\ti{\xi}_a^2 \ti{\xi}_b^2 (1+\ti{\xi}_a \ti{\xi}_b)}
    {(\ti{\xi}_a + \ti{\xi}_b)^2}  
	\frac{2z_{\ti{r},s}}{s_{\ti{r}s} z_{s,\ti{r}}}\bT_r^2
	\frac{1}{x_{\ti{a},\ti{r}} s_{\ti{a}\ti{r}}}
	\frac{\omega(ar)}{\omega(a)}
	P_{(ar) r}^{(0)}(x_{\ti{a},\ti{r}};\ep)\,,
\esp
 \label{eq:I_CrsFFSsCars0IFF2}
 \eeq
where the integrations over $\xi_{\ti{a},\ti{r}}$ and 
$\xi_{\ti{b},\ti{r}}$ were performed using the delta functions in 
\eqn{eq:int2pssiff}. Then, $x_{\ti{a},\ti{r}}$ and 
$s_{\ti{a}\ti{r}}$ can be expressed with $\ti{\xi}_a$ and 
$\ti{\xi}_b$ as in \eqns{eq:invertdaleotilde}{eq:xarevaltilde}.

%
% \cC{asr}{IFF (0,0)}\cC{as}{IF}\cS{s}{}
%

\subsubsection{
\texorpdfstring{$\cC{asr}{IFF (0,0)}\cC{as}{IF}\cS{s}{}$}
{CasrIFF00CasIFSs}}
\label{sec:CasrIFF00CasIFSs}

%%%%%
\paragraph{Subtraction term:} 

For an initial-state parton $a$, final-state parton $r$ and 
final-state gluon $s$ we define
\beq
\bsp
\cC{asr}{IFF (0,0)} \cC{as}{IF}\cS{s}{}(p_a,p_b;\mom{}_{m+X+2}) &= 
	(8\pi\as\mu^{2\ep})^2
	\frac{2}{x_{s,\ti{a}} s_{\ti{a}s}} \bT_a^2
\\&\times
	\frac{1}{x_{\ti{a},\ti{r}} s_{\ti{a}\ti{r}}}
	\hP_{(ar) r}^{(0)}(x_{\ti{a},\ti{r}},\kT{\ti{r},\ti{a}};\ep)
	\SME{(ar)b,m+X}{(0)}
    {(\ha{\ti{p}}_a,\ha{\ti{p}}_b;\momht{}_{m+X})}\,,
\label{eq:CasIFSsCars0IFF}
\esp
\eeq
where once more we obtain the reduced matrix element by removing 
the final-state parton $r$ and gluon $s$ and replacing the 
initial-state parton $a$ by a parton of flavor $(ar)$. The flavor 
of parton $(ar)$ is again fixed by requiring that $a=(ar)+r$ as in 
\fig{fig:Parflav}. The mapped momenta entering the reduced matrix 
element are given in \eqn{eq:IF-Smap}. Again, $\hP_{(ar) r}^{(0)}$ 
in \eqn{eq:CasIFSsCars0IFF} are the tree-level initial-final 
Altarelli-Parisi splitting functions of 
\eqnss{eq:Pqg0IF}{eq:Pgg0IF}, while the momentum fractions and 
transverse momentum entering \eqn{eq:CarsIFF00Ss} are defined by 
\eqns{eq:xja-def}{eq:kj-def}.

%%%%%
\paragraph{Momentum mapping and phase space factorization:}

This subtraction term is defined using the IF--S iterated momentum 
mapping of~\sect{sec:CarsIFF00Ss}, where the appropriate phase 
space convolution is also presented.

%%%%%
\paragraph{Integrated subtraction term:}

Using the definitions of the phase space convolution and the 
subtraction term, \eqns{eq:IF-Smap-PSconv}{eq:CasIFSsCars0IFF}, we 
write the integrated counterterm as
\beq
\bsp
&
\int_2{\cal N} \frac{1}{\omega(a)\omega(b)\Phi(p_a\cdot p_b)} 
	\PS{m+X+2}(\mom{}_{m+X+2};p_a+p_b)\,
	\cC{asr}{IFF (0,0)}\cC{as}{IF} \cS{s}{}(p_a,p_b;\mom{}_{m+X+2})
\\ &\qquad=
	\left[\frac{\as}{2\pi} S_\ep 
    \left(\frac{\mu^2}{s_{ab}}\right)^\ep\right]^2
	\Big( \left[\IcC{asr}{IFF (0,0)} \IcC{as}{IF} \IcS{s}{}\right] 
    \otimes \dsig{B}_{(ar)b,m+X}\Big)\,,
\esp
\label{eq:Int_CarsIFF00CasIFSs-fin}
\eeq
where
\beq
\left[\IcC{asr}{IFF (0,0)} \IcC{as}{IF} \IcS{s}{}\right] \otimes 
\dsig{B}_{(ar)b,m+X} = 
\int_{\lammin}^{\lammax} \rd \lam\, \int \rd \ti{\xi}_a\,
    \rd \ti{\xi}_b\, 
	\left[\IcC{asr}{IFF (0,0)} \IcC{as}{IF} \IcS{s}{}
    (\lam,\ti{\xi}_a, \ti{\xi}_b;\ep)\right]
	\dsig{B}_{(ar)b,m+X}(\ha{\ti{p}}_a, \ha{\ti{p}}_b)\,.
\eeq
The initial-state momenta in the Born cross section once again 
read $\ha{\ti{p}}_a = \lam \ti{\xi}_a p_a$ and $\ha{\ti{p}}_b = 
\lam \ti{\xi}_b p_b$ and the integrated counterterm is defined as
\beq
\bsp
\left[\IcC{asr}{IFF (0,0)}\IcC{as}{IF}\IcS{s}{}
(\lam,\ti{\xi}_a, \ti{\xi}_b; \ep)\right] &=
	\left( \frac{(4\pi)^2 }{S_\ep} s_{ab}^\ep \right)^2 
	\int \PS{II,F}(\ti{p}_{r},\ti{\xi}_a,\ti{\xi}_b)
	\frac{s_{ab}}{\pi}
	\PS{2}(p_s,K;p_a+p_b)
	\frac{2}{x_{s,\ti{a}} s_{\ti{a}s}} \bT_a^2
\\&\times
	\frac{\lam^3 \ti{\xi}_a \ti{\xi}_b}
    {x_{\ti{a},\ti{r}} s_{\ti{a}\ti{r}}}
	\frac{\omega(ar)}{\omega(a)}
	\hP_{(ar) r}^{(0)}(x_{\ti{a},\ti{r}},\kT{\ti{r},\ti{a}};\ep)\,.
\esp
\label{eq:I_CasIFSsCars0IFF}
\eeq
In order to evaluate $\left[\IcC{asr}{IFF (0,0)}\IcC{as}{IF}\IcS{s}
{}(\lam,\ti{\xi}_a, \ti{\xi}_b; \ep)\right]$, we exploit 
$\kT{\ti{r},\ti{a}}\cdot \ha{\ti{p}}_a=0$ and perform the 
azimuthal integration by passing to the azimuthally averaged 
splitting functions, $\hP_{(ar) r}^{(0)}
(x_{\ti{a},\ti{r}},\kT{\ti{r},\ti{a}};\ep) \to P_{(ar) r}^{(0)}
(x_{\ti{a},\ti{r}};\ep)$, see 
\eqnss{eq:Pqg-ave-IF}{eq:Pgg-ave-IF}. After azimuthal integration, 
the phase space measure $\PS{II,F}
(\ti{p}_{r},\ti{\xi}_a,\ti{\xi}_b)$ is fully fixed by the delta 
functions in \eqn{eq:ps2ar} and can be written as in 
\eqn{eq:int2pssiff}. Then, using $s_{\ti{a}\ti{b}} = \lam^2 
s_{ab}$, we find
\beq
\bsp
\left[\IcC{asr}{IFF (0,0)}\IcC{as}{IF}\IcS{s}{}
(\lam,\ti{\xi}_a, \ti{\xi}_b; \ep)\right] &=
	2 \frac{(4\pi)^2 }{S_\ep} s_{ab}^{1+\ep}
	\int \PS{2}(p_s,K;p_a+p_b)\, \frac{s_{ab}}{\pi} 
    \lam^{5-2\ep}
	\left[\frac{\ti{\xi}_a \ti{\xi}_b (1-\ti{\xi}_a^2) 
    (1-\ti{\xi}_b^2)}{(\ti{\xi}_a + \ti{\xi}_b)^2} \right]^{-\ep} 
\\&\times
	\frac{\ti{\xi}_a^2 \ti{\xi}_b^2 (1+\ti{\xi}_a \ti{\xi}_b)}
    {(\ti{\xi}_a + \ti{\xi}_b)^2}  
	\frac{2}{x_{s,\ti{a}} s_{\ti{a}s}}\bT_a^2
	\frac{1}{x_{\ti{a},\ti{r}} s_{\ti{a}\ti{r}}}
	\frac{\omega(ar)}{\omega(a)}
	P_{(ar) r}^{(0)}(x_{\ti{a},\ti{r}};\ep)\,.
\esp
\label{eq:I_CasIFSsCars0IFF2}
\eeq
As before, the integrations over $\xi_{\ti{a},\ti{r}}$ and 
$\xi_{\ti{b},\ti{r}}$ were performed using the delta functions in 
\eqn{eq:int2pssiff}, while $x_{\ti{a},\ti{r}}$ and 
$s_{\ti{a}\ti{r}}$ can be written in terms of $\ti{\xi}_a$ and 
$\ti{\xi}_b$ as in \eqns{eq:invertdaleotilde}{eq:xarevaltilde}.

%
% Final-state collinear limit of double soft-type 
% subtractions
%

\subsection{Final-state collinear limit of double soft-type 
subtractions}

%
% \cS{rs}{(0,0)}\cC{rs}{FF}
%

\subsubsection{
\texorpdfstring{$\cS{rs}{(0,0)}\cC{rs}{FF}$}{Srs00CrsFF}}
\label{sec:Srs00CrsFF}

%%%%%
\paragraph{Subtraction term:} 

For two final-state partons $r$ and $s$, either two gluons or a 
same-flavor quark-antiquark pair, we define
\beq
\bsp
&\cS{rs}{(0,0)}\cC{rs}{FF}(p_a,p_b;\mom{}_{m+X+2}) = 
	(8\pi\as\mu^{2\ep})^2
\\&\qquad\times	
	\sum_{\substack{i,k\in I \cup F \\ i,k\ne r,s}}
	\frac{1}{2} \calS_{\ti{\ha{i}}\ti{\ha{k}}}^{\mu \nu}(\wha{rs})
	\frac{1}{s_{rs}} 
    \la \mu |\hP^{(0)}_{r s}(z_{r,s},\kT{r,s};\ep)|\nu \ra
	\bT_i \bT_k
	\SME{ab,m+X}{(0)}
    {(\ti{\ha{p}}_a,\ti{\ha{p}}_b;\momth{}_{m+X})}\,,
\esp
\label{eq:CrsFFSrs0}
\eeq
where the reduced matrix element is obtained by simply removing 
the final-state partons $r$ and $s$. The set of momenta entering 
the reduced matrix element is defined below in \eqn{eq:S-FFmap}. 
In \eqn{eq:CrsFFSrs0}, the uncontracted eikonal factor 
$\calS_{\ti{\ha{i}}\ti{\ha{k}}}^{\mu \nu}(\wha{rs})$ is defined as
\beq
\calS_{\ti{\ha{i}}\ti{\ha{k}}}^{\mu \nu}(\wha{rs}) = 
	4 \frac{\ti{\ha{p}}_i^\mu \ti{\ha{p}}_k^\nu}
    {s_{\ti{\ha{i}}\wha{rs}} s_{\ti{\ha{k}}\wha{rs}}}\,,
	\label{eq:uneik}
\eeq
while $\hP^{(0)}_{r s}(z_{r,s},\kT{r,s};\ep)$ is the final-final 
single unresolved Altarelli-Parisi splitting kernel for $g \to 
q\qb$ or $g \to gg$ splitting, given in 
\eqns{eq:Pqq0FF}{eq:Pgg0FF}. The momentum fraction $z_{r,s}$ is  
defined by \eqn{eq:zjk-def}, while the transverse momentum 
$\kT{r,s}$ is constructed as in \eqn{eq:ktjk-ansatz}, with 
coefficients given in \eqns{eq:zirri}{eq:Zir} after the relabeling 
$i\to r$ and $r\to s$. Note that for $i=k$, \eqn{eq:CrsFFSrs0} 
does not vanish, since the product of the uncontracted eikonal 
factor with the transverse momentum-dependent part of the 
splitting kernels is non-zero. Of course, for color-singlet 
production, the sums in \eqn{eq:CrsFFSrs0} run only over $a$ and 
$b$. Then using color conservation, \eqn{eq:TaTb}, in this case we 
have simply
\beq
\bsp
&\cS{rs}{(0,0)}\cC{rs}{FF}(p_a,p_b;\mom{}_{X+2}) = 
	(8\pi\as\mu^{2\ep})^2
\\&\qquad\times	
    \frac12 \frac{1}{s_{rs}}
    \left(
    \calS_{\ti{\ha{a}}\ti{\ha{a}}}^{\mu \nu}(\wha{rs})
    + \calS_{\ti{\ha{b}}\ti{\ha{b}}}^{\mu \nu}(\wha{rs})
    - 2 \calS_{\ti{\ha{a}}\ti{\ha{b}}}^{\mu \nu}(\wha{rs})
    \right)
	\la \mu |\hP^{(0)}_{r s}(z_{r,s},\kT{r,s};\ep)|\nu \ra
	\bT^2_{\mathrm{ini}}
	\SME{ab,X}{(0)}{(\momth{}_{X};\ti{\ha{p}}_a,\ti{\ha{p}}_b)}\,.
\esp
\label{eq:CrsFFSrs0-colorsinglet}
\eeq
%

%%%%%
\paragraph{Momentum mapping:} 

The S--IF iterated momentum mapping is defined by the successive 
application of the final-state single collinear momentum mapping 
of \eqn{eq:cirFFmap} and the single soft momentum mapping of 
\eqn{eq:srmap},
\beq
\smp{\wha{rs}}{} \circ \cmp{rs}{FF}:\; 
	(p_a,p_b;\mom{}_{m+X+2}) \cmap{rs}{FF} 
    (\ha{p}_a,\ha{p}_b;\momh{}_{m+X+1}) \smap{\wha{rs}} 
    (\ti{\ha{p}}_a,\ti{\ha{p}}_b;\momth{}_{m+X})\,.
\eeq
Specifically, we first define the set of momenta $(\ha{p}_a, 
\ha{p}_b;\momh{}_{m+X+1})$ via the $\cmp{rs}{FF}$ mapping given in 
\eqn{eq:cirFFmap},
\beq
\bsp
\ha{p}_a^\mu &= (1-\al_{rs})p_a^\mu\,,
\\
\ha{p}_b^\mu &= (1-\al_{rs})p_b^\mu\,,
\\
\ha{p}_{rs}^\mu &= p_r^\mu + p_s^\mu - \al_{rs} (p_a+p_b)^\mu\,,
\\
\ha{p}_n^\mu &= p_n^\mu\,, \qquad n\in F\,,\quad n \ne r,s\,,
\\
\ha{p}_X^\mu &= p_X^\mu\,,
\esp
\label{eq:crsFFmap-2}
\eeq
with $\al_{rs}$ as in \eqn{eq:Cir0FF_al} after the relabeling 
$i\to r$ and $r\to s$. Then, the set $(\ti{\ha{p}}_a, 
\ti{\ha{p}}_b; \momth{}_{m+X})$ is constructed using the 
$\smp{\wha{rs}}{}$ mapping defined by \eqn{eq:srmap}, starting 
from the momenta in \eqn{eq:crsFFmap-2},
\beq
\bsp
\ti{\ha{p}}_a^\mu &= \lam_{\wha{rs}} \ha{p}_a^\mu\,,
\\
\ti{\ha{p}}_b^\mu &= \lam_{\wha{rs}} \ha{p}_b^\mu\,,
\\
\ti{\ha{p}}_n^\mu &= {\Lambda(\ha{P},\ti{\ha{P}})^\mu}_{\!\nu}\, 
\ha{p}_n^\nu\,, \qquad n \in F, \quad n\ne \wha{rs}\,,
\\
\ti{\ha{p}}_X^\mu &= {\Lambda(\ha{P},\ti{\ha{P}})^\mu}_{\!\nu}\, 
\ha{p}_X^\nu\,,
\esp
\label{eq:S-FFmap}
\eeq
where ${\Lambda(\ha{P},\ti{\ha{P}})^\mu}_{\!\nu}$ is the proper 
Lorentz transformation in \eqn{eq:Lambda_munu}, while
\beq
\ha{P}^\mu = (\ha{p}_a+\ha{p}_b)^\mu - \ha{p}_{rs}^\mu\,,
\qquad
\ti{\ha{P}}^\mu = (\ti{\ha{p}}_a+\ti{\ha{p}}_b)^\mu 
	= \lam_{\wha{rs}}(\ha{p}_a + \ha{p}_b)^\mu
	= \lam_{\wha{rs}}(1-\al_{rs})(p_a + p_b)^\mu\,.
\eeq
The quantity $\lam_{\wha{rs}}$ is computed as in \eqn{eq:lamr-def} 
but with hatted momenta, i.e., with the replacement $p_a \to 
\ha{p}_a$, $p_b \to \ha{p}_b$, $Q\to \ha{Q}=\ha{p}_a+\ha{p}_b$ and 
$p_r \to \ha{p}_{rs}$. The hat-tilde notation is used to indicate that 
the final set of mapped momenta is obtained by applying a collinear 
mapping first, followed by a soft one.

%%%%%
\paragraph{Phase space convolution:} 

Using the appropriate convolution formulae for the two mappings, 
\eqns{eq:CirFF-map-PSconv}{eq:Sr-map-PSconv}, we obtain the 
following representation of the $(m+X+2)$-particle phase space of 
\eqn{eq:psmx2} (with $k=2$),
\beq
\bsp
\PS{m+X+2}(\mom{}_{m+X+2};p_a+p_b) &=
	\int_{\almin}^{\almax} \rd\al\, 
	 \int_{\hlammin}^{\hlammax} \rd\ha{\lam}\,
	\PS{m+X}\big(\momth{}_{m+X}; \lam (1-\al)(p_a+p_b)\big)
\\&\times
	\frac{s_{\ha{a}\ha{b}}}{\pi} \ha{\lam}\,
	\PS{2}(\ha{p}_{rs},\ha{K},\ha{p}_a+\ha{p}_b)\,
	\frac{s_{\wha{rs}\ab}}{2\pi}
	\PS{2}(p_r,p_s; p_r+p_s)\,,
\esp
\label{eq:S-FFmap-PSconv} 
\eeq
were the momentum $\ha{K}^\mu$ is massive with mass $\ha{K}^2 = 
\ha{\lam}^2 s_{\ha{a}\ha{b}} = \ha{\lam}^2 (1-\al)^2 s_{ab}$. The 
limits of the $\al$ integration are given in \eqn{eq:alminmax}, 
while the limits of the $\ha{\lam}$ integration read
\beq
\hlammin = \frac{M}{(1-\al)\sqrt{s_{ab}}}\,,
\qquad
\hlammax = 1\,.
\label{eq:limitshlam1}
\eeq
Finally, the flux factor $\Phi(p_a\cdot p_b)$ can be written in 
terms of the mapped momenta as follows,
\beq
\Phi(p_a\cdot p_b) 
    = \frac{\Phi(\ti{\ha{p}}_a\cdot \ti{\ha{p}}_b)}
    {(1-\al)^2 \ha{\lam}^2}\,.
\label{eq:S-FFflux}
\eeq
%

%%%%%
\paragraph{Integrated counterterm:} 

Using the definitions of the subtraction term and the phase space 
convolution, \eqns{eq:CrsFFSrs0}{eq:S-FFmap-PSconv}, we write the 
integrated counterterm as
\beq
\bsp
&
\int_2{\cal N} \frac{1}{\omega(a)\omega(b)\Phi(p_a\cdot p_b)} 
	\PS{m+X+2}(\mom{}_{m+X+2};p_a+p_b)\,
	\cS{rs}{(0,0)}\cC{rs}{FF}(p_a,p_b;\mom{}_{m+X+2})
\\ &\qquad=
	\left[\frac{\as}{2\pi} S_\ep 
    \left(\frac{\mu^2}{s_{ab}}\right)^\ep\right]^2
	\Big( \left[\IcS{rs}{(0,0)} \IcC{rs}{FF}\right] \otimes 
    \dsig{B}_{ab,m+X}\Big)\,,
\esp
\label{eq:Int_Srs00CrsFF-fin}
\eeq
where 
\beq
\left[\IcS{rs}{(0,0)} \IcC{rs}{FF}\right] \otimes 
\dsig{B}_{ab,m+X} = 
\int_{\almin}^{\almax} \rd \al\, \int_{\hlammin}^{\hlammax} 
\rd \ha{\lam}\, 
	\sum_{\substack{i,k\in I \cup F \\ i,k\ne r,s}} 
	\frac{1}{2}
    \left[\IcS{rs}{(0,0)} \IcC{rs}{FF}
    (\al,\ha{\lam};\ep)\right]^{(i,k)}
	\bT_i \bT_k
	\dsig{B}_{ab,m+X}(\ti{\ha{p}}_a, \ti{\ha{p}}_b)\,.
\label{eq:ISrsCrsFF}
\eeq
The initial-state momenta entering the Born cross section are 
$\ti{\ha{p}}_a = (1-\al)\ha{\lam} p_a$ and $\ti{\ha{p}}_b = (1-
\al)\ha{\lam} p_b$ and the integrated counterterm is defined as
\beq
\bsp
\left[\IcS{rs}{(0,0)} \IcC{rs}{FF}
(\al,\ha{\lam};\ep)\right]^{(i,k)} &=
	\left( \frac{(4\pi)^2 }{S_\ep} s_{ab}^\ep \right)^2
	\int
	\frac{s_{\ha{a}\ha{b}}}{\pi}
	\PS{2}(\ha{p}_{rs},\ha{K};\ha{p}_a+\ha{p}_b)
	\frac{s_{\wha{rs}\ab}}{2\pi}
	\PS{2}(p_r,p_s; p_r+p_s)
	\frac{\ha{\lam}^3 (1-\alpha)^2}{s_{rs}}
\\&\times
	\calS_{\ti{\ha{i}}\ti{\ha{k}}}^{\mu \nu}(\wha{rs}) 
	 \la \mu |\hP^{(0)}_{r s}(z_{r,s},\kT{r,s};\ep)|\nu \ra \,.
\esp
\label{eq:I_cFFsoft}
\eeq
The evaluation of $\left[\IcS{rs}{(0,0)} \IcC{rs}{FF}
(\al,\ha{\lam};\ep)\right]^{(i,k)}$ requires some care and will be 
discussed in a dedicated publication. Of course, in the 
color-singlet case, we must compute the integrated counterterm 
only for $i,k=a,b$. Then, using \eqn{eq:TaTb}, we have 
\beq
\bsp
\left[\IcS{rs}{(0,0)} \IcC{rs}{FF}\right] \otimes 
\dsig{B}_{ab,X} = 
    \int_{\almin}^{\almax} \rd \al\, \int_{\hlammin}^{\hlammax} 
    \rd \ha{\lam}\, 
    \bigg\{&
	\frac{1}{2}
    \left[\IcS{rs}{(0,0)} \IcC{rs}{FF}
    (\al,\ha{\lam};\ep)\right]^{(a,a)}
    +\frac{1}{2}
    \left[\IcS{rs}{(0,0)} \IcC{rs}{FF}
    (\al,\ha{\lam};\ep)\right]^{(b,b)}
\\ &
    -\left[\IcS{rs}{(0,0)} \IcC{rs}{FF}
    (\al,\ha{\lam};\ep)\right]^{(a,b)}
    \bigg\}
	\bT^2_{\mathrm{ini}}
	\dsig{B}_{ab,X}(\ti{\ha{p}}_a, \ti{\ha{p}}_b)\,.
\esp
\label{eq:I_cFFsoft-colorsinglet}
\eeq
%

%
% \cC{ars}{IFF}\cS{rs}{(0,0)}\cC{rs}{FF}
%

\subsubsection{
\texorpdfstring{$\cC{ars}{IFF}\cS{rs}{(0,0)}\cC{rs}{FF}$}
{CarsIFFSrs00CrsFF}}
\label{sec:CarsIFFSrs00CrsFF}

%%%%%
\paragraph{Subtraction term:} 

For an initial-state parton $a$ and two final-state partons $r$ 
and $s$, either two gluons or a same-flavor quark-antiquark pair, 
we define
\beq
\bsp
&
\cC{a r_g s_g}{IFF}\cS{r_g s_g}{(0,0)}\cC{r_g s_g}{FF}
(p_a,p_b;\mom{}_{m+X+2}) = 
	(8\pi\as\mu^{2\ep})^2
	\frac{2}{s_{rs} s_{\ti{\ha{a}}\wha{rs}}} \bT_a^2 \CA 
\\ &\qquad \times
	\left[\frac{2}{x_{\wha{rs},\ti{\ha{a}}}} 
	\left(\frac{z_{r,s}}{1 - z_{r,s}} + 
    \frac{1 - z_{r,s}}{z_{r,s}}\right) 
	- (1-\ep) z_{r,s} (1 - z_{r,s}) 
    \frac{s_{\ti{\ha{a}}\kT{r,s}}^2}{\kT{r,s}^2
    s_{\ti{\ha{a}}\wha{rs}}}\right]
	\SME{ab,m+X}{(0)}
    {(\ti{\ha{p}}_a,\ti{\ha{p}}_b;\momth{}_{m+X})}\,,
\esp
\label{eq:CrsFFCarsIFFSrs0_gg}
\eeq
for two gluons, and
\beq
\bsp
& 
\cC{a r_q s_\qb}{IFF}\cS{r_q s_\qb}{(0,0)}\cC{r_q s_\qb}{FF}
(p_a,p_b;\mom{}_{m+X+2}) = 
\\&\qquad=
(8\pi\as\mu^{2\ep})^2
	\frac{2}{s_{rs} s_{\ti{\ha{a}}\wha{rs}}} \bT_a^2 \TR
	\left[\frac{1}{x_{\wha{rs},\ti{\ha{a}}}} 
	+ z_{r,s} (1 - z_{r,s}) \frac{s_{\ti{\ha{a}}\kT{r,s}}^2}
    {\kT{r,s}^2 s_{\ti{\ha{a}}\wha{rs}}}\right]
	\SME{ab,m+X}{(0)}
    {(\ti{\ha{p}}_a,\ti{\ha{p}}_b;\momth{}_{m+X})}\,,
\esp
\label{eq:CrsFFCarsIFFSrs0_qq}
\eeq
for a quark-antiquark pair. The reduced matrix elements are 
obtained by dropping the final-state partons $r$ and $s$, while 
the mapped momenta entering the reduced matrix elements are given 
in \eqn{eq:S-FFmap}. The momentum fraction $z_{r,s}$ is defined by 
\eqn{eq:zjk-def}, while $x_{\wha{rs},\ti{\ha{a}}}$ is given by 
\eqn{eq:xja-def}. The transverse momentum $\kT{r,s}^\mu$ is given 
by \eqn{eq:ktjk-ansatz} with coefficients as in 
\eqns{eq:zirri}{eq:Zir}, after the relabeling $i\to r$ and 
$r\to s$. Moreover, we used the notation
\beq
s_{\ti{\ha{a}}\kT{r,s}} = 2 \ti{\ha{p}}_a \cdot \kT{r,s}\,.
\eeq
%

%%%%%
\paragraph{Momentum mapping and phase space factorization:}

This subtraction term is defined using the S--IF iterated momentum 
mapping of~\sect{sec:Srs00CrsFF}, where the appropriate phase 
space convolution is also presented.

%%%%%
\paragraph{Integrated subtraction term:}

Using the definitions of the phase space convolution, 
\eqn{eq:S-FFmap-PSconv}, and the subtraction terms, 
\eqns{eq:CrsFFCarsIFFSrs0_gg}{eq:CrsFFCarsIFFSrs0_qq}, we write 
the integrated counterterm as
\beq
\bsp
&
\int_2{\cal N} \frac{1}{\omega(a)\omega(b)\Phi(p_a\cdot p_b)} 
	\PS{m+X+2}(\mom{}_{m+X+2};p_a+p_b)\,
	\cC{ars}{IFF}\cS{rs}{(0,0)}\cC{rs}{FF}(p_a,p_b;\mom{}_{m+X+2})
\\ &\qquad=
	\left[\frac{\as}{2\pi} S_\ep 
    \left(\frac{\mu^2}{s_{ab}}\right)^\ep\right]^2
	\Big( \left[\IcC{ars}{IFF} \IcS{rs}{(0,0)} \IcC{rs}{FF}\right] 
    \otimes \dsig{B}_{ab,m+X}\Big)\,,
\esp
\label{eq:Int_CarsSrs00CrsFF-fin}
\eeq
where
\beq
\left[\IcC{ars}{IFF} \IcS{rs}{(0,0)} \IcC{rs}{FF}\right] \otimes 
\dsig{B}_{ab,m+X} = 
\int_{\almin}^{\almax} \rd \al\, \int_{\hlammin}^{\hlammax} 
\rd \ha{\lam}\, 
	\left[\IcC{ars}{IFF} \IcS{rs}{(0,0)} \IcC{rs}{FF}
    (\al,\ha{\lam};\ep)\right]
	\dsig{B}_{ab,m+X}(\ti{\ha{p}}_a,\ti{\ha{p}}_b)\,.
\eeq
The initial-state momenta entering the Born cross section are 
$\ti{\ha{p}}_a = (1-\al)\ha{\lam} p_a$ and $\ti{\ha{p}}_b = (1-
\al)\ha{\lam} p_b$ and the integrated counterterm is defined as
\beq
\bsp
&
\left[\IcC{a r_g s_g}{IFF} \IcS{r_g s_g}{(0,0)} \IcC{r_g s_g}{FF}
(\al,\ha{\lam};\ep) \right] = 
	\left( \frac{(4\pi)^2 }{S_\ep} s_{ab}^\ep \right)^2
	\int 
	\frac{s_{\ha{a}\ha{b}}}{\pi}
	\PS{2}(\ha{p}_{rs},\ha{K};\ha{p}_a+\ha{p}_b)
	\frac{s_{\wha{rs}\ab}}{2\pi}
	\PS{2}(p_r,p_s; p_r+p_s) 
\\&\qquad\times
	\frac{2 \ha{\lam}^3 (1-\alpha)^2}
    {s_{rs} s_{\ti{\ha{a}}\wha{rs}}} 
	\bT_a^2 \CA \left[ \frac{2}{x_{\wha{rs},\ti{\ha{a}}}} 
    \left(\frac{z_{r,s}}{1 - z_{r,s}} 
    + \frac{1 - z_{r,s}}{z_{r,s}}\right) 
	- (1-\ep) z_{r,s} (1 - z_{r,s}) 
    \frac{s_{\ti{\ha{a}}\kT{r,s}}^2}
    {\kT{r,s}^2 s_{\ti{\ha{a}}\wha{rs}}} \right]\,,
\esp
\label{eq:I-CrsFFCarsIFFSrs0gg}
\eeq
for $r$ and $s$ both gluons and
\beq
\bsp
&
\left[\IcC{a r_q s_\qb}{IFF} \IcS{r_q s_\qb}{(0,0)} 
\IcC{r_q s_\qb}{FF}(\al,\ha{\lam};\ep) \right] = 
	\left( \frac{(4\pi)^2 }{S_\ep} s_{ab}^\ep \right)^2
	\int 
	\frac{s_{\ha{a}\ha{b}}}{\pi}
	\PS{2}(\ha{p}_{rs},\ha{K};\ha{p}_a+\ha{p}_b)
	\frac{s_{\wha{rs}\ab}}{2\pi}
	\PS{2}(p_r,p_s; p_r+p_s) 
\\&\qquad\times
	\frac{2 \ha{\lam}^3 (1-\alpha)^2}
    {s_{rs} s_{\ti{\ha{a}}\wha{rs}}} 
	\bT_a^2 \TR
	\left[ \frac{1}{x_{\wha{rs},\ti{\ha{a}}}}
	+ z_{r,s} (1 - z_{r,s}) \frac{s_{\ti{\ha{a}}\kT{r,s}}^2}
    {\kT{r,s}^2 s_{\ti{\ha{a}}\wha{rs}}} \right]\,,
\esp
\label{eq:I-CrsFFCarsIFFSrs0qq}
\eeq
for $r$ and $s$ a same-flavor quark-antiquark pair. The 
evaluation of $\left[\IcC{ars}{IFF} \IcS{rs}{(0,0)} \IcC{rs}{FF}
(\al,\ha{\lam};\ep)\right]$ requires some care due to the presence 
of the structure $s_{\ti{\ha{a}}\kT{r,s}}$ and will be discussed 
elsewhere. 

%
% Soft-type limits of double soft-type subtractions
%

\subsection{Soft-type limits of double soft-type subtractions}

%
% \cS{rs}{(0,0)}\cS{s}{}
%

\subsubsection{\texorpdfstring{$\cS{rs}{(0,0)}\cS{s}{}$}{Srs00Ss}}
\label{sec:Srs00Ss}

%%%%%
\paragraph{Subtraction term:} 

For final-state gluons $r$ and $s$ we define
\beq
\bsp
&
\cS{r s}{(0,0)}\cS{s}{}(p_a,p_b;\mom{}_{m+X+2}) = 
(8\pi\as\mu^{2\ep})^2
\\&\qquad\times
	\bigg[\frac18 
	\sum_{\substack{i,k,j,\ell \in I \cup F 
    \\ i, k, j, \ell \ne r, s}} 
	\calS_{\ti{\ti{i}}\ti{\ti{k}}}(\ti{r}) 
    \calS_{\ti{j}\ti{\ell}}(s)
	\{\bT_i \bT_k,\bT_j \bT_\ell\}
	\SME{ab,m+X}{(0)}{(\ti{\ti{p}}_a,\ti{\ti{p}}_b;\momtt{}_{m+X})}
\\&\qquad\qquad
	-\frac14 \CA
	\sum_{\substack{i,k \in I \cup F \\ i, k \ne r, s}} 
	\calS_{\ti{\ti{i}}\ti{\ti{k}}}(\ti{r})
	\left(
		\calS_{\ti{\ti{i}}\ti{r}}(s)
		+ \calS_{\ti{\ti{k}}\ti{r}}(s)
		- \calS_{\ti{\ti{i}}\ti{\ti{k}}}(s)
	\right)
	\bT_i \bT_k
	\SME{ab,m+X}{(0)}{(\ti{\ti{p}}_a,\ti{\ti{p}}_b;\momtt{}_{m+X})}
	\bigg]\,,
\esp
\label{eq:SsSrs00}
\eeq
where the reduced matrix elements are obtained by simply removing 
the final-state gluons $r$ and $s$. The set of momenta entering 
the reduced matrix elements is defined in \eqn{eq:S-Smap}. In 
\eqn{eq:SsSrs00}, the eikonal factors are given by 
\eqn{eq:eikfact} with appropriate relabeling as necessary. In the 
color-singlet case, the summations in \eqn{eq:SsSrs00} run only 
over the initial-state partons $a$ and $b$. Using color 
conservation (\eqns{eq:TaTb}{eq:TaTbTaTb}) we then find
\beq
\bsp
&
\cS{r s}{(0,0)}\cS{s}{}(p_a,p_b;\mom{}_{X+2}) = 
(8\pi\as\mu^{2\ep})^2
\\&\qquad\times
	\bigg[
	\calS_{\ti{\ti{a}}\ti{\ti{b}}}(\ti{r}) \calS_{\ti{a}\ti{b}}(s)
    \bT^2_{\mathrm{ini}}
	+\frac12
	\calS_{\ti{\ti{a}}\ti{\ti{b}}}(\ti{r})
	\left(
		\calS_{\ti{\ti{a}}\ti{r}}(s)
		+ \calS_{\ti{\ti{b}}\ti{r}}(s)
		- \calS_{\ti{\ti{a}}\ti{\ti{b}}}(s)
	\right)\CA\bigg]
	\bT^2_{\mathrm{ini}}
	\SME{ab,X}{(0)}{(\momtt{}_{X};\ti{\ti{p}}_a,\ti{\ti{p}}_b)}\,.
\esp
\label{eq:SsSrs00-colorsinglet}
\eeq
%

%%%%%
\paragraph{Momentum mapping:} 

The S--S momentum mapping is defined by  successive application of 
the single soft momentum mapping of \eqn{eq:srmap},
\beq
\smp{\ti{r}}{} \circ \smp{s}{}:\; 
	(p_a, p_b; \mom{}_{m+X+2}) 
	\smap{s}{} 
	(\ti{p}_a, \ti{p}_b; \momt{}_{m+X+1}) 
	\smap{\ha{r}}{} 
	(\ti{\ti{p}}_a,\ti{\ti{p}}_b;\momtt{}_{m+X}) \,.
\label{eq:srhssmap}
\eeq
Specifically, we first define the set of momenta $(\ti{p}_a, 
\ti{p}_b;\momt{}_{m+X+1})$ via the $\smp{s}{}$ mapping of 
\eqn{eq:ssmap},
\beq
\bsp
\ti{p}_a^\mu &= \lam_s p_a^\mu\,,
\\
\ti{p}_b^\mu &= \lam_s p_b^\mu\,,
\\
\ti{p}_n^\mu &= {\Lambda(P,\ti{P})^\mu}_{\!\nu}\, p_n^\nu\,, 
\qquad n \in F, \quad n\ne s\,,
\\
\ti{p}_X^\mu &= {\Lambda(P,\ti{P})^\mu}_{\!\nu}\, p_X^\nu\,,
\esp
\label{eq:ssmap-2}
\eeq
where ${\Lambda(P,\ti{P})^\mu}_{\!\nu}$ is the proper Lorentz 
transformation in \eqn{eq:Lambda_munu} and
\beq
P^\mu = (p_a+p_b)^\mu - p_s^\mu\,,
\qquad
\ti{P}^\mu = (\ti{p}_a+\ti{p}_b)^\mu 
	= \lam_s(p_a + p_b)^\mu\,,
\eeq
with $\lam_s$ defined by \eqn{eq:lamr-def} with the relabeling 
$r\to s$. Then, the set 
$(\ti{\ti{p}}_a,\ti{\ti{p}}_b;\momtt{}_{m+X})$ is constructed 
using the $\smp{\ti{r}}{}$ mapping of \eqn{eq:srmap} once again, 
starting from the momenta in \eqn{eq:ssmap-2},
\beq
\bsp
\ti{\ti{p}}_a^\mu &= \lam_{\ti{r}} \ti{p}_a^\mu\,,
\\
\ti{\ti{p}}_b^\mu &= \lam_{\ti{r}} \ti{p}_b^\mu\,,
\\
\ti{\ti{p}}_n^\mu &= {\Lambda(\ti{P},\ti{\ti{P}})^\mu}_{\!\nu}\, 
\ti{p}_n^\nu\,, \qquad n \in F, \quad n\ne \ti{r}\,,
\\
\ti{\ti{p}}_X^\mu &= {\Lambda(\ti{P},\ti{\ti{P}})^\mu}_{\!\nu}\, 
\ti{p}_X^\nu\,,
\esp
\label{eq:S-Smap}
\eeq
where ${\Lambda(\ti{P},\ti{\ti{P}})^\mu}_{\!\nu}$ is the proper 
Lorentz transformation in \eqn{eq:Lambda_munu}, while
\beq
\ti{P}^\mu = (\ti{p}_a+\ti{p}_b)^\mu - \ti{p}_{r}^\mu\,,
\qquad
\ti{\ti{P}}^\mu = (\ti{\ti{p}}_a+\ti{\ti{p}}_b)^\mu 
	= \lam_{\ti{r}}(\ti{p}_a + \ti{p}_b)^\mu
	= \lam_s \lam_{\ti{r}} (p_a + p_b)^\mu\,.
\eeq
The quantity $\lam_{\ti{r}}$ is computed as in \eqn{eq:lamr-def} 
but with tilded momenta, i.e., with the replacement $p_a \to 
\ti{p}_a$, $p_b \to \ti{p}_b$, $Q\to \ti{Q}=\ti{p}_a+\ti{p}_b$ and 
$p_r \to \ti{p}_{r}$. We employ the double tilde notation to indicate 
that the final set of mapped momenta are obtained by the successive 
application of two soft mappings.

%%%%%
\paragraph{Phase space convolution:} 

Using the appropriate convolution formula for the single soft 
mapping, \eqn{eq:Sr-map-PSconv}, we obtain the following 
representation of the $(m+X+2)$-particle phase space of 
\eqn{eq:psmx2} (with $k=2$),
\beq
\bsp
\PS{m+X+2}(\mom{}_{m+X+2};p_a+p_b) &=
	\int_{\lammin}^{\lammax} \rd \lam\,
	\int_{\tlammin}^{\tlammax} \rd \ti{\lam}\,
	\PS{m+X}\big(\momtt{}_{m+X}; \lam \ti{\lam}(p_a+p_b)\big)
\\&\times
    \frac{s_{\ti{a}\ti{b}}}{\pi} \ti{\lam}\,
	\PS{2}(\ti{p}_r,\ti{K};\ti{p}_a+\ti{p}_b)\,
	\frac{s_{ab}}{\pi} \lam\,
	\PS{2}(p_s,K;p_a+p_b)\,,
\esp
\label{eq:S-Smap-PSconv} 
\eeq
where $K^\mu$ and $\ti{K}^\mu$ are massive momenta with masses 
$K^2 = \lam^2 s_{ab}$ and $\ti{K}^2 = \ti{\lam}^2 s_{\ti{a}\ti{b}} = 
\lam^2 \ti{\lam}^2 s_{ab}$. The limits of the $\lam$ integration are 
given in \eqn{eq:limitslam}, while the limits of the $\ha{\lam}$ 
integration read
\beq
\hlammin = \frac{M}{\lam \sqrt{s_{ab}}}\,,
\qquad
\hlammax = 1\,.
\label{eq:limitshlam2}
\eeq
Finally, the flux factor $\Phi(p_a\cdot p_b)$ can be written in 
terms of the mapped momenta as follows,
\beq
\Phi(p_a\cdot p_b) 
= \frac{\Phi(\ti{\ti{p}}_a\cdot \ti{\ti{p}}_b)}
    {\lsq \ti{\lam}^2}\,.
\label{eq:SSSflux}
\eeq

%%%%%
\paragraph{Integrated subtraction term:}

Using the definitions of the subtraction term and the phase space 
convolution, \eqns{eq:SsSrs00}{eq:S-Smap-PSconv}, we write the 
integrated counterterm as
\beq
\bsp
& 
\int_2{\cal N} \frac{1}{\omega(a)\omega(b)\Phi(p_a\cdot p_b)} 
	\PS{m+X+2}(\mom{}_{m+X+2};p_a+p_b)
	\cS{rs}{(0,0)}\cS{s}{}(p_a,p_b;\mom{}_{m+X+2})
\\ &\qquad=
	\left[\frac{\as}{2\pi} S_\ep 
    \left(\frac{\mu^2}{s_{ab}}\right)^{\ep}\right]^2
	\Big(\left[\IcS{rs}{(0,0)}\IcS{s}{}\right] \otimes 
    \dsig{B}_{ab,m+X}\Big)
\esp
\label{eq:Int_Srs00Ss-fin}
\eeq
where 
\beq
\bsp
&
\left[\IcS{rs}{(0,0)}\IcS{s}{}\right] \otimes \dsig{B}_{ab,m+X} 
\\ &\qquad = 
\int_{\lammin}^{\lammax} \rd \lam\, \int_{\tlammin}^{\tlammax} 
\rd\ti{\lam}\, 
	\bigg\{
	\frac{1}{8} \sum_{\substack{i,k,j,\ell \in I \cup F 
    \\ i, k, j, \ell \ne r, s}} 
	\left[\IcS{rs}{(0,0)}\IcS{s}{}
    (\lam,\ti{\lam};\ep)\right]^{(ik,j\ell)}
	\{\bT_i \bT_k,\bT_j \bT_\ell\} \dsig{B}_{ab,m+X}
    (\ti{\ti{p}}_a, \ti{\ti{p}}_b)
\\&\qquad\qquad\qquad\qquad\qquad\qquad
	-\frac14 \CA 
	\sum_{\substack{i,k \in I \cup F \\ i, k \ne r, s}} 
	\left[\IcS{rs}{(0,0)}\IcS{s}{}
    (\lam,\ti{\lam};\ep)\right]^{(i,k)}
	\bT_i \bT_k \dsig{B}_{ab,m+X}(\ti{\ti{p}}_a, \ti{\ti{p}}_b)
	\bigg\}\,. 
\esp
\label{eq:ISggSg00}
\eeq
The initial-state momenta entering the Born cross sections in 
\eqn{eq:ISggSg00} are $\ti{\ti{p}}_a = \lam \ti{\lam} p_a$ and 
$\ti{\ti{p}}_b = \lam \ti{\lam} p_b$ and the integrated 
counterterms are defined as
\beq
\left[\IcS{rs}{(0,0)}\IcS{s}{}
(\lam,\ti{\lam};\ep)\right]^{(ik,j\ell)} =
	\left( \frac{s_{ab}}{\pi} \right)^2
	\left( \frac{(4\pi)^2 }{S_\ep} s_{ab}^\ep \right)^2 
	\int \PS{2}(p_s,K;p_a+p_b) \PS{2}
    (\ti{p}_r,\ti{K};\ti{p}_a+\ti{p}_b)\,
	\lam^5 \ti{\lam}^3  \calS_{\ti{\ti{i}}\ti{\ti{k}}}(\ti{r}) 
    \calS_{\ti{j}\ti{\ell}}(s)\,,
\label{eq:ISSS_ikjl} 
\eeq
and
\beq
\bsp
\left[\IcS{rs}{(0,0)}\IcS{s}{}(\lam,\ti{\lam};\ep)\right]^{(i,k)} &=
	\left( \frac{s_{ab}}{\pi} \right)^2
	\left( \frac{(4\pi)^2 }{S_\ep} s_{ab}^\ep \right)^2  
	\int \PS{2}(p_s,K;p_a+p_b) \PS{2}
    (\ti{p}_r,\ti{K};\ti{p}_a+\ti{p}_b)\,
\\&\times
	\lam^5 \ti{\lam}^3  \calS_{\ti{\ti{i}}\ti{\ti{k}}}(\ti{r})
	\left(
		\calS_{\ti{\ti{i}}\ti{r}}(s)
		+ \calS_{\ti{\ti{k}}\ti{r}}(s)
		- \calS_{\ti{\ti{i}}\ti{\ti{k}}}(s)
	\right)\,.
\label{eq:ISSS_ik}
\esp
\eeq
Of course, for color-singlet production, we only need to evaluate 
\eqns{eq:ISSS_ikjl}{eq:ISSS_ik} for $i,k,j,\ell=a,b$. Then, using 
\eqns{eq:TaTb}{eq:TaTbTaTb} we obtain
\beq
\bsp
\left[\IcS{rs}{(0,0)}\IcS{s}{}\right] \otimes \dsig{B}_{ab,X} = 
\int_{\lammin}^{\lammax} \rd \lam\, \int_{\tlammin}^{\tlammax} 
\rd\ti{\lam}\, 
	\bigg\{&
	\left[\IcS{rs}{(0,0)}\IcS{s}{}
    (\lam,\ti{\lam};\ep)\right]^{(ab,ab)}
    \bT^2_{\mathrm{ini}}
\\&
	+\frac12 \CA
	\left[\IcS{rs}{(0,0)}\IcS{s}{}
    (\lam,\ti{\lam};\ep)\right]^{(a,b)}\bigg\}
	\bT^2_{\mathrm{ini}}
    \dsig{B}_{ab,X}(\ti{\ti{p}}_a, \ti{\ti{p}}_b)\,. 
\esp
\label{eq:ISggSg00-colorsinglet}
\eeq
%

%
% \cC{ars}{IFF}\cS{rs}{(0,0)}\cS{s}{}
%

\subsubsection{
\texorpdfstring{$\cC{ars}{IFF}\cS{rs}{(0,0)}\cS{s}{}$}
{CarsIFFSrs00Ss}}
\label{sec:CarsIFFSrs00Ss}

%%%%%
\paragraph{Subtraction term:} 

For an initial-state parton $a$ and final-state gluons $r$ and $s$ 
we define
\beq
\bsp
&
\cC{ars}{IFF}\cS{rs}{(0,0)}\cS{s}{}(p_a,p_b;\mom{}_{m+X+2})
\\&\qquad=
	(8\pi\as\mu^{2\ep})^2
	P^{(\mathrm S), (0)}_{(ar) r s}
	(x_{\ti{a},\ti{r}s}, x_{\ti{r},\ti{a}}, x_{s,\ti{a}},
	s_{\ti{a}\ti{r}}, s_{\ti{a}s}, s_{\ti{r}s};\ep)
	\frac{1}{s_{\ti{\ti{a}}\ti{r}}}
    \frac{2}{x_{\ti{r},\ti{\ti{a}}}} \bT_a^2
	\SME{ab,m+X}{(0)}
    {(\ti{\ti{p}}_a,\ti{\ti{p}}_b;\momtt{}_{m+X})}\,,
\esp
\label{eq:SsCarsIFFSrs0}
\eeq
where once more the reduced matrix element is obtained by removing 
the final-state gluons $r$ and $s$, while the mapped momenta are 
given in \eqn{eq:S-Smap}. The initial-final-final soft function 
$P^{(\mathrm S), (0)}_{a g_r g_s}$ is the same as in 
\sect{sec:CarsIFF00Ss}. When both $r$ and $s$ are gluons, it takes 
the form,
\beq
P^{(\mathrm S), (0)}_{a g_r g_s}
(x_{\ti{a},\ti{r}s}, x_{\ti{r},\ti{a}}, x_{s,\ti{a}},
	s_{\ti{a}\ti{r}}, s_{\ti{a}s}, s_{\ti{r}s};\ep)
	 = \bT_a^2 \frac2{x_{s,\ti{a}} s_{\ti{a}s}}
	 + \CA \left(
	 	\frac{s_{\ti{a}\ti{r}}}{s_{\ti{a}s} s_{\ti{r}s}}
		- \frac{1}{x_{s,\ti{a}} s_{\ti{a}s}}
		+ \frac{x_{\ti{r},\ti{a}}}{x_{s,\ti{a}} s_{\ti{r}s}}
		\right)\,. 
\label{eq:SsPfgg0IFF}
\eeq
The momentum fractions entering \eqn{eq:SsCarsIFFSrs0} are defined 
by \eqns{eq:xja-def}{eq:xajk-def}.

%%%%%
\paragraph{Momentum mapping and phase space factorization:}

This subtraction term is defined using the S--S iterated momentum 
mapping of~\sect{sec:Srs00Ss}, where the appropriate phase space 
convolution is also presented.

%%%%%
\paragraph{Integrated subtraction term:}

Using the definitions of the phase space convolution and the 
subtraction term, \eqns{eq:S-Smap-PSconv}{eq:SsCarsIFFSrs0}, we 
write the integrated counterterm as
\beq
\bsp
&
\int_2{\cal N} \frac{1}{\omega(a)\omega(b)\Phi(p_a\cdot p_b)} 
	\PS{m+X+2}(\mom{}_{m+X+2};p_a+p_b)\,
	\cC{ars}{IFF}\cS{rs}{(0,0)}\cS{s}{}(p_a,p_b;\mom{}_{m+X+2})
\\ &\qquad=
	\left[\frac{\as}{2\pi} S_\ep 
    \left(\frac{\mu^2}{s_{ab}}\right)^\ep\right]^2
	\Big( \left[\IcC{ars}{IFF} \IcS{rs}{(0,0)} \IcS{s}{}\right] 
    \otimes \dsig{B}_{ab,m+X}\Big)\,,
\esp
\eeq
where
\beq
\left[\IcC{ars}{IFF} \IcS{rs}{(0,0)} \IcS{s}{}\right] \otimes 
\dsig{B}_{ab,m+X} = 
\int_{\lammin}^{\lammax} \rd \lam\, \int_{\tlammin}^{\tlammax} 
\rd \ti{\lam}\, 
	\left[\IcC{ars}{IFF} \IcS{rs}{(0,0)} \IcS{s}{}
    (\lam,\ti{\lam};\ep)\right]
	\dsig{B}_{ab,m+X}(\ti{\ti{p}}_a, \ti{\ti{p}}_b)\,.
\eeq
The initial-state momenta in the Born cross section are 
$\ti{\ti{p}}_a = \lam \ti{\lam} p_a$ and $\ti{\ti{p}}_b = \lam 
\ti{\lam} p_b$ as before and the integrated counterterm reads
\beq
\bsp
\left[\IcC{ars}{IFF}\IcS{rs}{(0,0)}\IcS{s}{}
(\lam,\ti{\lam};\ep)\right] &= 
	\left( \frac{s_{ab}}{\pi} \right)^2
	\left( \frac{(4\pi)^2 }{S_\ep} s_{ab}^\ep \right)^2
	\int \PS{2}(p_s,K;p_a+p_b) \PS{2}
    (\ti{p}_r,\ti{K};\ti{p}_a+\ti{p}_b)\,
\\&\times 
	\lam^5 \ti{\lam}^3
	P^{(\mathrm S), (0)}_{(ar) r s}
	(x_{\ti{a},\ti{r}s}, x_{\ti{r},\ti{a}}, x_{s,\ti{a}},
	s_{\ti{a}\ti{r}}, s_{\ti{a}s}, s_{\ti{r}s};\ep)
	\frac{1}{s_{\ti{\ti{a}}\ti{r}}}
    \frac{2}{x_{\ti{r},\ti{\ti{a}}}} \bT_a^2\,.
\esp
\label{eq:ISsCarsIFFSrs0}
\eeq

%
% \cS{rs}{(0,0)}\cC{as}{IF}\cS{s}{}
%

\subsubsection{
\texorpdfstring{$\cS{rs}{(0,0)}\cC{as}{IF}\cS{s}{}$}{Srs00CasIFSs}}
\label{sec:Srs00CasIFSs}

%%%%%
\paragraph{Subtraction term:} 

For an initial-state parton $a$ and final-state gluons $r$ and $s$ 
we define
\beq
\bsp
&
\cS{rs}{(0,0)}\cC{as}{IF}\cS{s}{}(p_a,p_b;\mom{}_{m+X+2})
\\&\qquad= 
	-(8\pi\as\mu^{2\ep})^2 \frac{1}{s_{\ti{a}s}} 
    \frac{2}{x_{s,\ti{a}}} \bT_a^2
	\sum_{\substack{j,k \in I \cup F \\ j, k \ne r, s}} 
	\frac{1}{2}\calS_{\ti{\ti{j}}\ti{\ti{k}}}(\ti{r}) \bT_j \bT_k
	\SME{ab,m+X}{(0)}
    {(\ti{\ti{p}}_a,\ti{\ti{p}}_b;\momtt{}_{m+X})} \,,
\esp
\label{eq:CasIFSsSrs0}
\eeq
where the reduced matrix element is obtained by removing the 
final-state gluons $r$ and $s$. The mapped momenta appearing in 
\eqn{eq:CasIFSsSrs0} are given in \eqn{eq:S-Smap}. The momentum 
fraction $x_{s,\ti{a}}$ is defined as in \eqn{eq:xja-def}, while 
the eikonal factor is given by \eqn{eq:eikfact}. Obviously, for 
color-singlet production we have $j,k=a,b$ and using color 
conservation (see \eqn{eq:TaTb}) we find simply
\beq
\cS{rs}{(0,0)}\cC{as}{IF}\cS{s}{}(p_a,p_b;\mom{}_{X+2}) = 
	(8\pi\as\mu^{2\ep})^2 \frac{1}{s_{\ti{a}s}} 
    \frac{2}{x_{s,\ti{a}}} 
	\calS_{\ti{\ti{a}}\ti{\ti{b}}}(\ti{r}) 
    \left(\bT^2_{\mathrm{ini}}\right)^2
    \SME{ab,X}{(0)}{(\ti{\ti{p}}_a,\ti{\ti{p}}_b;\momtt{}_{X})} \,.
\label{eq:CasIFSsSrs0-colorsinglet}
\eeq
%

%%%%%
\paragraph{Momentum mapping and phase space factorization:}

This subtraction term is defined using the S--S iterated momentum 
mapping of~\sect{sec:Srs00Ss}, where the appropriate phase space 
convolution is also presented.

%%%%%
\paragraph{Integrated subtraction term:}

Using the definitions of the phase space convolution and the 
subtraction term, \eqns{eq:S-Smap-PSconv}{eq:CasIFSsSrs0}, we 
write the integrated counterterm as 
\beq
\bsp
&
\int_2{\cal N} \frac{1}{\omega(a)\omega(b)\Phi(p_a\cdot p_b)} 
	\PS{m+X+2}(\mom{}_{m+X+2};p_a+p_b)\,
	\cS{rs}{(0,0)}\cC{as}{IF}\cS{s}{}(p_a,p_b;\mom{}_{m+X+2})
\\ &\qquad=
	\left[\frac{\as}{2\pi} S_\ep 
    \left(\frac{\mu^2}{s_{ab}}\right)^\ep\right]^2
	\Big( \left[\IcS{rs}{(0,0)} \IcC{as}{IF} \IcS{s}{}\right] 
    \otimes \dsig{B}_{ab,m+X}\Big)\,,
\esp
\eeq
where 
\beq
\bsp
&
\left[\IcS{rs}{(0,0)} \IcC{as}{IF} \IcS{s}{}\right] \otimes 
\dsig{B}_{ab,m+X} 
\\&\qquad = 
\int_{\lammin}^{\lammax} \rd \lam\, \int_{\tlammin}^{\tlammax} 
\rd \ti{\lam}\, 
	\sum_{\substack{j,k \in I \cup F \\ j, \ell \ne r, s}} 
    \frac{1}{2} 
	\left[\IcS{rs}{(0,0)} \IcC{as}{IF} \IcS{s}{}
    (\lam,\ti{\lam};\ep)\right]^{(j,k)}
	\bT_j \bT_k
	\dsig{B}_{ab,m+X}(\ti{\ti{p}}_a, \ti{\ti{p}}_b)\,.
\esp
\label{eq:ISrsCcsSs}
\eeq
The initial-state momenta in the Born cross section are 
$\ti{\ti{p}}_a = \lam \ti{\lam} p_a$ and $\ti{\ti{p}}_b = \lam 
\ti{\lam} p_b$ and the integrated counterterm is then defined as 
\beq
\bsp
\left[\IcS{rs}{(0,0)} \IcC{as}{IF} \IcS{s}{}
(\lam,\ti{\lam};\ep)\right]^{(j,k)} &= 
	- \left( \frac{s_{ab}}{\pi} \right)^2
	\left( \frac{(4\pi)^2 }{S_\ep} s_{ab}^\ep \right)^2
	\int \PS{2}(p_s,K;p_a+p_b) \PS{2}
    (\ti{p}_r,\ti{K};\ti{p}_a+\ti{p}_b)\,
\\ &\times
	\lam^5 \ti{\lam}^3  \frac{1}{s_{\ti{a}s}} 
    \frac{2}{x_{s,\ti{a}}} \bT_a^2 
	\calS_{\ti{\ti{j}}\ti{\ti{k}}}(\ti{r})\,.
\label{eq:ICasIFSsSrs0}
\esp
\eeq
For color-singlet production, the sums in \eqn{eq:ISrsCcsSs} run 
only over the initial-state partons $a$ and $b$. Then, using color 
conservation, we find 
\beq
\left[\IcS{rs}{(0,0)} \IcC{as}{IF} \IcS{s}{}\right] \otimes 
\dsig{B}_{ab,X} = 
    -\int_{\lammin}^{\lammax} \rd \lam\, 
    \int_{\tlammin}^{\tlammax} \rd \ti{\lam}\, 
	\left[\IcS{rs}{(0,0)} \IcC{as}{IF} \IcS{s}{}
    (\lam,\ti{\lam};\ep)\right]^{(a,b)}
	\bT^2_{\mathrm{ini}}
	\dsig{B}_{ab,X}(\ti{\ti{p}}_a, \ti{\ti{p}}_b)\,.
\label{eq:ISrsCcsSs-colorsinglet}
\eeq
%

%
% \cC{ars}{IFF}\cS{rs}{(0,0)}\cC{as}{IF}\cS{s}{}
%

\subsubsection{
\texorpdfstring{$\cC{ars}{IFF}\cS{rs}{(0,0)}\cC{as}{IF}\cS{s}{}$}
{CarsIFFSrs00CasIFSs}}
\label{sec:CarsIFFSrs00CasIFSs}

%%%%%
\paragraph{Subtraction term:} 

For an initial-state parton $a$ and final-state gluons $r$ and $s$ 
we define
\beq
\cC{ars}{IFF}\cS{rs}{(0,0)}\cC{as}{IF}\cS{s}{}
(p_a,p_b;\mom{}_{m+X+2}) = 
	(8\pi\as\mu^{2\ep})^2
	\frac{1}{s_{\ti{a}s}} \frac{2}{x_{s,\ti{a}}} \bT_a^2
	\frac{1}{s_{\ti{\ti{a}}\ti{r}}} 
    \frac{2}{x_{\ti{r},\ti{\ti{a}}}} \bT_a^2
	\SME{ab,m+X}{(0)}
    {(\ti{\ti{p}}_a,\ti{\ti{p}}_b;\momtt{}_{m+X})} \,,
\label{eq:CasIFSsCarsIFFSrs0}
\eeq
where once more we obtain the reduced matrix element by removing 
the final-state gluons $r$ and $s$ and the mapped momenta are 
defined in \eqn{eq:S-Smap}. The momentum fractions $x_{s,\ti{a}}$ 
and $x_{\ti{r},\ti{\ti{a}}}$ entering \eqn{eq:CasIFSsCarsIFFSrs0} 
are given by \eqn{eq:xja-def}.

%%%%%
\paragraph{Momentum mapping and phase space factorization:}

This subtraction term is defined using the S--S iterated momentum 
mapping of~\sect{sec:Srs00Ss}, where the appropriate phase space 
convolution is also presented.

%%%%%
\paragraph{Integrated subtraction term:}

Using the definitions of the phase space convolution and the 
subtraction term, \eqns{eq:S-Smap-PSconv}{eq:CasIFSsCarsIFFSrs0}, 
we write the integrated counterterm as
\beq
\bsp
&
\int_2{\cal N} \frac{1}{\omega(a)\omega(b)\Phi(p_a\cdot p_b)} 
	\PS{m+X+2}(\mom{}_{m+X+2};p_a+p_b)\,
	\cC{ars}{IFF}\cS{rs}{(0,0)}\cC{as}{IF}\cS{s}{}
    (p_a,p_b;\mom{}_{m+X+2})
\\ &\qquad=
	\left[\frac{\as}{2\pi} S_\ep 
    \left(\frac{\mu^2}{s_{ab}}\right)^\ep\right]^2
	\Big( \left[\IcC{ars}{IFF} \IcS{rs}{(0,0)} 
    \IcC{as}{IF} \IcS{s}{}\right] 
    \otimes \dsig{B}_{ab,m+X}\Big)\,,
\esp
\eeq
where
\beq
\left[\IcC{ars}{IFF} \IcS{rs}{(0,0)} \IcC{as}{IF} \IcS{s}{}\right] 
\otimes \dsig{B}_{ab,m+X} = 
\int_{\lammin}^{\lammax} \rd \lam\, \int_{\tlammin}^{\tlammax} 
\rd \ti{\lam}\, 
	\left[\IcC{ars}{IFF} \IcS{rs}{(0,0)} \IcC{as}{IF} \IcS{s}{}
    (\lam,\ti{\lam};\ep)\right]
	\dsig{B}_{ab,m+X}(\ti{\ti{p}}_a, \ti{\ti{p}}_b)\,.
\eeq
The initial-state momenta in the Born cross section are again 
$\ti{\ti{p}}_a = \lam \ti{\lam} p_a$ and $\ti{\ti{p}}_b = \lam 
\ti{\lam} p_b$ and the integrated counterterm is defined as
\beq
\bsp
\left[\IcC{ars}{IFF} \IcS{rs}{(0,0)} \IcC{as}{IF} \IcS{s}{}
(\lam,\ti{\lam};\ep)\right] &= 
	\left( \frac{s_{ab}}{\pi} \right)^2
	\left( \frac{(4\pi)^2 }{S_\ep} s_{ab}^\ep \right)^2
	\int \PS{2}(p_s,K;p_a+p_b) \PS{2}
    (\ti{p}_r,\ti{K};\ti{p}_a+\ti{p}_b)\,
\\&\times
	\lam^5 \ti{\lam}^3  
	\frac{1}{s_{\ti{a}s}} \frac{2}{x_{s,\ti{a}}}\bT_a^2
	\frac{1}{s_{\ti{\ti{a}}\ti{r}}} \frac{2}
    {x_{\ti{r},\ti{\ti{a}}}} \bT_a^2\,.
\esp
\label{eq:ICasIFSsCarsIFFSrs0}
\eeq

%%%
%%% Validation
%%%

\section{Validation}
\label{sec:validation}

In order to validate the definitions of the subtraction terms, we 
consider the production of a Higgs boson in gluon-gluon fusion in 
the Higgs effective field theory. In particular, we concentrate on 
the 
\beq
g(p_1) + g(p_2) \to H(p_3) + g(p_4) + g(p_5)
\eeq
fully gluonic subprocess which exhibits all IR singularities and 
spin correlations which may arise at NNLO accuracy in 
color-singlet production. This process has been implemented in the 
publicly available Monte Carlo program 
{\tt NNLOCAL}~\cite{DelDuca:2024ovc}, which includes a full set of 
dedicated phase space routines that allow the user to check the 
cancellation of kinematic singularities in any IR limit. However 
here we investigate these cancellations using an implementation of 
the matrix element and subtraction terms in the {\tt Mathematica} 
computer algebra system. The arbitrary precision capabilities of 
the latter allow us to assess the numerical behavior of the 
subtractions without having to concern ourselves with questions of 
numerical stability. In the following we present some 
representative results and draw particular attention to the main 
features regarding the structure of cancellations among the 
various subtraction terms.

To begin, we generate sequences of phase space points tending to all 
unique singular kinematic configurations. All radiation variables that 
are not directly relevant for approaching the chosen singular 
configuration, such as azimuthal angles, are kept fixed. The approach 
to any given unresolved configuration is measured by the value of some 
suitably chosen kinematic invariants that vanish in the exact limit. 
In all cases, we scale the invariants by the total squared momentum of 
the event and introduce the notation
\beq
y_{jk} = \frac{s_{jk}}{s_{12}} 
    = \frac{2p_j\cdot p_k}{2p_1\cdot p_2}\,,
\qquad 
y_{jQ} = \frac{s_{j(12)}}{s_{12}}
    = \frac{2p_j\cdot (p_1+p_2)}{2p_1\cdot p_2}\,,
\qquad j,k=1,\ldots,5\,.
\eeq
Normalizing the invariants in this way makes comparisons between 
the various cases more straightforward. The double real emission 
squared matrix element of course diverges in all limits. Hence, we 
examine the behavior of the {\em ratios} of the various 
subtraction terms to the full $gg\to Hgg$ squared matrix element.

\begin{figure}
    \centering
    \includegraphics[scale=1]{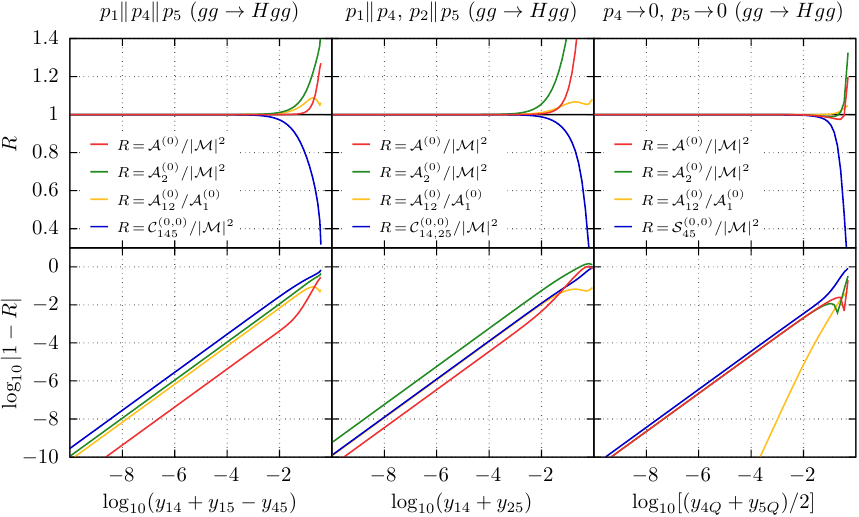}
    \caption{The behavior of ratios of subtraction terms and 
    the squared matrix element in double unresolved limits. 
    Here $|{\cal M}|^2$ denotes the tree-level squared matrix 
    element for the $g(p_1) + g(p_2) \to H(p_3) + g(p_4) + g(p_5)$
    subprocess, while ${\cal A}^{(0)}={\cal A}^{(0)}_1+{\cal A}^{(0)}_2-{\cal 
    A}_{12}$ is the sum of all subtraction terms.}
    \label{fig:A2lim}
\end{figure}

The results of such a study are reported in \fig{fig:A2lim} for 
the triple collinear, double collinear and double soft limits. In 
each case, we plot the ratio of the appropriate subtraction term to 
the squared matrix element (blue), the ratio of the total double 
unresolved subtraction term ${\cal A}^{(0)}_2$ to the squared matrix 
element (green), as well as the ratio of the iterated single 
unresolved subtraction term ${\cal A}^{(0)}_{12}$ to the single 
unresolved subtraction term ${\cal A}^{(0)}_1$ (yellow). Finally, the 
ratio of the sum of all subtraction terms, 
${\cal A}^{(0)} = {\cal A}^{(0)}_1 + {\cal A}^{(0)}_2 
- {\cal A}^{(0)}_{12}$ to the squared matrix element is also 
shown (red). Looking at the top panels, we observe first of all 
that in each case, the ratio of the appropriate subtraction term 
to the squared matrix element goes to one in the limit, 
establishing that the subtraction terms do indeed reproduce the 
singular behavior of the squared matrix element point-by-point in 
phase space. Moreover, we see that the ratio of the full double 
unresolved subtraction term ${\cal A}^{(0)}_2$ to the squared matrix 
element is also one in the limit. Hence, each double unresolved 
configuration is indeed counted precisely once and the appropriate 
cancellations necessary to avoid multiple subtraction take place 
as required. Furthermore, the ratio of the iterated single 
unresolved subtraction term ${\cal A}^{(0)}_{12}$ and the single 
unresolved subtraction term ${\cal A}^{(0)}_1$ approaches one in the 
limit as well. These two terms appear in the sum of all 
subtraction terms with opposite sign, so the fact that this ratio 
goes to one establishes that ${\cal A}^{(0)}_{12}$ correctly cancels 
${\cal A}^{(0)}_1$ in double unresolved limits. This then implies that 
the sum of all subtraction terms properly cancels the 
singularities of the squared matrix element point-by-point in 
phase space in these limits. In order to make this cancellation 
more apparent as the limit is approached, in the bottom panels we 
plot the absolute distance of each ratio from one on a logarithmic 
scale. The `spikes' visible in some of these graphs signal that 
the quantity $(1-R)$ has changed sign. We note that the structure 
of cancellations observed here between the squared matrix element 
and the various subtraction terms is expected to hold also for 
more general processes in our setup.

\begin{figure}
    \centering
    \includegraphics[scale=1]{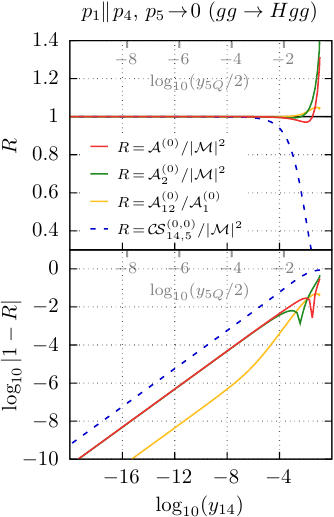}
    \caption{The behavior of ratios of subtraction terms and 
    the squared matrix element in the double unresolved 
    collinear-soft limit. $|{\cal M}|^2$ and ${\cal A}^{(0)}$ are the 
    same as in \fig{fig:A2lim}.}
    \label{fig:CSlim}
\end{figure}

Before moving on to examine the single unresolved limits, let us 
highlight specifically the case of the double unresolved 
collinear-soft singularity. The same ratios that were plotted for 
the other double unresolved limits are shown in \fig{fig:CSlim}. 
The notable difference with respect to the other cases discussed 
above is that in this case the collinear-soft subtraction term 
itself does not directly appear in ${\cal A}^{(0)}_2$, as explained in 
\sect{sec:doubleunresolv} (see \eqn{eq:CScancel1} in particular). 
Hence, the ratio of the would-be collinear-soft subtraction term 
to the matrix element is represented by a dashed blue line. 
Nevertheless, as expected, we see that ${\cal A}^{(0)}_2$ does 
reproduce the behavior of the squared matrix element and 
${\cal A}^{(0)}_{12}$ continues to cancel ${\cal A}^{(0)}_1$ also in 
this limit. Thus the sum of all subtraction terms has the correct 
behavior to cancel the double unresolved singularity of the 
squared matrix element also in the collinear-soft limit.

\begin{figure}
    \centering
    \includegraphics[scale=1]{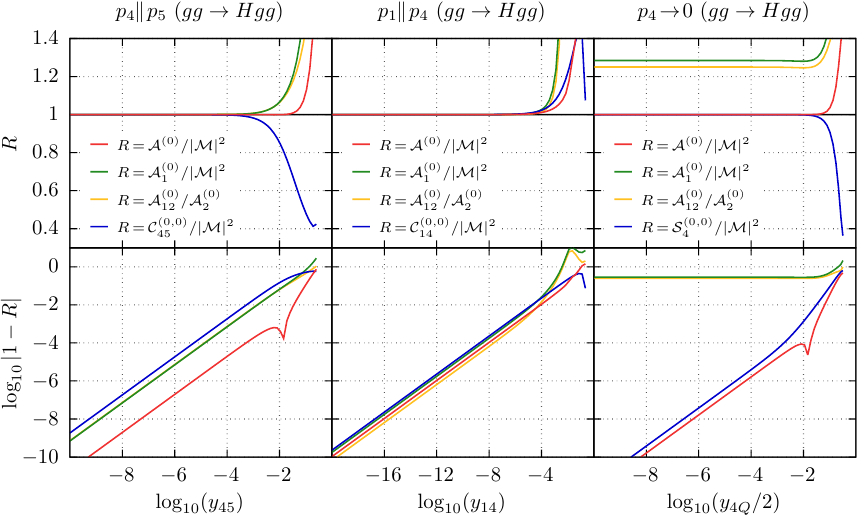}
    \caption{The behavior of ratios of subtraction terms and 
    the squared matrix element in single unresolved limits. 
    $|{\cal M}|^2$ and ${\cal A}^{(0)}$ are the same as in 
    \fig{fig:A2lim}.}
    \label{fig:A1lim}
\end{figure}

Finally, we turn to the case of single unresolved limits. In 
\fig{fig:A1lim}, we present results for the final-final single 
collinear, the initial-final single collinear and the single soft 
limits. This time, the relevant quantities are the ratio of the 
appropriate subtraction term to the squared matrix element (blue), 
the ratio of the total single unresolved subtraction term 
${\cal A}_1$ to the squared matrix element (green), the ratio of the 
iterated single unresolved subtraction term ${\cal A}^{(0)}_{12}$ to 
the double unresolved subtraction term ${\cal A}^{(0)}_2$ (yellow) and 
finally the ratio of the sum of all subtraction terms ${\cal A}^{(0)}$ 
to the squared matrix element (red). In the two collinear limits, 
we observe a behavior that is very much analogous to what we found 
in the double unresolved limits. The individual subtraction terms, 
as well as the total single unresolved subtraction term ${\cal 
A}_1$ reproduce the squared matrix element in the limits, while the 
full iterated single unresolved subtraction term ${\cal A}^{(0)}_{12}$ 
cancels the double unresolved subtraction term ${\cal A}^{(0)}_2$. 
Thus, the sum of all subtraction terms regularizes the single 
unresolved collinear singularities point-by-point in phase space. 
However, in the soft limit, a more intricate structure of 
cancellations is found. We see that while the individual soft 
subtraction term as well as the sum of all subtraction terms continue 
to correctly capture the singular structure of the squared matrix 
element in the limit, the ratio of the total single unresolved 
subtraction term ${\cal A}^{(0)}_1$ and the squared matrix element no 
longer goes to one. This behavior is however expected and arises due 
to singularities which appear in the {\em reduced matrix elements} 
entering certain single unresolved counterterms that are not 
relevant to canceling this soft limit. Similarly, the ratio of 
${\cal A}^{(0)}_{12}$ to ${\cal A}^{(0)}_{2}$ also does not approach 
one, hence these two terms do not cancel in the single soft limit. 
Again, this is expected. In fact, in this limit ${\cal A}^{(0)}_{12}$ 
must not only regularize the single unresolved singularities of 
${\cal A}^{(0)}_{2}$, but also those singularities in 
${\cal A}^{(0)}_{1}$ that originate from the diverging behavior of 
reduced matrix elements. This behavior is the typical one expected in 
our setup for single unresolved limits and the additional 
cancellations observed in the collinear cases (e.g., between 
${\cal A}^{(0)}_{12}$ and ${\cal A}^{(0)}_{2}$) are more or less 
accidental and due to the simplicity of the underlying Born process.

We finish this section by noting that using arbitrary precision 
arithmetic in {\tt Mathematica}, we have also checked the correct 
approach of the sum of all subtraction terms to the squared matrix 
element in very deep limits. We observe the required cancellation 
across several tens of orders of magnitude in all limits.

%%%
%%% Conclusions and outlook
%%%

\section{Conclusions and outlook}
\label{sec:conclusions}

In this work, we have presented the complete set of local 
counterterms necessary to regularize the double real emission 
contribution for perturbative calculations of hadronic collisions 
involving color-singlet production in the CoLoRFulNNLO method. 
This constitutes a crucial step towards the realization of this 
scheme as a general and practical framework for computing 
next-to-next-to-leading order predictions in Quantum 
Chromodynamics.

Explicit formulae, along with the fully detailed momentum 
mappings, were derived following the CoLoRFulNNLO approach. This 
systematic methodology constructs the counterterms based on the 
known universal behavior of the real-emission matrix elements in 
the soft and/or collinear emission limits. All necessary 
ingredients for an independent, {\it de novo} implementation have 
been meticulously reported herein, and the formulae have already 
been incorporated into existing software. Indeed, the full set of 
counterterms and mappings have been successfully implemented 
within the publicly available, general-purpose Monte Carlo program 
{\tt NNLOCAL}. This implementation provides a robust and flexible 
tool for the numerical evaluation of the regularized double real 
component of NNLO cross sections.

The construction of a complete subtraction formalism also mandates 
the integration of the local counterterms over the phase space of 
the unresolved emission(s). While the integration itself is beyond 
the scope of this paper and will be presented elsewhere, we have 
leveraged the freedom in the definition of the real-emission 
subtractions and the momentum mappings to satisfy simultaneously 
two critical requirements. First, to obtain expressions that allow 
for {\em analytic integration} over the phase space of unresolved 
radiation and second, to achieve a {\em small number} of 
counter-events necessary to represent the approximate cross 
sections that locally cancel the singularities of the real 
radiation matrix elements. Such an optimized construction 
significantly streamlines both the analytical and numerical 
aspects of future NNLO calculations. Crucially, the subtraction 
terms defined here can be used to regularize initial-state 
radiation also in processes with final-state jets, whose treatment 
will however require the definition of additional subtraction 
terms that correspond to infrared limits which do not occur in 
color-singlet production. In this sense, the counterterms 
presented here constitute a self-contained subset of the set of 
double unresolved subtraction terms for general hadronic processes.

The methods employed in this paper, which underpin the 
CoLoRFulNNLO approach, are highly general. Following the same 
procedures, this framework can be systematically extended to 
address final-state radiation in hadronic collisions and can also 
be generalized to even higher perturbative orders. Those 
developments are left for future work.

%%%
%%% Acknowledgments
%%%

\section*{Acknowledgments}
\label{sec:acknowledgments}

We are grateful to C.~Duhr, F.~Guadagni, P.~Mukherjee and 
especially S.~Van Thurenhout for their comments on the manuscript. 
G.S.~acknowledges the kind hospitality of the INFN Laboratori 
Nazionali di Frascati, where part of this work was completed. This 
work has been supported by grant K143451 of the National Research, 
Development and Innovation Fund in Hungary and the Bolyai 
Fellowship program of the Hungarian Academy of Sciences.

%%%
%%% Appendix
%%%

\appendix

%%%
%%% Spin and color correlated matrix elements
%%%

\section{Spin and color correlated matrix elements}
\label{appx:spincolor}

Consider the generic process of partons $a$ and $b$ scattering to 
produce a final state with $n$ partons and any number of non-QCD 
particles, collectively denoted by $X$, $a+b \to n\text{ partons} 
+ X$. The corresponding tree-level matrix element is denoted by
\beq
{\cal M}_{f_a,f_b,f_1,\ldots,f_n}^{c_a,c_b,c_1,\ldots,c_n;
s_a,s_b,s_1,\ldots,s_n}(p_a,p_b;\mom{}_{n+X})\,,
\label{eq:ME}
\eeq
where $(c_a,c_b,c_1,\ldots,c_n)$, $(s_a,s_b,s_1,\ldots,s_n)$ and 
$(f_a,f_b,f_1,\ldots,f_n)$ are the color, spin and flavor indices 
of the partons, while $(p_a,p_b;\mom{}_{n+X})$ are the particle 
momenta. In particular, the set of final-state momenta is denoted 
by $\mom{}_{n+X}$. The squared matrix element, summed over colors 
and spins, is then
\beq
\bsp
\SME{ab,n+X}{(0)}{(p_a,p_b;\mom{}_{n+X})} = 
    \sum_{\text{spins}} \sum_{\text{colors}}
    \Big[&
{\cal M}_{f_a,f_b,f_1,\ldots,f_n}^{c_a,c_b,c_1,\ldots,c_n;
s_a,s_b,s_1,\ldots,s_n}(p_a,p_b;\mom{}_{n+X})
    \Big]^*
\\ \times\, &
{\cal M}_{f_a,f_b,f_1,\ldots,f_n}^{c_a,c_b,c_1,\ldots,c_n;
s_a,s_b,s_1,\ldots,s_n}(p_a,p_b;\mom{}_{n+X})\,.
\esp
\label{eq:SME}
\eeq
Notice that $\SME{ab,n+X}{(0)}{(p_a,p_b;\mom{}_{n+X})}$ does not 
carry any normalization factor related to the number of spins or 
colors of the initial-state partons. 

In order to precisely define the product of Altarelli-Parisi 
splitting kernels and square matrix elements, we introduce the 
spin-polarization tensor as follows,
\beq
\bsp
{\cal T}_{ab,n+X}^{s_j s'_j}(p_a,p_b;\mom{}_{n+X}) = 
    \sum_{\text{spins } \ne s_j,s'_j} \sum_{\text{colors}}
    \Big[&
{\cal M}_{f_a,f_b,f_1,\ldots,f_n}^{c_a,c_b,c_1,\ldots,c_n;
s_a,s_b,s_1,\ldots,s'_j,\ldots,s_n}(p_a,p_b;\mom{}_{n+X})\Big]^*
\\ \times\, &
{\cal M}_{f_a,f_b,f_1,\ldots,f_n}^{c_a,c_b,c_1,\ldots,c_n;
s_a,s_b,s_1,\ldots,s_j,\ldots,s_n}(p_a,p_b;\mom{}_{n+X})\,.
\esp
\label{eq:Tspinpol}
\eeq
Thus, ${\cal T}_{ab,n+X}^{s_j s'_j}(p_a,p_b;\mom{}_{n+X})$ is 
defined similarly to the squared matrix element in \eqn{eq:SME}, 
however the sum over the spin polarizations of parton $j$ is not 
carried out. Then, products between splitting kernels and squared 
matrix elements are defined as
\beq
\hP^{(0)}_{i r} \SME{ab,n+X}{(0)}{(p_a,p_b;\mom{}_{n+X})} = 
    \sum_{s_{ir},s'_{ir}} \hP^{(0),s_{ir} s'_{ir}}_{i r} 
    {\cal T}_{ab,n+X}^{s_{ir} s'_{ir}}{(p_a,p_b;\mom{}_{n+X})}\,,
\label{eq:PxM}    
\eeq
where $s_{ir}$ and $s'_{ir}$ are the spin polarizations of the 
parent parton in the $(ir) \to i+r$ splitting. This definition 
generalizes  easily to the case when the squared matrix element is 
multiplied by two splitting kernels as happens in, e.g., 
\eqns{eq:CarbsIFIF00}{eq:CcsdrIFIF00CcsIF}. First, the appropriate 
spin polarization tensor is defined as
\beq
\bsp
{\cal T}_{ab,n+X}^{s_j s'_j, s_k s'_k}(p_a,p_b;\mom{}_{n+X}) = 
    \sum_{\substack{\text{spins } \ne \\ s_j,s'_j,s_k,s'_k}} 
    \sum_{\text{colors}}
    \Big[&
{\cal M}_{f_a,f_b,f_1,\ldots,f_n}^{c_a,c_b,c_1,\ldots,c_n;
s_a,s_b,s_1,\ldots,s'_j,\ldots,s'_k,\ldots,s_n}
(p_a,p_b;\mom{}_{n+X})\Big]^*
\\ \times\, &
{\cal M}_{f_a,f_b,f_1,\ldots,f_n}^{c_a,c_b,c_1,\ldots,c_n;
s_a,s_b,s_1,\ldots,s_j,\ldots,s_k,\ldots,s_n}
(p_a,p_b;\mom{}_{n+X})\,,
\esp
\label{eq:Tspinpol-2}
\eeq
i.e., the summations over the spin polarizations of partons $j$ 
and $k$ are not carried out. Then we have
\beq
\hP^{(0)}_{i r} \hP^{(0)}_{j s} 
\SME{ab,n+X}{(0)}{(p_a,p_b;\mom{}_{n+X})} = 
    \sum_{s_{ir},s'_{ir}} \sum_{s_{js},s'_{js}}
    \hP^{(0),s_{ir} s'_{ir}}_{i r} \hP^{(0),s_{js} s'_{js}}_{j s} 
    {\cal T}_{ab,n+X}^{s_{ir} s'_{ir}, s_{js} s'_{js}}
    {(p_a,p_b;\mom{}_{n+X})}\,,
\label{eq:PPxM}
\eeq
where now $s_{ir}$, $s'_{ir}$ and $s_{js}$, $s'_{js}$ are the spin 
polarizations of the parent partons in the $(ir) \to i+r$ and 
$(js) \to j+s$ splittings.

Turning to the definition of products of color-charge operators 
with squared matrix elements, we consider first the case of two 
color-charge operators and define
\beq
\bsp
\bT_j \bT_k \SME{ab,n+X}{(0)}{(p_a,p_b;\mom{}_{n+X})} &= 
    \sum_{\text{spins }} \sum_{\text{colors}}
    \Big[
{\cal M}_{f_a,f_b,f_1,\ldots,f_n}^{c_a,c_b,c_1,\ldots,c'_j,
\ldots,c_k,\ldots,c_n;s_a,s_b,s_1,\ldots,s_n}
(p_a,p_b;\mom{}_{n+X})\Big]^*
\\ &\times
T^{c}_{c'_j c_j} T^{c}_{c_k c'_k}
{\cal M}_{f_a,f_b,f_1,\ldots,f_n}^{c_a,c_b,c_1,\ldots,c_j,
\ldots,c'_k,\ldots,c_n;s_a,s_b,s_1,\ldots,s_n}
(p_a,p_b;\mom{}_{n+X})\,,
\esp
\label{eq:TjTkSME}
\eeq
where summation over repeated color indices is implied and the 
color-charge matrices are $T^{c}_{a_j b_j} = if_{ca_j b_j}$ (the 
color-charge matrix in the adjoint representation) if parton $j$ 
is a final-state gluon and $T^c_{\al_j \be_j} = t^c_{\al_j \be_j}$ 
(the color-charge matrix in the fundamental representation) if 
parton $j$ is a final-state quark. For a final-state antiquark 
$j$, we have $T^c_{\al_j \be_j} = \ba{t}^c_{\al_j \be_j} = -
t^c_{\be_j \al_j}$. The color-charge matrices for initial-state 
partons can be obtained by crossing and we have $T^c_{a_j b_j} = 
if_{a_j cb_j}$ for an initial-state gluon $j$, $T^c_{a_j b_j} = 
\ba{t}^c_{\al_j \be_j} = -t^c_{\al_j \be_j}$ for an initial-state 
quark $j$ and $T^c_{a_j b_j} = t^c_{\al_j \be_j}$ for an 
initial-state antiquark $j$. We must also consider the product of 
the squared matrix element with four color-charge operators as in 
\eqns{eq:Srsgg00}{eq:SsSrs00}. In this case we have
\beq
\bsp
&
\{\bT_i \bT_k,\bT_j \bT_\ell\} 
\SME{ab,n+X}{(0)}{(p_a,p_b;\mom{}_{n+X})} 
\\ &\qquad=
\sum_{\text{spins }} \sum_{\text{colors}}
    \Big[
{\cal M}_{f_a,f_b,f_1,\ldots,f_n}^{c_a,c_b,c_1,\ldots,c'_i,
\ldots,c'_j,\ldots,c_k,\ldots,c_\ell,\ldots,c_n;
s_a,s_b,s_1,\ldots,s_n}(p_a,p_b;\mom{}_{n+X})\Big]^*
\Big[T^{c}_{c'_i c_i} T^{c}_{c_k c'_k} 
T^{d}_{c'_j c_j} T^{d}_{c_\ell c'_\ell}
\\ &\qquad\qquad\qquad\qquad
+ T^{d}_{c'_j c_j} T^{d}_{c_\ell c'_\ell} 
T^{c}_{c'_i c_i} T^{c}_{c_k c'_k}\Big]
{\cal M}_{f_a,f_b,f_1,\ldots,f_n}^{c_a,c_b,c_1,\ldots,c_i,
\ldots,c_j,\ldots,c'_k,\ldots,c'_\ell,\ldots,c_n;
s_a,s_b,s_1,\ldots,s_n}(p_a,p_b;\mom{}_{n+X})\,.
\esp
\label{eq:TiTkTjTlSME}
\eeq
Once more, repeated color indices are understood to be summed 
over. We note furthermore that the color-charge algebra for the 
product $\bT_j \bT_k$ is
\beq
\bT_j \bT_k = \bT_k \bT_j\,,\quad\mbox{if }j\ne k
\qquad\mbox{and}\qquad
\bT_j^2 = C_j\,,
\eeq
where $C_j$ is the quadratic Casimir operator in the 
representation of parton $j$. Using the customary normalization of 
$\TR = 1/2$, we have $C_j = \CA = \Nc$ if $j$ is a gluon, while 
$C_j = \CF = (\Nc^2-1)/(2\Nc)$ if $j$ is a quark or antiquark.

Finally, in order to make contact with the color- and spin-space 
notation of~\refr{Catani:1996vz}, we note that by introducing a 
basis $|c_a,c_b,c_1,\ldots,c_n\ra \otimes 
|s_a,s_b,s_1,\ldots,s_n\ra$ in color- and spin-space, the matrix 
element in \eqn{eq:ME} can be represented by an abstract vector,
\beq
\bsp
\ket{ab,n+X}{(0)}{(p_a,p_b;\mom{}_{n+X})} &=
    \Big(|c_a,c_b,c_1,\ldots,c_n\ra \otimes 
    |s_a,s_b,s_1,\ldots,s_n\ra\Big)
\\&\times
    {\cal M}_{f_a,f_b,f_1,\ldots,f_n}^{c_a,c_b,c_1,\ldots,c_n;
    s_a,s_b,s_1,\ldots,s_n}(p_a,p_b;\mom{}_{n+X})\,.
\esp
\eeq
Notice that the flavors on the left-hand side are understood. 
Moreover, compared to~\refr{Catani:1996vz} (see their eq.~(3.11)), 
we {\em do not} include factors of $1/\sqrt{\Nc(f)}$ for each 
initial-state parton of flavor $f$ carrying $\Nc(f)$ colors. This 
corresponds to the normalization adopted in~\refr{Somogyi:2009ri}. 
Using this notation, the squared matrix element is simply
\beq
\SME{ab,n+X}{(0)}{(p_a,p_b;\mom{}_{n+X})} = 
\braket{ab,n+X}{(0)}{(p_a,p_b;\mom{}_{n+X})}{ab,n+X}{(0)}
{(p_a,p_b;\mom{}_{n+X})}\,.
\eeq
Then, products between splitting functions and squared matrix can 
be defined as
\beq
\hP^{(0)}_{i r} 
\SME{ab,n+X}{(0)}{(p_a,p_b;\mom{}_{n+X})} =
\bra{ab,n+X}{(0)}{(p_a,p_b;\mom{}_{n+X})}
\hP^{(0)}_{i r} 
\ket{ab,n+X}{(0)}{(p_a,p_b;\mom{}_{n+X})}
\eeq
and
\beq
\hP^{(0)}_{i r} \hP^{(0)}_{j s} 
\SME{ab,n+X}{(0)}{(p_a,p_b;\mom{}_{n+X})} =
\bra{ab,n+X}{(0)}{(p_a,p_b;\mom{}_{n+X})}
\hP^{(0)}_{i r} \hP^{(0)}_{j s} 
\ket{ab,n+X}{(0)}{(p_a,p_b;\mom{}_{n+X})}\,,
\eeq
where the splitting kernels $\hP^{(0)}_{i r}$ and $\hP^{(0)}_{j 
s}$ are matrices acting on the spin indices of the appropriate 
parent partons. In this notation, their matrix elements i.e., 
$\hP^{(0),s_{ir} s'_{ir}}_{i r}$ and $\hP^{(0),s_{js} s'_{js}}_{j 
s}$, are written as $\la s'_{ir}|\hP^{(0)}_{i r}|s_{ir}\ra$ and 
$\la s'_{js}|\hP^{(0)}_{j s}|s_{js}\ra$.\footnote{Here the generic 
symbol $s$ denotes the spin of the parent parton. This notation is 
appropriate if the parent is a quark or antiquark. However, when 
the parent is a gluon it is customary to use Greek letters, e.g., 
$\mu$ and $\nu$, to denote helicity indices. We will follow this 
convention below, see e.g., \eqnss{eq:Pqg0FF}{eq:Pgg0FF}.} 
Products of color-charge operators and matrix elements may be 
defined in a similar way. First, if the color-charge operator 
associated to the emission of a gluon from parton $i$ has color 
$c$, we can write
\beq
\bT_i = T_i^c |c\ra\,,
\eeq
and the action of this operator in color space is defined as
\beq
\la c'_a,c'_b,c'_1,\ldots,c'_i,\ldots,c'_n,c| \bT_i
|c_a,c_b,c_1,\ldots,c_i,\ldots,c_n\ra = 
\delta_{c'_a c_a}\delta_{c'_b c_b}\ldots T^c_{c'_i,c_i}\ldots 
\delta_{c'_n c_n}\,.
\eeq
Here the $T^c_{c'_i,c_i}$ are as explained above. Then, products 
between color-charge operators and matrix elements may be written 
as
\beq
\bT_j \bT_k \SME{ab,n+X}{(0)}{(p_a,p_b;\mom{}_{n+X})} = 
    \bra{ab,n+X}{(0)}{(p_a,p_b;\mom{}_{n+X})}
    \bT_j \bT_k
    \ket{ab,n+X}{(0)}{(p_a,p_b;\mom{}_{n+X})}\,,
\eeq
and
\beq
\bsp
&
\{\bT_i \bT_k,\bT_j \bT_\ell\} \SME{ab,n+X}{(0)}
{(p_a,p_b;\mom{}_{n+X})} 
\\&\qquad= 
    \bra{ab,n+X}{(0)}{(p_a,p_b;\mom{}_{n+X})}
    \{\bT_i \bT_k,\bT_j \bT_\ell\}
    \ket{ab,n+X}{(0)}{(p_a,p_b;\mom{}_{n+X})}\,,
\esp
\eeq
where $\{\bT_i \bT_k,\bT_j \bT_\ell\}$ denotes the anti-commutator 
of the products of color-charge operators.

%%%
%%% Momentum mappings
%%%

\section{Momentum mappings}
\label{appx:mommaps}

We gather here the definitions of the five elementary momentum 
mappings we have introduced.

In order to insure that this appendix is self-contained, we recall 
that throughout $s_{jk}$ denotes twice the dot-product of momenta 
$p_j$ and $p_k$. Also, indices in parentheses denote sums of the 
corresponding momenta so that, e.g.,
\beq
s_{jk} = 2p_j\cdot p_k\,,
\qquad
s_{j(k\ell)} = 2p_j\cdot (p_k+p_\ell)\,,
\qquad
s_{(jk)(\ell m)} = 2(p_j+p_k)\cdot (p_\ell+p_m)\,,
\eeq
and so on. Furthermore, $Q^\mu=p_a^\mu+p_b^\mu$ always denotes the 
total incoming {\em partonic} momentum. The various momentum mappings 
are then defined as follows.
\begin{itemize}
\item The final-state single collinear (FF) momentum mapping,
\beq
\cmp{ir}{FF}:\; (p_a, p_b; \mom{}_{m+X+2}) 
    \cmap{ir}{FF} (\ha{p}_a, \ha{p}_b; \momh{}_{m+X+1})\,,
\eeq
is defined as ($i,r \in F$)
\beq
\bsp
\ha{p}_a^\mu &= (1-\al_{ir})p_a^\mu\,,
\\
\ha{p}_b^\mu &= (1-\al_{ir})p_b^\mu\,,
\\
\ha{p}_{ir}^\mu &= p_i^\mu + p_r^\mu - \al_{ir} (p_a+p_b)^\mu\,,
\\
\ha{p}_n^\mu &= p_n^\mu\,, \qquad n\in F\,,\quad n \ne i,r\,,
\\
\ha{p}_X^\mu &= p_X^\mu\,.
\esp
\eeq
The value of $\al_{ir}$ is fixed by requiring that the parent 
momentum, $\ha{p}_{ir}^\mu$, be massless, $\ha{p}_{ir}^2 = 0$, 
which gives
\beq
\al_{ir} = \frac{1}{2}\left[\frac{s_{(ir)\ab}}{s_{ab}} 
    - \sqrt{\frac{s_{(ir)\ab}^2}{s_{ab}^2} 
    - \frac{4s_{ir}}{s_{ab}}}\,\right]\,.
\eeq

\item The initial-state single collinear (IF) momentum mapping,
\beq
\cmp{ab,r}{II,F}:\; (p_a, p_b; \mom{}_{m+X+2}) 
    \cmap{ab,r}{II,F} (\ha{p}_a, \ha{p}_b; \momh{}_{m+X+1})\,,
\eeq
is defined as ($a,b \in I$ and $r \in F$)
\beq
\bsp
\ha{p}_a^\mu &= \xi_{a,r} p_a^\mu\,,
\\
\ha{p}_b^\mu &= \xi_{b,r} p_b^\mu\,,
\\
\ha{p}_n^\mu &= {\Lambda(P,\ha{P})^\mu}_{\!\nu}\, p_n^\nu\,, 
\qquad n \in F, \quad n\ne r\,,
\\
\ha{p}_X^\mu &= {\Lambda(P,\ha{P})^\mu}_{\!\nu}\, p_X^\nu\,.
\esp
\eeq
Here ${\Lambda(P,\ha{P})^\mu}_{\!\nu}$ is a proper Lorentz 
transformation that takes the massive momentum $P^\mu$ into a 
momentum $\ha{P}^\mu$ of the same mass, 
\beq
{\Lambda(P,\ha{P})^\mu}_{\!\nu} = 
	{g^\mu}_{\!\nu} - 2\frac{(P+\ha{P})^\mu (P+\ha{P})_\nu}
    {(P+\ha{P})^2} + 2\frac{\ha{P}^\mu P_\nu}{P^2}\,,
\label{eq:Lambda_munu-appx}
\eeq
where 
\beq
P^\mu = (p_a+p_b)^\mu - p_r^\mu\,,
\qquad\qquad
\ha{P}^\mu = (\ha{p}_a + \ha{p}_b)^\mu 
    = (\xi_{a,r}p_a + \xi_{b,r} p_b)^\mu\,.
\eeq
The variables $\xi_{a,r}$ and $\xi_{b,r}$ are defined as follows,
\beq
\xi_{a,r} = \sqrt{\frac{s_{ab} - s_{br}}{s_{ab} - s_{ar}}
	\frac{s_{ab} - s_{r\ab}}{s_{ab}}}\,,
\qquad
\xi_{b,r} = \sqrt{\frac{s_{ab} - s_{ar}}{s_{ab} - s_{br}}
	\frac{s_{ab} - s_{r\ab}}{s_{ab}}}\,.
\eeq
These definitions ensure that indeed $P^2=\ha{P}^2$.

\item The single soft (S) mapping,
\beq
\smp{r}:\; (p_a,p_b;\mom{}_{m+X+2}) 
    \to (\ti{p}_a, \ti{p}_b; \momt{}_{m+X+1})\,,
\eeq
is defined as follows ($r \in F$) 
\beq
\bsp
\ti{p}_a^\mu &= \lambda_r p_a^\mu\,,
\\
\ti{p}_b^\mu &= \lambda_r p_b^\mu\,,
\\
\ti{p}_n^\mu &= {\Lambda(P,\ti{P})^\mu}_{\!\nu}\, p_n^\nu\,, 
\qquad n \in F, \quad n\ne r\,,
\\
\ti{p}_X^\mu &= {\Lambda(P,\ti{P})^\mu}_{\!\nu}\, p_X^\nu\,,
\esp
\eeq
where ${\Lambda(P,\ti{P})^\mu}_{\!\nu}$ is the proper Lorentz 
transformation of \eqn{eq:Lambda_munu-appx} that takes the massive 
momentum $P^\mu$ into a momentum $\ti{P}^\mu$ of the same mass, with
\beq
P^\mu = (p_a+p_b)^\mu - p_r^\mu
\qquad\mbox{and}\qquad
\ti{P}^\mu = (\ti{p}_a + \ti{p}_b)^\mu = \lambda_r (p_a+p_b)^\mu\,.
\eeq
The value of $\lambda_{r}$ is fixed by requiring that $P^2 = 
\ti{P}^2$,
\beq
\lambda_{r} = \sqrt{1 - \frac{s_{r\ab}}{s_{ab}}}\,.
\eeq

\item The initial-state double collinear momentum mapping,
\beq
\cmp{ab,rs}{II,FF}:\; (p_a, p_b; \mom{}_{m+X+2}) 
\cmap{ab,rs}{II,FF} (\ha{p}_a,\ha{p}_b;\momh{}_{m+X})\,,
\eeq
is defined as ($a,b \in I$ and $r,s \in F$)
\beq
\bsp
\ha{p}_a^\mu &= \xi_{a,rs} p_a^\mu\,,
\\
\ha{p}_b^\mu &= \xi_{b,rs} p_b^\mu\,,
\\
\ha{p}_n^\mu &= {\Lambda(P,\ha{P})^\mu}_{\!\nu}\, p_n^\nu\,, 
\qquad n \in F, \quad n\ne r\,,
\\
\ha{p}_X^\mu &= {\Lambda(P,\ha{P})^\mu}_{\!\nu}\, p_X^\nu\,,
\esp
\eeq
where ${\Lambda(P,\ha{P})^\mu}_{\!\nu}$ is the proper Lorentz 
transformation of \eqn{eq:Lambda_munu-appx} that takes the massive 
momentum $P^\mu$ into a momentum $\ha{P}^\mu$ of the same mass, with 
\beq
P^\mu = (p_a+p_b)^\mu - p_r^\mu - p_s^\mu \,,
\qquad
\ha{P}^\mu = (\ha{p}_a + \ha{p}_b)^\mu 
    = (\xi_{a,rs}p_a + \xi_{b,rs} p_b)^\mu\,.
\eeq
The variables $\xi_{a,rs}$ and $\xi_{b,rs}$ are defined as follows,
\beq
\xi_{a,rs} = \sqrt{\frac{s_{ab} - s_{b(rs)}}{s_{ab} - s_{a(rs)}}
	\frac{s_{ab} - s_{(rs)\ab} + s_{rs}}{s_{ab}}}\,,
\qquad\quad
\xi_{b,rs} = \sqrt{\frac{s_{ab} - s_{a(rs)}}{s_{ab} - s_{b(rs)}}
	\frac{s_{ab} - s_{(rs)\ab} + s_{rs}}{s_{ab}}}\,.
\eeq
These definitions ensure that indeed $P^2 = \ha{P}^2$.

\item The double soft mapping,
\beq
\smp{rs}:\; (p_a, p_b; \mom{}_{m+X+2}) 
    \smap{rs} (\ti{p}_a,\ti{p}_b;\momt{}_{m+X})\,,
\eeq
is defined as ($r,s \in F$),
\beq
\bsp
\ti{p}_a^\mu &= \lambda_{rs} p_a^\mu\,,
\\
\ti{p}_b^\mu &= \lambda_{rs} p_b^\mu\,,
\\
\ti{p}_n^\mu &= {\Lambda(P,\ha{P})^\mu}_{\!\nu}\, p_n^\nu\,,  
\qquad n \in F, \quad n\ne r\,,
\\
\ti{p}_X^\mu &= {\Lambda(P,\ha{P})^\mu}_{\!\nu}\, p_X^\nu\,,
\esp
\eeq
where ${\Lambda(P,\ha{P})^\mu}_{\!\nu}$ is the proper Lorentz 
transformation of \eqn{eq:Lambda_munu-appx} that takes the massive 
momentum $P^\mu$ into a momentum $\ti{P}^\mu$ of the same mass, 
where
\beq
P^\mu = (p_a+p_b)^\mu - p_r^\mu - p_s^\mu\,, 
\qquad
\ti{P}^\mu = (\ti{p}_a + \ti{p}_b)^\mu 
    = \lambda_{rs} (p_a+p_b)^\mu\,.
\eeq
The value of $\lambda_{rs}$ is fixed by requiring that 
$P^2 = \ti{P}^2$,
\beq
\lambda_{rs} = \sqrt{1 - \frac{s_{(rs)\ab}}{s_{ab}}  
    + \frac{s_{rs}}{s_{ab}}} \,.
\eeq

\end{itemize}

%%%
%%% Momentum fractions, transverse momenta and eikonal 
%%% factors
%%%

\section{Momentum fractions, transverse momenta and eikonal 
factors}
\label{appx:kinquant}

In this appendix we gather the definitions of the various 
kinematic quantities such as momentum fractions, transverse 
momenta and eikonal factors that we have introduced.

In order to ensure that this appendix is self-contained, we repeat our 
conventions for dot-products. Thus, throughout $s_{jk}$ denotes twice 
the dot-product of momenta $p_j$ and $p_k$, while indices in 
parentheses denote sums of the corresponding momenta so that, e.g.,
\beq
s_{jk} = 2p_j\cdot p_k\,,
\qquad
s_{j(k\ell)} = 2p_j\cdot (p_k+p_\ell)\,,
\qquad
s_{(jk)(\ell m)} = 2(p_j+p_k)\cdot (p_\ell+p_m)\,,
\eeq
and so on. Moreover, $Q^\mu=p_a^\mu+p_b^\mu$ always denotes the 
total incoming {\em partonic} momentum. Then, momentum fractions 
are defined using the following functions of momenta.
\begin{itemize}
\item For {\em final-state} partons $i$ and $r$ we define the 
kinematic function $z_{i,r}$ of the momenta $p_i^\mu$ and 
$p_r^\mu$ as
\beq
z_{i,r} = \frac{s_{i\ab}}{s_{(ir)\ab}}\,,
\qquad i,r\in F\,.
\eeq

\item For an {\em initial-state} parton $a$ and a 
{\em final-state} parton $r$, we define the kinematic functions
\beq
x_{r,a} = \frac{s_{r\ab}}{s_{a\ab}}\,,
\quad
x_{a,r} = 1 - x_{r,a} = 1 - \frac{s_{r\ab}}{s_{a\ab}}\,,
\qquad a\in I\,,\; r\in F\,.
\eeq

\item For an {\em initial-state} parton $a$ and {\em final-state} 
partons $r$ and $s$, we define the kinematic functions
\beq
x_{r,a} = \frac{s_{r\ab}}{s_{a\ab}}\,,
\quad
x_{s,a} = \frac{s_{s\ab}}{s_{a\ab}}\,,
\quad
x_{a,rs} = 1 - x_{r,a} - x_{s,a} = 
    1 - \frac{s_{r\ab}}{s_{a\ab}} - \frac{s_{s\ab}}{s_{a\ab}}\,,
\qquad a\in I\,,\; r,s\in F\,.
\eeq
\end{itemize}
Next, transverse momenta are defined as follows.
\begin{itemize}
\item For {\em final-state} partons $i$ and $r$ we set
\beq
\kT{i,r}^\mu = \zeta_{i,r} p_r^\mu - \zeta_{r,i} p_i^\mu 
    + Z_{ir} \ha{p}_{ir}^\mu\,,
\eeq
where
\beq
\zeta_{i,r} = z_{i,r} - \frac{s_{ir}}{\al_{ir} s_{(ir)\ab}}\,,
\quad
\zeta_{r,i} = z_{r,i} - \frac{s_{ir}}{\al_{ir} s_{(ir)\ab}}\,,
\quad
Z_{ir} = \frac{s_{ir}}{\al_{ir}s_{\wha{ir}\ab}}(z_{r,i}-z_{i,r})\,,
\eeq
with
\beq
\ha{p}_{ir}^\mu = p_i^\mu + p_r^\mu - \al_{ir}(p_a+p_b)^\mu
\eeq
and
\beq
\al_{ir} = \frac{1}{2}\left[\frac{s_{(ir)\ab}}{s_{ab}} - 
\sqrt{\frac{s_{(ir)\ab}^2}{s_{ab}^2} 
- \frac{4s_{ir}}{s_{ab}}}\,\right]\,.
\eeq

\item For an {\em initial-state} parton $a$ and a 
{\em final-state} parton $r$ we set
\beq
\kT{r,a}^\mu = p_{r,\perp}^\mu\,,
\quad
\kT{a,r}^\mu = -\kT{r,a}^\mu = -p_{r,\perp}^\mu\,,
\qquad a\in I\,,\; r\in F\,,
\eeq
where the orthogonal direction is defined with respect to the 
direction of $p_a^\mu$.

\item For an {\em initial-state} parton $a$ and {\em final-state} 
partons $r$ and $s$ we set
\beq
\kT{r,a}^\mu = p_{r,\perp}^\mu\,,
\quad
\kT{s,a}^\mu = p_{s,\perp}^\mu\,,
\quad
\kT{a,rs}^\mu = -\kT{r,a}^\mu -\kT{s,a}^\mu 
    = -p_{r,\perp}^\mu -p_{s,\perp}^\mu\,,
\qquad a\in I\,,\; r,s\in F\,,
\eeq
where the orthogonal direction is defined with respect to the 
direction of $p_a^\mu$.
\end{itemize}
Finally, for partons $j$ and $k$ (either initial-state or 
final-state) and {\em final-state} parton $r$, the eikonal factor 
is defined as
\beq
\calS_{jk}(r) = \frac{2s_{jk}}{s_{jr} s_{kr}}\,,
\qquad j,k\in I\cup F\,,\; r\in F\,.
\eeq
Recall that when the momenta in the arguments of any of these 
functions is drawn from a set of mapped momenta, e.g., 
$(\ti{p}_a,\ti{p}_b;\momt{})$, the index inherits the tilde (or 
corresponding) notation.

%%%
%%% Collinear splitting functions
%%%

\section{Collinear splitting functions}
\label{appx:APfunctions}

%
% Splitting functions
%

\subsection{Splitting functions}
\label{appx:AP-functions}

The tree-level Altarelli-Parisi splitting kernels for final-final 
collinear splitting are~\cite{Altarelli:1977zs}
\bal
\la s|\hP^{(0)}_{q g}(z,\kTm{};\ep)|s'\ra &= 
	\CF \left[\frac{1 + z^2}{1 - z} - \ep (1 - z)\right] 
    \delta_{ss'}\,,
\label{eq:Pqg0FF}
\\
\la s|\hP^{(0)}_{g q}(z,\kTm{};\ep)|s'\ra &= 
	\CF \left[\frac{1 + (1 - z)^2}{z} - \ep z\right] 
    \delta_{ss'}\,,	
\label{eq:Pgq0FF}
\\
\la \mu|\hP^{(0)}_{q \qb}(z,\kTm{};\ep)|\nu\ra &= 
	\TR \left[-g^{\mu\nu} + 4 z(1 - z) 
	\frac{\kTm{\mu} \kTm{\nu}}{\kTm{2}}\right]\,,	
\label{eq:Pqq0FF}
\\
\la \mu|\hP^{(0)}_{g g}(z,\kTm{};\ep)|\nu\ra &= 
	2\CA \left[
	-g^{\mu\nu}\left(\frac{z}{1 - z} + \frac{1 - z} {z}\right) 
	-2(1 - \ep)z(1 - z) 
	\frac{\kTm{\mu} \kTm{\nu}}{\kTm{2}}\right]\,,
\label{eq:Pgg0FF}
\eal
where we have used the color- and spin-state notation 
of~\cite{Catani:1996vz}, recalled in \appx{appx:spincolor} above. 
To avoid any ambiguity, we note that $z$ here refers to the 
momentum fraction of the first parton in the subscript. The 
azimuthally averaged splitting functions are obtained by 
contracting the splitting kernels with
\beq
\frac{1}{2} \delta_{ss'}
\label{eq:APq_ave}
\eeq
for fermion splitting or with
\beq
\frac{1}{d-2}d_{\mu\nu}(p,n) = \frac{1}{2(1-\ep)}
    \left(-g_{\mu\nu}
    + \frac{p^\mu n^\nu + p^\nu n^\mu}{p\cdot n}\right)
\label{eq:APg_ave}
\eeq
for gluon splitting. Here $p^\mu$ is the on-shell momentum of the 
parent gluon and $n$ is a reference vector required to fix the 
polarization vectors of the parent. Given that
\beq
\sum_{s,s'} \frac{1}{2} \delta_{ss'} \delta_{ss'} = 1
\eeq
and
\beq
\frac{1}{d-2}d_{\mu\nu}(p,n) (-g^{\mu\nu}) = 1\,,
\qquad
\frac{1}{d-2}d_{\mu\nu}(p,n)
\frac{\kTm{\mu} \kTm{\nu}}{\kTm{2}} = -\frac{1}{2(1-\ep)}\,,
\eeq
(we used $p\cdot \kT{}=0$) we see that the average of the 
splitting kernel $\hP_{ir}^{(0)}$ over the polarizations of the 
parent parton can be computed by a very simple prescription. For 
quark splitting we simply set
\beq
\delta_{ss'} \to 1\,,
\label{eq:APq_ave_presc}
\eeq
while for gluon splitting we replace
\beq
-g^{\mu\nu} \to 1\,,
\qquad
\frac{\kTm{\mu} \kTm{\nu}}{\kTm{2}} \to -\frac{1}{2(1-\ep)}\,.
\label{eq:APg_ave_presc}
\eeq
Thus, the averaged splitting functions $P_{ir}^{(0)}$ read
\bal
P_{qg}^{(0)}(z;\ep) &= 
	\CF\left[\frac{1+z^2}{1-z} - \ep(1-z)\right]\,,
\label{eq:Pqg-ave}
\\
P_{gq}^{(0)}(z;\ep) &= 
	\CF\left[\frac{1+(1-z)^2}{z} - \ep z\right]\,,
\label{eq:Pgq-ave}
\\
P_{q\qb}^{(0)}(z;\ep) &= 
	\TR\left[1 - \frac{2z(1-z)}{1-\ep}\right]\,,
\label{eq:Pqq-ave}
\\
P_{gg}^{(0)}(z;\ep) &= 
	2\CA\left[\frac{z}{1-z} + \frac{1-z}{z} + z(1-z)\right]\,.
\label{eq:Pgg-ave}
\eal

\begin{figure}
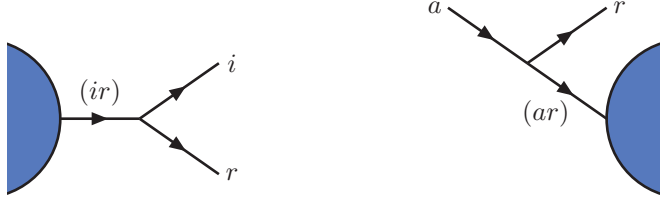

\setlength{\tabcolsep}{10pt}
\begin{center}
\includegraphics[scale=1]{Pir.pdf}
\hspace{5em}
\includegraphics[scale=1]{Par.pdf}
\end{center}
\caption{\label{fig:PirParflav}
The splitting configurations for $(ir)\to i+r$ final-state 
splitting (left) and $a\to (ar)+r$ initial-state splitting 
(right). The arrows in the figure denote momentum and flavor flow.
}
\end{figure}

The splitting kernels for initial-final collinear splitting can be 
obtained from \eqns{eq:Pqg0FF}{eq:Pgg0FF} by crossing. In order to 
derive the appropriate rules, consider crossing parton $i$ to the 
initial state, i.e., $i\to a$. Clearly, the parent parton $(ir)$ 
is then also crossed to the initial state, $(ir)\to (ar)$, and we 
have the corresponding reversal of momenta, $p_i^\mu\to -p_a^\mu$ 
and $p_{ir}^\mu \to -p_{ar}^\mu$. The appropriate splitting 
configurations and momentum flows were presented in 
\figs{fig:Pirflav}{fig:Parflav}, but are reproduced here in 
\fig{fig:PirParflav} for convenience.
Then we seek relations between the momentum fractions and 
transverse momenta describing the two collinear configurations. 
These can be obtained by noting that on the one hand, starting 
from the parametrization relevant to final-final splitting and 
reversing the momenta of partons $i$ and $(ir)$ gives
\bal
p_i^\mu = z_i p_{ir}^\mu + \kT{i}^\mu + \dots
&\;\;\xrightarrow{\;\;\mathrm{cross}\;\;}\;\;
p_a^\mu = z_i p_{ar}^\mu - \kT{i}^\mu + \dots\,,
\label{eq:picrossed}
\\
p_r^\mu = z_r p_{ir}^\mu + \kT{r}^{(F),\mu} + \dots
&\;\;\xrightarrow{\;\;\mathrm{cross}\;\;}\;\;
p_r^\mu = -z_r p_{ar}^\mu + \kT{r}^{(F),\mu} + \dots\,,
\label{eq:prcrossed}
\eal
where the $\ldots$ denote terms that vanish faster than 
$\kT{}^\mu$ in the collinear limit and which do not play a role in 
the following discussion. On the other hand, the corresponding 
parametrization in initial-final splitting reads
\bal
p_{ar}^\mu = x_a p_a^\mu + \kT{a}^\mu + \dots
&\;\;\Rightarrow\;\;
p_a^\mu = \frac{1}{x_a}p_{ar}^\mu - \frac{1}{x_a}\kT{a}^\mu 
+ \dots\,,
\label{eq:paexpr}
\\
p_r^\mu = x_r p_a + \kT{r}^{(I),\mu} + \dots
&\;\;\Rightarrow\;\;
p_r^\mu = \frac{x_r}{x_a}p_{ar}^\mu - \frac{x_r}{x_a}\kT{a}^\mu 
+ \kT{r}^{(I),\mu} + \dots\,.
\label{eq:prexpr}
\eal
Notice that we have used the superscripts $(F)$ and $(I)$ to 
distinguish the transverse momentum for parton $r$ in final-state 
splitting, $\kT{r}^{(F),\mu}$,  versus initial-state splitting, 
$\kT{r}^{(I),\mu}$. Then, comparing the ${\cal O}(\kT{}^0)$ and 
${\cal O}(\kT{})$ pieces of \eqns{eq:picrossed}{eq:paexpr} as well as 
\eqns{eq:prcrossed}{eq:prexpr}, we find
\beq
z_i \to \frac{1}{x_a}\,,
\qquad
z_r \to -\frac{x_r}{x_a}
\label{eq:crossIFx}
\eeq
and
\beq
\kT{i}^\mu\to \frac{1}{x_a}\kT{a}^\mu\,,
\qquad
\kT{r}^{(F),\mu} \to \frac{1}{x_a}\kT{r}^{(I),\mu}\,,
\label{eq:crossIFkT}
\eeq
where we used $\kT{a}^\mu = -\kT{r}^{(I),\mu}$. Notice that the 
relations $z_i+z_r=1$ and $\kT{i}^\mu + \kT{r}^{(F),\mu}=0$ are 
preserved under crossing as expected. Then, comparing 
\eqns{eq:CirFF00}{eq:CarIF00} and accounting for the signs 
associated with crossing a fermion from the final to the initial 
state, we find the relation 
\beq
\hP_{(ar) r}^{(0)}(x,k_\perp;\ep) = 
	-(-1)^{F(a)+F(ar)} x \hP_{\ba{a}r}(1/x,k_\perp/x;\ep)\,,
	\label{eq:Pifcross}
\eeq
where $\ba{a}$ is determined by $a+\ba{a}=g$ using the flavor 
summation rules of \eqn{eq:flsum}, while
\beq
F(q) = F(\qb) = 1\,,\qquad F(g) = 0\,.
\label{eq:Fsign}
\eeq
Note that factors which would account for the number of colors and 
spins of the incoming partons are not included because we define 
our matrix elements without those factors. Furthermore, the 
overall factor of $(-x)$ in \eqn{eq:Pifcross} accounts for the 
difference of kinematic factors that appear in 
\eqns{eq:CirFF00}{eq:CarIF00}. Indeed, the factor of $1/s_{ir}$ in 
\eqn{eq:CirFF00} becomes $-1/s_{ar}$ under crossing, however in 
\eqn{eq:CarIF00} we find instead $1/(x_a s_{ar})$. The ratio of 
these, i.e., $-x_a$, then appears in \eqn{eq:Pifcross}. Finally, 
we note that in \eqns{eq:Pqg0FF}{eq:Pgg0FF} the transverse 
momentum only appears in the form $\kT{}^\mu\kT{}^\nu/\kT{}^2$, 
hence the overall scale of $\kT{}^\mu$ is immaterial. Thus, the 
crossing rule in \eqn{eq:crossIFkT} and consequently 
\eqn{eq:Pifcross} could be simplified by setting $\kT{}^\mu/x \to 
\kT{}^\mu$ for the initial-state transverse momenta. The splitting 
functions for initial-final collinear splitting are given 
explicitly in \eqnss{eq:Pqg0IF}{eq:Pgg0IF} below.
\bal
\la s|\hP^{(0)}_{q g}(x,\kTm{};\ep)|s'\ra &= 
	\CF \left[\frac{1 + x^2}{1 - x} - \ep (1 - x)\right] 
    \delta_{ss'}\,,
\label{eq:Pqg0IF}
\\
\la \mu|\hP^{(0)}_{g q}(x,\kTm{};\ep)|\nu\ra &= 
	\TR \left[-g^{\mu\nu} x - 4\frac{1-x}{x} 
	\frac{\kTm{\mu} \kTm{\nu}}{\kTm{2}}\right] \,,	
\\
\la s|\hP^{(0)}_{q \qb}(x,\kTm{};\ep)|s'\ra &= 
	\CF \left[1 - \ep - 2 x(1  -x)\right] \delta_{ss'}\,,	
\\
\la \mu|\hP^{(0)}_{g g}(x,\kTm{};\ep)|\nu\ra &= 
	2\CA \left[
	-g^{\mu\nu}\left(x(1 - x) + \frac{x} {1-x}\right) 
	-2(1 - \ep)\frac{1 - x}{x} 
	\frac{\kTm{\mu} \kTm{\nu}}{\kTm{2}}\right]\,.
\label{eq:Pgg0IF}
\eal
Their azimuthally averaged counterparts read
\bal
P_{qg}^{(0)}(x;\ep) &= 
	\CF\left[\frac{1+x^2}{1-x} - \ep(1-x)\right]\,,
\label{eq:Pqg-ave-IF}
\\
P_{gq}^{(0)}(x;\ep) &= 
	\TR\left[x + 2\frac{1-x}{(1-\ep)x}\right]\,,
\label{eq:Pgq-ave-IF}
\\
P_{q\qb}^{(0)}(z;\ep) &= 
	\CF\left[1 - \ep - 2x(1-x)\right]\,,
\label{eq:Pqq-ave-IF}
\\
P_{gg}^{(0)}(z;\ep) &= 
	2\CA\left[x(1-x) + \frac{x}{1-x} + \frac{1-x}{x}\right]\,.
\label{eq:Pgg-ave-IF}
\eal
These expressions can be derived starting from 
\eqnss{eq:Pqg0IF}{eq:Pgg0IF} simply by using the replacements in 
\eqns{eq:APq_ave_presc}{eq:APg_ave_presc}.

Turning to the case of triple parton splitting, we first list the 
tree-level Altarelli-Parisi splitting functions for the $(irs) \to 
i+r+s$ final-state splitting. Introducing the notation,
\beq
s_{irs} = s_{ir}+s_{is}+s_{rs}\qquad\mbox{and}\qquad
t_{jk,l} = 2\frac{z_j s_{kl} - z_k s_{jl}}{z_j + z_k} 
	+ \frac{z_j - z_k}{z_j + z_k} s_{jk}\,,
\label{eq:tjkl}
\eeq
the splitting function for $q \to q \qb' q'$ is 
\beq
\bsp
&
\la s|\hP^{(0)}_{q_i \qb_r' q_s'}
(z_i, z_r, z_s, s_{ir}, s_{is}, s_{rs},
	\kT{i}, \kT{r}, \kT{s}; \ep)|s'\ra = 
\\&\qquad =
	\frac12 \CF\TR \frac{s_{irs}}{s_{rs}}\left[
	-\frac{t_{rs,i}^2}{s_{rs} s_{irs}} 
    + \frac{4 z_i + (z_r - z_s)^2}{z_r + z_s} 
	+ (1-2\ep)\left(z_r + z_s 
    - \frac{s_{rs}}{s_{irs}}\right)\right] \delta_{ss'}\,.
\label{eq:Pqqbpqp0FFF}
\esp
\eeq
If all quarks have identical flavors, we have
\beq
\bsp
&
\la s|\hP^{(0)}_{q_i \qb_r q_s}(z_i, z_r, z_s, s_{ir}, s_{is}, 
s_{rs}, \kT{i}, \kT{r}, \kT{s}; \ep)|s'\ra = 
\\&\qquad =
\la s|\hP^{(0)}_{q_i \qb_r' q_s'}(z_i, z_r, z_s, s_{ir}, s_{is}, 
s_{rs}, \kT{i}, \kT{r}, \kT{s}; \ep)|s'\ra
\\ &\qquad +
\la s|\hP^{(0)}_{q_i \qb_r' q_s'}(z_s, z_r, z_i, s_{rs}, s_{is}, 
s_{ir},	\kT{s}, \kT{r}, \kT{i}; \ep)|s'\ra
\\ &\qquad +
\la s|\hP^{(0),(\mathrm{id})}_{q_i \qb_r q_s}(z_i, z_r, z_s, 
s_{ir}, s_{is}, s_{rs}, \kT{i}, \kT{r}, \kT{s}; \ep)|s'\ra\,,
\label{eq:Pqqbq0FFF}
\esp
\eeq
where
\beq
\bsp
&
\la s|\hP^{(0),(\mathrm{id})}_{q_i \qb_r q_s}(z_i, z_r, z_s, 
s_{ir}, s_{is}, s_{rs}, \kT{i}, \kT{r}, \kT{s}; \ep)|s'\ra = 
\\&\qquad = 
	\CF\left(\CF - \frac{\CA}{2}\right)\bigg\{
	(1-\ep)\left(\frac{2s_{is}}{s_{rs}} - \ep\right) 
\\&\qquad
	+ \frac{s_{irs}}{s_{rs}}\left[\frac{1 + z_r^2}{1 - z_s} 
	- \frac{2 z_s}{1 - z_i}
	- \ep\left(\frac{(1-z_i)^2}{1-z_s} + 1 + z_r 
    - \frac{2 z_s}{1-z_i}\right)
	- \ep^2 (1-z_i)\right]
\\&\qquad
      - \frac{s_{irs}^2}{s_{rs} s_{ir}} \frac{z_r}{2}
      \left[\frac{1 + z_r^2}{(1 - z_s)(1 - z_i)}
      - \ep\left(1 + 2\frac{1-z_s}{1-z_i}\right)
      - \ep^2\right] + (i \leftrightarrow s)\,.
\label{eq:Pqqbq0idFFF}
\esp
\eeq
For $q\to qgg$ splitting we have
\beq
\bsp
&
\la s|\hP^{(0)}_{q_i g_r g_s}(z_i, z_r, z_s, s_{ir}, s_{is}, 
s_{rs}, \kT{i}, \kT{r}, \kT{s}; \ep)|s'\ra = 
\\&\qquad =
\CF^2
\la s|\hP^{(0),(\mathrm{ab})}_{q_i g_r g_s}(z_i, z_r, z_s, s_{ir}, 
s_{is}, s_{rs}, \kT{i}, \kT{r}, \kT{s}; \ep)|s'\ra
\\ &\qquad +
\CA \CF
\la s|\hP^{(0),(\mathrm{nab})}_{q_i g_r g_s}(z_i, z_r, z_s, 
s_{ir}, s_{is}, s_{rs}, \kT{i}, \kT{r}, \kT{s}; \ep)|s'\ra\,,
\label{eq:Pqgg0FFF}
\esp
\eeq
where
\beq
\bsp
&
\la s|\hP^{(0),(\mathrm{ab})}_{q_i g_r g_s}(z_i, z_r, z_s, s_{ir}, 
s_{is}, s_{rs}, \kT{i}, \kT{r}, \kT{s}; \ep)|s'\ra = 
\\&\qquad = 
	\bigg\{
	\frac{s_{irs}^2}{2 s_{ir} s_{is}} z_i 
	\left[\frac{1 + z_i^2}{z_r z_s}
	-\ep \frac{z_r^2 + z_s^2}{z_r z_s} - \ep(1+\ep)\right] 
\\& \qquad
      + \frac{s_{irs}}{s_{ir}}\left[\frac{z_i (1 - z_r) 
      + (1 - z_s)^3}{z_r z_s}
      + \ep^2 (1 + z_i) 
      - \ep \left(z_r^2 + z_r z_s + z_s^2\right) 
      \frac{1-z_s}{z_r z_s}\right] 
\\& \qquad
      + (1-\ep)\left[\ep 
      - (1-\ep)\frac{s_{is}}{s_{ir}}\right]\bigg\}
      + (r \leftrightarrow s)\,,
\label{eq:Pqgg0abFFF}
\esp
\eeq
and
\beq
\bsp
&
\la s|\hP^{(0),(\mathrm{nab})}_{q_i g_r g_s}(z_i, z_r, z_s, 
s_{ir}, s_{is}, s_{rs},	\kT{i}, \kT{r}, \kT{s}; \ep)|s'\ra = 
\\&\qquad = 
	\bigg\{
	(1-\ep)\left(\frac{t_{rs,i}^2}{4 s_{rs}^2} + \frac14 
    - \frac{\ep}{2}\right)
	+ \frac{s_{irs}^2}{2 s_{rs} s_{ir}} \bigg[
	\frac{(1 - z_i)^2(1-\ep) + 2z_i}{z_s}
\\&\qquad 
	+ \frac{z_s^2(1-\ep) + 2(1 - z_s)}{1 - z_i}\bigg] 
	- \frac{s_{irs}^2}{4 s_{ir} s_{is}} 
	z_i \left[\frac{(1 - z_i)^2(1-\ep) + 2 z_i}{z_r z_s} 
    + \ep(1-\ep)\right]
\\&\qquad 
	+ \frac{s_{irs}}{2s_{rs}} \left[
	(1-\ep)\frac{z_r (2 - 2 z_r + z_r^2) - z_s(6 - 6 z_s + z_s^2)}
    {z_s(1 - z_i)} 
	+ 2\ep\frac{z_i (z_r - 2 z_s) - z_s}{z_s (1 - z_i)}\right]
\\&\qquad
	+ \frac{s_{irs}}{2 s_{ir}} \bigg[
	(1-\ep)\frac{(1 - z_s)^3 + z_i^2 - z_s)}{z_s (1 - z_i)}
	-\ep\left(\frac{2(1 - z_s)(z_s - z_i)}
    {z_s (1 - z_i)} - z_r + z_s\right)
\\&\qquad
	- \frac{z_i (1 - z_r) + (1 - z_s)^3}{z_r z_s}
	+ \ep(1 - z_s)\left(\frac{z_r^2 + z_s^2}{z_r z_s} 
    - \ep\right)\bigg]
	\bigg\} 
	+ (r \leftrightarrow s)\,.
\label{eq:Pqgg0nabFFF}
\esp
\eeq
For $g \to gq\qb$ splitting we have
\beq
\bsp
&
\la \mu|\hP^{(0)}_{g_i q_r \qb_s}(z_i, z_r, z_s, s_{ir}, s_{is}, 
s_{rs}, \kT{i}, \kT{r}, \kT{s}; \ep)|\nu\ra = 
\\&\qquad =
\CF \TR
\la \mu|\hP^{(0),(\mathrm{ab})}_{g_i q_r \qb_s}(z_i, z_r, z_s, 
s_{ir}, s_{is}, s_{rs}, \kT{i}, \kT{r}, \kT{s}; \ep)|\nu\ra
\\ &\qquad +
\CA \TR
\la \mu|\hP^{(0),(\mathrm{nab})}_{g_i q_r \qb_s}(z_i, z_r, z_s, 
s_{ir}, s_{is}, s_{rs},	\kT{i}, \kT{r}, \kT{s}; \ep)|\nu\ra\,,
\label{eq:Pgqqb0FFF}
\esp
\eeq
where
\beq
\bsp
&
\la \mu|\hP^{(0),(\mathrm{ab})}_{g_i q_r \qb_s}(z_i, z_r, z_s, 
s_{ir}, s_{is}, s_{rs},	\kT{i}, \kT{r}, \kT{s}; \ep)|\nu\ra = 
\\&\qquad = 
	\Bigg\{
	-g^{\mu\nu}\left[
	-2 + \frac{2 s_{irs} s_{rs} + (1-\ep)(s_{irs} - s_{rs})^2}
    {s_{ir} s_{is}}\right]
	+ \frac{4s_{irs}}{s_{ir} s_{is}} \Bigg(
	\big(s_{is} - s_{irs}(1-z_r)\big) z_s
	\frac{\kT{r}^{\mu}\kT{r}^{\nu}}{\kT{r}^{2}}
\\& \qquad
	+ \big(s_{ir} - s_{irs}(1-z_s)\big) z_r
	\frac{\kT{s}^{\mu}\kT{s}^{\nu}}{\kT{s}^{2}}
	+ s_{rs} \frac{\kT{rs}^{\mu}\kT{rs}^{\nu}}{\kT{rs}^{2}}
	-(1-\ep) \big(s_{rs} - s_{irs}(1-z_i)\big) z_i
	\frac{\kT{i}^{\mu}\kT{i}^{\nu}}{\kT{i}^{2}}\Bigg)
	\Bigg\}\,,
\label{eq:Pgqqb0abFFF}
\esp
\eeq
and
\beq
\bsp
&
\la \mu|\hP^{(0),(\mathrm{nab})}_{g_i q_r \qb_s}(z_i, z_r, z_s, 
s_{ir}, s_{is}, s_{rs}, \kT{i}, \kT{r}, \kT{s}; \ep)|\nu\ra = 
\\&\qquad = 
	\frac14 \bigg\{\frac{s_{irs}}{s_{rs}^2} \left[
	g^{\mu\nu} \frac{t_{rs,i}^2}{s_{irs}} 
	- 16 \frac{z_r^2 z_s^2}{z_i (1 - z_i)} 
	\left(-\frac{s_{rs}}{z_r z_s} 
	\frac{\kT{rs}^{\mu}\kT{rs}^{\nu}}{\kT{rs}^{2}}\right)\right]
	+ \frac{s_{irs}}{s_{ir} s_{is}} \bigg[
	2 s_{irs} g^{\mu\nu}
\\&\qquad
	- 4 \Bigg(\big(s_{is} - s_{irs} (1 - z_r)\big) z_s 
	\frac{\kT{r}^{\mu}\kT{r}^{\nu}}{\kT{r}^{2}} 
	+ \big(s_{ir} - s_{irs} (1 - z_s)\big) z_r 
	\frac{\kT{s}^{\mu}\kT{s}^{\nu}}{\kT{s}^{2}}
\\&\qquad	
	+ s_{rs} 
	\frac{\kT{rs}^{\mu}\kT{rs}^{\nu}}{\kT{rs}^{2}}
	- (1 - \ep) \big(s_{rs} - s_{irs} (1 - z_i)) z_i 
	\frac{\kT{i}^{\mu}\kT{i}^{\nu}}{\kT{i}^{2}}\Bigg)\Bigg] 
\\&\qquad
	- g^{\mu\nu} \left[
	-(1 - 2 \ep) 
	+ 2 \frac{s_{irs}}{s_{ir}} \frac{1 - z_s}{z_i (1 - z_i)} 
	+ 2 \frac{s_{irs}}{s_{rs}} \frac{1 - z_i + 2 z_i^2}
    {z_i (1 - z_i)}\right] 
\\&\qquad
	+ \frac{s_{irs}}{s_{ir} s_{rs}} \Bigg[
	-2 s_{irs} g^{\mu\nu} \frac{z_r (1 - 2 z_i)}{z_i (1 - z_i)} 
	- 16 \big(s_{ir} - s_{irs} (1 - z_s)\big) z_s 
	\frac{\kT{s}^{\mu}\kT{s}^{\nu}}{\kT{s}^{2}} 
    \frac{z_r^2}{z_i (1 - z_i)} 
\\&\qquad
	+ 8 (1 - \ep) \big(s_{is} - s_{irs} (1 - z_r)\big) z_r 
	\frac{\kT{r}^{\mu}\kT{r}^{\nu}}{\kT{r}^{2}}
	+ 4 \Bigg(\big(s_{is} - s_{irs} (1 - z_r)\big) z_s 
	\frac{\kT{r}^{\mu}\kT{r}^{\nu}}{\kT{r}^{2}}
\\&\qquad
	+ \big(s_{ir} - s_{irs} (1 - z_s)\big) z_r 
	\frac{\kT{s}^{\mu}\kT{s}^{\nu}}{\kT{s}^{2}}
	+ s_{rs}
	\frac{\kT{rs}^{\mu}\kT{rs}^{\nu}}{\kT{rs}^{2}}\Bigg) 
	\left(\frac{2 z_r (z_s - z_i)}{z_i (1 - z_i)} 
    + (1 - \ep)\right)\Bigg]
	\bigg\}
	+ (r \leftrightarrow s)\,.
\label{eq:Pgqqb0nabFFF}
\esp
\eeq
Finally for $g \to ggg$ splitting, we have
\beq
\bsp
&
\la \mu|\hP^{(0)}_{g_i g_r g_s}(z_i, z_r, z_s, s_{ir}, s_{is}, 
s_{rs},	\kT{i}, \kT{r}, \kT{s}; \ep)|\nu\ra = 
\\&\qquad = \CA^2 \Bigg\{
\frac{(1 - \ep)}{4 s_{ir}^2} \left[
	-g^{\mu\nu} t_{ir,s}^2 
	+ 16 s_{irs} \frac{z_i^2 z_r^2}{z_s (1 - z_s)} 
	\left(-\frac{s_{ir}}{z_i z_r} 
	\frac{\kT{ir}^{\mu}\kT{ir}^{\nu}}{\kT{ir}^{2}}\right)\right] 
\\&\qquad
	- \frac34 (1 - \ep) g^{\mu\nu} 
	+ \frac{s_{irs}}{s_{ir}} g^{\mu\nu} \frac{1}{z_s} 
	\left[\frac{2 (1 - z_s) + 4 z_s^2}{1 - z_s} 
	- \frac{1 - 2 z_s (1 - z_s)}{z_i (1 - z_i)}\right] 
\\&\qquad
	+ \frac{s_{irs} (1 - \ep)}{s_{ir} s_{is}} 
	\Bigg[2 z_i \Bigg(
	\big(s_{is} - s_{irs} (1 - z_r)\big) z_r
	\frac{\kT{r}^{\mu}\kT{r}^{\nu}}{\kT{r}^{2}} 
	\frac{1 - 2 z_s}{z_s (1 - z_s)} 
\\&\qquad	
	+ \big(s_{ir} - s_{irs} (1 - z_s)\big) z_s 
	\frac{\kT{s}^{\mu}\kT{s}^{\nu}}{\kT{s}^{2}} 	
	\frac{1 - 2 z_r}{z_r (1 - z_r)}\Bigg) 
	+ \frac{s_{irs}}{2 (1 - \ep)} g^{\mu\nu} 
	\bigg(\frac{4 z_r z_s + 2 z_i (1 - z_i) - 1}
    {(1 - z_r) (1 - z_s)} 
\\&\qquad		
	- \frac{1 - 2 z_i (1 - z_i)}{z_r z_s}\bigg) 
	+ \Bigg(\big(s_{is} - s_{irs} (1 - z_r)\big) z_s 
	\frac{\kT{r}^{\mu}\kT{r}^{\nu}}{\kT{r}^{2}} 
	+ \big(s_{ir} - s_{irs} (1 - z_s)\big) z_r 
	\frac{\kT{s}^{\mu}\kT{s}^{\nu}}{\kT{s}^{2}} 
\\&\qquad		
	+ s_{rs} 
	\frac{\kT{rs}^{\mu}\kT{rs}^{\nu}}{\kT{rs}^{2}}\Bigg) 
	\left(\frac{2 z_r (1 - z_r)}{z_s (1 - z_s)} - 3\right)\Bigg]
	\Bigg\} 
	+ (\mbox{5 permutations})\,.
\label{eq:Pggg0FFF}
\esp
\eeq
In \eqnss{eq:Pgqqb0abFFF}{eq:Pggg0FFF} we used the notation,
\beq
\kT{jk}^\mu 
    = \frac{\kT{j}^{\mu}}{z_j} - \frac{\kT{k}^{\mu}}{z_k}\,.
\label{eq:kTjk-def}
\eeq
As discussed in \ref{sec:extension}, we prefer to write gluon 
splitting functions such that any transverse momentum $\kT{}^\mu$ 
only appears in the combination $\kT{}^\mu\kT{}^\nu/\kT{}^2$. 
Thus, the gluon splitting functions reported above are obtained 
from the corresponding expressions in~\refr{Catani:1999ss} by the 
following prescription:
\begin{enumerate}
\item First, all occurrences of $\kT{j}^\mu \kT{k}^\nu + 
\kT{k}^\mu \kT{j}^\nu$ are rewritten with the help of 
\eqn{eq:kTjk-def} such that only terms of the form $\kT{l}^\mu 
\kT{l}^\nu$ ($l=j,k,jk$) appear,
\beq
\kT{j}^\mu \kT{k}^\nu + \kT{k}^\mu \kT{j}^\nu = 
	\frac{z_k}{z_j} \kT{j}^\mu \kT{k}^\nu
	+ \frac{z_j}{z_k} \kT{k}^\mu \kT{j}^\nu
	- z_j z_k \kT{jk}^\mu \kT{jk}^\nu\,.
\label{eq:ktkt2ktktonkt2_0}
\eeq
\item Next, terms of the form $\kT{l}^\mu \kT{l}^\nu$ ($l=j,k,jk$) 
are replaced according to the rules,
\bal
\kT{j}^\mu \kT{j}^\nu &\to 
	\big(s_{kl} - s_{jkl} (1-z_j)\big) z_j
	\frac{\kT{j}^\mu \kT{j}^\nu}{\kT{j}^2}\,,
\label{eq:ktkt2ktktonkt2_1}
\\
\kT{jk}^\mu \kT{jk}^\nu &\to 
	-\frac{s_{jk}}{z_j z_k}
	\frac{\kT{jk}^\mu \kT{jk}^\nu}{\kT{jk}^2}\,.
\label{eq:ktkt2ktktonkt2_2}
\eal
\end{enumerate}
These rules ensure that the collinear behavior of the triple 
collinear subtraction can be matched with that of the single 
collinear subtraction in single unresolved regions.

\begin{figure}
\setlength{\tabcolsep}{10pt}
\begin{center}
\includegraphics[scale=1]{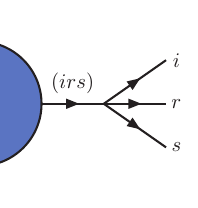}
\hspace{5em}
\includegraphics[scale=1]{Pars.pdf}
\end{center}
\caption{\label{fig:PirsParsflav}
    The triple collinear splitting configurations for 
    $(irs)\to i+r+s$ final-state splitting (left) and 
    $a\to (ars)+r+s$ initial-state splitting (right). 
    The arrows in the figure denote momentum and flavor flow.
}
\end{figure}

Splitting functions for initial-state triple collinear splitting 
can be obtained once more by crossing. In order to derive the 
appropriate rules, we again consider the crossing of parton $i$ to 
the initial state, which implies also that the parent parton 
$(irs)$ is crossed, $i\to a$ and $(irs)\to (ars)$. As before, the 
corresponding momenta are reversed, $p_i^\mu\to -p_a^\mu$ and 
$p_{irs}^\mu\to -p_{ars}^\mu$. The appropriate splitting 
configurations and momentum flows are shown in 
\fig{fig:PirsParsflav}.
Then repeating the analysis leading to 
\eqns{eq:crossIFx}{eq:crossIFkT}, we write first the crossed form 
of the final-final-final collinear parametrization,
\bal
p_i^\mu = z_i p_{irs}^\mu + \kT{i}^\mu + \dots
&\;\;\xrightarrow{\;\;\mathrm{cross}\;\;}\;\;
p_a^\mu = z_i p_{ars}^\mu - \kT{i}^\mu + \dots\,,
\label{eq:picrossed2}
\\
p_r^\mu = z_r p_{irs}^\mu + \kT{r}^{(F),\mu} 
+ \dots
&\;\;\xrightarrow{\;\;\mathrm{cross}\;\;}\;\;
p_r^\mu = -z_r p_{ars}^\mu + \kT{r}^{(F),\mu} + \dots\,,
\label{eq:prcrossed2}
\\
p_s^\mu = z_s p_{irs}^\mu + \kT{s}^{(F),\mu} 
+ \dots
&\;\;\xrightarrow{\;\;\mathrm{cross}\;\;}\;\;
p_s^\mu = -z_s p_{ars}^\mu + \kT{s}^{(F),\mu} + \dots\,.
\label{eq:pscrossed2}
\eal
The corresponding parametrization for initial-final-final 
splitting is
\bal
p_{ars}^\mu = x_a p_a^\mu + \kT{a}^\mu + \dots
&\;\;\Rightarrow\;\;
p_a^\mu = \frac{1}{x_a}p_{ars}^\mu - \frac{1}{x_a}\kT{a}^\mu 
+ \dots\,,
\label{eq:paexpr2}
\\
p_r^\mu = x_r p_a + \kT{r}^{(I),\mu} + \dots
&\;\;\Rightarrow\;\;
p_r^\mu = \frac{x_r}{x_a}p_{ars}^\mu - \frac{x_r}{x_a}\kT{a}^\mu 
+ \kT{r}^{(I),\mu} + \dots\,.
\label{eq:prexpr2}
\\
p_s^\mu = x_s p_a + \kT{s}^{(I),\mu} + \dots
&\;\;\Rightarrow\;\;
p_s^\mu = \frac{x_s}{x_a}p_{ars}^\mu - \frac{x_s}{x_a}\kT{a}^\mu 
+ \kT{s}^{(I),\mu} + \dots\,.
\label{eq:psexpr2}
\eal
Once more, the superscripts $(F)$ and $(I)$ distinguish transverse 
momenta in final-state and initial-state splitting. Then, 
comparing the ${\cal O}(\kT{}^0)$ and ${\cal O}(\kT{})$ pieces of 
\eqnss{eq:picrossed2}{eq:pscrossed2} and 
\eqnss{eq:paexpr2}{eq:psexpr2}, we find
\beq
z_i \to \frac{1}{x_a}\,,
\qquad
z_r \to -\frac{x_r}{x_a}\,,
\qquad
z_s \to -\frac{x_s}{x_a}
\label{eq:crossIFFx}
\eeq
and
\beq
\kT{i}^\mu\to \frac{1}{x_a}\kT{a}^\mu\,,
\qquad
\kT{r}^{(F),\mu} \to \frac{1-x_s}{x_a}\kT{r}^{(I),\mu} + 
\frac{x_r}{x_a}\kT{s}^{(I),\mu}\,,
\qquad
\kT{s}^{(F),\mu} \to \frac{1-x_r}{x_a}\kT{s}^{(I),\mu} + 
\frac{x_s}{x_a}\kT{r}^{(I),\mu}\,,
\label{eq:crossIFFkT}
\eeq
where we used $\kT{a}^\mu = -\kT{r}^{(I),\mu} -\kT{s}^{(I),\mu}$. 
The relations $z_i+z_r+z_s=1$ and $\kT{i}^\mu + \kT{r}^{(F),\mu} + 
\kT{s}^{(F),\mu}=0$ are preserved under crossing as expected. 
Then, accounting for the signs associated with crossing a fermion 
from the final to the initial state, we find the relation 
\beq
\bsp
&\hP_{(ars) r s}^{(0)}(x_a, x_r, x_s, s_{ar}, s_{as}, s_{rs},
	\kT{a}, \kT{r}, \kT{s}; \ep) = 
	(-1)^{F(a)+F(ars)} x_a 
\\&\qquad \times    
    \hP_{\ba{a} r s}
	\bigg(\frac{1}{x_a},-\frac{x_r}{x_a},-\frac{x_s}{x_a},
    -s_{ar},-s_{as},s_{rs},\frac{\kT{a}}{x_a},
    \frac{1-x_s}{x_a}\kT{r}+\frac{x_r}{x_a}\kT{s},
    \frac{1-x_r}{x_a}\kT{s}+\frac{x_s}{x_a}\kT{r};\ep\bigg)\,.
\esp
\label{eq:Piffcross}
\eeq
Comments similar to those below \eqn{eq:Pifcross} apply to 
\eqn{eq:Piffcross} as well. In particular, factors which would 
account for the numbers of colors and spins of the incoming 
partons are not included. Further, the overall factor of $x_a$ in 
\eqn{eq:Piffcross} appears because we choose to extract an overall 
factor of $1/x_a$ in \eqn{eq:CarsIFF00}. Last, since transverse 
momenta only appear in the combination 
$\kT{}^\mu\kT{}^\nu/\kT{}^2$, the crossing rule in 
\eqn{eq:crossIFFkT} and thus \eqn{eq:Piffcross} could be 
simplified by dropping the $x_a$ which appear in denominators in 
the definitions of initial-state transverse momenta.

Finally, we note that the azimuthally averaged triple splitting 
kernels can be obtained in the same way as before. In fact, the 
replacement rules of \eqns{eq:APq_ave_presc}{eq:APg_ave_presc} 
obviously continue to hold also in this case.

%
% Strongly-ordered splitting functions
%

\subsection{Strongly-ordered splitting functions}
\label{appx:SO-AP-functions}

Next, we recall the tree-level strongly-ordered final-state 
splitting functions for $(irs)\to i+(rs)\to i+r+s$ splitting 
from~\refrs{Somogyi:2005xz,Somogyi:2006da}. For quark splitting we 
have
\bal
&
\la s|\hP^{\mathrm{(C)}, (0)}_{q_i \qb_r' q_s'}
	(z_{r,s}, \kT{rs}, \kTt{rs}, z_{\ha{i},\wha{rs}},  
	\kT{\ha{i}\wha{rs}}, \ha{p}_i; \ep)|s'\ra
\label{eq:CFFPqqbq0FFF}
\\&\qquad= 
    \TR \Bigg[
	P^{(0)}_{q_{\ha{i}} g_{\wha{rs}}}(z_{\ha{i},\wha{rs}};\ep)
	- 2\CF z_{r,s} (1-z_{r,s})\left(1 - z_{\ha{i},\wha{rs}} - 
	\frac{s_{\ha{i}\kT{rs}}^2}
    {\kT{rs}^2 s_{\ha{i}\wha{rs}}}\right)\Bigg] 
	\delta_{ss'}\,,
\nt
\\
&
\la s|\hP^{\mathrm{(C)}, (0)}_{q_i g_r g_s}
	(z_{r,s}, \kT{rs}, \kTt{rs}, z_{\ha{i},\wha{rs}},  
	\kT{\ha{i}\wha{rs}}, \ha{p}_i; \ep)|s'\ra 
\label{eq:CFFPqgg0FFF}
\\&\qquad= 
    2\CA \Bigg[
	P^{(0)}_{q_{\ha{i}} g_{\wha{rs}}}(z_{\ha{i},\wha{rs}};\ep)
	\left(\frac{z_{r,s}}{1-z_{r,s}} 
    + \frac{1-z_{r,s}}{z_{r,s}}\right)
	+\CF (1-\ep) z_{r,s} (1-z_{r,s})
    \left(1 - z_{\ha{i},\wha{rs}} - \frac{s_{\ha{i}\kT{rs}}^2}
    {\kT{rs}^2 s_{\ha{i}\wha{rs}}}\right)\Bigg] 
	\delta_{ss'}\,,
\nt
\\
&
\la s|\hP^{\mathrm{(C)}, (0)}_{g_i q_r g_s}
	(z_{r,s}, \kT{rs}, \kTt{rs}, z_{\ha{i},\wha{rs}},  
	\kT{\ha{i}\wha{rs}}, \ha{p}_i; \ep)|s'\ra = 
	P^{(0)}_{q_{\wha{rs}} g_{\ha{i}}}(z_{\wha{rs},\ha{i}};\ep)
	P^{(0)}_{q_r g_s}(z_{r,s};\ep)
	\delta_{ss'}\,.
\label{eq:CFFPgqg0FFF}
\eal
Notice that the flavors of the partons involved in the most 
collinear splitting, i.e., $r$ and $s$, always appear as the 
second and third indices in the subscript. We will keep this 
convention throughout. For gluon splitting we have
\bal
&
\la \mu|\hP^{\mathrm{(C)}, (0)}_{\qb_i q_r g_s}
	(z_{r,s}, \kT{rs}, \kTt{rs}, z_{\ha{i},\wha{rs}},  
	\kT{\ha{i}\wha{rs}}, \ha{p}_i; \ep)|\nu\ra = 
	\la \mu |\hP^{(0)}_{\qb_{\ha{i}} q_{\wha{rs}}}
	(z_{\wha{rs},\ha{i}}, \kT{\wha{rs}\ha{i}}; \ep)| \nu \ra 
	P^{(0)}_{q_r g_s}(z_{r,s}; \ep)\,,
\label{eq:CFFPqbqg0FFF}
\\
&
\la \mu|\hP^{\mathrm{(C)}, (0)}_{g_i q_r \qb_s}
	(z_{r,s}, \kT{rs}, \kTt{rs}, z_{\ha{i},\wha{rs}},  
	\kT{\ha{i}\wha{rs}}, \ha{p}_i; \ep)| \nu \ra
\label{eq:CFFPgqqb0FFF}
\\&\qquad=
	2\CA \TR \Bigg[-g^{\mu\nu}\left(\frac{z_{\ha{i},\wha{rs}}}
    {1-z_{\ha{i},\wha{rs}}}
	+ \frac{1-z_{\ha{i},\wha{rs}}}{z_{\ha{i},\wha{rs}}}
	+ z_{r,s}(1-z_{r,s})\frac{s_{\ha{i}\kT{rs}}^2}
    {\kT{rs}^2 s_{\ha{i}\wha{rs}}}\right)
\nt\\&\qquad     
	+ 4 z_{r,s}(1-z_{r,s}) \frac{1-z_{\ha{i},\wha{rs}}}
    {z_{\ha{i},\wha{rs}}}
	\frac{\kTt{rs}^\mu \kTt{rs}^\nu}{\kTt{rs}^2}\Bigg] 
    -4 \CA (1-\ep) z_{\ha{i},\wha{rs}}(1-z_{\ha{i},\wha{rs}}) 
    P^{(0)}_{q_r \qb_s}(z_{r,s};\ep)
	\frac{\kT{\ha{i}\wha{rs}}^\mu \kT{\ha{i}\wha{rs}}^\nu}
    {\kT{\ha{i}\wha{rs}}^2}\,,
\nt
\\
&
\la \mu|\hP^{\mathrm{(C)}, (0)}_{g_i g_r g_s}
	(z_{r,s}, \kT{rs}, \kTt{rs}, z_{\ha{i},\wha{rs}},  
	\kT{\ha{i}\wha{rs}}, \ha{p}_i; \ep)| \nu \ra
\label{eq:CFFPggg0FFF}
\\&\qquad=
	4\CA^2 \Bigg[-g^{\mu\nu}\left(\frac{z_{\ha{i},\wha{rs}}}
    {1-z_{\ha{i},\wha{rs}}}
	+ \frac{1-z_{\ha{i},\wha{rs}}}{z_{\ha{i},\wha{rs}}}\right)
	\left(\frac{z_{r,s}}{1-z_{r,s}} 
    + \frac{1-z_{r,s}}{z_{r,s}}\right) 
\nt\\& \qquad 
    + g^{\mu\nu} z_{r,s}(1-z_{r,s}) \frac{1-\ep}{2}
	\frac{s_{\ha{i}\kT{rs}}^2}{\kT{rs}^2 s_{\ha{i}\wha{rs}}}
	-2 (1-\ep) z_{r,s}(1-z_{r,s})
	\frac{1-z_{\ha{i},\wha{rs}}}{z_{\ha{i},\wha{rs}}}
	\frac{\kTt{rs}^\mu \kTt{rs}^\nu}{\kTt{rs}^2}\Bigg] 
\nt\\& \qquad  
    - 4 \CA (1-\ep) z_{\ha{i},\wha{rs}}(1-z_{\ha{i},\wha{rs}}) 
    P^{(0)}_{g_r g_s}(z_{r,s};\ep)
	\frac{\kT{\ha{i}\wha{rs}}^\mu \kT{\ha{i}\wha{rs}}^\nu}
    {\kT{\ha{i}\wha{rs}}^2}\,,
\nt
\eal
where the azimuthally averaged tree-level splitting functions 
$P^{(0)}$ are given in \eqnss{eq:Pqg-ave}{eq:Pgg-ave}.

The strongly-ordered splitting functions involving initial-state 
splitting can be obtained from those for final-state splitting, 
\eqnss{eq:CFFPqqbq0FFF}{eq:CFFPggg0FFF}, by simple crossing 
relations. Consider first the case of $a\to a+(rs)\to a+r+s$ 
splitting. The appropriate strongly-ordered splitting function 
which appears in \eqn{eq:CarsIFF00CrsFF} is obtained by the 
crossing relation, 
\beq
\bsp
&
\hP^{\mathrm{(C)}, (0)}_{(ars) r s}
(z_r, \kT{r}, \kTh{rs}, x_{\ha{a}}, \kT{\wha{rs}}, \ha{p}_a;\ep)
\\ &\qquad= 
	-(-1)^{F(a)+F(ars)} x_{\ha{a}}
	\hP^{\mathrm{(C)}, (0)}_{\ba{a} rs}
	(z_r, \kT{r}, \kTh{rs}, 1/x_{\ha{a}}, 
    \kT{\wha{rs}}/x_{\ha{a}}, -\ha{p}_a;\ep)\,,
\label{eq:PCars-crossing}
\esp
\eeq
where the sign factors $F(a)$ and $F(ars)$ are given by 
\eqn{eq:Fsign}. Next, the $a\to (as)+r\to a+r+s$ strongly-ordered 
splitting appearing function appearing in \eqn{eq:CasrIFF00CasIF} 
is given by the following crossing formula, 
\beq
\bsp&
\hP^{\mathrm{(C)}, (0)}_{(ars) (as) s}
(x_a, \kT{s}, \kTh{s}, x_{\ha{a}}, \kT{\ha{r}}, \ha{p}_r;\ep) 
\\&\qquad= 
	(-1)^{F(as)+F(ars)} x_a x_{\ha{a}}
	\hP^{\mathrm{(C)}, (0)}_{r \ba{a} s}(1/x_a, \kT{s}/x_a, 
    \kTh{s}/x_a, 1/x_{\ha{a}}, \kT{\ha{r}}/x_{\ha{a}}, 
    \ha{p}_r;\ep)\,,
\esp	
\label{eq:PCars-crossing-IF}
\eeq
where the sign factors $F(as)$ and $F(ars)$ are given by 
\eqn{eq:Fsign}. Factors that would account for the numbers of 
colors and spins of the incoming partons are not included in 
\eqns{eq:PCars-crossing}{eq:PCars-crossing-IF}. We recall that the 
flavors of the partons involved in the most collinear splitting, 
i.e., $(as)$ and $s$, always appear as the second and third index 
in the subscript.

Finally, we mention that the azimuthally averaged forms of the 
strongly ordered splitting functions are computed in the same way 
as their non-strongly ordered counterparts. In particular, one may 
simply use the replacement rules 
in~\eqns{eq:APq_ave_presc}{eq:APg_ave_presc}.

%
% Soft limit of the splitting functions 
%

\subsection{Soft limit of the splitting functions}
\label{appx:Soft-AP-functions}

Finally, we recall the tree-level soft functions 
$P^{(\mathrm S),(0)}_{irs}$. These represent the single soft 
$p_s^\mu\to 0$ limit of a set of $n$ collinear final-state 
partons~\cite{DelDuca:2019ggv}, for $n=3$ and 
read~\cite{Somogyi:2005xz},
\bal
P^{(\mathrm S), (0)}_{f_i f_r q_s}
(z_i,z_r,z_s,s_{ir},s_{is},s_{rs}) &= 0\,,
\label{eq:SsPfifrq0FFF}
\\
P^{(\mathrm S), (0)}_{q_i g_r g_s}
(z_i,z_r,z_s,s_{ir},s_{is},s_{rs},) &= 
	\CF \frac{2}{s_{is}}\frac{z_i}{z_s}
	+ \CA\left(
		\frac{s_{ir}}{s_{is} s_{rs}} 
		- \frac{1}{s_{is}} \frac{z_i}{z_s}
		+ \frac{1}{s_{rs}} \frac{z_r}{z_s}
		\right)\,,
\label{eq:SsPqgg0FFF}
\\
P^{(\mathrm S), (0)}_{g_i q_r g_s}
(z_i,z_r,z_s,s_{ir},s_{is},s_{rs}) &= 
	\CF \frac{2}{s_{rs}}\frac{z_r}{z_s}
	+ \CA\left(
		\frac{s_{ir}}{s_{is} s_{rs}} 
		+ \frac{1}{s_{is}} \frac{z_i}{z_s}
		- \frac{1}{s_{rs}} \frac{z_r}{z_s}
		\right)\,,
\label{eq:SsPgqg0FFF}
\\
P^{(\mathrm S), (0)}_{q_i \qb_r g_s}
(z_i,z_r,z_s,s_{ir},s_{is},s_{rs}) &= 
	\CF \frac{2 s_{ir}}{s_{is} s_{rs}}
	+ \CA\left(
		-\frac{s_{ir}}{s_{is} s_{rs}} 
		+ \frac{1}{s_{is}} \frac{z_i}{z_s}
		+ \frac{1}{s_{rs}} \frac{z_r}{z_s}
		\right)\,,
\label{eq:SsPqqg0FFF}
\\
P^{(\mathrm S), (0)}_{g_i g_r g_s}
(z_i,z_r,z_s,s_{ir},s_{is},s_{rs}) &= 
	\CA\left(
		\frac{s_{ir}}{s_{is} s_{rs}} 
		+ \frac{1}{s_{is}} \frac{z_i}{z_s}
		+ \frac{1}{s_{rs}} \frac{z_r}{z_s}
		\right)\,.
\label{eq:SsPggg0FFF}
\eal
Notice that the index of the soft parton $s$ appears last in the 
subscript. We note in passing that 
\eqnss{eq:SsPgqg0FFF}{eq:SsPggg0FFF} can be written in a uniform 
manner~\cite{Bolzoni:2010bt},
\beq
\bsp
&
P^{(\mathrm S), (0)}_{f_i f_r g_s}
(z_i,z_r,z_s,s_{ir},s_{is},s_{rs}) 
\\&\qquad=
(\bT_{i}^2+\bT_r^2-\bT_{ir}^2) \frac{s_{ir}}{s_{is} s_{rs}} 
+(\bT_{ir}^2+\bT_i^2-\bT_r^2) \frac{1}{s_{is}} \frac{z_i}{z_s}
+(\bT_{ir}^2-\bT_i^2+\bT_r^2) \frac{1}{s_{rs}} \frac{z_r}{z_s}\,.
\esp
\eeq

The corresponding soft function for initial-state splitting 
appears in \eqn{eq:CarsIFF00Ss} and can be obtained by the  
following simple crossing relation,
\beq
P^{(\mathrm S), (0)}_{(ar) r s}
	(x_a,x_r,x_s,s_{ar}, s_{as}, s_{rs};\ep) =
    -(-1)^{F(a)+F(ar)}P^{(\mathrm S),(0)}_{\ba{a} r s}
	\left(\frac{1}{x_a}, 
	-\frac{x_r}{x_a}, 
	-\frac{x_s}{x_a},
	-s_{ar}, -s_{as}, s_{rs}\right)\,,
	\label{eq:sapspliIFF}
\eeq
where the sign factors $F(a)$ and $F(ar)$ are given by 
\eqn{eq:Fsign} and we have exploited the fact that parton $s$ is a 
gluon.

%%%
%%% Bibliography
%%%

% ========== ========== ========== ========== ==========

\bibliographystyle{JHEP}
\bibliography{nnlocal}

\end{document}